\documentclass[english,twocolumn]{revtex4}

\usepackage[T1]{fontenc}
\usepackage[utf8]{inputenc}
\usepackage{color}
\usepackage{babel}
\usepackage{verbatim}
\usepackage{refstyle}
\usepackage{amsmath}
\usepackage{amssymb}
\usepackage{cancel}
\usepackage{graphicx}
\usepackage{wasysym}
\usepackage{esint}
\usepackage{microtype}
\usepackage[unicode=true,
 bookmarks=true,bookmarksnumbered=false,bookmarksopen=false,
 breaklinks=false,pdfborder={0 0 0},pdfborderstyle={},backref=false,colorlinks=true]
 {hyperref}
\hypersetup{pdftitle={Distribution of the order parameter: analytic theory},
 pdfauthor={Anton Khvalyuk, Mikhail Feigel'man},
 pdfsubject={Superconductivity}}

\makeatletter

\AtBeginDocument{\providecommand\secref[1]{\ref{sec:#1}}}
\AtBeginDocument{\providecommand\subsecref[1]{\ref{subsec:#1}}}
\AtBeginDocument{\providecommand\eqref[1]{\ref{eq:#1}}}
\AtBeginDocument{\providecommand\appref[1]{\ref{app:#1}}}
\AtBeginDocument{\providecommand\figref[1]{\ref{fig:#1}}}
\AtBeginDocument{\providecommand\subsecappref[1]{\ref{subsecapp:#1}}}
\AtBeginDocument{\providecommand\eqappref[1]{\ref{eqapp:#1}}}
\AtBeginDocument{\providecommand\figappref[1]{\ref{figapp:#1}}}
\newcommand{\lyxdot}{.}

\RS@ifundefined{subsecref}
  {\newref{subsec}{name = \RSsectxt}}
  {}
\RS@ifundefined{thmref}
  {\def\RSthmtxt{theorem~}\newref{thm}{name = \RSthmtxt}}
  {}
\RS@ifundefined{lemref}
  {\def\RSlemtxt{lemma~}\newref{lem}{name = \RSlemtxt}}
  {}

\usepackage{chngcntr} 

\newref{fig}{ refcmd={Figure~\ref{#1}} }
\newref{figapp}{ refcmd={Figure~\ref{#1} of SM~\citep{SM}} }
\newref{tab}{ refcmd={Table~\ref{#1}} }
\newref{eq}{ refcmd={(\ref{#1})} }
\newref{eqapp}{ refcmd={(\ref*{#1}) of SM~\citep{SM}} }
\newref{subsec}{ refcmd={Subsection~\ref{#1}} }
\newref{sec}{ refcmd={Section~\ref{#1}} }
\newref{app}{ refcmd={Section~\ref{#1} of SM~\citep{SM}} }
\newref{subsecapp}{ refcmd={Subsection~\ref{#1} of SM~\citep{SM}} }
\newref{part}{ refcmd={Part~\ref{#1}} }

\makeatother

\begin{document}
\title{Distribution of the order parameter in strongly disordered superconductors:
analytic theory}
\author{Anton V. Khvalyuk}
\email{anton.k@itp.ac.ru}

\affiliation{Skolkovo Institute of Science and Technology, 143026 Skolkovo, Russia}
\affiliation{L. D. Landau Institute for Theoretical Physics, Kosygin Str. 2, Moscow
119334, Russia}
\author{Mikhail V. Feigel'man}
\email{feigel@landau.ac.ru}

\affiliation{L. D. Landau Institute for Theoretical Physics, Kosygin Str. 2, Moscow
119334, Russia}
\affiliation{Moscow Institute of Physics and Technology, Dolgoprudny, Russia}
\date{\today{}}
\begin{abstract}
We developed an analytic theory of inhomogeneous superconducting pairing
in strongly disordered materials, which are moderately close to superconducting-insulator
transition. Single-electron eigenstates are assumed to be Anderson-localized,
with a large localization volume. Superconductivity develops due to
coherent delocalization of originally localized preformed Cooper pairs.
The key assumption of the theory is that each such pair is coupled
to a large number $Z\gg1$ of similar neighboring pairs. We derived
integral equations for the probability distribution $P\left(\Delta\right)$
of local superconducting order parameter $\Delta\left(\boldsymbol{r}\right)$
and analyzed their solutions in the limit of small dimensionless Cooper
coupling constant $\lambda\ll1$. The shape of the order-parameter
distribution is found to depend crucially upon the effective number
of \textquotedbl nearest neighbors\textquotedbl{} $Z_{\text{eff}}=2\nu_{0}\Delta_{0}Z$,
with $\nu_{0}$ being the single-particle density of states at the
Fermi level. The solution we provide is valid both at large and small
$Z_{\text{eff}}$; the latter case is nontrivial as the function $P\left(\Delta\right)$
is heavily non-Gaussian. One of our key findings is the discovery
of a broad range of parameters where the distribution function $P\left(\Delta\right)$
is non-Gaussian but also non-critical (in the sense of SIT criticality).
The analytic results are supplemented by numerical data, and good
agreement between them is observed. 
\end{abstract}
\maketitle

\section{Introduction}

Strongly disordered superconductors are interesting both from fundamental
and practical perspectives. The fundamental problem of a quantum (zero-temperature)
phase transition between superconducting and insulating ground states
(Superconductor-Insulator transition, or SIT) attracted considerable
attention since mid-80's~\citep{Goldman89,Fisher90,Shahar97,Fazio2001,Gantmakher2010}
and got an additional burst of research during the last decade. Prominent
examples include various structurally different realizations of the
SIT, such as granular arrays of Josephson Junctions or thick homogeneous
films of amorphous Indium Oxide. The whole variety of phenomena collectively
labeled as SIT demonstrate a great deal of diversity in the underlying
physics and thus cannot be possibly explained by a single mechanism
(see the recent review \citep{SFK-review-2020} for further details).
In this paper, we theoretically demonstrate several rather persistent
properties of 3D materials with homogeneous structure and strong microscopic
disorder.

The practical side of interest to strongly disordered superconductors
stems from potential applications in quantum computing technologies
in the form of so called ``superinductors''~\citep{Doucot12,Kitaev13,Groszowski18,Mooij05,Mooij06}.
These are much wanted yet so far mostly hypothetical inductive devices
that combine nearly absent dissipation at low energies (in GHz range)
with high inductance and small spatial size such that kinetic inductance
per square $L_{\square}$ exceeds $10\text{ nH}$. The principal opportunity
to fabricate such a device is provided by the platform of thick films
of strongly disordered superconductors. Indeed, the latter feature
low superfluid density $\rho_{s}$ and the associated high kinetic
inductance per square $L_{\square}\sim1/\rho_{s}$, enabling one to
implement an superinductor within a compact geometry. Such extreme
values for these materials are a consequence of high normal state
resistance induced by disorder~\citep[sec. 3.10]{Tinkham_supercond-textbook}\citep{Sacepe_subgap-excitations_APS-Meeting-March-2021}\citep[ Fig. 3b, 3c in particular]{sherman2015higgs_for-disorder-driven-suppresion-of-rho-s}.
On the other hand, the necessity for the absence of low-energy dissipation
requires one to use materials with a well resolved gap in the optical
excitation spectrum --- a feature so natural for superconducting
materials.

However, it occurs that the two conditions mentioned above (low $\rho_{s}$
and absence of any low-energy excitations) come into conflict. Superconductors
which are \textit{too close to SIT} unavoidably contain some non-zero
density of low-lying collective modes even when single-electron density
of states (1-DoS) is fully gapped, as it is demonstrated by theoretical
analysis~\citep{Feigelman_Microwave_2018} and experimental observations~\citep{Sacepe_subgap-excitations_APS-Meeting-March-2021}.
Yet, the question of low-energy modes in strongly disordered superconductors
is by no means resolved qualitatively. The preliminary analysis performed
in paper~\citep{Feigelman_Microwave_2018} was based upon the approximation
of constant superconducting order parameter $\Delta(\mathbf{r})=\Delta$,
which is far from being obviously correct. Instead, a self-consistent
theory of the system's collective modes without the use of such a
drastic approximation is needed. Moreover, the spatial distribution
of superconducting order parameter can now be probed by means of modern
low-temperature Scanning Tunneling Microscopy methods \citep{Sacepe10,Sacepe_2011_for-pair-preformation,Kamlapure12,Ganguly17,Tamir_private}.
It is thus of both fundamental and practical interest to develop a
theory that would be able to: 1)~describe realistic spatial distributions
of the order parameter, and 2)~describe the behavior of collective
modes on top of the spatially inhomogeneous superconducting state.
In the present paper we deal with the first of these problems only,
leaving the second one for the near future.

The local probability distribution function $P(\Delta)$ of superconducting
order parameter has already been addressed in several important limiting
cases of disorder strength. The limit of small disorder corresponds
to usual dirty superconductors with diffusive transport in normal
state. For this regime, the structure of statistical fluctuations
of the order parameter was analyzed in the seminal paper~\citep[see Sec. 3 in particular]{L0_1972}
by means of semiclassical theory of superconductivity, demonstrating
a narrow purely Gaussian $P\left(\Delta\right)$. In the opposite
limit of small disorder, the single-electron wave functions suffer
Anderson localization transition, rendering the conventional semiclassical
approach inapplicable. To describe this regime, the work~\citep{Ioffe_2010_ref-about-Onsager}
substantiated the model on the Bethe lattice, while the subsequent
paper~\citep{Feigelman_SIT_2010} showed that the resulting $P\left(\Delta\right)$
exhibits critical features, such as ``fat tails'' extended to the
region of large $\Delta$, much larger than the typical value $\Delta_{\text{typ}}$.
However, realistic experiments usually deal with superconducting samples
which fall within neither of the two limiting cases described above;
it is especially so for superconductors which may serve as candidates
for construction of superinductors. On the one hand, superconducting
materials discussed in the work are \emph{much more disordered} than
usual dirty superconductors, to the extent where neither the standard
semiclassical theory of Ref.~\citep{L0_1972} nor even the mere Gaussian
approximation for $P\left(\Delta\right)$ are applicable. As suggested
by numerical data~\citep{Randeria01,Bouadim11} and experimental
observations~\citep{Sacepe_2011_for-pair-preformation,Lemarie2013,Tamir_private},
this type of materials features heavily non-Gaussian profiles of the
order parameter distribution. On the other hand, the level of disorder,
the resulting non-critical distribution $P\left(\Delta\right)$ and
the requirement for the absence of low-energy excitations are all
suggesting that the samples of interest are \emph{somewhat away from
the SIT}, so that the the critical theory of Ref.~\citep{Feigelman_SIT_2010}
is also inapplicable. The present paper is devoted to the development
of analytical methods able to study the order parameter distribution
in the materials that belong to the region in between the two limiting
cases. The latter turns out to be parametrically broad, as we also
show below. While our approach is general and valid \emph{in principle}
at all temperatures, in this paper we consider $T=0$ limit only.

This paper is organized as follows. We formulate our theoretical model
in \secref{The-model}. Within it, we review the relevant phenomenology
of disordered superconductors and formulate the Hamiltonian of the
system. The corresponding static self-consistency equations for the
order parameter are then introduced along with a brief discussion
of applicability and several known limits. The section is closed by
a brief discussion of the methods used in previous works to analyze
problems similar to the one stated in the present work. \secref{Distribution-of-the-OP}
then presents the body of our theoretical approach. In~\subsecref{General-local-equations},
we start by deriving a general set of equations to describe the statistics
of solution to systems of local \emph{nonlinear} equations with disorder,
such as the self-consistency equations for the order parameter. Within
the following~\subsecref{The-limit-of-small-Delta}, those equations
are substantially simplified in the physically relevant limit of small
order parameter $\nu_{0}\Delta_{0}\ll1$ and large number of neighbors
$Z\gg1$ within the localization volume of a given single-particle
state. Such simplifications render the presented equations amenable
for both numerical and analytical analysis. In~\subsecref{Weak-coupling-approximation},
the reader can find an explicit \emph{analytical} solution to the
the proposed equations on the distribution function of the order parameter
and related quantities in terms of certain special functions. The
following~\subsecref{Numerical-analysis} then briefly describes
the numerical routines used to analyze both the original self-consistency
equations in a particular realization of disorder in the system and
the derived equations on the distributions of various physical quantities
across different disorder realizations. In~\subsecref{Overview-of-the-results}
we demonstrate the key outcomes of our theoretical analysis: the profile
of the distribution function as a function of the parameters of the
model and the asymptotic behavior of the distribution. The subsection
also contains some results for the distribution of other local physical
quantities. \subsecref{Extreme-value-statistics_fluctuating-coupling}
then introduces and analyzes several important extensions of our model
that allow us to draw conclusions about the robustness of our findings.
Finally, \secref{Discussion-and-Conclusions} summarizes the key theoretical
achievements and outlines several immediate developments. This paper
is accompanied by the Supplementary Materials (SM)~\citep{SM} that
contain additional technical information on various steps of theoretical
and numerical analysis employed in this work.

\section{The model\label{sec:The-model}}

\subsection{Phenomenology of strongly disordered superconductors\label{subsec:Phenomenology-overview}}

The physics of superconductor-insulator transition (SIT) owes its
rich phenomenology to the underlying complexity of the Anderson Localization
transition in the single-particle spectrum of the system. The paper~\citep{Feigelman_Fractal-SC_2010}
conducts an extensive research of the topic, building upon the seminal
paper~\citep{Ma_Lee_1985_Ref-to-pseudospins} and early numerical
studies~\citep{Randeria01}; here we employ a simplified description
proposed and substantiated in Ref.-s~\citep{Ioffe_2010_ref-about-Onsager,Feigelman_SIT_2010}

The single-particle electron states are described by spatially localized
wave-functions~$\psi_{i}\left(r\right)$ with large localization
volume~$V_{\text{loc}}$ and complex spatial structure~\citep[Sec. 2]{Feigelman_Fractal-SC_2010}.
The single-particle eigenenergies $\xi_{i}$ of these states can be
approximated as randomly distributed independent variables, with the
typical width of the distribution $\nu\left(\xi\right)$ being of
order of the Fermi energy $E_{F}$. We assume that this distribution
arranges a finite density of states \emph{per spin projection} $\nu_{0}=\nu\left(\xi=0\right)\sim1/E_{F}$
at the Fermi level.

Even prior to the emergence of the global superconducting coherence,
the systems in question are known to favor the formation of localized
Cooper pairs~\citep[Sec. 3 and ref. therein]{Feigelman_Fractal-SC_2010}.
This phenomenon can be delineated by an additional energy~$E_{\text{PG}}$
per each unpaired electron in the system. For the systems of interest,
the typical scale of $E_{\text{PG}}$ is significantly larger than
all superconducting energy scales~\citep[Sec. 4.3]{Feigelman_Fractal-SC_2010}.
Consequently, single-particle excitations barely contribute to low-energy
physics. One is thus able to describe the relevant physics by considering
only the states corresponding to presence or absence of a local Cooper
pair on a given single-particle state~$i$, effectively halving the
Hilbert space, as described in~\citep[Sec. 6]{Feigelman_Fractal-SC_2010}.

The superconducting order in the system then corresponds to coherent
delocalization of preformed Cooper pairs, as demonstrated experimentally
in Ref.~\citep{Sacepe_2011_for-pair-preformation} and supported
by numerical data~\citep{Bouadim11}. Such behavior results from
attractive Cooper-like pairwise interaction between the Cooper pairs.
This interaction is assumed to be local, so that it only connects
single-particle states with a finite spatial overlap. As a result,
each single-particle state~$i$ is effectively interacting with other
states located within the localization volume of~$i$. However, the
particular subset of those states is rather nontrivial due to both
the complex structure of the single-particle wave-functions~$\psi_{i}\left(r\right)$
and explicit dependence of the matrix element of the interaction on
energy difference~$\xi_{i}-\xi_{j}$ between the interacting states.
To describe the emerging phenomenology, we employ a simplistic model
of the spatial structure of matrix elements that assumes each single-particle
state~$i$ to be effectively connected to a \emph{constant} number~$Z$
of states chosen at random from within the localization volume of~$i$.
The value of~$Z$ can be estimated as a small fraction of the total
number of states within the localization volume that has significant
spatial overlap with a given state~$i$, so that $Z\sim nV_{\text{loc}}\cdot\eta$,
where $n$ is the electron concentration and $\eta$ is a small numerical
factor. Due to the proximity to the Anderson transition, the localization
volume~$V_{\text{loc}}$ is large~\citep[Sec. 2]{Feigelman_Fractal-SC_2010},
thus also rendering $Z\gg1$, even despite the smallness provided
by~$\eta$. We note, however, that for the analysis presented below
it is only important that $Z$ itself is a large quantity. In particular,
the analysis of a model where each site has the value of $Z$ distributed
according to Poisson distribution suggest that the fluctuations of
$Z$ do not play a significant role in the observed behavior.

In what follows, we will also retain the information about the energy
dependence~$D\left(\xi_{i}-\xi_{j}\right)$ of the matrix elements
of the interaction. This energy dependence is primarily characterized
by the large energy cutoff $\varepsilon_{D}$ that is typically of
the order of the Debye energy of phonons. Due to this energy scale,
the interaction between the states with energy difference~$\left|\xi_{i}-\xi_{j}\right|$
larger than $\varepsilon_{D}$ is essentially absent. On the other
hand, we assume that the localization volume of single-particle electron
states is large enough to secure the continuity of phonon spectrum,
i.e. $\delta_{\text{loc}}\ll\varepsilon_{D}$, with $\delta_{\text{loc}}$
being the characteristic phonon level spacing in the localization
volume. It is worth mentioning that the actual profile of $D$ for
dirty superconductors with pseudogap is known to exhibit substantial
dependence at small energies due to the underlying phenomenology of
Anderson insulator~\citep[Sec. 4]{Feigelman_Fractal-SC_2010}. This
feature presents an additional complication which does not seem to
be universally relevant. We will thus simplify the model below by
assuming that $D$ is smooth in the vicinity of the zero energy difference
and arranges a small static coupling constant~$D\left(0\right)$.
The latter is then conventionally parametrized by small dimensionless
Cooper constant~$\lambda\ll1$ as $D\left(0\right)=\lambda/\left(2\nu_{0}Z\right)$,
where the multiplier~$Z$ in the denominator ensures proper normalization
of the matrix element.

An important issue is related to the spatial geometry of the manifold
spanned by the indices of eigenstates~$i,j,...$, etc. On the one
hand, the eigenstates~$\psi_{i}\left(\boldsymbol{r}\right)$ are
supposed to be localized in the physical 3D space (or in the effectively
2D space in case of very thin films), and the locations~$R_{i}$
of the maxima in the absolute values~$\left|\psi_{i}\left(\boldsymbol{r}\right)\right|$
constitute a set of points in real 3D (or 2D) space. On the other
hand, the major role in the formation of the superconducting state
is played specifically by the eigenstates close to the Fermi-level
and \emph{in addition} also sufficiently strongly coupled to each
other. Since coupling amplitudes between eigenstates near the mobility
edge strongly vary in magnitude, only small fraction of all eigenstates~$\psi_{j}\left(\boldsymbol{r}\right)$
that can be found around the selected one --- $\psi_{i}\left(\boldsymbol{r}\right)$
--- is coupled to $\psi_{i}\left(\boldsymbol{r}\right)$ considerably.
The resulting spatial structure of \emph{interacting} eigenstates
can be considered, in some approximation, as a strongly diluted random
graph with some large but finite number of neighbors $Z$ per each
participating ``site''. The crucial feature of this graph --- as
opposed to the usual Euclidean lattice --- is its loopless structure.
More exactly, a random graph with coordination number $Z$ that is
much smaller than the total number of sites~$N$, does contain loops,
but their typical size grows with system size as~$\sim\ln N/\ln\left(Z-1\right)$,
while small loops are nearly absent~\citep{bollobas2001random}.
This, in turn, suppresses infra-red fluctuations of the order parameter,
which are known to be crucial for the adequate description of thermal
phase transitions in low-dimensional systems. On the other hand, in
the present problem we are interested in statistical properties of
the order parameter at lowest temperatures, where thermal fluctuations
are absent anyway. The most important effects to be studied here are
due to strong statistical fluctuations (of quenched disorder), which
can be considered within the loop-less approximation.

\subsection{The model Hamiltonian\label{subsec:The-model-Hamiltonian}}

The presented phenomenological picture allows us to adopt the following
model Hamiltonian of a strongly disordered superconductor on the verge
of localization transition and with a well developed pseudogap:
\begin{align}
H & =\sum_{i}\xi_{i}\left(a_{i\downarrow}^{\dagger}a_{i\downarrow}+a_{i\uparrow}^{\dagger}a_{i\uparrow}\right)\nonumber \\
 & -\sum_{\left\langle ij\right\rangle }D_{ij}\left(a_{i\downarrow}^{\dagger}a_{i\uparrow}^{\dagger}a_{j\uparrow}a_{j\downarrow}+\text{Herm. conj.}\right).
\label{eq:Model-Hamiltonian}
\end{align}
Here, $a_{i\sigma}^{\dagger},a_{i\sigma}$ are fermionic creation
and annihilation operators of single-particles states $\psi_{i\sigma}$
obeying standard commutation relations, with $\sigma\in\left\{ \uparrow,\downarrow\right\} $
denoting the spin of the electron. The discussed preformation of Cooper
pairs reduces the Hilbert space to eigenstates of Cooper pair occupation
number 
\begin{equation}
n_{i}=\frac{1}{2}\left(a_{i\downarrow}^{\dagger}a_{i\downarrow}+a_{i\uparrow}^{\dagger}a_{i\uparrow}\right)=\left\{ 0,1\right\} ,
\end{equation}
which is obviously conserved by the Hamiltonian. The first term in
Eq.~\eqref{Model-Hamiltonian} then reproduces the randomly distributed
independent single-particle energies~$\xi_{i}$. The corresponding
distribution $\nu\left(\xi\right)$ has a typical width of order of
the Fermi energy~$E_{F}$. The particular profile of $\nu\left(\xi\right)$
is of little importance for the low-energy physics as long as the
single-particle density of states~$\nu_{0}=\nu\left(\xi=0\right)$
is finite, i. e.~$\nu_{0}\sim1/E_{F}$. The second term in Eq.~\eqref{Model-Hamiltonian}
represents local Cooper-like interaction, with the summation going
over all pairs~$\left\langle ij\right\rangle $ of effectively interacting
single-particle states. We assume that each state~$i$ is effectively
coupled to a large number~$Z\gg1$ of other localized states. Importantly,
the pairs of coupled states are chosen completely at random, so that
the resulting structure bears no information about the original 3D
nature of the system (as opposed to similar models that are formulated
on a lattice, see e.g. the 2D-CMF model of Ref.~\citep{Lemarie2013}),
while also preserving some notion of the translation symmetry (in
contrast to the models on a portion of the Bethe lattice, as e.g.
the one of Ref~\citep{Feigelman_SIT_2010}). The matrix element~$D_{ij}$
of the interaction is determined by the energy dependence of the interaction
and is modeled by a smooth function with the following asymptotic
properties
\begin{equation}
D_{ij}=D\left(\xi_{i}-\xi_{j}\right)\approx\begin{cases}
\frac{\lambda}{2\nu_{0}Z}, & \left|\xi_{i}-\xi_{j}\right|\apprle\varepsilon_{D},\\
0, & \left|\xi_{i}-\xi_{j}\right|\apprge\varepsilon_{D},
\end{cases}
\end{equation}
where $\lambda\ll1$ is the dimensionless Cooper constant and $\varepsilon_{D}\ll W$
is the characteristic scale of energy dependence of the Cooper interaction. 

\subsection{The self-consistency equation\label{subsec:The-saddle-point-equation}}

The superconducting transition for the Hamiltonian~\eqref{Model-Hamiltonian}
is captured by the saddle-point (Bogolyubov) approach. According to
it, one approximates the Cooper interaction with coupling to the field
of the complex order parameter~$\Delta$. The latter is then found
as a minimum of the self-consistent free energy. In the absence of
time reversal symmetry breaking factors, such as magnetic field or
external current, the field of the order parameter~$\Delta_{i}$
can be chosen to be real and positive. One then determines the zero
temperature configuration of the order parameter as a positive solution
to the following self-consistency equation~\citep[Sec. 4.3 and Sec. 6.1]{Feigelman_Fractal-SC_2010}:
\begin{equation}
\Delta_{i}=\sum_{j\in\partial i}D\left(\xi_{i}-\xi_{j}\right)\frac{\Delta_{j}}{\sqrt{\Delta_{j}^{2}+\xi_{j}^{2}}},
\label{eq:saddle-point_order-parameter}
\end{equation}
where the summation in the right hand side goes over $Z$~states
labeled with index~$j$ that interact with a given state $i$. The
reader can find the derivation of this equation for the original Hamiltonian~\eqref{Model-Hamiltonian}
in~\appref{Saddle-point-equation_derivation}. One then has to solve
the equation~\eqref{saddle-point_order-parameter} for a given realization
of random energies $\xi_{i}$ and subsequently analyze the statistical
properties of the resulting ensemble of $\Delta_{i}$, such as the
local probability distribution and the structure of spatial correlations. 

However, the conventional self-consistent approach fails to describe
the Superconductor Insulator Transition (SIT) itself. Namely, Eq.~\eqref{saddle-point_order-parameter}
posses nontrivial solutions for arbitrary weak Cooper coupling strength,
while in reality one observes destruction of the global superconducting
order at a certain value of the coupling constant~\citep{Feigelman_SIT_2010}.
The correct description of the SIT requires careful treatment of the
self-action of the order parameter in a form of so-called Onsager
reaction term. The papers~\citep{Ioffe_2010_ref-about-Onsager,Feigelman_SIT_2010}
provide a consistent account for this effect by means of the cavity
method~\citep{MezardParisi_cavity0,MezardParisi_cavity1} and demonstrate
the emergence of broad probability distributions of the order parameter
with slow power-law decay at large values, thus revealing the defining
role of extreme values in the corresponding quantum phase transition.
However, the paper~\citep{Feigelman_SIT_2010} also demonstrates
that the effects of self-action are only relevant for $Z\apprle Z_{1}$,
where 
\begin{equation}
Z_{1}=\lambda\exp\left\{ \frac{1}{2\lambda}\right\} ,
\label{eq:Z1_definition}
\end{equation}
with $\lambda\ll1$ being the dimensionless Cooper coupling constant.
Away from this region the reaction term constitutes only a small correction,
rendering the self-consistency equation~\eqref{saddle-point_order-parameter}
applicable. We will thus limit our analysis to the case~$Z\apprge Z_{1}$,
although our technique could be extended to include the Onsager reaction
term. Despite the introduced limitation, we report a broad region
of $Z$ values for which the distribution of the order parameter still
assumes substantially non-Gaussian profile indicative of the competition
between strong fluctuations and global superconducting order.

\subsection{Mean-field solution\label{subsec:mean-field-solution}}

The typical scale of the order parameter in Eq.~\eqref{saddle-point_order-parameter}
can be established by a simple mean-field approach. Namely, one seeks
a spatially uniform solution $\Delta_{i}=\Delta_{0}=\text{const}$,
approximating the right-hand side of the self-consistency equation~\eqref{saddle-point_order-parameter}
by its statistical average. This substitution is justified \emph{a
priori} for sufficiently large values of $Z$ by virtue of the central
limit theorem. As suggested by the seminal paper~\citep{Ma_Lee_1985_Ref-to-pseudospins},
a physical estimate for the relevant range of $Z$ could be obtained
by demanding that each single-particle state has at least one other
resonant state within the energy interval of size $\Delta_{0}$. This
results on the following criteria:
\begin{equation}
Z\apprge Z_{2}=\frac{1}{2\nu_{0}\Delta_{0}}\sim2\nu_{0}\varepsilon_{D}\cdot e^{1/\lambda}.
\label{eq:Z2}
\end{equation}
In this case, one can neglect the fluctuations of the right hand side
of Eq.~\eqref{saddle-point_order-parameter} around its mean value
and obtain:
\begin{equation}
\Delta_{0}\left(\xi_{0}\right)=Z\left\langle \frac{\Delta_{0}\left(\xi\right)}{\sqrt{\Delta_{0}^{2}\left(\xi\right)+\xi^{2}}}\cdot D\left(\xi-\xi_{0}\right)\right\rangle _{\xi},
\label{eq:mean-field_order-parameter_eq}
\end{equation}
where $\left\langle \bullet\right\rangle _{\xi}$ denotes the statistical
distribution w.r.t the distribution of $\xi$. The equation still
contains the value~$\xi_{0}$ of the disorder field at a given site,
reflecting the fact that the order parameter is itself a function
of onsite energy~$\xi_{0}$. 

The value of $\Delta_{0}$ is found self-consistently by solving the
resulting integral equation. The smallness of the coupling $D\left(\xi\right)\sim\lambda/\left(2\nu_{0}Z\right)$
at small energies $\left|\xi\right|\ll\varepsilon_{D}$ enables one
to provide an analytical solution for the order parameter close to
the Fermi surface in a form of the celebrated BCS expression:
\begin{equation}
\Delta_{0}\left(\xi_{0}=0\right)=2E_{0}\cdot\exp\left\{ -\frac{1}{\lambda}\right\} ,
\label{eq:mean-field-delta_zero-temp}
\end{equation}
where the value of $E_{0}\sim\varepsilon_{D}$ is expressed via the
single-particle density of states~$\nu\left(\xi\right)$ and the
exact profile of the $D$~function. The explicit form for $E_{0}$
is presented in \appref{Saddle-point-equation_derivation}.

As this point, it is worth introducing one more microscopic parameter
that turn out to play the defining role for the distribution of the
order parameter:
\begin{equation}
\kappa=\frac{\lambda}{Z/Z_{2}}\equiv\frac{D\left(0\right)}{\Delta_{0}}
\label{eq:Z-eff_and_kappa_definition}
\end{equation}
Qualitatively, this parameter combines the information about the criteria~\eqref{Z2}
and the strength of the attractive interaction in the form of the
dimensionless coupling constant $\lambda$. Otherwise, the value of
$\kappa$ bears the meaning of properly rescaled matrix element of
bare attractive interaction. A particularly important aspect of this
parameter is that it quantifies the competition between the superconducting
pairing and the disorder. The former enters the expression via the
value of the bare matrix element of the attraction, and the latter
is represented by the mean field of the order parameter which is defined
by the distribution of the onsite disorder according to Eq.~\eqref{mean-field_order-parameter_eq}.

While our analysis shows that the mean-field result~\eqref{mean-field-delta_zero-temp}
is only justified for $Z\apprge Z_{2}$, the exponential smallness
of the actual order parameter rests solely on the smallness of the
coupling constant~$\lambda$. This makes $\Delta_{0}$ a valid scale
to describe the typical magnitude of the true solution to the self-consistency
equation~\eqref{saddle-point_order-parameter} in the whole range
$Z\apprge Z_{1}$ we are interested in. Below we find distribution
function~$P\left(\Delta\right)$ and show that it can be strongly
non-Gaussian in general, while narrow Gaussian shape is realized if
the inequality~\eqref{Z2} is satisfied.

\subsection{Relation to previous studies}

Our analytical approach presented below in \secref{Distribution-of-the-OP}
borrows certain features from the methods that are widely used to
analyze statistical physics of disordered systems on the Bethe lattice.
The latter is defined as an infinite tree with all but one vertices
having $Z-1$ descendants and one ancestor, while the root site has
$Z$ descendants and no ancestor, so that each vertex has exactly
$Z$ neighbors in total. One of the key properties of the Bethe lattice
is the absence of loops which enables one to derive recursive relations
for both a given local quantity itself and distribution function of
this quantity across various disorder realizations. Qualitatively,
such possibility can be perceived as a consequence of the fact that
in a system with no loops any two non-overlapping subsystems are connected
by a single chain of sites that arranges the exchange of statistical
information and thus induces statistical correlations. This allows
one to analyze the statistical properties of the system by considering
the state of just a single site. A prominent exploitation of this
feature was provided by M.~Mezard and G.~Parisi within their analysis
of spin glass problems on the Bethe lattice~\citep{MezardParisi_cavity0,MezardParisi_cavity1}
by means of the ``cavity method''. A similar approach was used in
Ref.~\citep{Feigelman_SIT_2010} for a model of strongly disordered
superconductor that is structurally similar to the one used in the
present work. 

However, one should be careful when using a finite portion of the
Bethe lattice as a model for any physical system. The issue is that
truncating the Bethe lattice explicitly breaks the equivalence between
different vertices in the system and thus induces a certain preferred
direction in the system. Precisely for this reason we use the ensemble
of Random Regular Graphs (RRGs) and its generalizations as a finite
size approximation to the Bethe lattice. The important difference
between the two structures is that a typical RRG inevitably contains
large loops with lengths of the order of the graph's diameter $D\sim\ln N/\ln\left(Z-1\right)$
that serve to restore the translation symmetry in the system~\citep{bollobas2001random}.
Remarkably, our theoretical and numerical analysis shows that as long
as the number of neighbors $Z$ of each site is large enough and the
disorder is not critically strong (in the sense of the vicinity of
the SIT), neither the presence of even short loops nor even the irregularity
of the base graph (in the sense that each site might have different
number of neighbors) have any noticeable influence on the distribution
of the order parameter.

It is worth discussing two more subtle differences between our present
approach and the one used previously in Ref.~\citep{Feigelman_SIT_2010}.
The cavity method~\citep{MezardParisi_cavity0,MezardParisi_cavity1}
was developed originally for Ising-type problems. Relying on the exact
recursive relation for the conditional partition function, it derives
its power from the possibility to parametrize the latter in terms
a ``local field''~$h_{i}$ defined for each site of the problem.
This is possible for the classical Ising problem where only two classical
states per site are present. Upon taking into account the normalization
condition we are left with only one real number~$h_{i}$ that parametrizes
the conditional partition function. Our superconducting problem is
different in two aspects. One of them is due to the quantum nature
of local degrees of freedom, as it was already discussed in~\citep{Feigelman_SIT_2010}.
Namely, the Hamiltonian~\eqref{Model-Hamiltonian} can be exactly
mapped on the spin~$1/2$ $XY$~model in transverse field, with
the corresponding spin degrees of freedom termed pseudospins~\citep{Ma_Lee_1985_Ref-to-pseudospins}.
Ref.~\citep{Feigelman_SIT_2010} then uses the ``static approximation''
that neglects dynamic correlations between pseudospins. The second
feature (left unnoticed in Ref.~\citep{Feigelman_SIT_2010}) is that,
even with quantum effects neglected, the conditional partition function
for a spin~$1/2$ degree of freedom with $XY$~symmetry cannot be
parametrized, in general, by a single complex field~$\Delta_{i}$.

A generalization of the cavity method is certainly possible for this
type of order parameter as well, but it is more involved. The difference
between the cavity mapping used in Ref.~\citep{Feigelman_SIT_2010}
and the exact one becomes important once the terms nonlinear in the
magnitude of the order parameter become essential for physics. We
expect that the recursive equations derived and analyzed in Ref.~\citep{Feigelman_SIT_2010}
are exact (leaving aside the additional problem with the accuracy
of ``the static approximation'') as long as the amplitude of the
order parameter is small in some appropriate sense. For example, the
linearized form of these equations is perfectly suitable, e. g., for
the analysis of the temperature-driven transition. It is also correct
to use the recursive equations of Ref.~\citep{Feigelman_SIT_2010}
for the analysis of the long tail of the order parameter distribution,
as the effects of nonlinearity are also weak in this case. In the
present paper we are interested in the shape of the complete distribution
function $P\left(\Delta\right)$ at $T=0$, where the effects of nonlinearity
are strong. Thus here we prefer to employ classical form of the self-consistency
equations~\eqref{saddle-point_order-parameter}; as explained in
the previous Subsection, the related inaccuracy (as long as we do
not include Onsager reaction term) is small as the ratio $Z_{1}/Z\ll1$.

\section{Distribution of the order parameter\label{sec:Distribution-of-the-OP}}

In this section, we present both analytical and numerical results
for the onsite joint probability distribution of fields $\xi$ and
$\Delta$ on a given site. The latter is defined as 
\begin{equation}
P_{i}\left(\xi,\Delta\right)=\left\langle \delta\left(\xi-\xi_{i}\right)\delta\left(\Delta-\Delta_{i}\left(\left\{ \xi\right\} \right)\right)\right\rangle ,
\end{equation}
where $\delta\left(x\right)$ is the Dirac $\delta$-function, $\Delta_{i}\left(\left\{ \xi\right\} \right)$
is the exact solution of the self-consistency equation~\eqref{saddle-point_order-parameter}
for a given realization of the disorder field~$\xi$, and the average
$\left\langle \bullet\right\rangle $ is performed over all configurations
of the $\xi$ field. The distribution is normalized by definition:
\begin{equation}
\intop_{0}^{\infty}d\Delta'\,P\left(\xi,\Delta'\right)=\nu\left(\xi\right),
\end{equation}
where $\nu\left(\xi\right)$ is the distribution of the original onsite
disorder field~$\xi$.

\subsection{Equation on the distribution in a locally tree-like system\label{subsec:General-local-equations}}

\begin{figure}
\begin{centering}
\includegraphics[scale=0.6]{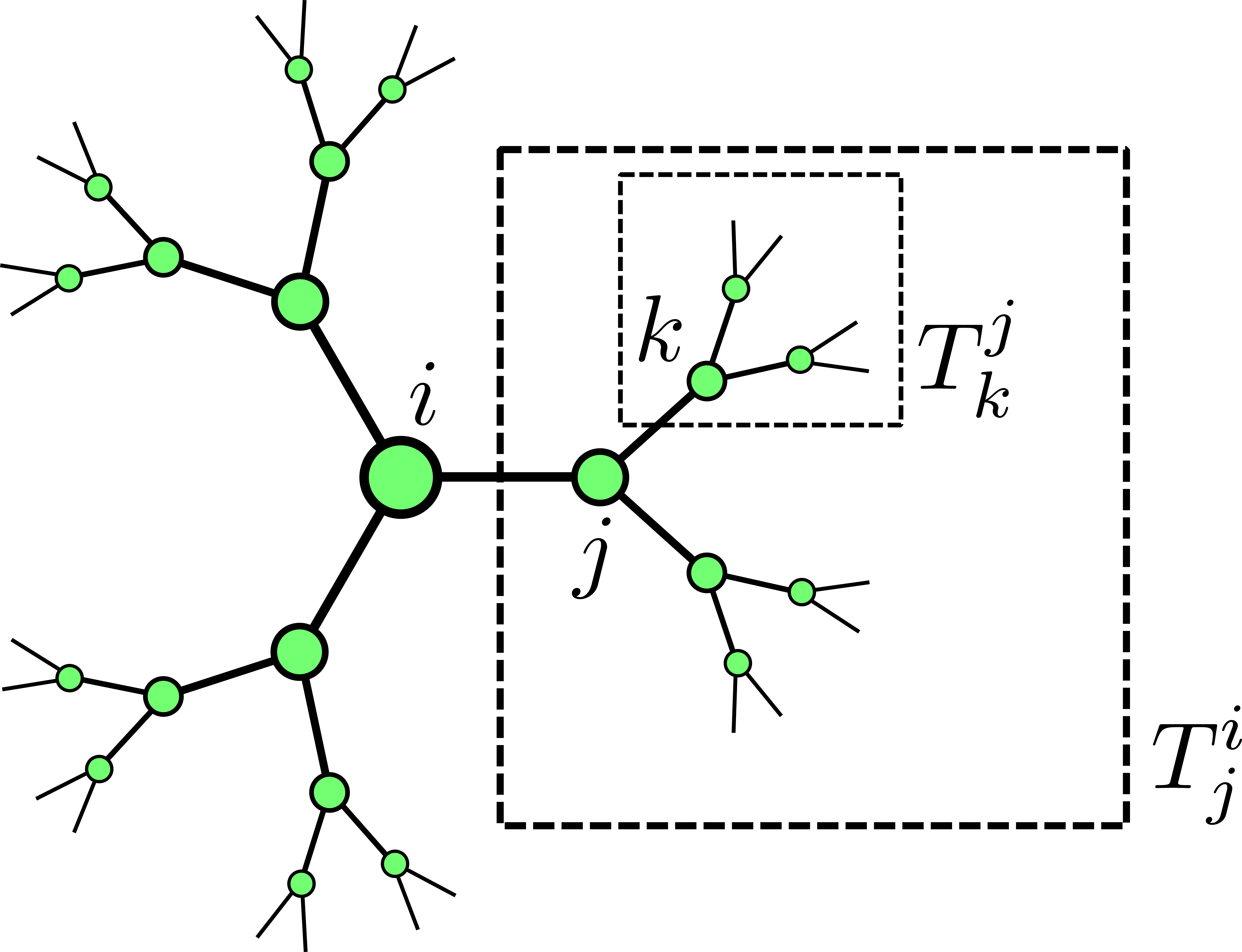}
\par\end{centering}
\caption{A schematic illustration of the neighborhood of radius~$d=3$ of
a particular vertex~$i$ of a Random Regular Graph (RRG) of degree~$Z=3$,
i. e with each vertex having exactly three neighbors. Large RRGs are
known to exhibit vanishing concentration of short loops~\citep{McKay_1981},
so that up to some large distance~$d$ the neighborhood of~$i$
represents a loop-free structure, i. e. a tree. In particular, each
neighboring vertex~$j$ is a root of the corresponding branch~$T_{j}^{i}$
consisting of all vertices that can be reached from~$i$ by a path
containing at most $d$~edges. Because the whole neighborhood is
a tree, such path is unique. Similarly, each nearest neighbor of~$j$
except $i$~itself is also a root of a tree~$T_{k}^{j}$ nested
in~$T_{j}^{i}$. Such a nested structure is convenient for various
recursive considerations. \label{fig:tree-like_structure}}
\end{figure}

Within our model, each single-particle state~$i$ is effectively
interacting with $Z$ other single-particle states selected at random.
The corresponding structure of the matrix elements can be represented
by an instance of so called Random Regular Graphs. The latter are
known to exhibit vanishing concentration of finite loops in the thermodynamical
limit~\citep{McKay_1981}. In other words, the sites at distances
up to some large distance~$d$ from any chosen site~$i$ form a
regular loop-free structure rooted at $i$ with probability approaching
unity as the total number of sites~$N$ tends to infinity. A fragment
of the corresponding structure termed locally tree-like is illustrated
on \figref{tree-like_structure}.

For the physical system in question, one expects that the spatial
distribution of the order parameter exhibits a finite correlation
radius, at least away from the SIT. This implies that the value of
the order parameter at a given site is only sensitive to the characteristics
of neighboring sites up to some finite correlation distance~$d_{0}$
away from the chosen site. In conjunction with the locally tree-like
structure, this property suggests that for each site~$i$ the neighboring
sites~$j\in\partial i$ are only correlated via the site~$i$ itself.
Indeed, the underlying graph only contains large loops that are much
longer than the correlation length~$d_{0}$, and thus cannot influence
distributions of any local quantities.

To make use of the described properties, we consider the system where
the values of both $\xi$ and $\Delta$ at a given site~$i$ are
fixed externally, i. e. $\Delta_{i_{0}}=\Delta_{0}$ and $\xi_{i_{0}}=\xi_{0}$,
as opposed to finding $\Delta_{i}$ from the self-consistency equations~\eqref{saddle-point_order-parameter}
for site~$i$. Now, consider a nearest neighbor~$j\in\partial i$
of the ``quenched'' site~$i$. Due to the aforementioned structure
of spatial correlations, the exact solution $\Delta_{j}^{i}\left(\left\{ \xi\right\} |\xi_{0},\Delta_{0}\right)$
to the modified version of the self-consistency equations~\eqref{saddle-point_order-parameter}
depends considerably only on the values of the disorder field~$\xi$
within some finite region~$T_{j}^{i}$ rooted at~$j$, see \figref{tree-like_structure}.
Crucially, the described locally tree-like structure implies that
for different~$j$ the corresponding ``essential'' regions~$T_{j}^{0}$
are non-overlapping. This translates to the fact that the pairs~$\left(\xi_{j},\Delta_{j}\right)$
for various~$j\in\partial i$ are rendered uncorrelated in the modified
problem, as they are determined by non-overlapping regions. 

Similarly to the initial problem, we are interested in the joint distribution
of $\Delta$ and $\xi$ for site $j$ in the nearest neighborhood
of $i$ for the case when both $\Delta$ and $\xi$ at site $i$ itself
are fixed externally. The corresponding distribution function is defined
as
\begin{align}
 & P_{j}^{i}\left(\xi_{1},\Delta_{1}|\xi_{0},\Delta_{0}\right)\nonumber \\
 & =\left\langle \delta\left(\xi_{1}-\xi_{j}\right)\delta\left(\Delta_{1}-\Delta_{j}^{i}\left(\left\{ \xi\right\} |\xi_{0},\Delta_{0}\right)\right)\right\rangle ,
\label{eq:P1-definition}
\end{align}

\noindent where $\Delta_{j}^{i}\left(\left\{ \xi\right\} |\xi_{0},\Delta_{0}\right)$
is the exact solution of the self-consistency equation~\eqref{saddle-point_order-parameter}
for a given realization of the disorder field $\xi$ and a fixed value
$\Delta_{0}$ of the order parameter at site $i$. The average $\left\langle \bullet\right\rangle $
is now performed over the values of $\xi$ at all sites except $i$,
where the disorder field assumes the value of $\xi_{0}$. The new
distribution function is properly normalized, i. e.
\begin{equation}
\intop_{0}^{\infty}d\Delta_{1}'\,P_{j}^{i}\left(\xi_{1},\Delta_{1}'|\xi_{0},\Delta_{0}\right)=\nu\left(\xi_{1}\right),
\label{eq:P1-normalization}
\end{equation}
valid for any $\xi_{0},\Delta_{0},\xi_{1}$. The aforementioned partition
of the neighborhood of $i$ into non-overlapping tree-like structures
$T_{i}^{j}$ then translates to the fact that the averaging in~\eqref{P1-definition}
only reflects the statistical fluctuations of $\xi$ in the corresponding
region $T_{i}^{j}$ originating from the site $j$ of interest. 

The local structure of the problem along with the outlined above statistical
independence of different neighbors $j\in\partial i$ in the modified
problem allows one to connect the onsite distribution $P_{i}\left(\xi_{0},\Delta_{0}\right)$
at site $i$ with the distributions $P_{j}^{i}$ in the modified problem.
To this end, one uses the self-consistency equation~\eqref{saddle-point_order-parameter}
for site $i$. On the one hand, it is trivially satisfied by the exact
solution $\Delta_{i}\left(\left\{ \xi\right\} \right)$ to the \emph{original}
problem. On the other hand, the values of $\Delta_{j}$ are given
by the solutions $\Delta_{j}^{i}\left(\left\{ \xi\right\} |\xi_{0},\Delta_{0}\right)$
to the \emph{modified} problem for a consistent choice of the values
$\xi_{0},\Delta_{0}$. In other words, letting $\Delta_{j}=\Delta_{j}^{i}\left(\left\{ \xi\right\} |\xi_{0},\Delta_{0}\right)$
with $\xi_{0}=\xi_{i}$, $\Delta_{0}=\Delta_{i}$ produces an equation
on the value of $\Delta_{i}$ itself. These two observations valid
for any disorder realization can be translated to the following relation
between the two problems:
\begin{widetext}
\begin{equation}
P_{i}\left(\xi,\Delta\right)=P_{i}\left(\xi\right)\cdot\intop_{-\infty}^{\infty}\frac{d\tau}{2\pi}\cdot\frac{\partial}{\partial\Delta}\left\{ \left(\intop_{E}^{\Delta}d\Delta'\cdot e^{-i\tau\Delta'}\right)\prod_{j\in\partial i}\left(\intop d\xi_{j}d\Delta_{j}\cdot P_{j}^{i}\left(\xi_{j},\Delta_{j}|\xi,\Delta\right)\cdot e^{i\tau f\left(\xi_{j},\Delta_{j}|\xi\right)}\right)\right\} .
\label{eq:P-via-P1_1}
\end{equation}
\end{widetext}

\noindent Here, $P_{i}\left(\xi\right)$ is the distribution of the
onsite disorder, $f\left(\xi_{j},\Delta_{j}|\xi\right)$ represents
a shorthand for the right hand side of the self-consistency equation~\eqref{saddle-point_order-parameter}:
\begin{equation}
f\left(\xi_{j},\Delta_{j}|\xi\right)=\frac{\Delta_{j}}{\sqrt{\Delta_{j}^{2}+\xi_{j}^{2}}}\cdot D\left(\xi_{j}-\xi\right).
\label{eq:saddle-point-rhs_notation}
\end{equation}
The lower integration limit $E$ in the integral over $\Delta'$ can
be set to an arbitrary positive constant. While the value of the whole
expression does not depend on $E$ due to normalization of the probability
distribution $P_{j}^{i}$, one can use various values of $E$ to simplify
the calculations. The specific structure of the equation is due to
the fact that computing a distributions of solutions to a given equation
with disorder requires taking into account the Jacobian resulting
from replacing the $\delta$-function of the solution with a $\delta$-function
of the corresponding equation. The detailed derivation of Eq.~\eqref{P-via-P1_1}
is presented in \appref{equation-on-distribution}.

In a similar fashion, one can formally consider quenching the site
$j$ as well and determining the resulting onsite distribution $P_{k}^{j}\left(\xi_{2},\Delta_{2}|\xi_{1},\Delta_{1}\right)$
for some $k\in\partial j\backslash\left\{ i\right\} $, i. e. next-to-nearest
neighbor of the initial site $i$. It is important, that due to the
tree-like structure, the distribution $P_{k}^{j}$ receives no information
about the values of field $\xi$ and $\Delta$ at the initial site
$i$. The same considerations as the one that lead to Eq.~\eqref{P-via-P1_1}
then allow one to connect the onsite distribution $P_{j}^{i}$ of
the site $j$ with those on all nearest neighbors of $j$ except $i$
itself:
\begin{widetext}
\begin{align}
P_{j}^{i}\left(\xi_{1},\Delta_{1}|\xi_{0},\Delta_{0}\right) & =\nu\left(\xi_{1}\right)\cdot\intop_{-\infty}^{\infty}\frac{d\tau}{2\pi}\cdot\frac{\partial}{\partial\Delta_{1}}\left\{ \left(\intop_{E}^{\Delta_{1}}d\Delta'_{1}e^{-i\tau\Delta'_{1}+i\tau f\left(\xi_{0},\Delta_{0}|\xi_{1}\right)}\right)\right.\nonumber \\
 & \times\left.\prod_{k\in\partial j\backslash\left\{ i\right\} }\left(\intop d\xi_{k}d\Delta_{k}\cdot P_{k}^{j}\left(\xi_{k},\Delta_{k}|\xi_{1},\Delta_{1}\right)\cdot e^{i\tau f\left(\xi_{k},\Delta_{k}|\xi_{j}\right)}\right)\right\} .
\end{align}
\end{widetext}

The final step of the derivation is to exploit translational and rotational
symmetries of the underlying graph, as the latter are restored after
averaging over disorder. In other words, the choice of $i$ and $j\in\partial i$
is arbitrary, so that translational invariance implies independence
of both the original $P_{i}$ and the modified $P_{j}^{i}$ distributions
on the choice of $i$, while rotational invariance suggests that $P_{j}^{i}$
is the same for all $j\in\partial i$. This allows one to replace
all $P_{j}^{i}$ with just a single function $P_{1}$, arriving at
the central results of this section:
\begin{widetext}
\begin{equation}
P\left(\xi,\Delta\right)=\nu\left(\xi\right)\intop_{-\infty}^{\infty}\frac{d\tau}{2\pi}\,\frac{\partial}{\partial\Delta}\left\{ \left(\intop_{E}^{\Delta}d\Delta'\,e^{-i\tau\Delta'}\right)\left(\intop d\xi_{1}d\Delta_{1}\cdot P_{1}\left(\xi_{1},\Delta_{1}|\xi,\Delta\right)e^{i\tau f\left(\xi_{1},\Delta_{1}|\xi\right)}\right)^{Z}\right\} ,
\label{eq:exact-equation-on-P}
\end{equation}

\begin{align}
P_{1}\left(\xi_{1},\Delta_{1}|\xi_{0},\Delta_{0}\right) & =\nu\left(\xi_{0}\right)\intop_{-\infty}^{\infty}\frac{d\tau}{2\pi}\,\frac{\partial}{\partial\Delta_{1}}\left\{ \intop_{E}^{\Delta_{1}}d\Delta_{1}'\,\exp\left\{ -i\tau\Delta_{1}'+i\tau f\left(\xi_{0},\Delta_{0}|\xi_{1}\right)\right\} \right.\nonumber \\
 & \times\left.\left(\intop d\xi_{2}d\Delta_{2}\cdot P_{1}\left(\xi_{2},\Delta_{2}|\xi_{1},\Delta_{1}\right)e^{i\tau f\left(\xi_{2},\Delta_{2}|\xi_{1}\right)}\right)^{Z-1}\right\} .
\label{eq:exact-equation-on-P1}
\end{align}
\end{widetext}

\noindent Both expressions~(\hphantom{}\ref{eq:exact-equation-on-P}\nobreakdash-\ref{eq:exact-equation-on-P1}\hphantom{})
preserve the normalization of the distributions, as can be checked
by direct computation. 

The accuracy of equations~(\hphantom{}\ref{eq:exact-equation-on-P}\nobreakdash-\ref{eq:exact-equation-on-P1}\hphantom{})
is governed by the presence of small loops in the system. However,
the relative magnitude of the corresponding corrections is estimated
as $\sim Z^{-l}$. This estimation originates from the fact that correlations
in the distribution of $\Delta$ can be shown to decay as $Z^{-d}$.
Because of the aforementioned loopless structure of large regular
graphs, the equations~(\hphantom{}\ref{eq:exact-equation-on-P}\nobreakdash-\ref{eq:exact-equation-on-P1}\hphantom{})
become exact in the thermodynamical limit. In reality, however, finite
loops are present in the system, but their concentration is typically
small~\citep{McKay_1981}, rendering their physical effect insignificant.
Our additional numerical experiments show that for sufficiently large
$Z$ even the shortest loops of length three do not cause any noticeable
deformation of the onsite distribution functions. Namely, the empirical
distribution of the order parameter on those sites that are members
of any cycle of length three in the graph is statistically indistinguishable
from the probability distribution for the remaining fraction of sites.

We also note that our approach allows a systematic computation of
any other joint probability distribution functions for any group of
sites of finite spatial size. In particular, a joint probability distribution
$P_{ij}\left(\xi_{i},\Delta_{i};\xi_{j},\Delta_{j}\right)$ for any
two sites at some finite distance $d$ is expressible in terms of
certain integro-differential transform of the product of two $P_{1}$
functions. It is worth noting at this point, that both the direct
inspection of our approach and the answer for the joint probability
distribution for the two neighboring sites $i$ and $j\in\partial i$
suggests that $P_{1}$ \emph{does not }coincide with a conditional
distribution function of the form $P_{ij}\left(\xi_{i},\Delta_{i};\xi_{j},\Delta_{j}\right)/P_{i}\left(\xi_{i},\Delta_{i}\right)$.
Although the two objects share some qualitative properties, they are
in fact quite different quantitatively. The difference can be traced
down to the aforementioned Jacobian originating from representing
the $\delta$-function of solution in terms of the $\delta$-function
of the original equations. 

We conclude this subsection by noting that the developed formalism
allows numerous extensions of the form of the $f$~function. As long
as the underlying physical assumptions of conditional statistical
decoupling (i. e. the locality of correlations) hold true, the exact
form of the right-hand side of the analyzed equation~\eqref{saddle-point_order-parameter}
is of little importance. Possible generalizations include the effects
of finite temperature and other types of uncorrelated disorder. In
particular, \appref{Effect-of-fluctuating-coupling} presents analysis
of a more general model that reflects mesoscopic fluctuation in the
values of the matrix elements between localized electron states. The
key qualitative changes to our results due to such fluctuations are
summarized in \subsecref{Extreme-value-statistics_fluctuating-coupling}.

\subsection{The limit of small $\Delta$ and large $Z$\label{subsec:The-limit-of-small-Delta}}

Having equations~(\hphantom{}\ref{eq:exact-equation-on-P}\nobreakdash-\ref{eq:exact-equation-on-P1}\hphantom{})
at hand, it is now our aim to simplify the equations in order to reflect
the fact that the typical scale of the order parameter is the only
relevant energy scale in the problem. In other words, we want to exploit
the hierarchy of scales of the form $\Delta\ll\varepsilon_{D},E_{F}$
that is naturally present in the problem. By carefully expanding the
equations~(\hphantom{}\ref{eq:exact-equation-on-P}\nobreakdash-\ref{eq:exact-equation-on-P1}\hphantom{})
according to this relation of scales, we will eventually be able to
solve the equation~\eqref{exact-equation-on-P1} for $P_{1}$ and
calculate the resulting distribution $P\left(\xi,\Delta\right)$ by
means of~\eqref{exact-equation-on-P}. 

We start by introducing the following dimensionless quantities:
\begin{equation}
x_{i}=\frac{\xi_{i}}{\Delta_{0}},\,\,\,\,\,y_{i}=\frac{\Delta_{i}}{\Delta_{0}},
\end{equation}
where $\Delta_{0}$ is the mean field value of the order parameter
defined in \subsecref{mean-field-solution}. Similarly to the conventional
theory of superconductivity, we then expect that the high-energy physics
playing out at scales $\varepsilon_{D},E_{F}$ does not find its way
in the low-energy physics, as the sole role of higher energies is
to dictate the overall scale of superconducting correlations.

The equation~\eqref{exact-equation-on-P1} suggests the following
quantity as a proper object in the limit of small $\Delta$:
\begin{widetext}
\begin{equation}
m\left(S|x,y\right):=\ln\left\{ \left[\intop d\xi_{1}d\Delta_{1}\cdot P_{1}\left(\xi_{1},\Delta_{1}|\xi,\Delta\right)\cdot\exp\left\{ iS\cdot f\left(\xi_{1},\Delta_{1}|\xi\right)/\Delta_{0}\right\} \right]^{Z-1}\right\} ,\,\,\,\,\,\xi=\Delta_{0}x,\,\,\,\,\Delta=\Delta_{0}y.
\label{eq:m-function_definition}
\end{equation}
\end{widetext}

\noindent It represents a dimensionless form of the cumulant generating
function for the right hand side of the self-consistency equation~\eqref{saddle-point_order-parameter}
for site $j$ in the modified version of the problem, see the detailed
description in the preceding \subsecref{General-local-equations}.
In particular, the normalization condition~\eqref{P1-normalization}
translates to the following trivial identity:
\begin{equation}
m\left(0|x,y\right)=0,
\end{equation}
valid for any $x,y$.

The integro-differential equation~\eqref{exact-equation-on-P1} can
be reformulated in terms of $m$ function in a straightforward fashion.
The proper low-energy limit of this equation consists of formally
retaining only the leading orders in powers of small parameters $\nu_{0}\Delta_{0},\,1/Z\ll1$
while treating their product as a finite constant $Z_{\text{eff}}=2\nu_{0}\Delta_{0}\cdot\left(Z-1\right)$
that may attain any numerical value, either large or small. The physical
meaning of $Z_{\text{eff}}$ is the \emph{effective} number of interacting
neighbors, that is, pairs with local energies within the energy stripe
of width $\sim\Delta$. Evidently, local fluctuations of the order
parameter will be small if $Z_{\text{eff}}\gg1$. A proper reduction
of Eq.~\eqref{exact-equation-on-P1} to the low-energy sector of
the theory should be implemented with care due to logarithmic divergency
at high energies, with the latter being typical for any kind of BCS-like
theory. Working out a proper cutoff for this divergence requires certain
technical effort. The corresponding technical details are described
in \appref{equation-on-distribution_small-Delta} for a simple case
of trivial energy dependence of the matrix element, i. e. $D\left(\xi\right)=D\left(0\right)=\text{const}$.
Although not exactly physical, the latter case showcases all insights
necessary to obtain a controlled limit of small $\Delta_{0}$. \appref{Effect-of-energy-dependence-of-matrix-element}
then describes the generalization of the approach to the case of smooth
$D\left(\xi\right)$ with some finite energy scale of the order of
the Debye energy $\varepsilon_{D}$. Below we formulate the outcome
of this procedure.

The $m\left(S|x,y\right)$ function possesses the following parametrization
that is natural to describe the effects resulting from carefully processing
the aforementioned logarithmic behavior in the theory:
\begin{equation}
m\left(S|x,y\right)=iSm_{1}\left(w\right)+m_{2}\left(S|w\right),
\label{eq:m_split-form}
\end{equation}
\begin{equation}
w=\omega\left(z=x/y\right)=\frac{1}{\sqrt{1+z^{2}}},
\end{equation}
valid for $\left|x\right|\leq x_{\max}$, where $x_{\max}\sim\varepsilon_{D}/\Delta_{0}\gg1$
by assumption. The function $m_{2}$ is constructed in such a way
that its expansion in powers of small $S$ starts from the second
order, i. e. $m_{2}\left(S|w,x\right)=O\left(S^{2}\right)$ for $S\ll1$.
For both $m_{1},\,m_{2}$, the $w$ arguments assumes values in $\left[0,1\right]$.
The functions $m_{1},\,m_{2}$ then satisfy the following pair of
integro-differential equations:
\begin{widetext}
\begin{align}
m_{1}\left(w\right) & =m_{1}\left(0\right)+\kappa w\cdot\alpha+\lambda\intop_{0}^{1}dw_{1}\cdot\sqrt{1-w_{1}^{2}}\cdot\frac{m_{1}\left(w_{1}\right)-m_{1}\left(0\right)}{w_{1}}\nonumber \\
 & +\lambda\intop_{0}^{\infty}dy_{1}\cdot y_{1}\ln\frac{1}{y_{1}}\cdot\intop_{\mathbb{R}-i0}\frac{ds}{2\pi}\cdot\exp\left\{ is\kappa\cdot w\right\} \cdot\exp\left\{ m\left(s|0\right)-isy_{1}\right\} ,
\label{eq:eq-on-m1}
\end{align}
\begin{align}
m_{2}\left(S|w\right) & =\lambda\cdot\intop_{0}^{1}dw_{1}\cdot\frac{\exp\left\{ iS\kappa\cdot w_{1}\right\} -1-iS\kappa\cdot w_{1}}{w_{1}^{2}\sqrt{1-w_{1}^{2}}}\cdot\left[1-w_{1}\left(1-w_{1}^{2}\right)\frac{\partial}{\partial w_{1}}\right]\cdot\left[\frac{\kappa w+m_{1}\left(w_{1}\right)}{\kappa}\right].
\label{eq:eq-on-m2}
\end{align}
\end{widetext}

\noindent These equations constitute a proper low-energy limit of
equation~\eqref{exact-equation-on-P1}. The result contains three
controlling parameters $\lambda,\,\kappa,\,\alpha$ that define the
form of the solution and are themselves defined by high-energy physics.
By definition, $\lambda=2\nu_{0}ZD\left(0\right)$ is the dimensionless
Cooper attraction constant, the parameter $\kappa$ is defined as
\begin{equation}
\kappa=\frac{\lambda}{2\nu_{0}\Delta_{0}\left(Z-1\right)}=\frac{\lambda}{Z_{\text{eff}}},
\label{eq:kappa-def}
\end{equation}
and the value of $\alpha$ is given by the following expression:
\begin{equation}
\alpha=1+\lambda\intop_{\mathbb{R}}\frac{d\xi\cdot\nu\left(\xi\right)}{2\nu_{0}}\cdot\frac{D\left(\xi\right)}{D\left(0\right)}\cdot\frac{D\left(\xi\right)-\mathcal{D}\left(\xi\right)}{D\left(0\right)\left|\xi\right|},
\label{eq:def-of-alpha}
\end{equation}
where the $\mathcal{D}$ function is the solution to the following
integral equation:
\begin{align}
 & \mathcal{D}\left(\xi_{0}\right)=D\left(\xi_{0}\right)+\nonumber \\
 & \lambda\intop\frac{d\xi\nu\left(\xi\right)}{2\nu_{0}}\frac{D\left(0\right)D\left(\xi-\xi_{0}\right)-D\left(\xi_{0}\right)D\left(\xi\right)}{D^{2}\left(0\right)\left|\xi\right|}\mathcal{D}\left(\xi\right).
\label{eq:def-of-energy-dependence-of-the-OP}
\end{align}
The physical sense of $\mathcal{D}$ is to reflect the mean-field
energy dependence of the order parameter at scales $\xi\sim\varepsilon_{D}$.
Namely, it describes the behavior of the solution $\Delta\left(\xi\right)=\Delta_{0}\mathcal{D}\left(\xi\right)$
to the mean-field equation~\eqref{mean-field_order-parameter_eq},
see \appref{Saddle-point-equation_derivation} for details. As already
mentioned above, the derivation of these results is presented in \appref{equation-on-distribution_small-Delta}
for the simple case with $D\left(\xi\right)=D\left(0\right)=\text{const}$
and in \appref{Effect-of-energy-dependence-of-matrix-element} for
the case of smooth $D$. The resulting expressions are applicable
as long as the actual value of the order parameter $\Delta\sim\Delta_{0}$
is much smaller than any other typical scale in the problem.

The solution to~(\hphantom{}\ref{eq:def-of-alpha}\nobreakdash-\ref{eq:def-of-energy-dependence-of-the-OP}\hphantom{})
renders the value of $\alpha$ that is close to unity as long as the
coupling constant $\lambda$ is small enough:
\begin{equation}
\alpha\approx1+\lambda^{2}c,\,\,\,\,\,c\sim1.
\end{equation}
Furthermore, the exact values of both $\alpha$ and $\lambda$ provide
only a certain quantitative effect, while the only essential role
in the statistics of the order parameter belongs to the parameter
$\kappa$. In particular, in the following \subsecref{Extreme-value-statistics}
it is shown that large values of $\kappa$ correspond to heavily non-Gaussian
regime of the distribution, while the region $\kappa\ll1$ reproduces
the Gaussian statistics as it corresponds to the region defined by~\eqref{Z2}. 

Once the solution to equations~(\hphantom{}\ref{eq:eq-on-m1}\nobreakdash-\ref{eq:eq-on-m2}\hphantom{})
is obtained, one uses the expression~\eqref{exact-equation-on-P}
to calculate the joint probability distribution $P\left(x,y\right)$
of the fields $x=\xi/\Delta_{0}$ and $y=\Delta/\Delta_{0}$:
\begin{widetext}
\begin{equation}
P\left(x,y\right)=P\left(x\right)\cdot\intop_{\mathbb{R}}\frac{ds}{2\pi}\cdot\frac{\partial}{\partial y}\left\{ \left[\intop_{0}^{y}dy^{'}\exp\left\{ -isy^{'}\right\} \right]\cdot\exp\left\{ m\left(s|\omega\left(x/y\right)\right)\right\} \right\} ,\,\,\,\,\omega\left(z=x/y\right)=\frac{1}{\sqrt{1+z^{2}}},
\label{eq:joint-distribution_expression-via-m}
\end{equation}
\end{widetext}

\noindent where all probability distributions are understood in their
dimensionless form, so that the probability measure is defined as
$P\left(x\right)dx$, $P\left(x,y\right)dxdy$, etc. In particular,
the value of $P\left(x\right)$ is given by $P\left(x\right)=\Delta_{0}\cdot\nu\left(\xi=\Delta_{0}x\right)$.
The expression is valid for $\left|x\right|\ll\varepsilon_{D}/\Delta_{0}$,
while the remaining region is covered in \appref{Effect-of-energy-dependence-of-matrix-element}.
At this point, a comment is in order regarding the qualitative behavior
of $P\left(x,y\right)$ with respect to the first argument $x=\xi/\Delta_{0}$.
From general physics reasoning one expects that there are two important
regions: $\left|x\right|\sim1$ and $\left|x\right|\apprge\varepsilon_{D}/\Delta_{0}\gg1$.
In the former, the joint distribution is expected to exhibit nontrivial
behavior that is the central topic of this paper. On the contrary,
the region of large $\left|x\right|$ describes the situation when
the Cooper attraction is not effective anymore because the corresponding
single-particle state is two far away from the Fermi surface and thus
does not contribute to the global superconducting order. As a result,
one expects that for $\left|x\right|\apprge\varepsilon_{D}/\Delta_{0}$
the joint probability distribution is concentrated around $y=0$ and
thus bears no physical meaning whatsoever. 

The distribution $P\left(y\right)$ of the order parameter is then
obtained by integrating the joint distribution $P\left(x,y\right)$
over $x$. According to the discussion above, the upper limit of this
integration is $x_{\max}\sim\varepsilon_{D}/\Delta_{0}$ which corresponds
to local site energies close to Fermi level, i. e. $\left|\xi\right|\apprle\varepsilon_{D}$.
The result has the following simple form:
\begin{equation}
P_{0}\left(y\right)=\intop_{\mathbb{R}}\frac{ds}{2\pi}\cdot\exp\left\{ m\left(s|0\right)-isy\right\} .
\label{eq:expr-for-P0-via-m}
\end{equation}
It is now evident that the quantity $m\left(s|0\right)$ represents
the cumulant generating function of the order parameter, that is
\begin{equation}
m\left(s|0\right)=\ln\left[\left\langle \exp\left\{ is\cdot\frac{\Delta}{\Delta_{0}}\right\} \right\rangle \right],
\end{equation}
where the average $\left\langle \bullet\right\rangle $ is taken over
the distribution $P_{0}$, i. e. only takes into account physically
relevant states close to the Fermi surface. 

The theoretical approach developed thus far can be summarized as follows.
Given the values of the parameters $\kappa,\lambda,\alpha$ defined
by high energy physics according to equations~(\hphantom{}\ref{eq:kappa-def}\nobreakdash-\ref{eq:def-of-energy-dependence-of-the-OP}\hphantom{}),
one solves the system of equations~(\hphantom{}\ref{eq:eq-on-m1}\nobreakdash-\ref{eq:eq-on-m2}\hphantom{})
for the $m$ function. This function alone contains complete information
about the statistical properties of the self-consistency equations~\eqref{saddle-point_order-parameter}.
In particular, the very definition \eqref{m-function_definition}
of the $m$~function implies that the modified distribution $P_{1}\left(x_{1},y_{1}|x,y\right)$
is directly restored from $m\left(S|x,y\right)$ by computing the
right-hand side of \eqref{exact-equation-on-P1}, with the latter
being expressible in terms of $m$ alone. One then uses expression~\eqref{expr-for-P0-via-m}
to calculate the onsite probability distribution of the order parameter
close to the Fermi surface or a similar expression for joint probability
distributions of interest. The latter can be systematically expressed
in terms of the $P_{1}\left(x_{1},y_{1}|x,y\right)$ distribution
according to the procedure delineated in \subsecref{General-local-equations}.

\subsection{Weak coupling approximation $\lambda\ll1$\label{subsec:Weak-coupling-approximation}}

It turns out that the equations~(\hphantom{}\ref{eq:eq-on-m1}\nobreakdash-\ref{eq:eq-on-m2}\hphantom{})
admit a complete analytical solution for the case of small coupling
$\lambda$. While we have already used the smallness of the coupling
constant in the form of the corresponding exponential smallness of
the order parameter to derive the equations~(\hphantom{}\ref{eq:eq-on-m1}\nobreakdash-\ref{eq:eq-on-m2}\hphantom{})
themselves, the value of $\lambda$ in the resulting low-energy theory
is not restricted to small values and can itself assume values of
the order of unity. For the case of small values of $\lambda$, however,
we now present a consistent expansion of the $m$ function in powers
of small $\lambda$ that constitutes a full solution to the system~(\hphantom{}\ref{eq:eq-on-m1}\nobreakdash-\ref{eq:eq-on-m2}\hphantom{}).
A detailed procedure is presented in \appref{Solution_small-lambda},
while this Subsection demonstrates the final results.

The leading term of the $m_{2}$ function reads:
\begin{equation}
m_{2}\left(S|w\right)=\lambda\cdot\left[\left(w+w_{0}\right)\Phi_{0}\left(\kappa S\right)+\Phi_{1}\left(\kappa S\right)\right],
\label{eq:m2-small-lambda-limit}
\end{equation}
where $\Phi_{0}$ and $\Phi_{1}$ are special functions with the following
integral representations:
\begin{equation}
\Phi_{0}\left(\sigma\right)=\intop_{0}^{1}\frac{dw_{1}}{w_{1}^{2}\sqrt{1-w_{1}^{2}}}\left\{ e^{i\sigma w_{1}}-1-i\sigma w_{1}\right\} ,
\end{equation}
\begin{equation}
\Phi_{1}\left(\sigma\right)=\intop_{0}^{1}\frac{w_{1}dw_{1}}{\sqrt{1-w_{1}^{2}}}\left\{ e^{i\sigma w_{1}}-1-i\sigma w_{1}\right\} ,
\end{equation}
and $w_{0}$ is a constant that is determined below in a self-consistent
fashion. The special functions can be expressed in terms of generalized
hypergeometric series, see \appref{Solution_small-lambda}. One then
substitutes this form of the $m_{2}$~function in Eq.~\eqref{eq-on-m1}
for the remaining $m_{1}$ term. Restoring the functional form of
the $w$-dependence up to the same precision as the expression~\eqref{m2-small-lambda-limit}
for $m_{2}$ then renders:
\begin{align}
m_{1}\left(w\right) & =\kappa\left(w+w_{0}\right)\nonumber \\
 & +\lambda\left[\left(w_{0}+w\right)\ln\frac{1}{w_{0}+w}-w_{0}\ln\frac{1}{w_{0}}\right].
\end{align}
Finally, equation~\eqref{eq-on-m1} also produces a self-consistency
equation for $m_{1}\left(0\right)$, which allows one to determine
the value of $w_{0}$:
\begin{equation}
w_{0}=w_{0}^{\left(0\right)}+\lambda\cdot w_{0}^{\left(1\right)},
\end{equation}
\begin{equation}
w_{0}^{\left(0\right)}=\frac{\pi/4}{W\left(\pi\kappa/4\right)},\,\,\,\,\,w_{0}^{\left(1\right)}=\frac{\frac{\pi}{4}\ln\frac{1}{\kappa}+F\left(w_{0}^{\left(0\right)}\right)}{\ln\kappa w_{0}^{\left(0\right)}+1}.
\label{eq:w0-expr-for-leading-terms}
\end{equation}
where $W\left(z\right)$ is the principal branch of the Lambert's
$W$-function, and $F\left(w\right)$ is a special function with the
following integral representation:
\begin{align}
F\left(w\right) & =\intop_{0}^{1}dw_{1}\cdot\frac{w_{1}^{2}+\left(1-w_{1}^{2}\right)\ln\frac{1}{w}}{\sqrt{1-w_{1}^{2}}}\nonumber \\
 & +\intop_{0}^{1}dw_{1}\cdot\frac{\left(w+w_{1}\right)^{2}\ln\frac{w}{w+w_{1}}+ww_{1}}{w_{1}^{2}\sqrt{1-w_{1}^{2}}}.
\label{eq:F-function-def}
\end{align}
\appref{Solution_small-lambda} contains an explicit expression for
the $F$~function in terms of polylogarithm function $\text{Li}_{2}\left(z\right)$.
Equations~(\hphantom{}\ref{eq:m2-small-lambda-limit}\nobreakdash-\ref{eq:F-function-def}\hphantom{})
thus constitute a complete solution for $m$ function that is restored
from $m_{1}$ and $m_{2}$ contributions according to Eq.~\eqref{m_split-form}.
The obtained expressions are then to be used to compute the value
of the distribution function $P_{0}\left(y\right)$ by means of Eq.~\eqref{expr-for-P0-via-m}.
\figref{numerics-vs-theory_delta-distrib} features the resulting
theoretical curves along with the ones obtained with the use the exact
solution to the equations~(\hphantom{}\ref{eq:eq-on-m1}\nobreakdash-\ref{eq:eq-on-m2}\hphantom{})
and with a histogram of direct numerical solution to the original
self-consistency equations~\eqref{saddle-point_order-parameter}.

The applicability of the presented expansion is limited by the subleading
terms in $\lambda$. The corresponding control parameter is given
by
\begin{equation}
\frac{\lambda}{w_{0}^{\left(0\right)}}=\lambda\cdot\frac{4}{\pi}W\left(\frac{\pi\kappa}{4}\right)\ll1,
\end{equation}
which, in turn, limits the value of the microscopic parameter $Z$
of our model as
\begin{equation}
Z\gg Z^{*}=\frac{\pi}{4}\cdot\frac{\lambda}{2\nu_{0}\cdot2\varepsilon_{D}}\cdot\exp\left\{ \frac{1}{\lambda}\left(1-\frac{\pi}{4}\right)\right\} .
\label{eq:small-lambda-condition_Z-criteria}
\end{equation}
Remarkably, the resulting scale of $Z$ is exponentially smaller than
the value of $Z_{1}=\lambda\exp\left\{ 1/2\lambda\right\} $, which
limits the applicability of the original self-consistency equations~\eqref{saddle-point_order-parameter}
due to the neglect of the Onsager reaction terms, as explained in
the discussion after Eq.~\eqref{Z1_definition}.

We have thus obtained a set of expressions that fully describe the
statistics of the order parameter in the entire region of applicability
of the original self-consistency equations~\eqref{saddle-point_order-parameter}.
Namely, expressions \eqref{m2-small-lambda-limit}~through~\eqref{F-function-def}
explicitly describe the $m$~function, which, in turn, contains full
information about the joint statistics of the order parameter $\Delta$
and the disorder field $\xi$, as explained in \subsecref{General-local-equations}.

\subsection{Extreme value statistics\label{subsec:Extreme-value-statistics}}

The exact equations~(\hphantom{}\ref{eq:eq-on-m1}\nobreakdash-\ref{eq:eq-on-m2}\hphantom{})
presented earlier admit asymptotic analysis that allows one to extract
the behavior of the probability density function $P_{0}\left(y\right)$
of the dimensionless order parameter $y$ in several important limiting
cases. These include the limit of Gaussian distribution of the order
parameter that connects our model to the conventional weak disorder
limit as well as the the extreme value statistics in the regime of
non-Gaussian distribution of the order parameter corresponding to
moderate and large values of $\kappa$. 

\subsubsection{Gaussian regime of weak disorder $\kappa\ll\lambda$}

We start by formally considering the limit of large number of neighbors
that corresponds to the regime of weak fluctuations. Within our theory,
this regime is realized at $\kappa\apprle\lambda$, in consistence
with the physical criteria articulated in \subsecref{mean-field-solution}.
For small values of $\kappa$, the integral over $s$ in Eq.~\eqref{expr-for-P0-via-m}
for the probability distribution $P_{0}\left(y\right)$ gains its
value near the trivial saddle point $s=0$, as the $m$ function depends
on $s$ only via a combination $\kappa s$. This, in turn, implies
that only the two leading terms in the expansion of the $m$ function
in powers of small $s$ are important for the value of the integral~\eqref{expr-for-P0-via-m}.
As it is shown in \subsecappref{Gaussian-limit}, these leading terms
are straightforwardly extracted from the system~(\hphantom{}\ref{eq:eq-on-m1}\nobreakdash-\ref{eq:eq-on-m2}\hphantom{})
and read:
\begin{align}
 & m\left(S\ll\kappa^{-1}|w\right)=\left\{ 1+\left(\frac{\pi}{4}+w\right)\left(1-\lambda\right)\kappa\right\} \left(iS\right),\nonumber \\
 & +\frac{1}{2}\lambda\kappa\cdot\left\{ \frac{\pi}{2}+\left[\frac{\pi^{2}}{8}+\frac{2}{3}\right]\left(1-\lambda\right)\kappa+\frac{\pi}{2}\kappa w\right\} \left(iS\right)^{2}.
\label{eq:m-Gaussian-regime}
\end{align}
The higher order corrections are negligible for $\kappa S\ll1$. With
this expression at hand, one obtains the following approximate expressions
for the probability density function of the order parameter:
\begin{equation}
P_{0}\left(y\right)\approx\frac{1}{\sqrt{2\pi\sigma^{2}}}\exp\left\{ -\frac{\left(y-\left\langle y\right\rangle \right)^{2}}{2\sigma^{2}}\right\} ,
\label{eq:P0-Gaussian-regime}
\end{equation}
\begin{equation}
\left\langle y\right\rangle =1+\frac{\pi}{4}\left(1-\lambda\right)\kappa,
\label{eq:Gaussian-regime_mean-value}
\end{equation}
\begin{equation}
\sigma^{2}=\frac{\pi}{2}\lambda\kappa\cdot\left\{ 1+\left[\frac{\pi}{4}+\frac{4}{3\pi}\right]\left(1-\lambda\right)\kappa\right\} .
\label{eq:Gaussian-regime_variance}
\end{equation}
As already mentioned, the discussed approximation is valid for $\kappa\apprle\lambda$,
as follows from analysis of higher order corrections to the expansion~\eqref{m-Gaussian-regime},
see \subsecappref{Gaussian-limit} for details. The presented results~(\hphantom{}\ref{eq:P0-Gaussian-regime}\nobreakdash-\ref{eq:Gaussian-regime_variance}\hphantom{})
are otherwise accessible by a direct averaging of the original self-consistency
equations~\eqref{saddle-point_order-parameter}. Indeed, upon applying
the central limit theorem to the right hand side of Eq.~\eqref{saddle-point_order-parameter},
one concludes that the order parameter in the left hand side obeys
a Gaussian distribution~\eqref{P0-Gaussian-regime} with the parameters
given by equations~\eqref{Gaussian-regime_mean-value}~and~\eqref{Gaussian-regime_variance}.
The region $\kappa\apprle\lambda$ is thus consistent with the basic
expectations in the regime of weak disorder.

\subsubsection{Strong disorder $\kappa\apprge\lambda$, small-$y$ tail}

In case $\kappa\ge\lambda$ the full shape of the distribution function
$P_{0}\left(y\right)$ cannot be computed analytically in general
case. However, its behavior at both large and small values of $y$
is reproduced by the saddle-point analysis of the corresponding integral~\eqref{expr-for-P0-via-m}.
The latter, in turn, requires asymptotic analysis for the $m$~function
at large purely imaginary arguments. This asymptotic behavior can
be extracted from~\eqref{eq-on-m2}. A detailed exposition of the
procedure is presented in \appref{Extreme-value-statistics}, while
here we only quote the results. 

For small values of $y$ one finds the following asymptotic expression
for the probability:
\begin{equation}
P_{0}\left(y\apprle1\right)\approx\sqrt{\frac{\zeta\left(y\right)}{2\pi\cdot\left[\lambda\left\langle y\right\rangle \right]^{2}}}\cdot\exp\left\{ -\zeta\left(y\right)\right\} ,
\label{eq:P0-small-y-tail}
\end{equation}
with the exponent $\zeta\left(y\right)$ given by
\begin{equation}
\zeta\left(y\right)=\frac{\lambda\left\langle y\right\rangle }{2\kappa}\exp\left\{ \frac{1}{\lambda}\left(1-\frac{y}{\left\langle y\right\rangle }\right)-\frac{\left\langle y\ln y\right\rangle }{\left\langle y\right\rangle }-\gamma\right\} ,
\label{eq:zeta_def}
\end{equation}
where $\left\langle \bullet\right\rangle $ denotes the mean value
with respect to the full distribution $P_{0}\left(y\right)$ itself,
and $\gamma=0.577...$ is the Euler-Mascheroni constant. The expressions~(\hphantom{}\ref{eq:P0-small-y-tail}\nobreakdash-\ref{eq:zeta_def}\hphantom{})
are valid as long as the value of $\zeta$ is sufficiently large,
viz.
\begin{equation}
\zeta\left(y\right)\gg\max\left\{ \frac{\lambda\left\langle y\right\rangle }{\kappa},1\right\} .
\label{eq:small-y-region_criteria}
\end{equation}
For the case $\kappa\apprle\lambda\ll1$ considered, the condition
above reduces to $\zeta\gg1$. We choose to retain the more general
form for the discussion relevant to the case $\kappa\apprle\lambda$
below.

\subsubsection{Strong disorder $\kappa\apprge1$, large-$y$ tail}

In the limit of large values of $y$, the following asymptotic expression
takes place:
\begin{align}
\ln P_{0}^{\text{lead}}\left(y\right) & \sim-\frac{y-\left\langle y\right\rangle }{\kappa}\cdot\left[\ln\psi+\frac{1}{2}\ln\ln\psi-1\right]\nonumber \\
 & +\ln\left[\frac{1}{\sqrt{2\pi\kappa\left(y-\left\langle y\right\rangle \right)}}\right],
\label{eq:P0-large-y-tail}
\end{align}
where $\psi$ is a rescaled distance to the mean value:
\begin{equation}
\psi=\frac{y-\left\langle y\right\rangle }{\lambda m_{1}\left(1\right)\sqrt{\frac{\pi}{2}}},
\label{eq:P0-large-y-tail_psi-definition}
\end{equation}
with $m_{1}\left(1\right)$ being the exact value of the $m_{1}$
function at $w=1$ given by
\begin{equation}
m_{1}\left(1\right)=\left\langle y\right\rangle +\kappa+\lambda\left\langle \left(y+\kappa\right)\ln\frac{1}{y+\kappa}-y\ln\frac{1}{y}\right\rangle .
\end{equation}
The similarity sign ``$\sim$'' in Eq.~\eqref{P0-large-y-tail}
expresses the fact that the logarithm of the distribution function
$\ln P_{0}\left(y\right)$ can be evaluated explicitly only up to
subleading corrections of the order $\left(y-\left\langle y\right\rangle \right)/\ln\left(y-\left\langle y\right\rangle \right)$.
The latter are themselves growing functions of $y$, which prevents
us from evaluating a proper asymptotic form of the $P_{0}$ function.
A correct expression can only be formulated in terms of the saddle-point
approximation that uses the exact form of the $m$~function to estimate
the value of the integral~\eqref{expr-for-P0-via-m}. The applicability
of the asymptotic form~\eqref{P0-large-y-tail} is controlled by
the following condition:
\begin{equation}
y-\left\langle y\right\rangle \gg\kappa.
\label{eq:large-y-region_criteria}
\end{equation}

We note that while the asymptotic expressions~\eqref{P0-small-y-tail}~and~\eqref{P0-large-y-tail}
can be used for any value of $\kappa$, the corresponding behavior
is essentially unobservable for $\kappa\ll1$. Indeed, in the latter
case, the criteria of applicability for the limiting expressions presented
above correspond to Eq.~\eqref{large-y-region_criteria} for large
$y$ and to 
\begin{equation}
1-\frac{y}{\left\langle y\right\rangle }\apprge\lambda\cdot\left[\ln2+\gamma-\frac{\left\langle y\ln\frac{1}{y}\right\rangle }{\left\langle y\right\rangle }\right]\sim\lambda.
\end{equation}
for small $y$. On the other hand, the Gaussian probability distribution
\eqref{P0-Gaussian-regime} assumes exponentially small values for
\begin{equation}
\left|y-\left\langle y\right\rangle \right|\gg\sigma\sim\lambda\kappa.
\end{equation}
This implies that for the Gaussian regime $\kappa\apprle\lambda$
the asymptotic expressions~\eqref{P0-small-y-tail}~and~\eqref{P0-large-y-tail}
only become applicable in the region where the the absolute value
of the probability is already exponentially small.

\subsubsection{Strong disorder $\kappa\apprge\lambda$, oscillatory behavior at
large $y$}

The asymptotic expression~\eqref{P0-large-y-tail} does not account
for the subleading saddle points in the integral~\eqref{expr-for-P0-via-m}
over $s$ that are present for the case $y>\left\langle y\right\rangle $
(as discussed in detail in \subsecappref{Probability-large-y-asymptotic}).
The total probability is given by a sum over contributions from all
saddle points:
\begin{equation}
P_{0}\left(y\right)=P_{0}^{\text{lead}}\left(y\right)+\sum_{n=-\infty}^{\infty}P_{0}^{\left(n\right)}\left(y\right),
\label{eq:P0-large-y-tail_sum-over-saddle-points}
\end{equation}
where $P_{0}^{\text{lead}}\left(y\right)$ is the leading contribution
described by~\eqref{P0-large-y-tail}, and $P_{0}^{\left(n\right)}\left(y\right)$
is the subleading term produced by a pair of complex secondary saddle
points $z_{-n}=\overline{z_{n}}$ enumerated by $n\in\mathbb{Z}$.
Similarly to the quality of estimation~\eqref{P0-large-y-tail},
a proper asymptotic expression for each subleading contribution requires
the exact form of the $m$~function. One can provide only the leading
log-accurate expression for each of the subleading contributions:
\begin{equation}
\ln\frac{P_{0}^{\left(n\right)}\left(y\right)}{P_{0}^{\text{lead}}\left(y\right)}\sim-\frac{y-\left\langle y\right\rangle }{\kappa}\cdot2\pi in\cdot\left(1+\frac{1}{2\ln\psi}\right),
\label{eq:P0-large-y_secondary-contribution-estimation}
\end{equation}
with $\psi$ defined in Eq.~\eqref{P0-large-y-tail_psi-definition}.
While we are not able to provide an asymptotic expression for the
result of the summation due to the poor accuracy of the estimation
of the summation terms, even at the level of Eq.~\eqref{P0-large-y_secondary-contribution-estimation}
one can observe that the resulting probability distribution exhibits
oscillations. Indeed, the estimation~\eqref{P0-large-y_secondary-contribution-estimation}
indicates that each secondary contribution is close to a periodic
function with period $\Delta y=\kappa$. The sum~\eqref{P0-large-y-tail_sum-over-saddle-points}
thus features constructive interference from all contributions at
values of $y$ described by
\begin{equation}
y^{\left(n+1\right)}-y^{\left(n\right)}\approx\kappa,\,\,\,\,y^{\left(0\right)}=\left\langle y\right\rangle 
\end{equation}
where $n\in\mathbb{N}$ enumerates the secondary peak that emerges
from the such an interference.

\subsection{Numerical analysis of the problem\label{subsec:Numerical-analysis}}

In this section, we briefly describe the numerical routines used to
analyze both the original self-consistency equation~\eqref{saddle-point_order-parameter}
and the integral equations~(\hphantom{}\ref{eq:eq-on-m1}\nobreakdash-\ref{eq:eq-on-m2}\hphantom{})
that constitute the core outcome of the theoretical analysis. 

One immediate way to gather the statistics of the solution of the
self-consistency equation~\eqref{saddle-point_order-parameter} is
to solve it directly for the values of $\Delta_{i}$ in a number of
sufficiently large realizations of the system. To this end, we generate
an instance of Random Regular Graph along with a random set of values
$\xi_{i}$ for each site and then solve the system~\eqref{saddle-point_order-parameter}
by a suitable iterative procedure. The size of the base graph reaches
$N=2^{23}\approx8.4\cdot10^{6}$, which allowed us to ensure that
thermodynamic limit in all quantities of interest was achieved. The
distribution of onsite disorder field $\nu\left(\xi\right)$ only
determines the overall superconducting scale and otherwise has little
to no effect on any of properly rescaled distributions of the order
parameter, in full agreement with the general physics as well as our
theory. For this reason, all numerical data quoted below uses the
box distribution of the form $\nu\left(\xi\right)=\theta\left(\left|\xi\right|-1\right)/2$
with $\nu_{0}=1/2$, although other distributions have also been considered
and observed to behave in accord with our theoretical expectations.
The Fermi energy $E_{F}$, being the characteristic scale of the distribution,
is always used as the energy unit, so all dimensionfull quantities
such as $D\left(\xi\right)$ are measured in units of $E_{F}$. The
numerical routine uses the version of the model with a trivial energy
dependence of the interaction matrix element $D\left(\xi\right)=D\left(0\right)=\text{const}$,
and other models are immediately available. However, both the general
physics reasoning and our theoretical analysis (see \appref{Effect-of-energy-dependence-of-matrix-element}
for details) indicate that there is no practical difference between
various profiles of $D\left(\xi\right)$ as long as the they are smooth
on superconducting energy scales, i. e. $D\left(\xi\sim\Delta\right)\approx D\left(0\right)$.

The key focus of this work, however, is to use the derived equations
to describe the statistics of the order parameter analytically. The
remaining technical challenge at this point is to solve the pair of
integro-differential equations~(\hphantom{}\ref{eq:eq-on-m1}\nobreakdash-\ref{eq:eq-on-m2}\hphantom{})
for the $m$~function. While \subsecref{Weak-coupling-approximation}
provides an approximate analytical solution in terms of special functions,
it is still important to verify the numerical accuracy of this approximation.
We designed a certain numerical procedure that iteratively constructs
the solution to the integro-differential equations~(\hphantom{}\ref{eq:eq-on-m1}\nobreakdash-\ref{eq:eq-on-m2}\hphantom{}).
The implementation can be found at \citep{m-function_numerics_implementation};
it allows one to obtain the solution in several minutes on a usual
laptop. Once the solution is determined either numerically or analytically
by means of equations (\hphantom{}\ref{eq:m2-small-lambda-limit}\nobreakdash-\ref{eq:F-function-def}\hphantom{}),
our routine then provides an efficient way to perform the numerical
integration of Eq.~\eqref{expr-for-P0-via-m} to calculate the probability
distribution $P\left(y\right)$ and other objects of interest, such
as the joint probability distribution $P\left(x,y\right)$ given by~\eqref{joint-distribution_expression-via-m}.
Various averages over the resulting distribution are then available
via either yet another numerical integration or by exploiting the
fact that the function~$m\left(S|0\right)$ represents the cumulant
generating function of the $P\left(y\right)$ distribution, with the
both methods being optimized within the routine.

We emphasize that the primary outcomes of our analysis are analytical,
while the developed numerical routines are mainly used to confirm
the analytical results.

\subsection{Overview of the main results\label{subsec:Overview-of-the-results}}

\subsubsection{The shape of the distribution at various values of disorder}

\begin{figure}
\begin{centering}
\includegraphics[viewport=0bp 82bp 1200bp 818bp,clip,scale=0.105]{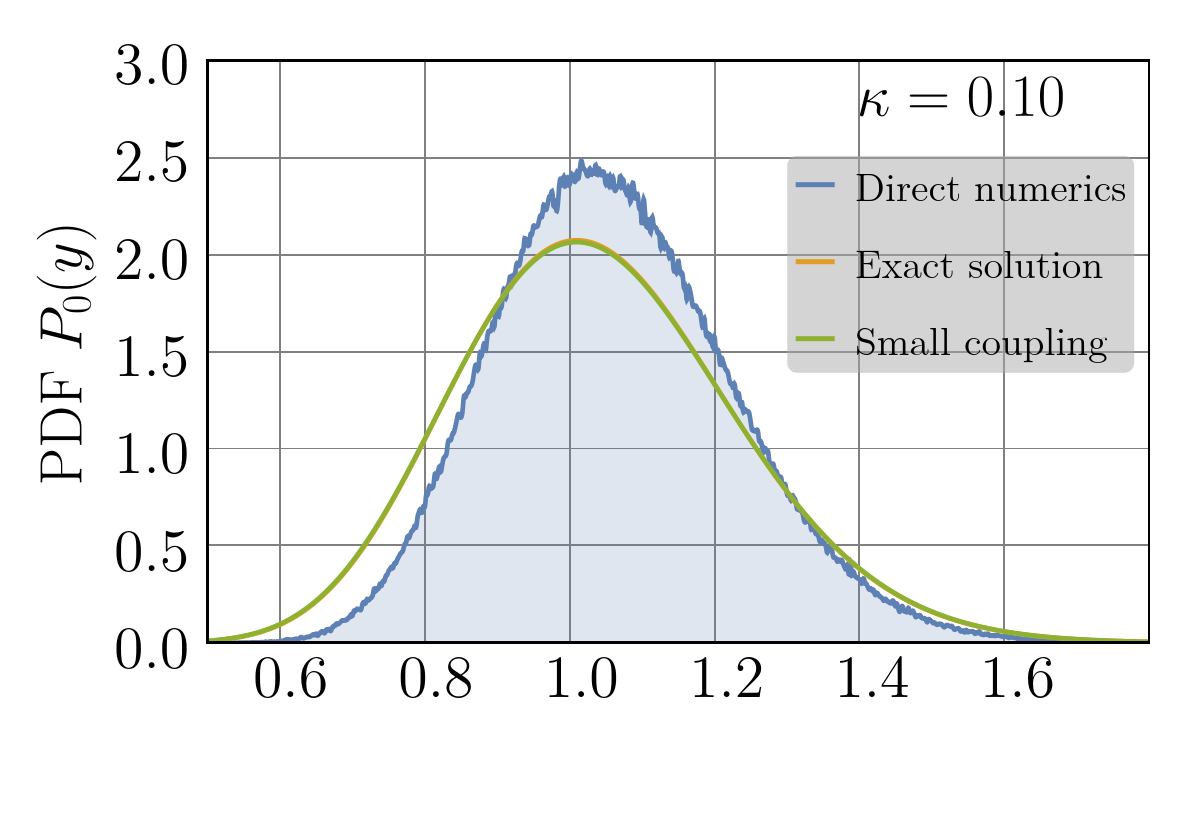}\hspace*{\fill}\includegraphics[viewport=80bp 82bp 1200bp 818bp,clip,scale=0.105]{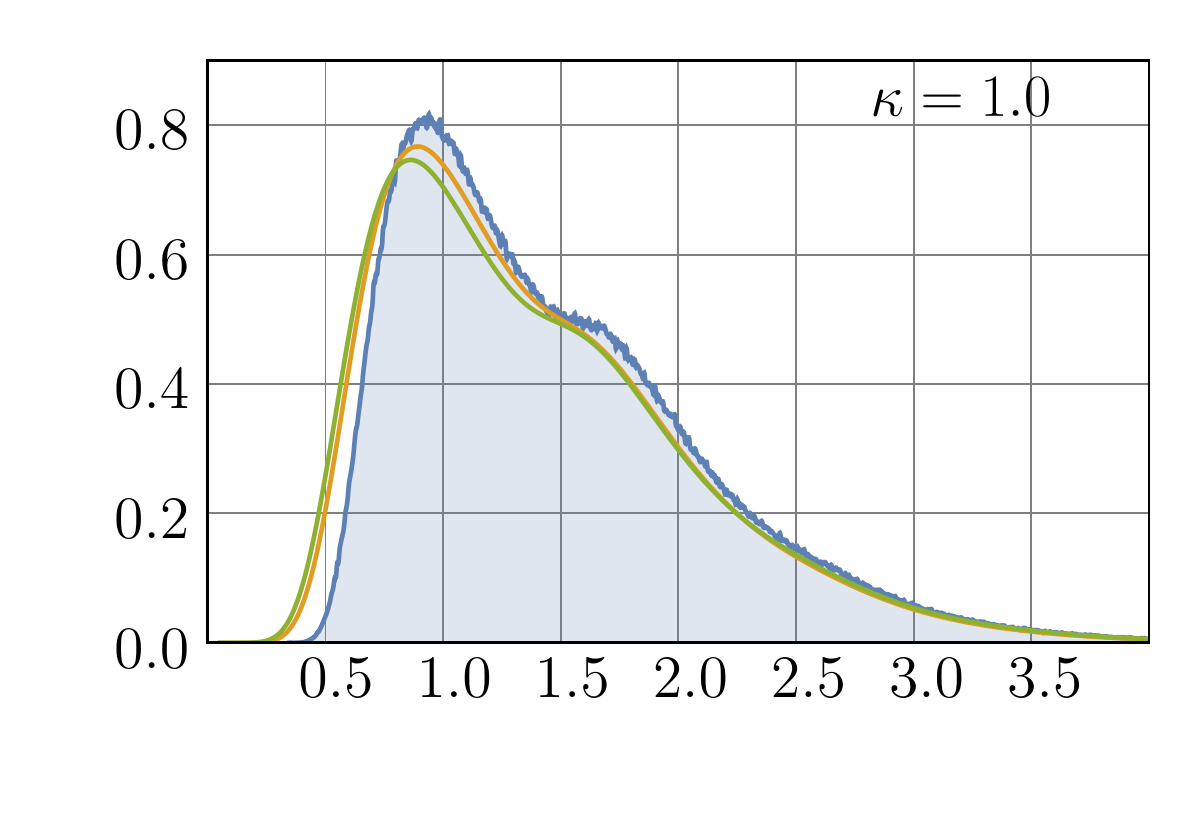}
\par\end{centering}
\begin{centering}
\includegraphics[viewport=12bp 82bp 1200bp 818bp,clip,scale=0.105]{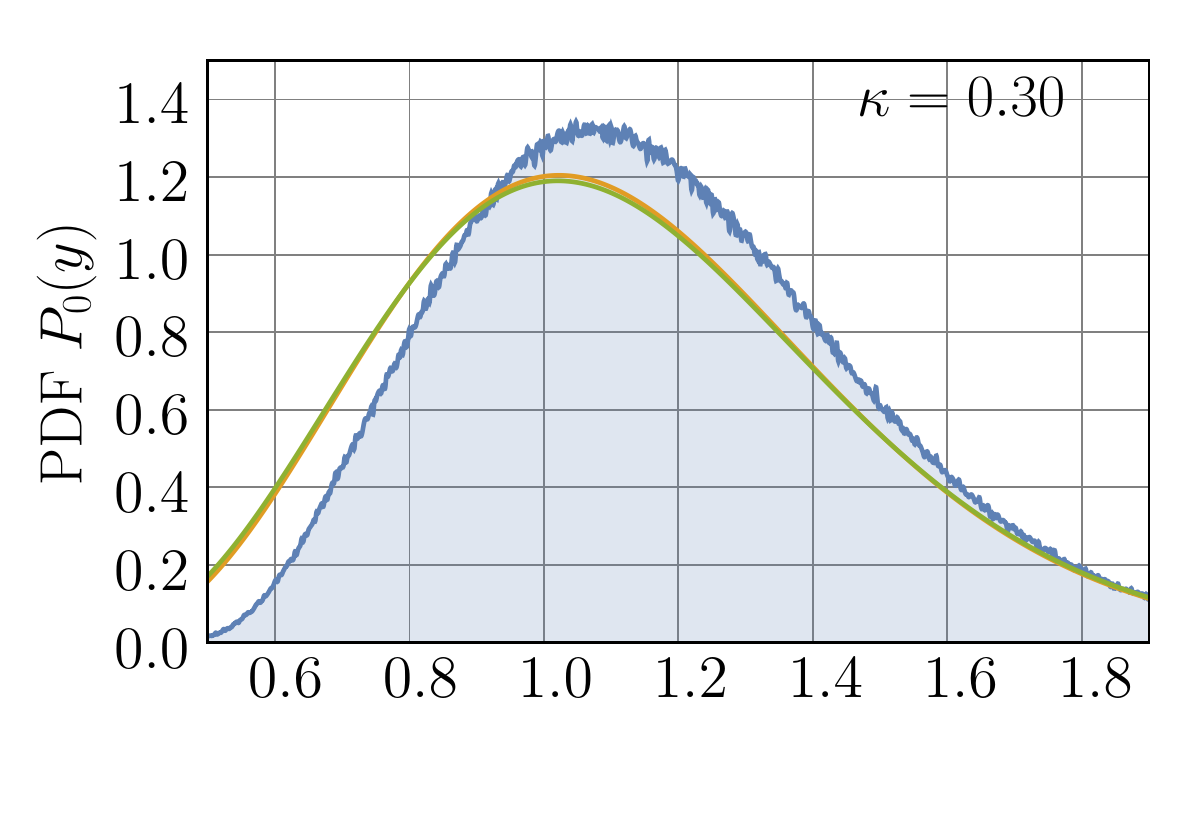}\hspace*{\fill}\includegraphics[viewport=82bp 82bp 1200bp 818bp,clip,scale=0.105]{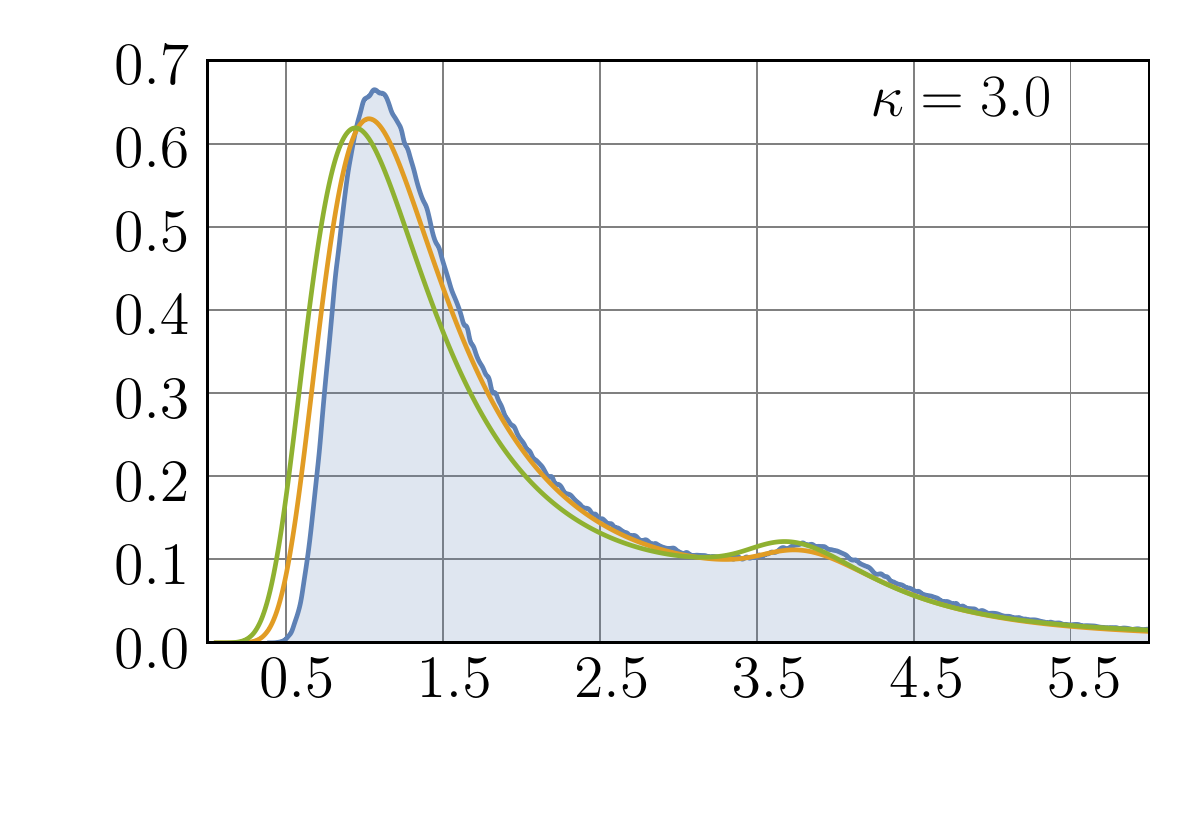}
\par\end{centering}
\begin{centering}
\includegraphics[viewport=12bp 0bp 1200bp 818bp,scale=0.105]{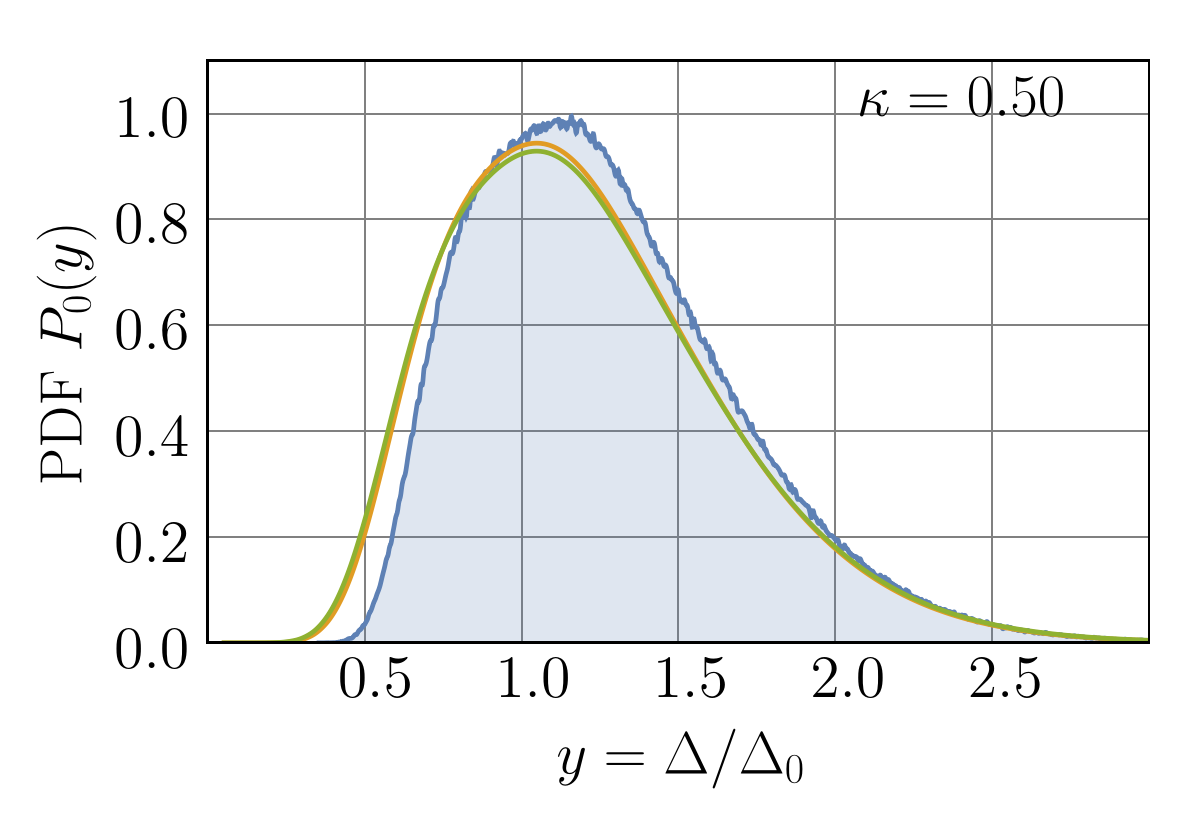}\hspace*{\fill}\includegraphics[viewport=82bp 0bp 1200bp 818bp,clip,scale=0.105]{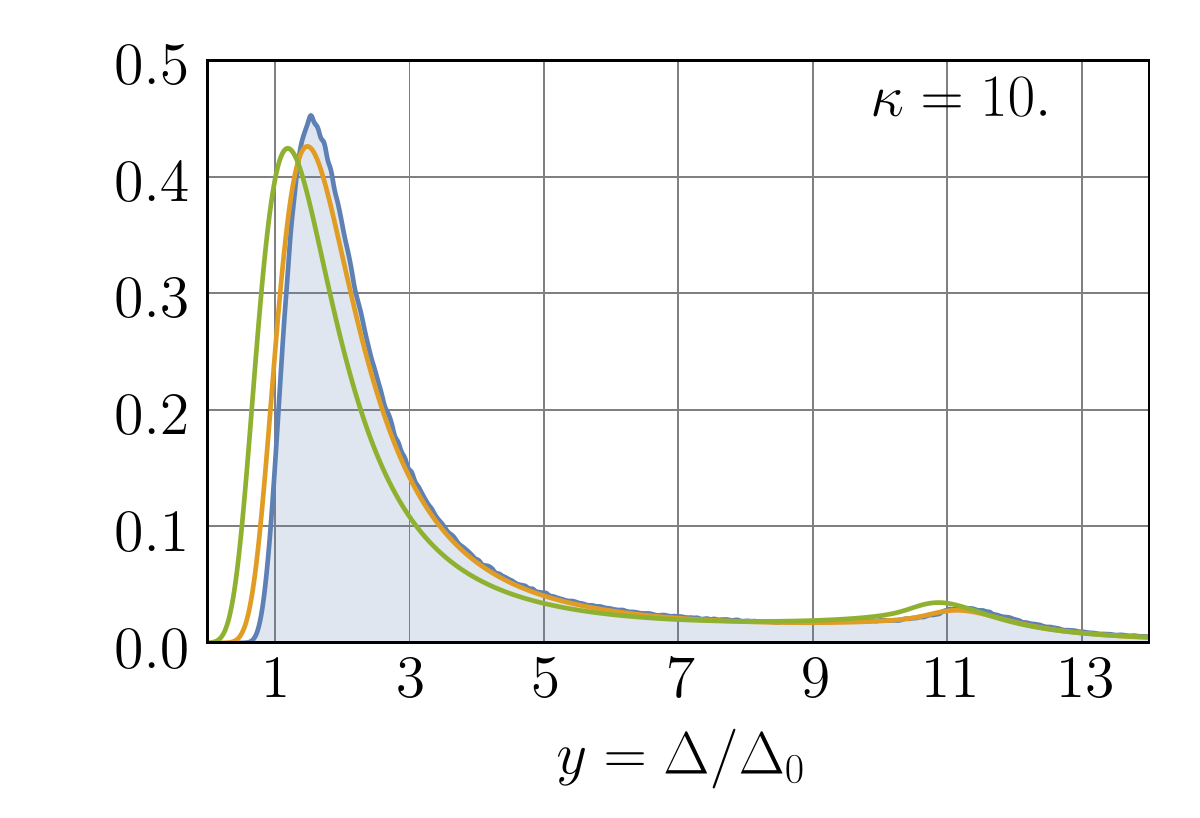}
\par\end{centering}
\caption{A series of plots for the probability density function (PDF) of the
dimensionless order parameter $P\left(\Delta/\Delta_{0}\right)$ for
various values of the parameter $\kappa$. The filled blue line is
the histogram obtained from direct numerical solution of the self-consistency
equations~\eqref{saddle-point_order-parameter} on a Random Regular
Graph of size $N=2^{22}\approx4.2\cdot10^{6}$. The orange line is
obtained by solving the equations~(\protect\hphantom{}\ref{eq:eq-on-m1}\protect\nobreakdash-\ref{eq:eq-on-m2}\protect\hphantom{})
for the function $m\left(S|w\right)$ and subsequently evaluating
the integral~\eqref{expr-for-P0-via-m} for the distribution function.
The green line uses the analytical expressions~(\protect\hphantom{}\ref{eq:m2-small-lambda-limit}\protect\nobreakdash-\ref{eq:F-function-def}\protect\hphantom{})
of~\subsecref{Weak-coupling-approximation} to approximate the value
of the $m$~function used to compute the integral~\eqref{expr-for-P0-via-m}
for the PDF. For simplicity, the model with $D\left(\xi\right)=\text{const}$
is used. Values of $\kappa=\left\{ 0.3,0.5,1.0,3.0,10\right\} $ are
realized in the system with $Z=51$ and $\lambda\approx\left\{ 0.199,0.177,0.154,0.129,0.110\right\} $
respectively, and $\kappa=0.1$ corresponds to $Z=101$ and $\lambda\approx0.222$.
The last pair of values for $Z,\lambda$ is motivated by the fact
that larger values of $\lambda$ render large values of $\Delta$,
while our theory corresponds to the limit $\nu_{0}\Delta\ll1$, with
the leading correction being of order $2\nu_{0}\Delta_{0}/\lambda\approx\left(\kappa Z\right)^{-1}$.
That is why in order to obtain small $\kappa$ one has to use larger
$Z$ so as to keep the value of $\lambda$ small enough. The aforementioned
corrections to small $\Delta_{0}$ limit are also responsible for
the mismatch between the theory and numerical data that is pronounced
for $\kappa=0.1,0.3$ and is also somewhat observable for larger values
of $\kappa$ with an apparent decreasing trend (the theoretical curves
have no fitting parameters). The mismatch between the two instances
of the theoretical descriptions originates from corrections of order
$\sim\lambda^{2}$ neglected in the approximate analytical solution
(green line), see \subsecref{Weak-coupling-approximation} for details.
One can observe the defining role of $\kappa$ for the profile of
he distribution: small $\kappa$ produce Gaussian regime, while large
$\kappa$ render nontrivial distribution function, whose asymptotic
behavior is discussed in \subsecref{Extreme-value-statistics}. \label{fig:numerics-vs-theory_delta-distrib}}
\end{figure}

\figref{numerics-vs-theory_delta-distrib} showcases the results of
both procedures for various values of microscopic parameters of the
model corresponding to qualitatively different profiles of the distribution
function $P_{0}\left(y\right)$. As it is evident from both the numerical
studies and the analytical solution presented below, the parameter
$\kappa$ plays the defining role in the qualitative form of the solution.
Indeed, small values of $\kappa\ll1$ correspond to the regime of
small disorder with a Gaussian distribution of the order parameter,
while the opposite case of $\kappa\apprge1$ implies a rather involved
non-Gaussian profile of the distribution. The exact form and asymptotic
behavior of this strong-disorder profile is described in \subsecref{Extreme-value-statistics}.
In particular, a proper discussion of the apparent secondary maximum
in the distribution $P_{0}\left(y\right)$ observed for $\kappa\apprge1$
is provided.

The physical reason behind the existence of diverse profiles of the
distribution function $P_{0}\left(y\right)$ is related to the smallness
of the Cooper coupling constant $\lambda$. As was explained in \subsecref{The-saddle-point-equation},
the bare \textquotedbl number of neighbors\textquotedbl{} $Z$ in
our model must be above $Z_{1}=\lambda\cdot e^{1/2\lambda}$ in order
to substantiate our disregard for the Onsager reaction terms in the
original self-consistency equation~\eqref{saddle-point_order-parameter}.
On the other hand, it is only at $Z\apprge Z_{2}\sim e^{1/\lambda}$
when one observes suppression of local fluctuations of the order parameter
due to statistical self-averaging, see Eq.~\eqref{Z2} and the associated
discussion. The smallness of $\lambda$ then renders an exponentially
large region $Z_{1}\ll Z\ll Z_{2}$ where the distribution of the
order parameter assumes a complicated profile presented. Taking for
the sake of example $\lambda=0.2$ we find that $Z_{1}\approx2.5$
and $Z_{2}\approx30$; in terms of the $\kappa$ parameter defined
in Eq.~\eqref{Z-eff_and_kappa_definition}, the accessible values
range from arbitrarily small $\kappa$ up to $\kappa\apprle10$.

\subsubsection{Asymptotic behavior of the distribution}

\begin{figure}
\begin{centering}
\includegraphics[viewport=37.4844bp 0bp 1200.252bp 820bp,clip,scale=0.21]{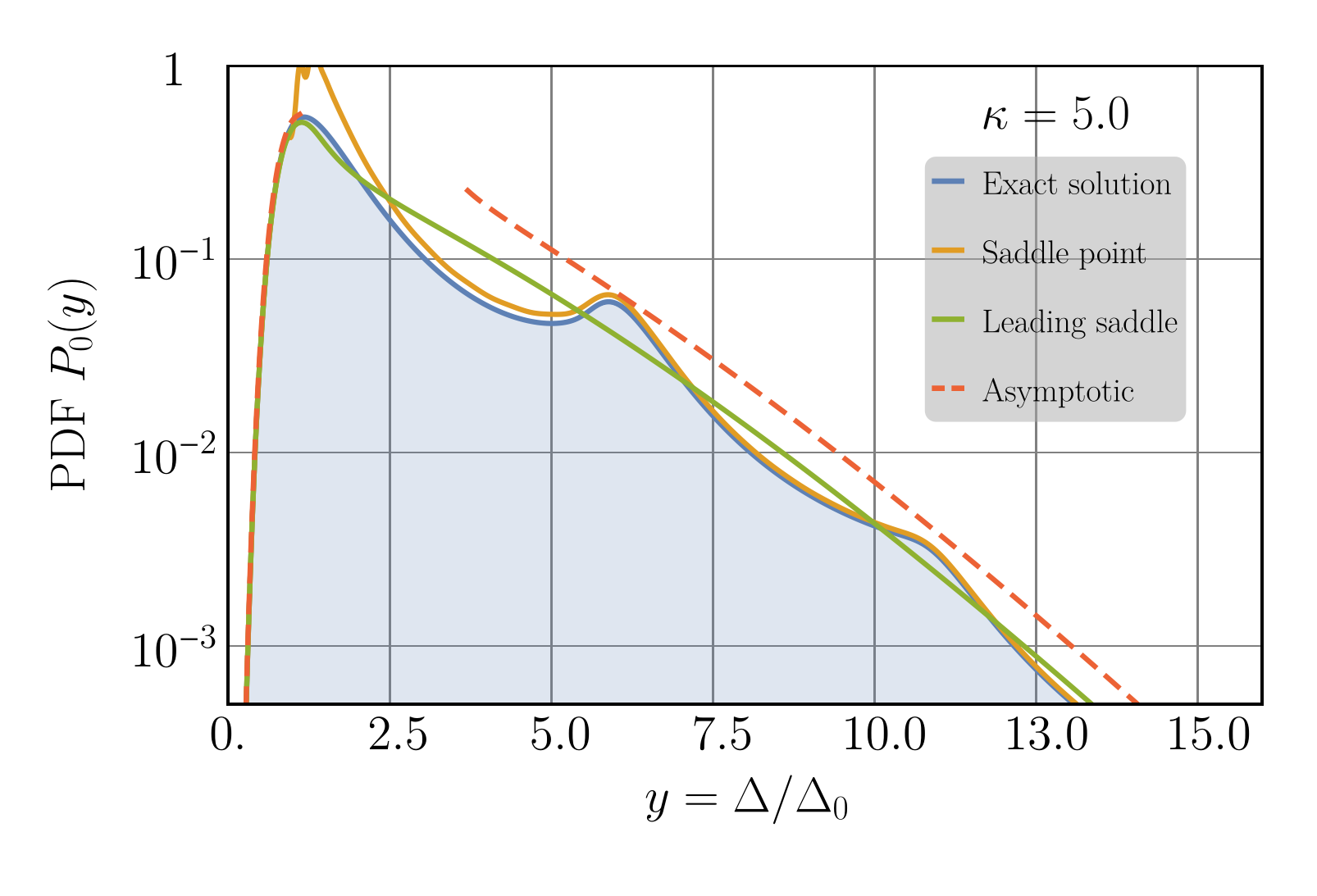}
\par\end{centering}
\caption{A log-scale plot reflecting the asymptotic behavior of the probability
density function (PDF) of the dimensionless order parameter $y=\Delta/\Delta_{0}$.
The filled blue curve represents the value of the integral~\eqref{expr-for-P0-via-m}
obtained by direct numerical integration. The orange line corresponds
to saddle-point approximation of the integral~\eqref{expr-for-P0-via-m}
with all saddle points taken into account for $y>\left\langle y\right\rangle $.
The green line reflects contribution of the leading purely imaginary
saddle point only. When required, the $m$~function is determined
from the numerical solution of equations~(\protect\hphantom{}\ref{eq:eq-on-m1}\protect\nobreakdash-\ref{eq:eq-on-m2}\protect\hphantom{}),
see \subsecref{Numerical-analysis} for details on the numerical routine.
Finally, the dashed red line corresponds to approximate analytic expressions
presented in the Main Text: Eq.~\eqref{P0-large-y-tail} for large
values of $y>\left\langle y\right\rangle $ and Eq.\protect\nobreakdash-s~(\protect\hphantom{}\ref{eq:P0-small-y-tail}\protect\nobreakdash-\ref{eq:zeta_def}\protect\hphantom{})
for $y<\left\langle y\right\rangle $. The microscopic parameters
of the model are $D\left(\xi\right)=\text{const}$, $\lambda\approx0.120$,
$Z=51$ and $\kappa=5.0$. All saddle-point-type approximations naturally
fail in the region $y\sim\left\langle y\right\rangle $ due to vanishing
second derivative at the saddle point. On the other hand, all of them
show reasonable agreement with the exact value for both large and
small values of $y$. The discussion of the secondary peaks at large
values of $y$ in given in the Main Text. \label{fig:P0-log-plot_with-asymptotics}}
\end{figure}

\figref{P0-log-plot_with-asymptotics} provides a demonstration of
the approximate behavior described by the asymptotic equations~\eqref{P0-small-y-tail}~and~\eqref{P0-large-y-tail}
superimposed on the distribution obtained by exact numerical solution
of the equations~(\hphantom{}\ref{eq:eq-on-m1}\nobreakdash-\ref{eq:eq-on-m2}\hphantom{})
with respect to $m_{1}\left(w\right),m_{1}\left(S|w\right)$~functions
(the numerical procedure is explained in \subsecref{Numerical-analysis}).
In addition to that, this Figure also features the estimations obtained
from using the \emph{exact} form of the $m$~function determine the
position of the saddle points and evaluate the resulting approximation
of the integral~\eqref{expr-for-P0-via-m} for the probability density. 

We note that the asymptotic form given by Eq.\nobreakdash-s~(\hphantom{}\ref{eq:P0-small-y-tail}\nobreakdash-\ref{eq:zeta_def}\hphantom{})
for $y<1$ demonstrates excellent agreement with the exact result.
However, the situation is more involved in the opposite limit of large
$y$. The provided approximation~\eqref{P0-large-y-tail} for $y\apprge1$
does describe the asymptotic behavior of the distribution function
$P_{0}\left(y\right)$ up to a constant of order unity, in accordance
with the quoted accuracy of the corresponding calculation, see the
discussion under \eqref{P0-large-y-tail}. On top of that, the oscillations
with period $\Delta y=\kappa$ proposed by estimations~(\hphantom{}\ref{eq:P0-large-y-tail_sum-over-saddle-points}\nobreakdash-\ref{eq:P0-large-y_secondary-contribution-estimation}\hphantom{})
are also observed.

The observed double-exponential behavior of the probability at $y\le\left\langle y\right\rangle $
is secured by a certain type of local disorder configurations. Indeed,
one can observe directly from the self-consistency equation~\eqref{saddle-point_order-parameter}
that the only feasible way to produce anomalously low value of the
order parameter on a given site is to have the values of the disorder
fields $\xi_{j}$ \emph{on all nearest neighbors }larger (in absolute
value) than a certain threshold $\xi_{\min}\gg\Delta$. The value
of the threshold can be estimated from the mean-field-like treatment
of the self-consistency equation and renders $\xi_{\min}\sim\frac{\left\langle \Delta\right\rangle }{2}\exp\left\{ \frac{1}{\lambda}\left(1-\frac{\Delta}{\left\langle \Delta\right\rangle }\right)\right\} $,
and the probability of the such an event to occur in the statistics
of $\xi$ is estimated as $P\left(\min\left|\xi\right|>\xi_{\min}\right)\approx\exp\left\{ -2\nu_{0}Z\xi_{\min}\right\} $
for $Z\gg1$ and $\xi_{\min}\ll E_{F}$. Combining these two estimations
correctly reproduces the exponential part of Eq.\nobreakdash-s~(\hphantom{}\ref{eq:P0-small-y-tail}\nobreakdash-\ref{eq:zeta_def}\hphantom{}).
A more detailed version of this reasoning is given in \subsecappref{Probability-small-y-asymptotic}.

The secondary maxima in the probability distribution $P_{0}\left(y\right)$
also admit a decent physical interpretation in each particular realization
of the disorder fields~$\xi$. Namely, the $n$\nobreakdash-th secondary
maximum of the distribution corresponds to the sites with exactly
$n$ neighbors with small value of onsite disorder $\left|\xi_{i}\right|\sim\Delta_{0}$.
The apparent sharpness of the peaks can be perceived as a consequence
of Van Hove-type singularity in the probability distribution of the
terms in the right hand side of the self-consistency equation~\eqref{saddle-point_order-parameter}.
The latter exhibit a quadratic maximum at $\xi=0$, and thus posses
the probability density that features a square root singularity as
$\xi\rightarrow0$, viz.
\begin{equation}
P\left(\epsilon=\frac{\Delta_{0}}{\sqrt{\Delta_{0}^{2}+\xi^{2}}}\right)\approx\frac{\Delta_{0}}{\sqrt{2\left(1-\epsilon\right)}},\,\,\,\,\,\epsilon\rightarrow1.
\end{equation}
\subsecappref{Probability_large-y-asymptotic_secondary-peaks} describes
several quantitative tests to verify this hypothesis at the level
of an individual disorder realization. The results are of unequivocal
support to the proposed interpretation. 

This explanation also suggests that the observed features of the distribution
originate from an unphysical assumption that the matrix element of
interaction is constant, so that the described singularity of Van
Hove type is well pronounced. On the other hand, in real system one
naturally expects fluctuations in the coupling matrix element. In
the following \subsecref{Extreme-value-statistics_fluctuating-coupling},
we analyze an extension of our model that includes these fluctuations.
Our conclusions clearly reflect that the described secondary maxima
in the distribution of the order parameter are smeared by fluctuations
of the coupling constant.

\subsubsection{Joint probability distribution}

\begin{figure}
\begin{centering}
\includegraphics[viewport=25bp 5bp 950bp 430bp,clip,scale=0.27]{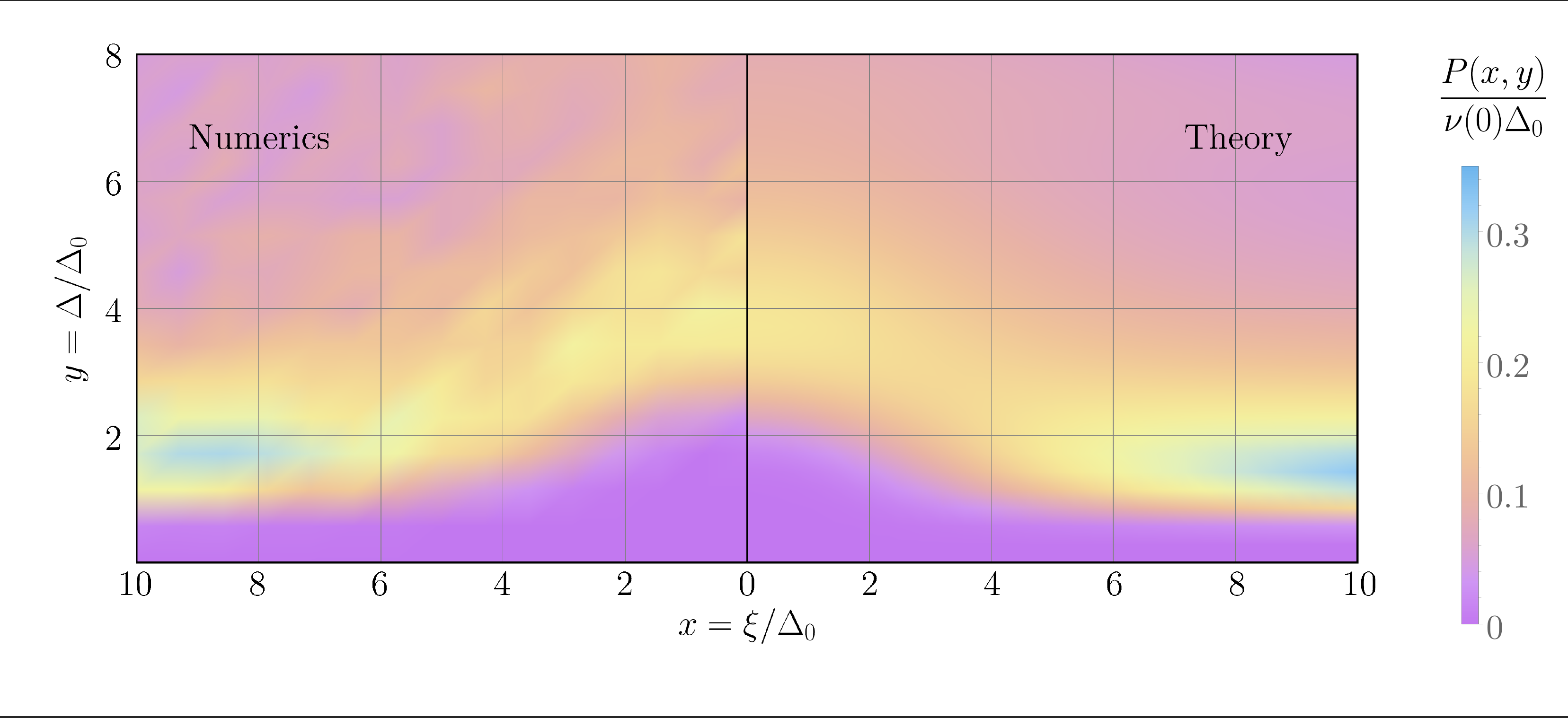}
\par\end{centering}
\caption{A color map of the joint probability density function $P\left(x,y\right)$
of the onsite values of dimensionless disorder field $x=\xi/\Delta_{0}$
and dimensionless order parameter $y=\Delta/\Delta_{0}$ in the vicinity
of the Fermi surface corresponding to $\xi=0$. The color encodes
the value of the probability density according to the legend to the
right. The left color map corresponds to (smoothened) histogram obtained
from direct numerical solution of the original self-consistency equations~\eqref{saddle-point_order-parameter},
and the right color map reflects the result of the theoretical calculation
performed according to expression~\eqref{joint-distribution_expression-via-m}
with the $m$~function determined from the numerical solution of
equations~(\protect\hphantom{}\ref{eq:eq-on-m1}\protect\nobreakdash-\ref{eq:eq-on-m2}\protect\hphantom{}).
For simplicity, the model with $D\left(\xi\right)=\text{const}$ is
used. The parameters of the model in both cases are $\lambda\approx0.120$
and $Z=51$, which corresponds to $\kappa=5.0$. The observed jitter
in the results of the direct numerical solution (left plot) is due
to the finite size of the corresponding sample: even despite the fact
that a system with $N=2^{22}\approx4.2\cdot10^{6}$ sites is used,
only $\sim N\cdot\left(2\nu_{0}\Delta_{0}\right)\sim6\cdot10^{4}$
of them contribute to the presented histogram, resulting in an average
of just $\sim250$ points contributing to each bin of the histogram
for the chosen bin size $\Delta x=\Delta y\approx0.71$. The two color
maps demonstrate reasonable agreement, simultaneously reproducing
several important qualitative features of the joint PDF. In particular,
one observes a considerable deformation of the conditional distribution
$P_{\text{c}}\left(\Delta\right):=P\left(\xi,\Delta\right)/P\left(\xi\right)$
as $\left|\xi\right|$ decreases. See the Main Text for a detailed
discussion. \label{fig:numerics-vs-theory_joint-distrib}}
\end{figure}

We also present the results for the joint probability distribution
$P\left(x,y\right)$ of the dimensionless order parameter $y=\Delta/\Delta_{0}$
and the corresponding onsite local field $x=\xi/\Delta_{0}$. \figref{numerics-vs-theory_joint-distrib}
shows the color maps of the distribution as found from the theoretical
approach presented above along with the data obtained from exact numerical
solution of the original self-consistency equations~\eqref{saddle-point_order-parameter},
as explained earlier. The two pictures indicate a clear agreement
up to statistical noise present in the numerical data due to finite
sample size.

\begin{figure}
\begin{centering}
\includegraphics[viewport=75bp 37.5114bp 1200bp 820bp,clip,scale=0.21]{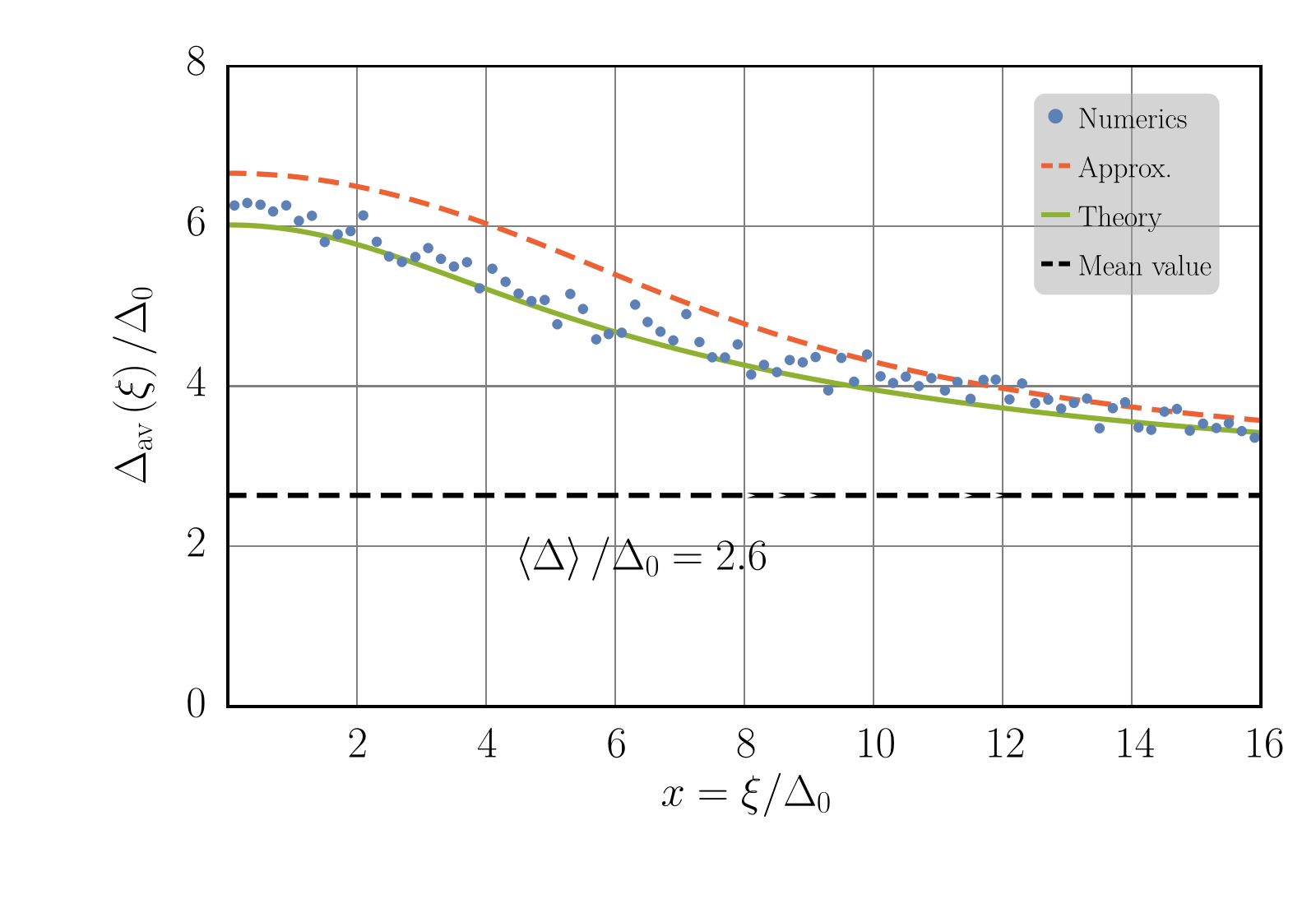}
\par\end{centering}
\caption{The plot of the conditional average of the dimensionless order parameter
$\Delta_{\text{av}}\left(\xi\right)/\Delta_{0}=\intop d\Delta\cdot\Delta/\Delta_{0}\cdot P\left(\Delta,\xi\right)/P\left(\xi\right)$
as a function of onsite value of the dimensionless disorder field
$x=\xi/\Delta_{0}$. The blue points corresponds to the result calculated
from the direct numerical solution of the self-consistency equations~\eqref{saddle-point_order-parameter}.
The solid green line corresponds to the conditional average computed
by direct integration from the theoretical joint probability distribution
given by Eq.~\eqref{joint-distribution_expression-via-m}, with the
$m$~function determined from the numerical solution of equations~(\protect\hphantom{}\ref{eq:eq-on-m1}\protect\nobreakdash-\ref{eq:eq-on-m2}\protect\hphantom{}).
The red dashed line corresponds to physically relevant solution of
the approximate equations~(\protect\hphantom{}\ref{eq:approximate-conditional-average_1}\protect\nobreakdash-\ref{eq:approximate-conditional-average_2}\protect\hphantom{}).
Finally, the black dashed line denotes the value of the total average
$\left\langle \Delta\right\rangle /\Delta_{0}$ of the dimensionless
order parameter as found from both the numerical data and analytic
theory. The microscopic parameters of the model are $D\left(\xi\right)=\text{const}$,
$\lambda\approx0.120$, $Z=51$ and $\kappa=5.0$, so that a direct
comparison with \figref{numerics-vs-theory_joint-distrib} is appropriate.
\label{fig:conditional-mean-value_behavior}}
\end{figure}

While the distribution quickly approaches the profile corresponding
to the factorized distribution of the form $P_{0}\left(y\right)\cdot P\left(x\right)$
at sufficiently large values of $\xi$, there is a noticeable deformation
in the region $\xi/\Delta_{0}\apprle5$ indicative of the strong correlation
between the onsite values of $\xi$ and $\Delta$. As can be seen
from the original self-consistency equation~\eqref{saddle-point_order-parameter},
such behavior is a secondary consequence of the fact that a low value
of $\xi$ at a given site $i$ results in an increase of the order
parameter \emph{at all neighboring sites} $j\in\partial i$ by a contribution
of the order $D\left(\xi_{j}\right)/Z\sim\Delta_{0}\cdot\kappa$.
This, in turn, leads to the enhancement of the value of the order
parameter on the chosen site $i$. These \emph{qualitative} considerations
allow one to estimate the position of the conditional distribution
average $\Delta_{\text{av}}\left(\xi\right)=\intop d\Delta\cdot\Delta\cdot P\left(\Delta,\xi\right)/P\left(\xi\right)$
as an appropriate solution to the following system of equations:
\begin{align}
 & \Delta_{\text{av}}\approx\Delta_{\text{neighb}}\left(1+\lambda\cdot\ln\frac{\left\langle \Delta\right\rangle }{\Delta_{\text{neighb}}}\right),
\label{eq:approximate-conditional-average_1}\\
 & \Delta_{\text{neighb}}\approx\left\langle \Delta\right\rangle +\kappa\Delta_{0}\cdot\frac{\Delta_{\text{av}}}{\sqrt{\Delta_{\text{av}}^{2}+\xi^{2}}}.
\label{eq:approximate-conditional-average_2}
\end{align}
At large values of $\xi$ the solution $\Delta_{\text{av}}$ approaches
the total expectation $\left\langle \Delta\right\rangle $, while
at $\xi\rightarrow0$ the result behaves as $\Delta_{\text{av}}\approx\left\langle \Delta\right\rangle +\kappa\Delta_{0}$,
in full agreement to what is observed on \figref{numerics-vs-theory_joint-distrib}.
A plot of the full dependence $\Delta_{\text{av}}\left(\xi\right)$
is presented on~\figref{conditional-mean-value_behavior} and shows
a reasonable agreement with both data obtained from the direct numerical
solutions of the self-consistency equations and the curve calculated
by appropriate numerical integration of the theoretical expression~\eqref{joint-distribution_expression-via-m}.

We would like to emphasize, however, that this behavior is subject
to revision upon introduction of the Onsager reaction term discussed
in \subsecref{The-saddle-point-equation}. While we expect that for
$Z\ge Z_{1}$ this term is of little importance for the distribution
function of the order parameter, the profile of the \emph{onsite joint}
distribution function $P\left(\Delta,\xi\right)$ at $\left|\xi\right|\sim\Delta$
can potentially experience noticeable deformations from the described
behavior. Indeed, the physical interpretation of the reaction term
is to mediate the self-action of the order parameter, that is, the
indirect response of a given quantity to its own change through the
corresponding responses of the neighboring fields. The latter mechanism
is precisely what leads to the described profile of the joint probability
function at small values of $\left|\xi\right|$. That is why even
for sufficiently large values of $Z$ the Onsager reaction term might
have a significant effect on the shape of the onsite joint distribution
function $P\left(\Delta,\xi\right)$ for $\left|\xi\right|\sim\Delta$. 

It is also worth mentioning that the joint probability distribution
$P\left(\Delta,\xi\right)$ is of more physical significance than
the distribution $P\left(\Delta\right)$ of the order parameter alone.
Indeed, computation of various physical observables for the given
configuration of the order parameter involves values both $\xi$ and
$\Delta$ for states close to Fermi level, i. e. with~$\left|\xi\right|\sim\Delta$.
As~Figures \ref{fig:numerics-vs-theory_joint-distrib}~and~\ref{fig:conditional-mean-value_behavior}
suggest, treating fields $\xi$ and $\Delta$ as independent would
thus result in qualitatively incorrect results. One particular example
of this is the spectrum of collective low-energy excitations discussed
in~Ref.~\citep{Feigelman_Microwave_2018}: the inverse Green's Function
of those modes is sensitive to onsite values of $\xi$ and $\Delta$
in equal measures, so that computing the average Green's Function
actually demands the aforementioned joint distribution close to the
Fermi surface. Another important question yet to be analyzed is the
connection between the field of the order parameter $\Delta$ discussed
in this work and experimentally measurable quantities. While the order
parameter in weakly disorder superconductors can be probed e.g. via
the single-particle density of states~\citep{L0_1972}, no theory
exists to our knowledge of a similar connection in the case of strong
disorder with a pseudogap. We believe such a theory will inevitably
require the knowledge of joint distribution functions of both $\xi$
and $\Delta$. 

\subsection{The effect of weak fluctuations of the coupling amplitudes\label{subsec:Extreme-value-statistics_fluctuating-coupling}}

In this Subsection, we analyze a generalization of our model that
allows for the fluctuations of the interaction matrix element between
each pair of interacting single-particle states. We model these fluctuations
by assigning a random magnitude to the bare matrix element $D_{ij}$
of the interaction between each pair of interacting states on top
of its smooth dependence on the energy difference $\xi_{i}-\xi_{j}$
of the two states. This corresponds to the following generalization
of the self-consistency equation~\eqref{saddle-point_order-parameter}:
\begin{equation}
\Delta_{i}=\sum_{j\in\partial i}c_{ij}D\left(\xi_{i}-\xi_{j}\right)\frac{\Delta_{j}}{\sqrt{\Delta_{j}^{2}+\xi_{j}^{2}}},
\label{eq:fluctuating-coupling_saddle-point-equation}
\end{equation}
where $D\left(\xi\right)$ is the energy dependence of the interaction
described previously, and $c_{ij}$ are independent random variables
distributed according to some distribution $P\left(c\right)$. In
particular, letting $P\left(c\right)=\delta\left(c-1\right)$ leads
one back to the self-consistency equation~\eqref{saddle-point_order-parameter}
analyzed earlier. The new equation~\eqref{fluctuating-coupling_saddle-point-equation}
now includes two sources of disorder: the randomness of the single-particle
energies $\xi_{i}$ and the one from the distribution of the coupling
matrix elements $D_{ij}=c_{ij}\cdot D\left(\xi_{i}-\xi_{j}\right)$.

One can conduct the mean-field analysis of Eq.~\eqref{fluctuating-coupling_saddle-point-equation}
similar to that of \subsecref{mean-field-solution}. The latter is
still valid for sufficiently large number of neighbors, i. e. $\left\langle c\right\rangle Z\cdot2\nu_{0}\Delta\gg1$.
One can then assert a spatially uniform order parameter for energies
close to the to Fermi surface and obtain
\begin{equation}
\Delta_{R}=2E_{0}\cdot\exp\left\{ -\frac{1}{\lambda_{R}}\right\} ,\,\,\,\,\,\,\lambda_{R}=\left\langle c\right\rangle 2\nu_{0}D\left(0\right)Z,
\label{eq:fluctuating-coupling_mean-field}
\end{equation}
where $\lambda_{R}$ is the new dimensionless Cooper attraction constant,
and the value of $E_{0}\sim\varepsilon_{D}$ is still determined by
higher energy scales, but with the new value of the mean matrix element.

Our theoretical approach can be generalized to describe the model
above, as explained in detail in \appref{Effect-of-fluctuating-coupling}.
In particular, the $m$~function retains its role of the central
object in the theory. Here, we only present the proper counterpart
of Eq.\nobreakdash-s~(\hphantom{}\ref{eq:eq-on-m1}\nobreakdash-\ref{eq:eq-on-m2}\hphantom{})
valid for $x\apprle\varepsilon_{D}/\Delta_{0}$:
\begin{widetext}
\begin{align}
m_{1}\left(w\right) & =m_{1}\left(0\right)+\boxed{\left\langle c^{2}\right\rangle }\cdot\kappa w\alpha+\boxed{\left\langle c\right\rangle }\cdot\lambda\intop_{0}^{1}dw_{1}\cdot\sqrt{1-w_{1}^{2}}\cdot\frac{m_{1}\left(w_{1},0\right)-m_{1}\left(0,0\right)}{w_{1}}\nonumber \\
 & +\lambda\intop_{0}^{\infty}dy_{1}\cdot y_{1}\ln\frac{1}{y_{1}}\cdot\intop_{\mathbb{R}-i0}\frac{ds}{2\pi}\cdot\boxed{\intop dcP\left(c\right)}\cdot\boxed{c}\exp\left\{ i\boxed{c}s\kappa w\right\} \cdot\exp\left\{ m\left(s|0,0\right)-isy_{1}\right\} ,
\label{eq:fluctuating-coupling_m1-equation}
\end{align}
\begin{equation}
m_{2}\left(S|w\right)=\lambda\cdot\boxed{\intop dcP\left(c\right)}\cdot\intop_{0}^{1}dw_{1}\cdot\frac{\exp\left\{ iS\kappa\boxed{c}w_{1}\right\} -1-iS\kappa\boxed{c}w_{1}}{w_{1}^{2}\sqrt{1-w_{1}^{2}}}\cdot\left[1-w_{1}\left(1-w_{1}^{2}\right)\frac{\partial}{\partial w_{1}}\right]\cdot\left[\frac{\boxed{c}\kappa w+m_{1}\left(w_{1}\right)}{\kappa}\right].
\label{eq:fluctuating-coupling_m2-equation}
\end{equation}
\end{widetext}

In these equations, the boxes highlight the difference brought in
by the fluctuations of the matrix element in comparison with equations~(\hphantom{}\ref{eq:eq-on-m1}\nobreakdash-\ref{eq:eq-on-m2}\hphantom{}).
Once the solution to these equations is found, expressions \eqref{joint-distribution_expression-via-m}~and~\eqref{expr-for-P0-via-m}
for the probability density of the dimensionless order parameter $P_{0}\left(y\right)$
and the joint probability density $P\left(x,y\right)$ of onsite values
of $x=\xi_{i}/\Delta_{0}$ and $y=\Delta_{i}/\Delta_{0}$ are applicable
without modifications.

\subsubsection{Generalization to fluctuating number of neighbors $Z$}

We first note that these equations allow one to effortlessly analyze
the effect of the fluctuating number of neighbors $Z$. To this end,
one lets $P\left(c\right)=p\cdot\delta\left(1-c\right)+\left(1-p\right)\delta\left(c\right)$,
so that each edge is either ``turned on'' with probability $p\in\left[0,1\right]$,
or ``turned off'' with probability $1-p$. As a result, each site
has a fluctuating number of neighbors with Poisson distribution characterized
by mean value $\left\langle Z\right\rangle =pZ$. With such choice
of the distribution function $P\left(c\right)$ one can explicitly
perform all the averages in Eq.\nobreakdash-s~(\hphantom{}\ref{eq:fluctuating-coupling_m1-equation}\nobreakdash-\ref{eq:fluctuating-coupling_m2-equation}\hphantom{}).
Remarkably, the outcome \emph{is identical to} the equations~(\hphantom{}\ref{eq:eq-on-m1}\nobreakdash-\ref{eq:eq-on-m2}\hphantom{})
for the case without fluctuations of the number of neighbors upon
proper renormalization of the microscopical constants $\lambda,\alpha,Z,\Delta_{0},\kappa$.
Namely, one simply has to replace
\begin{equation}
\lambda\mapsto\lambda_{R}=p\lambda,\,\,\,\alpha\mapsto\alpha_{R}=p\alpha
\end{equation}
and calculate all other low-energy quantities in the theory using
these modified values. One particular example of this is the mean-field
value of the order parameter~\eqref{fluctuating-coupling_mean-field}
that now contains precisely $\lambda_{R}$ in both the exponent and
the prefactor $E_{0}$ defined by higher energies. Consequently, the
remaining microscopical constants are renormalized as
\begin{equation}
Z_{R}=pZ,\,\,\,\kappa_{R}=\frac{\lambda_{R}}{\Delta_{R}\cdot Z_{R}}.
\end{equation}
The derivation of these results is presented in \subsecappref{fluctuating-coupling_extreme-value-statistics}.
We once again underscore that such a picture implies absence of any
practical significance of the fluctuations of the number of neighbors
in our model. 

\subsubsection{Weak fluctuations of the coupling constant $\lambda$}

A more complicated situation arises, however, if one introduces disorder
in the value of $c$ itself. For this calculation, we choose $c$
to be distributed according to a narrow distribution with mean value
$\left\langle c\right\rangle =1$, variance $\left\langle \left(c-1\right)^{2}\right\rangle =\delta^{2}$
and exponentially decaying tails. One can then repeat the asymptotic
analysis of \subsecref{Extreme-value-statistics} to extract the influence
of the introduced fluctuations of the coupling matrix elements on
the extreme value statistics. A detailed exposition is presented in
\appref{Effect-of-fluctuating-coupling}, while here we summarize
the key results and qualitative conclusions.

In the region of small value of $y$, that corresponds to a unique
saddle point of the form $S=+it,\,\,t\gg1$, one can expand the Eq.~\eqref{fluctuating-coupling_m1-equation}
w.r.t small deviations of $c$ from its mean value. Upon estimating
the probability~\eqref{expr-for-P0-via-m} with the help of the resulting
asymptotic expression, the double-exponential asymptotic behavior
described by Eq.\nobreakdash-s~(\hphantom{}\ref{eq:P0-small-y-tail}\nobreakdash-\ref{eq:zeta_def}\hphantom{})
remains valid with only a slight modification of the form
\begin{equation}
\zeta\left(y\right)\mapsto\zeta\left(y\right)\exp\left\{ \delta^{2}/2\right\} .
\end{equation}
However, with finite $\delta$ this regime now extends only to a finite
lower value of the probability density:
\begin{equation}
P_{0}\left(y\right)\apprge\frac{1}{\sqrt{2\pi\cdot\lambda\left\langle y\right\rangle \kappa\delta^{2}}}\exp\left\{ -\frac{\lambda\left\langle y\right\rangle }{\kappa\delta^{2}}\right\} .
\end{equation}
This also implies that the double-exponential regime is only present
while
\begin{equation}
\delta\apprle\sqrt{\lambda/\kappa}=\sqrt{Z_{\text{eff}}}.
\end{equation}
The value of $P_{0}\left(y\right)$ for larger values of $\delta$
is described by a different asymptotic behavior with much slower decay
in the region of small~$y/\left\langle y\right\rangle $. It can
be interpreted as a change in the type of the dominating optimal fluctuation
that delivers the body of the distribution for low values of the order
parameter. Indeed, for the case with $\delta=0$ the only way to render
a small value of the order parameter was to have all neighboring values
of $\left|\xi\right|$ large enough, as explained in~\subsecref{Extreme-value-statistics}.
However, sufficiently strong fluctuations of the coupling constant
provide a finite probability of a region with a diminished values
of the coupling constant to neighboring sites with relatively small
values of $\xi$. The behavior of the distribution would thus reflect
the competition between these two sets of configurations. As a consequence,
one expects that in this case the answer will be sensitive to the
particular form of the distribution $P\left(c\right)$ as well as
any local correlations present in the joint distribution of the coupling
matrix elements $c_{ij}$ and the onsite energies $\xi_{i}$. 

The asymptotic behavior of the distribution for large values of the
order parameter can also be analyzed within the perturbative expansion
of Eq.~\eqref{fluctuating-coupling_m1-equation} w.r.t small deviation
of $c$ from its mean value. One obtains that each of the multiple
saddles point of the integral~\eqref{expr-for-P0-via-m} for the
probability acquire an additional multiplier that can be estimated
as
\begin{equation}
P_{0}^{\left(n\right)}\left(y\right)\sim P_{0}^{\left(n\right)}\left(y,\delta=0\right)\cdot\exp\left\{ \frac{\left(z_{n}\delta\right)^{2}}{2}\frac{y-\left\langle y\right\rangle }{\kappa}\right\} ,
\end{equation}
where $z_{n}=iS_{n}\kappa$ describes the position of the corresponding
saddle point, and $P_{n}\left(y,\delta=0\right)$ stands for the magnitude
of the contribution without fluctuations of the matrix element. This
result implies that the asymptotic expression~\eqref{P0-large-y-tail}
delivered by the main saddle point with $n=0$ remains \emph{qualitatively}
intact up to $\delta\sim1$, at which point the perturbative expansion
w.r.t small $\delta$ ceases to be applicable. Furthermore, each secondary
saddle point acquires an extra multiplier of the form $\exp\left\{ -\frac{\left(2\pi n\delta\right)^{2}}{2}\frac{y-\left\langle y\right\rangle }{\kappa}\right\} $
due to the imaginary part $z_{n}$ which is close to $2\pi n$. As
a result, the oscillations produced by these secondary saddle points
are suppressed at $2\pi\delta\sim1$. 

\figref{P0-log-plot_with-weak-coupling-fluctuations} below presents
the demonstration of the qualitative picture presented above in the
form of both theoretical curves and histograms obtained from direct
numerical solution of the modified self-consistency equations~\eqref{fluctuating-coupling_saddle-point-equation}
for several realizations of the disorder. In particular, it clearly
illustrates the persistence of both asymptotic trends observed in
\subsecref{Extreme-value-statistics}, while also demonstrating how
the secondary maxima are smeared as the value of $\delta$ is growing. 

\begin{figure}
\begin{centering}
\includegraphics[viewport=0bp 0bp 1200.75bp 820bp,clip,scale=0.21]{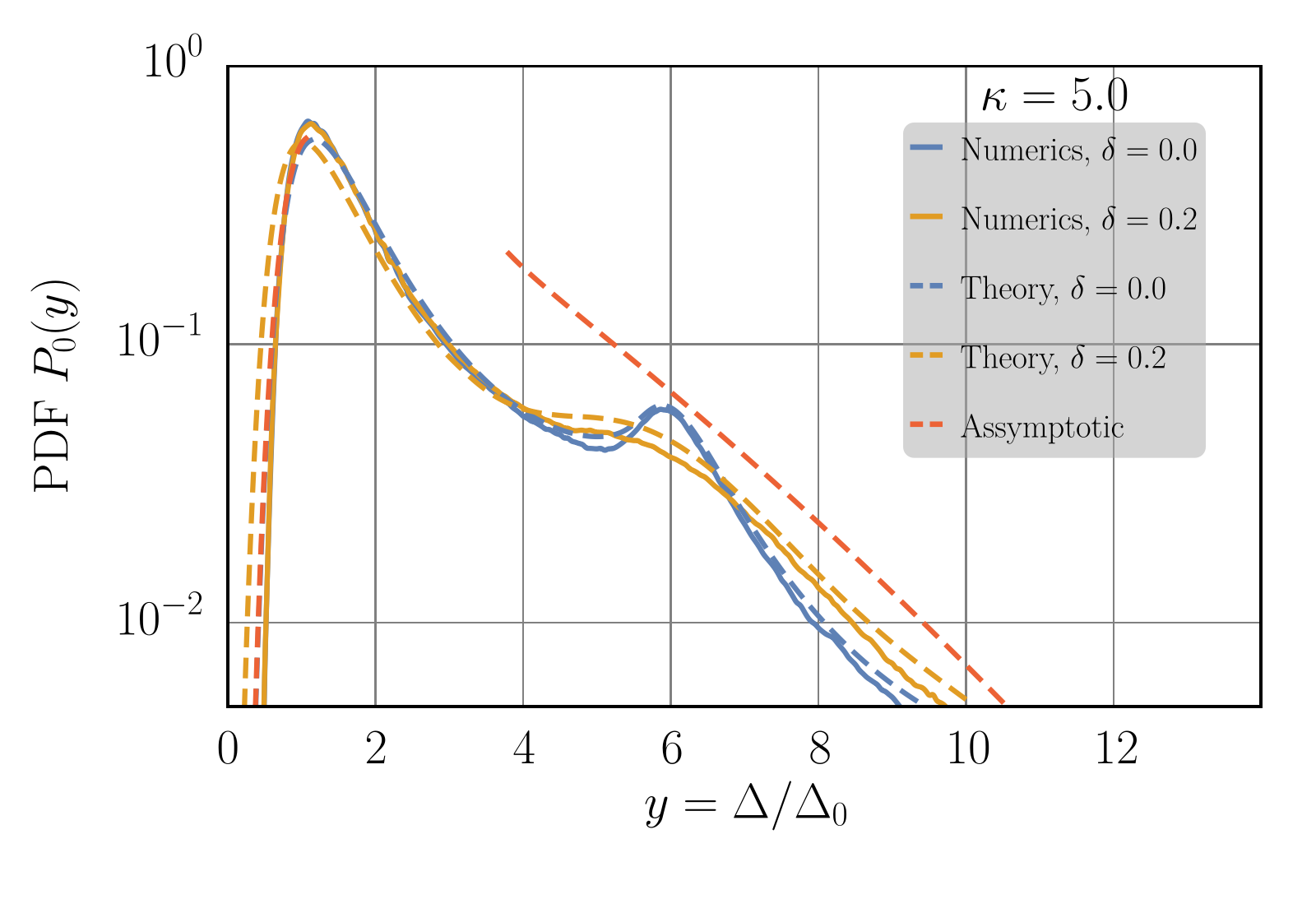}
\par\end{centering}
\caption{A log-scale plot for the PDF of the dimensionless order parameter
$P\left(\Delta/\Delta_{0}\right)$ for various strength of the fluctuations
of the interaction matrix element $D_{ij}=c_{ij}D\left(\xi_{i}-\xi_{j}\right)$.
The distribution of $c$ is log-normal with parameters that ensure
$\left\langle c\right\rangle =1$, $\left\langle \left(c-1\right)^{2}\right\rangle =\delta^{2}$.
The solid lines represent the smoothened histogram obtained from direct
numerical solution of the self-consistency equations~\eqref{fluctuating-coupling_saddle-point-equation}
on 3 instances of Random Regular Graph of size $N=2^{17}\approx1,3\cdot10^{5}$.
The dashed lines uses the proper generalization of the weak coupling
approximations of~\subsecref{Weak-coupling-approximation} to approximate
the value of the $m$~function used to compute the integral~\eqref{expr-for-P0-via-m}
for the PDF. Finally, the dashed red line corresponds to approximate
analytic expressions for the case without fluctuations of coupling
matrix element: Eq.~\eqref{P0-large-y-tail} for large values of
$y>\left\langle y\right\rangle $ and Eq.\protect\nobreakdash-s~(\protect\hphantom{}\ref{eq:P0-small-y-tail}\protect\nobreakdash-\ref{eq:zeta_def}\protect\hphantom{})
for $y<\left\langle y\right\rangle $. The microscopic parameters
of the model are $D\left(\xi\right)=\text{const}$, $\lambda\approx0.120$,
$Z=51$ and $\kappa=5.0$. The mismatch between the theoretical description
and the numerical histogram originates from subleading corrections
of order $O\left(\Delta_{0}/\lambda\right)$ and $O\left(\lambda^{2}\right)$,
see also notes on this under \figref{numerics-vs-theory_delta-distrib}
\label{fig:P0-log-plot_with-weak-coupling-fluctuations}}
\end{figure}

\section{Discussion and Conclusions\label{sec:Discussion-and-Conclusions}}

\begin{figure}
\begin{centering}
\includegraphics[scale=0.4]{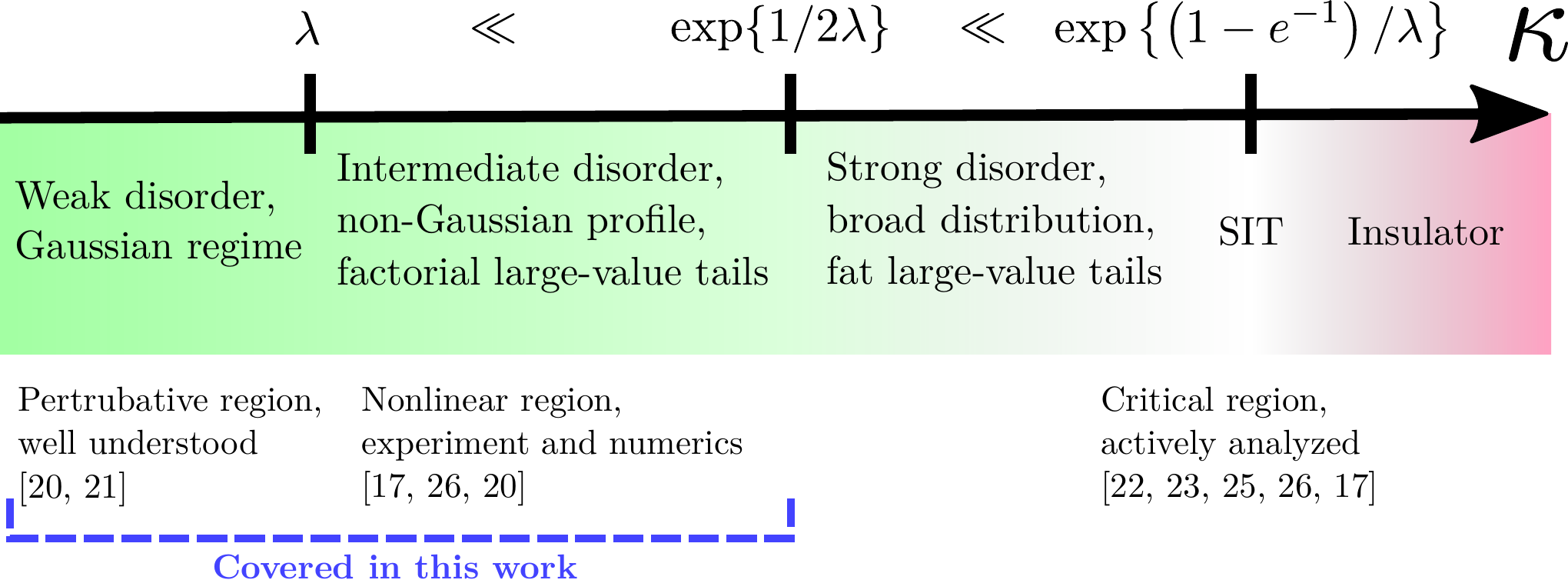}
\par\end{centering}
\caption{Qualitative phase diagram describing the behavior of the distribution
of the order parameter $P\left(\Delta\right)$ in strongly disordered
superconductors. The horizontal axis schematically describes the strength
of disorder measured by the parameter $\kappa$ defined in Eq.~\eqref{Z-eff_and_kappa_definition}.
Various colors indicate the perceived level of inhomogeneity of the
superconducting state: the green color corresponds to a well pronounced
superconducting state with nearly uniform value of the order parameter,
light-green and white colors represent a manifestly nonuniform superconducting
state with strongly non-Gaussian or even critical distributions of
the order parameter, and the red color stands for the insulating state
of the system with no superconducting order parameter. The blue dashed
line highlights the range of parameters available to our theoretical
description. \label{fig:OP-distribution_qualitative-phase-diagram}}
\end{figure}

In the present paper we developed systematic theory able to describe
statistics of superconducting order parameter in strongly disordered
pseudo-gaped superconductors. We have discovered the existence of
a wide region of parameters where usual semiclassical approach to
dirty superconductors is not valid, but, at the same time, the universal
behavior typical for the close proximity to SIT~\citep{Feigelman_SIT_2010}
does not take place either. In this wide range of parameters, the
shape of the distribution function~$P\left(\Delta\right)$ is controlled
by the single parameter~$\kappa$ defined in Eq.~\eqref{Z-eff_and_kappa_definition}.
Small~$\kappa$ corresponds to limit of weak disorder that is typical
for usual dirty superconductors. This limit is characterized by narrow
Gaussian distribution of the order parameter is observed, see Eq.~\eqref{P0-Gaussian-regime}.
On the other hand, at $\kappa\apprge\lambda$, with $\lambda$ being
the dimensionless Cooper constant, the distribution becomes highly
non-trivial. We are able to calculate its explicit form for all values
of $\Delta/\Delta_{0}$ in terms of certain special functions, as
presented in \subsecref{Weak-coupling-approximation}. The asymptotic
behavior of the distribution density $P\left(\Delta\right)$ is given
by equations (\hphantom{}\ref{eq:P0-small-y-tail}\nobreakdash-\ref{eq:zeta_def}\hphantom{})~and~\eqref{P0-large-y-tail}
for small and large values of $\Delta/\Delta_{0}$ respectively. These
functions do depend on the value of $\kappa$; in principle, it opens
the possibility to extract the value of $\kappa$ for specific disordered
superconductor via measuring the local distribution $P\left(\Delta\right)$
by means of scanning tunneling methods. Our model, however, breaks
down in a small vicinity of the~SIT described by exponentially large
values of $\kappa\apprge\exp\left\{ \frac{1}{2\lambda}\right\} \gg1$.
The phase diagram following from our findings is sketched on \figref{OP-distribution_qualitative-phase-diagram}.

We emphasize that the very existence of a separate region with a broad
range of disorder strengths featuring a non-trivial profile of the
distribution function $P\left(\Delta\right)$ is related to the smallness
of Cooper attraction constant $\lambda\ll1$. Until recently, small-$\lambda$
region was not attainable for direct numerical simulations of real
2D and 3D systems due to size restrictions. Advances in this field~\citep{Evers2019,PhysRevB.102.184507,Trivedi2021}
seem to make such a study possible.

The shape of distribution function $P\left(\Delta\right)$ was found
to differ considerably from the fat-tail distributions obtained previously
in Ref.\nobreakdash-s~\citep{Feigelman_SIT_2010,Lemarie2013} by
different analytic and numerical methods. Concerning available experimental
data, we note, first of all, that the interpretation of the tunneling
conductance $dI/dV$ in terms of the theoretical order parameter is
not straightforward in the case of large spatial fluctuations $\Delta\left(\boldsymbol{r}\right)$.
Indeed, in such case the half-width of the gap defined as the energy
distance between the peaks in $dI/dV$ is not just given by the order
parameter $\Delta$ itself, as it is the case in the classical superconductor
with constant $\Delta$. In fact the shape of $dI/dV$ is controlled
by the local Density of States (DoS) $\nu\left(E\right)$ which should
be obtained, in principle, via the solution of the generalized Usadel
equation for the local electron Green function in the background of
spatially fluctuating order parameter $\Delta\left(\boldsymbol{r}\right)$
as well as in presence of a pseudo-gap. Such a program had never been
implemented yet, to our knowledge.

Qualitatively, it seems evident that more direct access to the local
values of $\Delta\left(\boldsymbol{r}\right)$ is provided by the
heights $R$ of the \textquotedbl coherence peaks\textquotedbl{}
in local tunneling conductance~$dI/dV\left(\boldsymbol{r}\right)$.
Early experimental data~\citep{Sacepe_2011_for-pair-preformation}
demonstrates substantial change in the distribution of peak heights
$\mathcal{P}\left(R\right)$ with the increase of disorder, similarly
to the effect of increasing our theoretical parameter $\kappa$ upon
the shape of $P_{0}\left(\Delta\right)$, see \figref{numerics-vs-theory_delta-distrib}.
Another type of theoretical analysis provided in Ref.~\citep{Lemarie2013}
predicts extremely broad distribution of the Tracy-Widom universal
shape in terms of the logarithmic variable $R_{s}=\ln R/\left\langle R\right\rangle $;
however, their experimental data on Fig.~6 leaves space for different
interpretations as well. A recent study~\citep{PhysRevB.102.184507}
of strongly disordered 3D superconductor by means of numerical solution
of Bogolyubov-De~Gennes equations provides a number of various distribution
functions for $P\left(\Delta\right)$, which could be analyzed in
terms of our theory; for now we can say that the generic feature ---
an increase of normalized width of the distribution with disorder
--- is reproduced there as well. 

The model we have studied here is limited in several regards. First
of all, our initial model approximates the matrix element of the Cooper
attraction by a constant value, that is further endowed with a weak
dependence energy difference. However, the actual amplitude of the
interaction in each disorder realization is proportional with the
overlap of the corresponding wave-functions $\intop\psi_{i}^{2}\left(\boldsymbol{r}\right)\psi_{j}^{2}\left(\boldsymbol{r}\right)d\boldsymbol{r}$
and thus exhibits direct statistical fluctuations at least of the
order of its mean value. In \subsecref{Extreme-value-statistics_fluctuating-coupling}
we have briefly analyzed an extended model that incorporates this
effect in the simplest fashion possible. Our analysis indicates that
these direct statistical fluctuations do not alter our conclusions
about the large-value asymptotic behavior of the distribution of the
order parameter, while only removing several unphysical features such
as secondary maxima. However, it also follows from our results that
even relatively small fluctuations of the interaction matrix element
can distort the low-value asymptotic behavior of the distribution
of the order parameter. The character of this distortion is generally
sensitive to the local structure of the distribution of the matrix
elements and requires further analysis.

Secondly, the energy dependence of the matrix element $D\left(\omega=\left|\xi_{i}-\xi_{j}\right|\right)$
is assumed to be smooth at the relevant energy scale of Debye energy
$\varepsilon_{D}$. It is not necessarily the case for strongly disordered
superconductor with the Fermi energy located inside the localized
band; the point is that the relevant matrix elements between localized
eigenstates contain~\citep[Subsec. 2.2.5]{Feigelman_Fractal-SC_2010}
the Mott resonances leading to a singular behavior $D\left(\omega\right)\propto\left|\ln\omega/\delta_{L}\right|^{d-1}$.
This feature can be incorporated in our approach as long as the overall
separation of scales $\Delta\ll\varepsilon_{D},E_{F}$ is maintained.

Thirdly, we have analyzed the mean-field equations for $T=0$ only.
Non-zero temperatures can be included into our formalism simply by
multiplying the function $f\left(\xi_{j},\Delta_{j}|\xi_{i}\right)$
defined in Eq.~\eqref{saddle-point-rhs_notation} by $\tanh\frac{\sqrt{\Delta_{j}^{2}+\xi_{j}^{2}}}{2T}$.
It will complicate further analysis, but low-$T$ corrections to the
obtained results are possible to derive.

The nearest extensions of the developed theory will contain study
of low-energy collective modes in strongly disordered superconductors.
The aim is to revisit this subject, considered originally in Ref.~\citep{Feigelman_Microwave_2018}
with the presently developed understanding about the order parameter
distribution. Another important subject is to include the Onsager
reaction term in our free energy functional; it would allow to consider
the region closer to~SIT by our methods. Finally, it is of practical
importance to establish a reliable connection between the order parameter~$\Delta$
studied in this work and experimentally measurable quantities, as
none such connections exist to date for strongly disordered superconductors.
\begin{acknowledgments}
We are grateful to Yan Fyodorov, Lev Ioffe, and Igor Poboiko for many
useful discussions. A. K. is also grateful to Vladislav P. Serebrennikov
for his assistance with the hardware for numerical computations. This
research was supported by the Russian Science Foundation grant \#
20-12-00361.
\end{acknowledgments}

\bibliographystyle{unsrt}
\bibliography{Bibliography}

\newpage{}

\appendix
\onecolumngrid 
\counterwithin{figure}{section} 
\renewcommand\thefigure{\thesection\arabic{figure}} 
\renewcommand\appendixname{Supplementary Material} 
\newref{eqapp}{ refcmd={(\ref{#1})} }
\newref{eq}{ refcmd={(\ref{#1}) of the Main Text} }
\newref{app}{ refcmd={Section~\ref{#1}} }
\newref{subsecapp}{ refcmd={Subsection~\ref{#1}} }
\newref{subsec}{ refcmd={Subsection~\ref{#1} of the Main Text} }
\newref{sec}{ refcmd={Section~\ref{#1} of the Main Text} }
\newref{figapp}{ refcmd={Figure~\ref{#1}} }
\newref{fig}{ refcmd={Figure~\ref{#1} of the Main Text} }

\section{The Saddle-point Approximation\label{app:Saddle-point-equation_derivation}}

In this Appendix, we present a concise description of the saddle-point
approximation for the model Hamiltonian \eqref{Model-Hamiltonian}.
The partition function~$\mathcal{Z}$ and the free energy~$F$ of
the model are defined as
\begin{equation}
-\beta F=\ln\mathcal{Z}=\ln\text{Tr}\left\{ \exp\left\{ -\beta H\right\} \right\} ,\label{eqapp:partition-function}
\end{equation}
where the trace is taken over all fermionic degrees of freedom that
correspond to eigenstates of Cooper pair occupation operator:
\begin{equation}
n_{i}=\frac{1}{2}\left(a_{i\downarrow}^{\dagger}a_{i\downarrow}+a_{i\uparrow}^{\dagger}a_{i\uparrow}\right)=\left\{ 0,1\right\} .
\end{equation}
This limitation arises due to the presence of a large pseudogap in
the system, see \subsecref{Phenomenology-overview} for more details.

\subsection{Saddle point free energy}

It is convenient to rewrite the partition function \eqappref{partition-function}
in terms of imaginary time interaction picture with respect to the
non-interacting part of the Hamiltonian. The latter is defined as
\begin{equation}
H_{0}:=\sum_{i}\xi_{i}\left(a_{i\downarrow}^{\dagger}a_{i\downarrow}+a_{i\uparrow}^{\dagger}a_{i\uparrow}\right).
\end{equation}
Time-dependent operators in the interaction picture then reads
\[
X\left(\tau\right)=e^{\tau H_{0}}Xe^{-\tau H_{0}}.
\]
One can then represent the partition function in the following way:
\begin{equation}
-\beta F=\ln\text{Tr}\left\{ \text{Texp}\left\{ -\beta H_{\text{int}}\left(\tau\right)\right\} e^{-\beta H_{0}}\right\} .
\end{equation}
Here, the symbol $\text{Texp}$ stands for the imaginary-time-ordered
exponent, and $H_{\text{int}}$ is the interaction term:
\begin{equation}
H_{\text{int}}\left(\tau\right)=-\sum_{\left\langle ij\right\rangle }D\left(\xi_{i}-\xi_{j}\right)\left(a_{i\downarrow}^{\dagger}\left(\tau\right)a_{i\uparrow}^{\dagger}\left(\tau\right)a_{j\uparrow}\left(\tau\right)a_{j\downarrow}\left(\tau\right)+\text{Herm. conj.}\right).
\end{equation}
Under the sign of the time ordering, one can decouple the interaction
term by means of the functional Hubbard-Stratanovich transformation
\citep{Altland_Simons}:
\begin{equation}
\text{Texp}\left\{ -\beta H_{\text{int}}\left(\tau\right)\right\} =\frac{1}{\mathcal{Z}_{\Delta}}\intop_{\Delta\left(0\right)=\Delta\left(\tau\right)}D\Delta\left(\tau\right)\cdot\exp\left\{ -S_{\Delta}\left[\Delta\right]\right\} \cdot\text{Texp}\left\{ +\sum_{i}\intop d\tau\left(a_{i\downarrow}^{\dagger}\left(\tau\right)a_{i\uparrow}^{\dagger}\left(\tau\right)\Delta_{i}\left(\tau\right)+\text{Herm. conj.}\right)\right\} ,\label{eqapp:partition-function_functional-representation}
\end{equation}
\begin{equation}
\mathcal{Z}_{\Delta}=\intop_{\Delta\left(0\right)=\Delta\left(\tau\right)}D\Delta\left(\tau\right)\cdot\exp\left\{ -S_{\Delta}\left[\Delta\right]\right\} ,
\end{equation}
where the functional integration is done over all complex fields $\Delta_{i}\left(\tau\right)$
obeying periodic boundary conditions w.r.t $\tau$. The integration
weight is given by the following imaginary time action:
\begin{equation}
S_{\Delta}=\frac{1}{2}\sum_{ij}\cdot\intop_{0}^{\beta}d\tau\cdot\Delta_{i}^{*}\left(\tau\right)\cdot\left(V^{-1}\right)_{ij}\cdot\Delta_{j}\left(\tau\right),
\end{equation}
where $V^{-1}$ is the matrix inverse to the interaction matrix element,
viz.
\begin{equation}
\sum_{j:\left\langle jk\right\rangle }\left(V^{-1}\right)_{ij}\cdot D\left(\xi_{i}-\xi_{k}\right)=\delta_{ik}.
\end{equation}

One typically expects the mean field analysis to be perfectly applicable
for a well developed superconductivity away from the transition temperature,
i. e. for $1-T/T_{c}\gg\text{Gi}$. For the functional representation
\eqappref{partition-function_functional-representation}, this corresponds
to evaluating the functional integral by means of the saddle-point
approximation. The saddle-point configuration of the order parameter
is time-independent, so that the saddle-point value of the free energy
corresponds to the minimum of the following free energy function with
respect to the order parameter:
\begin{equation}
-\beta F\left[\Delta\right]=-\frac{\beta}{2}\sum_{ij}\cdot\Delta_{i}^{*}\cdot\left(V^{-1}\right)_{ij}\cdot\Delta_{j}+\ln\text{Tr}\left\{ \exp\left\{ -\beta H_{\text{MF}}\right\} \right\} ,\label{eqapp:saddle-point-free-energy}
\end{equation}
\begin{equation}
H_{\text{MF}}=\sum_{i}\xi_{i}\left(a_{i\downarrow}^{\dagger}a_{i\downarrow}+a_{i\uparrow}^{\dagger}a_{i\uparrow}\right)-\sum_{i}\left(\Delta_{i}a_{i\downarrow}^{\dagger}a_{i\uparrow}^{\dagger}+\text{Herm. conj.}\right).
\end{equation}
The second term in \eqappref{saddle-point-free-energy} is obtained
after using the fact that for the case of time-independent $\Delta$~field,
the trace of the time-ordered exponent can be rewritten in terms of
a trace over fermionic degrees of freedom of the mean field Hamiltonian~$H_{\text{MF}}$. 

The second term in \eqappref{saddle-point-free-energy} can be evaluated
explicitly:
\begin{equation}
\ln\text{Tr}\left\{ \exp\left\{ -\beta H_{\text{MF}}\right\} \right\} =\sum_{i}\ln\left[2e^{-\beta\xi_{i}}\cosh\left\{ \beta\sqrt{\xi_{i}^{2}+\left|\Delta_{i}\right|^{2}}\right\} \right].
\end{equation}
Note that this expression differs from a similar term in the conventional
theory of superconductivity by absence of the quasi-particle contribution,
as the latter is exponentially suppressed due to the well-developed
pseudogap. The final expression for the free energy then reads:
\begin{equation}
F\left[\Delta\right]=\frac{1}{2}\sum_{ij}\cdot\Delta_{i}^{*}\cdot\left(V^{-1}\right)_{ij}\cdot\Delta_{j}-T\sum_{i}\ln\left[2\cosh\left\{ \beta\sqrt{\xi_{i}^{2}+\left|\Delta_{i}\right|^{2}}\right\} \right]+\sum_{i}\xi_{i}.\label{eqapp:saddle-point-free-energy_final}
\end{equation}

\subsection{Saddle point equation for the order parameter}

The order parameter is determined as the minimum of the saddle-point
free energy \eqappref{saddle-point-free-energy_final}. The corresponding
saddle point equation reads:
\begin{equation}
\frac{\partial F\left[\Delta\right]}{\partial\Delta_{i}^{*}}=0\Leftrightarrow\sum_{j}\cdot\left(V^{-1}\right)_{ij}\cdot\Delta_{j}-\frac{\Delta_{i}}{\sqrt{\left|\Delta_{i}\right|^{2}+\xi_{i}^{2}}}\cdot\tanh\left\{ \beta\sqrt{\xi_{i}^{2}+\left|\Delta_{i}\right|^{2}}\right\} .
\end{equation}
Reverting back the $V$~matrix renders the celebrated self-consistency
equation:
\begin{equation}
\Delta_{i}=\sum_{j\in\partial i}f\left(\xi_{j},\Delta_{j}|\xi_{i}\right),
\end{equation}
\begin{equation}
f\left(\xi,\Delta|\xi_{0}\right)=D\left(\xi-\xi_{0}\right)\frac{\Delta}{\sqrt{\left|\Delta\right|^{2}+\xi^{2}}}\tanh\left\{ \beta\sqrt{\left|\Delta\right|^{2}+\xi^{2}}\right\} .
\end{equation}
In the absence of magnetic field and similar time-reversal symmetry
breaking factors, the order parameter can be chosen to be real and
positive. Finally, the zero temperature case corresponds to $\beta\rightarrow\infty$,
which results in the equation~\eqref{saddle-point_order-parameter}:
\begin{equation}
\Delta_{i}=\sum_{j\in\partial i}D\left(\xi_{j}-\xi_{i}\right)\frac{\Delta_{j}}{\sqrt{\Delta_{j}^{2}+\xi_{j}^{2}}}.\label{eqapp:zero-temperature-saddle-point-equation}
\end{equation}

\subsection{Mean field solution at zero temperature}

It is informative to analyze the resulting self-consistency equation~\eqappref{zero-temperature-saddle-point-equation}
in the regime of weak disorder, when the order parameter is nearly
homogeneous. As already discussed in \subsecref{mean-field-solution},
this approach is justified for sufficiently large number of neighbors~$Z$
by virtue of the central limit theorem. In this case, one can simplify
the self-consistency equation at zero temperature~\eqappref{zero-temperature-saddle-point-equation}
to the following form:
\begin{equation}
\Delta\left(\xi_{0}\right)=Z\cdot\intop d\xi\cdot\nu\left(\xi\right)\cdot D\left(\xi-\xi_{0}\right)\frac{\Delta\left(\xi\right)}{\sqrt{\Delta^{2}\left(\xi\right)+\xi^{2}}}.\label{eqapp:saddle-point_mean-field}
\end{equation}
This expression represents an integral equation on the value of the
order parameter $\Delta$ for a site with a given value of the onsite
disorder $\xi_{0}$.

Let us now take into account that the~$D$~function describes some
weak attraction with a typical energy scale being the Debye energy
$\varepsilon_{D}$. Similarly to the conventional theory of superconductivity,
the resulting value of the order parameter then appears to be exponentially
small with respect to the dimensionless coupling constant. For our
model, the latter is defined is
\begin{equation}
\lambda=2\nu_{0}\cdot ZD\left(0\right).
\end{equation}
The exponential smallness then follows from the fact that the integral
over $\xi$ in the right hand side of \eqappref{saddle-point_mean-field}
is logarithmic due to the $1/\xi$ asymptotic of the expression with
a square root. Secondly, because the superconducting scale $\Delta$
is exponentially smaller than the Debye energy $\varepsilon_{D}$,
the key role of $D\left(\xi\right)$ is to provide an upper cut-off
for the otherwise logarithmically diverging integral over $\xi$ in~\eqappref{saddle-point_mean-field}.
The integral itself can thus be estimated as:
\[
\intop d\xi\cdot\nu\left(\xi\right)\cdot D\left(\xi-\xi_{0}\right)\frac{\Delta\left(\xi\right)}{\sqrt{\Delta^{2}\left(\xi\right)+\xi^{2}}}\sim2\nu\left(0\right)\cdot D\left(0\right)\ln\frac{\varepsilon_{D}}{\Delta\left(0\right)}+\text{const},
\]
where the constant term of order unity is controlled by high energies,
as we will demonstrate below. This also implies that the $\xi_{0}$-dependence
of the order parameter approximately replicates that of $D\left(\xi_{0}\right)$,
thus also suggesting the scale of order $\xi_{0}\sim\varepsilon_{D}$
for the dependence of $\Delta$ on $\xi_{0}$.

To conduct a more quantitative analysis, let us introduce the following
notation:
\begin{equation}
u\left(\xi\right):=\frac{D\left(\xi\right)}{D\left(0\right)},\,\,\,\,\,d\left(\xi_{0}\right):=\frac{\Delta\left(\xi_{0}\right)}{\Delta_{0}},\,\,\,\,\,\Delta\left(0\right)=\Delta_{0},\,\,\,\,\,\nu_{0}=\nu\left(0\right).\label{eqapp:saddle-point-equation-derivation_envelope-functions-defs}
\end{equation}
As discussed above, both $u\left(\xi\right)$ and $d\left(\xi_{0}\right)$
are expected to have $\varepsilon_{D}$ as the energy scale of the
$\xi$-dependence. Note also that both $u$ and $d$ functions are
normalized as $u\left(0\right)=d\left(0\right)=1$ by construction.
In this notation, the mean field equation \eqappref{saddle-point_mean-field}
reads
\[
d\left(\xi_{0}\right)=\lambda\cdot\intop d\xi\cdot\frac{\nu\left(\xi\right)}{2\nu_{0}}\cdot u\left(\xi-\xi_{0}\right)\frac{d\left(\xi\right)}{\sqrt{\Delta_{0}^{2}\cdot d^{2}\left(\xi\right)+\xi^{2}}}.
\]
Assuming that the value of $\Delta_{0}$ is the smallest energy scale
in the problem, one can perform two important simplifications. First,
one neglects the $\xi$-dependence of the expression under the square
root, as it rendered irrelevant already for $\left|\xi\right|\apprge\Delta_{0}$,
well below the region where $u\left(\xi\right)$ deviates from unity
considerably. Secondly, one can split the integral over $\xi$ into
two contributions: the low energy part gaining its value at $\left|\xi\right|\sim\Delta_{0}$
and the high-energy part collecting its value from a large region
$\Delta_{0}\ll\left|\xi\right|\apprle\varepsilon_{D}$. The result
reads:
\begin{align}
d\left(\xi_{0}\right) & =\lambda\cdot\intop d\xi\cdot\frac{\nu\left(\xi\right)}{2\nu_{0}}\cdot u\left(\xi-\xi_{0}\right)\frac{d\left(\xi\right)}{\sqrt{\Delta_{0}^{2}+\xi^{2}}}\nonumber \\
 & =\lambda\eta\left(\xi_{0}\right)\cdot\intop d\xi\cdot\frac{\nu\left(\xi\right)}{2\nu_{0}}\cdot\frac{u\left(\xi\right)d\left(\xi\right)}{\sqrt{\Delta_{0}^{2}+\xi^{2}}}+\lambda\cdot\intop d\xi\cdot\frac{\nu\left(\xi\right)}{2\nu_{0}}\cdot\frac{u\left(\xi-\xi_{0}\right)d\left(\xi\right)-u\left(\xi_{0}\right)u\left(\xi\right)d\left(\xi\right)}{\sqrt{\Delta_{0}^{2}+\xi^{2}}}.
\end{align}
The second term now gains its value from the aforementioned large
region $\Delta_{0}\ll\left|\xi\right|\apprle\varepsilon_{D}$ and
thus the $\Delta_{0}^{2}$ term in the denominator can be neglected,
rendering:
\begin{equation}
d\left(\xi_{0}\right)=\lambda u\left(\xi_{0}\right)\cdot\intop d\xi\cdot\frac{\nu\left(\xi\right)}{2\nu_{0}}\cdot\frac{u\left(\xi\right)d\left(\xi\right)}{\sqrt{\Delta_{0}^{2}+\xi^{2}}}+\lambda\cdot\intop d\xi\cdot\frac{\nu\left(\xi\right)}{2\nu_{0}}\cdot\frac{u\left(\xi-\xi_{0}\right)d\left(\xi\right)-u\left(\xi_{0}\right)u\left(\xi\right)d\left(\xi\right)}{\left|\xi\right|}.\label{eqapp:mean-field-equation_dimensionless-form}
\end{equation}
The normalization condition $d\left(0\right)=1$ fixes the exact value
for the first term:
\begin{equation}
1=\lambda\cdot\intop d\xi\cdot\frac{\nu\left(\xi\right)}{2\nu_{0}}\cdot\frac{u\left(\xi\right)d\left(\xi\right)}{\sqrt{\Delta_{0}^{2}+\xi^{2}}},\label{eqapp:mean-field_order-parameter-value}
\end{equation}
which then allows one to simplify the equation~\eqappref{mean-field-equation_dimensionless-form}
to:
\begin{equation}
d\left(\xi_{0}\right)=u\left(\xi_{0}\right)+\lambda\cdot\intop d\xi\cdot\frac{\nu\left(\xi\right)}{2\nu_{0}}\cdot\frac{u\left(\xi-\xi_{0}\right)d\left(\xi\right)-u\left(\xi_{0}\right)u\left(\xi\right)d\left(\xi\right)}{\left|\xi\right|}.\label{eqapp:mean-field_d-function-equation}
\end{equation}
This equation represents an integral equation on the $d$~function.
As expected, it does not contain any information about the order parameter
whatsoever, reflecting the fact that the behavior of $d$ is determined
solely by higher energies.

One then turns to the low-energy part represented by Eq.~\eqappref{mean-field_order-parameter-value}.
In order to extract the value of $\Delta_{0}$, one uses the following
integral representation for the root function:
\begin{equation}
\frac{1}{\sqrt{\Delta_{0}^{2}+\xi^{2}}}=\intop\frac{dt}{2\pi}\cdot2K_{0}\left(\Delta_{0}\left|t\right|\right)e^{it\xi},
\end{equation}
where $K_{0}$ is the modified Bessel function. The expression~\eqappref{mean-field_order-parameter-value}
then reads:
\begin{equation}
1=\lambda\cdot\intop\frac{dt}{2\pi}\cdot2K_{0}\left(\Delta_{0}\left|t\right|\right)\cdot\intop d\xi\cdot\frac{\nu\left(\xi\right)}{2\nu_{0}}\cdot u\left(\xi\right)d\left(\xi\right)\cdot e^{it\xi}.
\end{equation}
After integrating over $\xi$ the resulting function of $t$ decays
quickly beyond $\left|t\right|\apprge\varepsilon_{D}^{-1}$, as governed
by the behavior of both $d$ and $u$ functions. The resulting integral
over $t$ then also converges at $\left|t\right|\apprle\varepsilon_{D}^{-1}$,
allowing one to formally expand the Bessel function in the limit $\Delta_{0}\left|t\right|\ll1$:
\begin{equation}
1=\lambda\cdot\intop\frac{dt}{2\pi}\cdot2\left[\ln\frac{2}{\Delta_{0}\left|t\right|}-\gamma+O\left(\Delta_{0}^{2}\ln\Delta_{0}\right)\right]\cdot\intop d\xi\cdot\frac{\nu\left(\xi\right)}{2\nu_{0}}\cdot u\left(\xi\right)d\left(\xi\right)\cdot e^{it\xi},\label{eqapp:mean-field-equation-1}
\end{equation}
where $\gamma=0.577...$ is the Euler-Mascheroni constant. It is convenient
introduce the following notation
\begin{equation}
2\ln\frac{E_{0}}{\varepsilon_{D}}=\intop\frac{dt}{2\pi}\cdot2\left[\ln\frac{1}{\varepsilon_{D}\left|t\right|}-\gamma\right]\cdot\intop d\xi\cdot\frac{\nu\left(\xi\right)}{2\nu_{0}}\cdot u\left(\xi\right)d\left(\xi\right)\cdot e^{it\xi}.\label{eqapp:E0-definition}
\end{equation}
The value of $E_{0}$ is of the order of Debye energy $\varepsilon_{D}$,
as will be demonstrated in a moment. One can then simplify equation
\eqappref{mean-field-equation-1} to
\begin{align}
1 & =\lambda\cdot\intop\frac{dt}{2\pi}\cdot2\left[\ln\frac{2\varepsilon_{D}}{\Delta_{0}}+\ln\frac{1}{\varepsilon_{D}\left|t\right|}-\gamma\right]\cdot\intop d\xi\cdot\frac{\nu\left(\xi\right)}{2\nu_{0}}\cdot u\left(\xi\right)d\left(\xi\right)\cdot e^{it\xi}\nonumber \\
 & =\lambda\cdot\left[2\ln\frac{2\varepsilon_{D}}{\Delta_{0}}\cdot\intop d\xi\cdot\frac{\nu\left(\xi\right)}{2\nu_{0}}\cdot u\left(\xi\right)d\left(\xi\right)\cdot\delta\left(\xi\right)+\intop\frac{dt}{2\pi}\cdot2\left[\ln\frac{1}{\varepsilon_{D}\left|t\right|}-\gamma\right]\cdot\intop d\xi\cdot\frac{\nu\left(\xi\right)}{2\nu_{0}}\cdot u\left(\xi\right)d\left(\xi\right)\cdot e^{it\xi}\right]\nonumber \\
 & =\boxed{\lambda\cdot\ln\frac{2E_{0}}{\Delta_{0}}=1}.
\end{align}
The last equation highlighted with a box finally renders the BCS solution
\eqref{mean-field-delta_zero-temp} for the order parameter:
\begin{equation}
\Delta_{0}=2E_{0}\exp\left\{ -\frac{1}{\lambda}\right\} ,
\end{equation}
where the exact energy scale $E_{0}\sim\varepsilon_{D}$ is determined
by high energies, as evident from Eq.~\eqappref{E0-definition}.

It is worth noting that one can build perturbative expansion for both
$d\left(\xi\right)$ and $E_{0}$ in powers of small $\lambda$. As
it is obvious from Eq.~\eqappref{mean-field_d-function-equation},
the leading order for $d\left(\xi_{0}\right)$ is given by
\begin{equation}
d\left(\xi_{0}\right)=u\left(\xi_{0}\right)+O\left(\lambda\right),
\end{equation}
which confirms the qualitative expectation that $\Delta\left(\xi_{0}\right)$
resembles the profile of $D\left(\xi\right)$. The value of $E_{0}$
is then read off from Eq.~\eqappref{E0-definition}:
\begin{equation}
E_{0}=\varepsilon_{D}\cdot\exp\left\{ +\frac{1}{2}\cdot\intop\frac{dt}{2\pi}\cdot2\left[\ln\frac{1}{\varepsilon_{D}\left|t\right|}-\gamma\right]\cdot\intop d\xi\cdot\frac{\nu\left(\xi\right)}{2\nu_{0}}\cdot u^{2}\left(\xi\right)\cdot e^{it\xi}\right\} .
\end{equation}
For instance, a simplistic model of the form $u\left(\xi\right)=\theta\left(\varepsilon_{D}-\left|\xi\right|\right)$
renders:
\[
E_{0}=\varepsilon_{D}\cdot\exp\left\{ +\frac{1}{2}\intop\frac{dt}{2\pi}\cdot\left[\ln\frac{1}{\varepsilon_{D}\left|t\right|}-\gamma\right]\cdot\frac{2\sin\varepsilon_{D}t}{t}\right\} =\varepsilon_{D}\cdot\exp\left\{ +\frac{1}{2}\intop\frac{dx}{2\pi}\cdot\left[\ln\frac{1}{\left|x\right|}-\gamma\right]\cdot\frac{2\sin x}{x}\right\} =\varepsilon_{D},
\]
in full agreement with the textbook results \citep{Altland_Simons}.

\section{Equation on the distribution of the order parameter\label{app:equation-on-distribution}}

In this appendix, we present the derivation of the equation on the
distribution of the order parameter obeying the self-consistency equation~\eqref{saddle-point_order-parameter}.
In what follows, it is convenient to represent the equation in the
following general form:
\begin{equation}
\Delta_{i}=\sum_{j\in\partial i}f\left(\xi_{j},\Delta_{j}|\xi_{i}\right),\label{eqapp:saddle-point_functional-form}
\end{equation}
where $f$ represents the functional form of the equation. For instance,
the zero temperature case reads
\begin{equation}
f\left(\xi_{j},\Delta_{j}|\xi_{i}\right)=\frac{\Delta_{j}}{\sqrt{\Delta_{j}^{2}+\xi_{j}^{2}}}\cdot D\left(\xi_{j}-\xi_{i}\right),
\end{equation}
as read off directly from the self-consistency equation~\eqref{saddle-point_order-parameter}
itself. We assume that for every particular realization of the disorder
field $\xi_{i}$ there exists a unique \emph{stable} solution w.r.t
the order parameter field $\Delta_{i}$. Let us denote this solution
as~$S_{i}\left(\left\{ \xi_{i}\right\} _{i\in G}\right)$. In this
context, the term stability essentially means that the solution is
a minimum of the free energy.%
{} Note that the configuration of the order parameter explicitly depends
on the disorder fields $\xi_{i}$.

We are also interested in the following modification of the problem.
Consider the system~\eqappref{saddle-point_functional-form}, in
which the equation for the value of $\Delta$ at site $i_{0}$ was
replaced by manually specifying the value of the order parameter,
so that~$\Delta_{i_{0}}=\Delta_{0}$. For brevity, we will denote
the value of $\xi_{i_{0}}$ at the corresponding site as~$\xi_{0}$.
For an arbitrary choice of $\Delta_{0}$ and $\xi_{0}$, this new
problem is not identical to the initial one, hence the solution to
the modified system of self-consistency equations represents a different
function of disorder on the remaining sites $j\in G\backslash\left\{ i_{0}\right\} $.
Let us denote this function as
\begin{equation}
S_{j}^{i_{0}}\left(\left\{ \xi_{j}\right\} _{j\neq i}|\xi_{0},\Delta_{0}\right).
\end{equation}
Just as the original problem, this modified problem does contain explicit
dependence on the disorder field~$\xi_{i}$ in the remaining system.
However, it now also depends on the choice of $\Delta_{0}$ and $\xi_{0}$.
The key observation at this point is that solution to the modified
problem $S^{i_{0}}$ coincides with the solution to the original problem
$S$ if and only if one chooses the value of $\Delta_{0}$ consistent
with the configuration of both $\Delta$ and $\xi$ fields in the
remaining system. In other words, the following identity holds
\begin{equation}
\forall j\neq i_{0}:\,\,\,S_{j}^{i_{0}}\left(\left\{ \xi_{j}\right\} |\xi_{0},\Delta_{0}\right)\equiv S_{j}\left(\left\{ \xi_{i}\right\} \right)\Leftrightarrow\begin{cases}
\xi_{0}=\xi_{i_{0}},\\
\Delta_{0}=S_{i}\left(\left\{ \xi_{i}\right\} \right),\\
\forall j\neq i_{0}:\,\,\,\,\Delta_{j}=S_{j}\left(\left\{ \xi_{i}\right\} \right).
\end{cases}
\end{equation}
Equivalently, one must demand the value of $\Delta_{0}$ itself to
satisfy the self-consistency equation \eqappref{saddle-point_functional-form}
at $i_{0}$:
\begin{equation}
\forall j\neq i_{0}:\,\,\,S_{j}^{i_{0}}\left(\left\{ \xi_{j}\right\} |\xi_{0},\Delta_{0}\right)\equiv S_{j}\left(\left\{ \xi_{i}\right\} \right)\Leftrightarrow\begin{cases}
\xi_{0}=\xi_{i_{0}},\\
\Delta_{0}=\sum\limits _{j\in\partial i_{0}}f\left(\xi_{j},\Delta_{j}|\xi_{0}\right),\\
\forall j\neq i_{0}:\,\,\,\,\Delta_{j}=S_{j}\left(\left\{ \xi_{i}\right\} \right).
\end{cases}\label{eqapp:equivalence-of-original-and-modified-systems}
\end{equation}
One particularly important interpretation of the Eq.~\eqappref{equivalence-of-original-and-modified-systems}
is that solving the full system of self-consistency equations~\eqappref{saddle-point_functional-form}
can be performed in two steps. First, one solves the modified problem
for some externally specified value of $\Delta_{0}$ and thus restores
the $S_{j}^{i_{0}}$ function. Then one plugs the result into the
self-consistency equation for site $i_{0}$ itself and solves the
resulting equation on $\Delta_{0}$. Because each value of each $\Delta_{j}$
now implicitly depends on $\Delta_{0}$ via the $S_{j}^{i_{0}}$ function,
the second step is by no means simpler than solving the original systems
of equations. Nevertheless, this two-step procedure formalizes the
concept of locality in the original self-consistency equation in a
sense that in order to restore the solution in a given finite region,
one only has to specify the values at the boundary of this region
as a function of the values inside the region.

This interpretation allows one to come up with a relation between
the ensembles of solutions in the two versions of the problem. Consider
the joint probability distribution of the values of $\xi$ and $\Delta$
at $i_{0}$ and its nearest neighbors $j\in\partial i_{0}$ in the
original problem. This function is defined as:
\begin{equation}
P\left(\left\{ \overline{\xi}_{k},\Delta_{k}\right\} \right):=\left\langle \prod_{k}\delta\left(\overline{\xi}_{k}-\xi_{k}\right)\delta\left(\Delta_{k}-S_{k}\right)\right\rangle _{\xi},\label{eqapp:joint-probability_definition}
\end{equation}
where $k$ runs through $\partial i_{0}$ and $i_{0}$ itself, and
the average $\left\langle \bullet\right\rangle $ is performed over
all values of $\xi_{j}$. Because of the equivalence~\eqappref{equivalence-of-original-and-modified-systems},
we can perform a change of variables in the argument of the $\delta$-function
resulting in:
\begin{equation}
\prod_{k}\delta\left(\Delta_{k}-S_{k}\right)=\prod_{k\in\partial i_{0}}\delta\left(\Delta_{k}-S_{k}^{i_{0}}\right)\cdot\delta\left(\Delta_{0}-s\right)\cdot\left|\det\hat{M}\right|.\label{eqapp:expression-for-delta-function-of-equations}
\end{equation}
Note that the right hand side now contains the solution of the modified
problem $S_{k}^{i_{0}}$ as well as a new function $s$, which represents
the explicit expression for the value of $\Delta_{0}$ as a function
of all neighboring values:
\begin{equation}
s\left(\xi_{i_{0}}|\left\{ \xi_{j},\Delta_{j}\right\} _{j\in\partial i_{0}}\right)=\sum\limits _{j\in\partial i_{0}}f\left(\xi_{j},\Delta_{j}|\xi_{i_{0}}\right).
\end{equation}
In addition, the expression \eqappref{expression-for-delta-function-of-equations}
includes the Jacobian of the transformation~$M$ resulting from the
aforementioned change of variables:
\begin{equation}
\hat{M}:=\frac{\delta\left(s,\left\{ S_{j}^{i_{0}}\right\} _{j\in\partial i_{0}}\right)}{\delta\left(S_{i_{0}},\left\{ S_{j}\right\} _{j\in\partial i_{0}}\right)}.
\end{equation}
Its matrix elements have the following block structure:
\begin{equation}
M_{jk}=\begin{pmatrix}1 & -\frac{\partial s}{\partial\Delta_{k}}\\
-\frac{\partial S_{j}^{i_{0}}}{\partial\Delta_{0}} & \delta_{jk}
\end{pmatrix},
\end{equation}
where the first index correspond to site $i_{0}$ itself, and the
remaining indices enumerate neighbors $j\in\partial i_{0}$. The determinant
of this matrix can easily be computed:
\begin{equation}
\det\hat{M}=1-\sum_{j\in\partial i_{0}}\frac{\partial s}{\partial\Delta_{j}}\frac{\partial S_{j}^{i_{0}}}{\partial\Delta_{0}}.\label{eqapp:expression-for-Jacobian}
\end{equation}
Also, because the original solution $S_{j}$ is assumed to represent
a minimum of the free energy, it can be shown that the matrix $\hat{M}$
is positive definite, thus rendering its determinant also positive.
This allows one to drop the absolute value sign in \eqappref{expression-for-delta-function-of-equations}
and further rewrite it in the following form:

\begin{align}
\prod_{k\in\partial i_{0}\cup\left\{ i_{0}\right\} }\delta\left(\Delta_{k}-S_{k}\right) & =\prod_{k\in\partial i_{0}}\delta\left(\Delta_{k}-S_{k}^{i_{0}}\right)\cdot\delta\left(\Delta_{i_{0}}-s\right)\nonumber \\
 & -\sum_{j}\prod_{k\in\partial i_{0},k\neq j}\delta\left(\Delta_{k}-S_{k}^{i_{0}}\right)\cdot\left[\frac{\partial}{\partial\Delta_{0}}\intop^{\Delta_{j}}d\Delta_{j}^{'}\cdot\delta\left(\Delta_{j}^{'}-S_{j}^{i_{0}}\right)\right]\cdot\left[\frac{\partial}{\partial\Delta_{j}}\intop^{\Delta_{0}}d\Delta_{0}^{'}\cdot\delta\left(\Delta_{0}^{'}-s\right)\right].\label{eqapp:expression-for-delta-function-of-equations-2}
\end{align}
The first term in this expression corresponds to unitary term in the
Jacobian \eqappref{expression-for-Jacobian}, and the second term
reproduces the part with derivatives by exploiting the fact that
\begin{equation}
\frac{\partial f}{\partial y}\delta\left(x-f\left(y\right)\right)=\frac{\partial}{\partial y}\intop^{x}dx'\cdot\delta\left(x'-f\left(y\right)\right).
\end{equation}
The lower limit of the integral is of little importance as long as
it does not depend on $y$, so that its influence vanishes upon differentiation.

\begin{figure}
\begin{centering}
\includegraphics[scale=0.9]{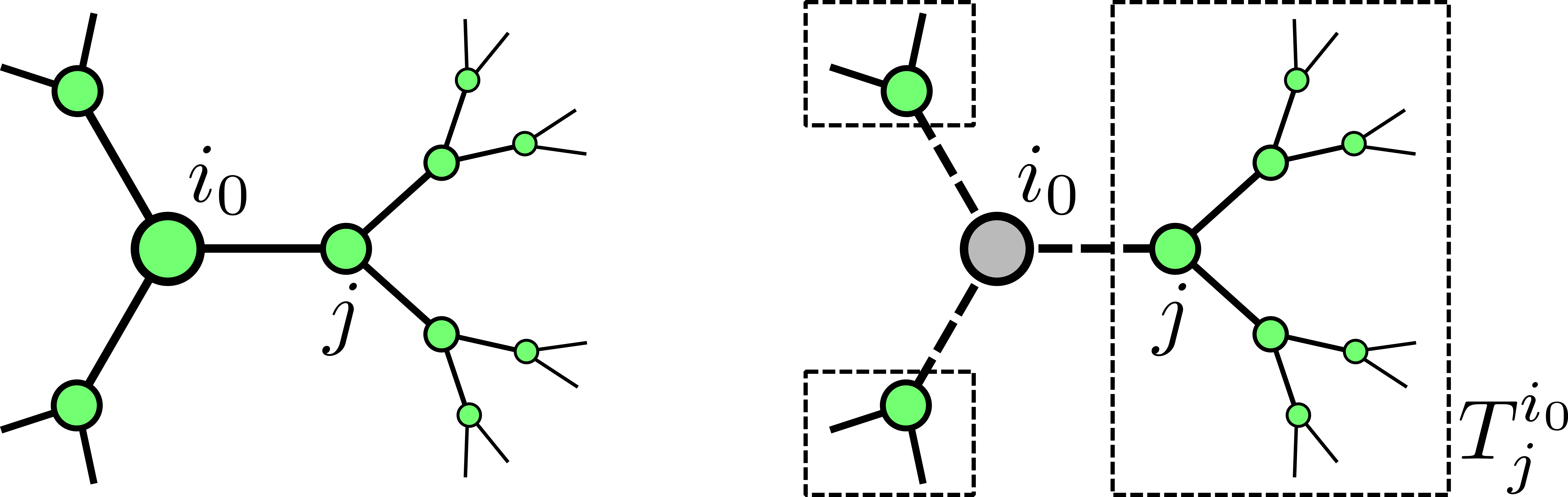}
\par\end{centering}
\caption{A neighborhood of the chosen site $i_{0}$ with $Z=3$ is depicted.
The left figure represents the original problem, and the right figure
corresponds to the modified problem with the site $i_{0}$ (in gray)
and all its edges (dashed lines) being \textquotedblleft quenched\textquotedblright{}
to some externally specified values. For a given neighbor of $j$
of the chosen site, the value of the order parameter $\Delta_{j}$
is determined by the values of $\xi$ within the corresponding branch
denoted as $T_{j}^{i_{0}}$ (highlighted with a dashed rectangle).
Because of locality and the fact that the original equation contains
a summation over all neighbors, the influence of each particular site
in $T_{j}^{i_{0}}$ decreases with the distance away from the discussed
site $j$. This is schematically represented by the size of each site.
In particular, one expects that beyond some finite distance $d$ each
particular site has virtually no effect on the value of the order
parameter at site $j$.\label{figapp:neighborhood-of-a-chosen-site}}

\end{figure}

As a final step, we exploit the locally tree-like structure of the
graph. Consider sites $j\in\partial i_{0}$, that is, the nearest
neighborhood of the fixed site $i_{0}$. Let us also denote the local
tree of some large depth $d$ originating at $j$ and spreading away
from $i_{0}$ as $T_{j}^{i_{0}}$, see also \figappref{neighborhood-of-a-chosen-site}.
Locally tree-like structure of the graph implies that the local trees
for different $j$ start to overlap only when $d$ approaches the
diameter of the entire system, with the latter diverging in the thermodynamical
limit. On the other hand, the self-consistency equations~\eqappref{saddle-point_functional-form}
involve a sum of large number of fluctuating variables. Consequently,
$S_{j}^{i_{0}}$ for a given $j$ is essentially sensitives only to
the values of $\xi$ within a tree $T_{j}^{i_{0}}$ of some finite
depth $d_{0}$, with the latter playing the role of the correlation
length. Such an argumentation is known to be valid only for $Z\gg Z_{1}=\lambda\exp\left\{ \frac{1}{2\lambda}\right\} $
\citep{Feigelman_SIT_2010}, as discussed in the main text. The key
conclusion from the observations above is that the functions $S_{j}^{i_{0}}$
for various $j$ depend on non-overlapping sets of $\xi_{j}$ values,
thus leading to statistical independence of $S_{j}^{i_{0}}$ w.r.t
the ensemble of independent $\xi_{j}$ in the thermodynamical limit.
Note that this does not imply the same behavior for $S_{j}$ in the
full system, where the neighborhood of $i_{0}$ is correlated precisely
due to presence of $i_{0}$ itself. 

The described decoupling between the values of the order parameter
in the modified problem allows one to average each $\delta$-function
in the right-hand side of Eq.~\eqappref{expression-for-delta-function-of-equations-2}
independently. That is why, the original expression for the joint
probability~\eqappref{joint-probability_definition} can be rewritten
in the following form:

\begin{align}
P\left(\left\{ \xi_{k},\Delta_{k}\right\} \right) & =\prod_{j\in\partial i_{0}}P_{j}^{i_{0}}\left(\xi_{j},\Delta_{j}|\xi_{0},\Delta_{0}\right)\cdot P\left(\xi_{0}\right)\cdot\delta\left(\Delta_{0}-s\right)\nonumber \\
 & -\sum_{j}\prod_{k\in\partial i_{0}\neq j}P_{k}^{i_{0}}\left(\xi_{k},\Delta_{k}|\xi_{0},\Delta_{0}\right)\cdot\left[\frac{\partial}{\partial\Delta_{0}}\intop^{\Delta_{j}}d\Delta_{j}^{'}\cdot P_{j}^{i_{0}}\left(\xi_{j},\Delta_{j}^{'}|\xi_{0},\Delta_{0}\right)\right]\cdot\left[P\left(\xi_{0}\right)\frac{\partial}{\partial\Delta_{j}}\intop^{\Delta_{0}}d\Delta_{0}^{'}\cdot\delta\left(\Delta_{0}^{'}-s\right)\right],
\end{align}
where the function $P_{j}^{i_{0}}$ is the probability density function
of a given neighbor $j$ of site $i_{0}$ in the modified problem:
\begin{equation}
P_{j}^{i_{0}}\left(\overline{\xi}_{j},\Delta_{j}|\xi_{0},\Delta_{0}\right):=\left\langle \delta\left(\overline{\xi}_{j}-\xi_{j}\right)\delta\left(\Delta_{j}-S_{j}^{i_{0}}\right)\right\rangle _{\xi}.
\end{equation}
The average $\left\langle \bullet\right\rangle $ in this expression
is now performed over all configurations of disorder with $\xi_{i_{0}}=\xi_{0}$.
One is then interested in integrating out the neighboring sites to
obtain the expression for the onsite probability density:

\begin{align}
P_{i_{0}}\left(\xi_{0},\Delta_{0}\right) & =\prod_{j\in\partial i_{0}}\intop d\xi_{j}d\Delta_{j}\cdot P_{j}^{i_{0}}\left(\xi_{j},\Delta_{j}|\xi_{0},\Delta_{0}\right)\cdot P\left(\xi_{0}\right)\delta\left(\Delta_{0}-s\right)\nonumber \\
 & -\prod_{k\in\partial i_{0}}\intop d\xi_{k}d\Delta_{k}\cdot\sum_{j}\prod_{k\in\partial i_{0},k\neq j}P_{k}^{i_{0}}\left(\xi_{k},\Delta_{k}|\xi_{0},\Delta_{0}\right)\nonumber \\
 & \times\left[\frac{\partial}{\partial\Delta_{0}}\intop^{\Delta_{j}}d\Delta_{j}^{'}\cdot P_{j}^{i_{0}}\left(\xi_{j},\Delta_{j}^{'}|\xi_{0},\Delta_{0}\right)\right]\cdot\left[P\left(\xi_{0}\right)\frac{\partial}{\partial\Delta_{j}}\intop^{\Delta_{0}}d\Delta_{0}^{'}\cdot\delta\left(\Delta_{0}^{'}-s\right)\right].\label{eqapp:connection-between-onsite-probabilities}
\end{align}
This equation connects the onsite probability distribution in the
original problem with a similar object in the modified problem. One
can derive similar expression for a joint probability distribution
of any local set of sites in the full system. 

Now, the argumentation that lead to the relation~\eqappref{connection-between-onsite-probabilities}
between original and modified problems remains entirely valid if one
formally performs the same steps one more time for any site $j\in\partial i_{0}$.
Namely, fixing the value of the order parameter on $j$ as well results
in a new function $S^{j,i_{0}}$ of the remaining $\xi$ values. This
function is connected with the previous iteration as
\begin{equation}
\forall k\neq i_{0},j_{0}:\,\,\,S_{k}^{j,i_{0}}\left(\left\{ \xi_{k}\right\} |\Delta_{1},\xi_{1},\Delta_{0},\xi_{0}\right)\equiv S_{k}^{i_{0}}\left(\left\{ \xi_{j}\right\} |\Delta_{0},\xi_{0}\right)\Leftrightarrow\begin{cases}
\xi_{1}=\xi_{j},\\
\Delta_{1}=s^{i_{0}}\left(\xi_{1}|\left\{ \xi_{k},\Delta_{k}\right\} _{k\in\partial j}\right),\\
\forall k\neq i_{0},j_{0}:\,\,\,\,\Delta_{k}=S_{k}^{i_{0}}\left(\left\{ \xi_{j}\right\} |\Delta_{0},\xi_{0}\right),
\end{cases}\label{eqapp:equivalence-of-two-modified-systems}
\end{equation}
where $\Delta_{1}$ and $\xi_{1}$ are the values of fields at site
$j$, and $s^{i_{0}}$ stands for the following expression:
\begin{equation}
s^{i_{0}}\left(\xi_{1}|\left\{ \xi_{k},\Delta_{k}\right\} _{k\in\partial j}\right):=\sum\limits _{k\in\partial j\backslash\left\{ i_{0}\right\} }f\left(\xi_{k},\Delta_{k}|\xi_{1}\right)+f\left(\Delta_{0},\xi_{0}|\xi_{1}\right).
\end{equation}
Note that we have explicitly split the summation over the nearest
neighbors of $j$ into the contribution the ``fixed'' site $i_{0}$
and all other neighbors of $j$. Performing the same type of reasoning
as the one already discussed, one proceeds to derive an expression
similar to Eq.~\eqappref{connection-between-onsite-probabilities},
which expresses $P_{j}^{i_{0}}$ in terms of $P_{k}^{j}$ for $k\neq i_{0}$
and $s^{i_{0}}$:

\begin{align}
P_{j}^{i_{0}}\left(\xi_{1},\Delta_{1}|\xi_{0},\Delta_{0}\right) & =\prod_{k\in\partial j,k\neq i_{0}}\intop d\xi_{k}d\Delta_{k}\cdot P_{k}^{j}\left(\xi_{k},\Delta_{k}|\xi_{1},\Delta_{1}\right)\cdot P\left(\xi_{1}\right)\delta\left(\Delta_{1}-s^{i_{0}}\right)\nonumber \\
 & -\prod_{k\in\partial j,k\neq i_{0}}\intop d\xi_{k}d\Delta_{k}\cdot\sum_{k}\prod_{r\in\partial j,r\neq i_{0},k}P_{r}^{j}\left(\xi_{r},\Delta_{r}|\xi_{1},\Delta_{1}\right)\nonumber \\
 & \times\left[\frac{\partial}{\partial\Delta_{1}}\intop^{\Delta_{k}}d\Delta_{k}^{'}\cdot P_{k}^{j}\left(\xi_{k},\Delta_{k}^{'}|\xi_{1},\Delta_{1}\right)\right]\cdot\left[P\left(\xi_{1}\right)\frac{\partial}{\partial\Delta_{k}}\intop^{\Delta_{1}}d\Delta_{1}^{'}\cdot\delta\left(\Delta_{1}^{'}-s^{i_{0}}\right)\right].\label{eqapp:connection-between-modified-onsite-probabilities}
\end{align}
There are two main differences between this equation and the one for
the onsite probability density. First of all, the right-hand side
now includes the solutions to the modified problem for all neighbors
of $j$ except $i_{0}$. Secondly, the function $s^{i_{0}}$ in the
right-hand side contains the explicit dependence on the arguments
$\Delta_{0}$, $\xi_{0}$ of the target function $P_{j}^{i_{0}}$
in the left-hand side. 

At this point, one can recall that all local distributions assume
translationally invariant form in the thermodynamical limit, so that
onsite distributions in both original and modified problems are expected
to be independent on the actual position of $i_{0}$ and $k\in\partial i_{0}$.
Let us denote the corresponding functions as $P\left(\xi_{0},\Delta_{0}\right)$
and $\mathcal{P}\left(\xi_{1},\Delta_{1}|\xi_{0},\Delta_{0}\right)$,
respectively. Equation~\eqappref{connection-between-modified-onsite-probabilities}
then turns into a closed equation on $\mathcal{P}$, and Eq.~\eqappref{connection-between-onsite-probabilities}
expresses the onsite probability distribution $P$ of the original
problem via $\mathcal{P}$. After some transformations, the result
can be rewritten as:

\begin{equation}
P\left(\xi_{0},\Delta_{0}\right)=P\left(\xi_{0}\right)\cdot\intop_{\mathbb{R}}\frac{dt}{2\pi}\cdot\frac{\partial}{\partial\Delta_{0}}\left\{ \left[\intop^{\Delta_{0}}d\Delta^{'}\exp\left\{ -it\Delta^{'}\right\} \right]\cdot\left[\intop d\xi d\Delta\cdot\mathcal{P}\left(\xi,\Delta|\xi_{0},\Delta_{0}\right)\cdot\exp\left\{ itf\left(\xi,\Delta|\xi_{0}\right)\right\} \right]^{Z}\right\} ,\label{eqapp:equation-on-onsite-distribution}
\end{equation}
\begin{align}
\mathcal{P}\left(\xi_{1},\Delta_{1}|\xi_{0},\Delta_{0}\right) & =P\left(\xi_{1}\right)\cdot\intop_{\mathbb{R}}\frac{dt}{2\pi}\cdot\exp\left\{ itf\left(\xi_{0},\Delta_{0}|\xi_{1}\right)\right\} \nonumber \\
 & \times\frac{\partial}{\partial\Delta_{1}}\left\{ \left[\intop^{\Delta_{1}}d\Delta_{1}^{'}\exp\left\{ -it\Delta_{1}^{'}\right\} \right]\cdot\left[\intop d\xi d\Delta\cdot\mathcal{P}\left(\xi,\Delta|\xi_{1},\Delta_{1}\right)\cdot\exp\left\{ itf\left(\xi,\Delta|\xi_{1}\right)\right\} \right]^{Z-1}\right\} ,\label{eqapp:equation-on-modified-distribution}
\end{align}
where we made use of the integral representation of the $\delta$-function
in terms of a Fourier integral as well as the additive form of both
$s$ and $s^{i_{0}}$ functions. One then has to solve Eq.~\eqappref{equation-on-modified-distribution}
with respect to $\mathcal{P}$ and use it to calculate the actual
onsite distribution $P$ by means of Eq.~\eqappref{equation-on-onsite-distribution}.
As a final remark, we note that expressions similar to Eq.~\eqappref{equation-on-onsite-distribution}
can be obtained for joint probability distributions of fields on several
neighboring sites, allowing one to study correlations in the distribution
of the order parameter. For instance, the joint probability distribution
for a pair of two neighboring sites is expressed as
\begin{align}
P\left(\xi_{1},\Delta_{1};\xi_{2},\Delta_{2}\right) & =\mathcal{P}\left(\xi_{1},\Delta_{1}|\xi_{2},\Delta_{2}\right)\cdot\mathcal{P}\left(\xi_{2},\Delta_{2}|\xi_{1},\Delta_{1}\right)\nonumber \\
 & -\left[\frac{\partial}{\partial\Delta_{2}}\intop^{\Delta_{1}}d\Delta_{1}^{'}\mathcal{P}\left(\xi_{1},\Delta_{1}^{'}|\xi_{2},\Delta_{2}\right)\right]\left[\frac{\partial}{\partial\Delta_{1}}\intop^{\Delta_{2}}d\Delta_{2}^{'}\mathcal{P}\left(\xi_{2},\Delta_{2}^{'}|\xi_{1},\Delta_{1}\right)\right].
\end{align}

\section{Equation on distribution in the limit $\nu_{0}\Delta_{0}\ll1$, $Z\gg1$\label{app:equation-on-distribution_small-Delta}}

\subsection{Notation and relevant assumptions}

In this Appendix we analyze the distribution of the order parameter
in the limit of small $\nu_{0}\Delta$ and large $Z$. The limit is
controlled by a finite value of 
\begin{equation}
Z_{\text{eff}}=2\nu_{0}\Delta_{0}\cdot\left(Z-1\right),
\end{equation}
where $\Delta_{0}$ is the naive mean-field value of the order parameter
as defined by Eq.~\eqref{mean-field_order-parameter_eq}, and $\nu_{0}=\nu\left(0\right)$
is the single-particle density of states at the Fermi level per spin
projection. The value of $Z_{\text{eff}}$ represents the \emph{effective}
average number of neighbors for each site, that is, the number of
neighbors that have their $\xi_{i}$ values within the stripe of width
$2\Delta_{0}$ around the value of $\xi_{i}$ at the given site.

In what follows, we limit the consideration to the case of zero temperature.
We will also assume the attractive interaction to be energy-independent,
while the influence of a smooth dependence of the matrix element on
energy is discussed \appref{Effect-of-fluctuating-coupling}. The
self-consistency equation then simply reads:
\begin{equation}
\Delta_{i}=\sum_{j\in\partial i}\omega\left(z_{j}=\frac{\xi_{j}}{\Delta_{j}}\right)\cdot D\left(0\right),\,\,\,\,\omega\left(z\right)=\frac{1}{\sqrt{1+z^{2}}}.\label{eqapp:saddle-point-equation_no-D-dependence}
\end{equation}
One of the most profound consequences of the employed simplification
is that the typical scale of the order parameter is altered, as evident
already at the level of the mean-field equation~\eqappref{mean-field_order-parameter-value}.
Indeed, in the absence of the $D$ function, the single-particle density
of states $\nu$ becomes the only function to provide a high-energy
cut-off of order of $E_{F}$ for otherwise logarithmically diverging
integral in \eqappref{mean-field_order-parameter-value}. This change,
however, does not influence major low-energy properties, as described
in \appref{Effect-of-fluctuating-coupling}.

We also assert the following relation between the typical energy scales:
\begin{equation}
\Delta_{0}\ll\varepsilon_{D},E_{F}.\label{eqapp:small-delta-assumption}
\end{equation}
Here, $\Delta_{0}$ is an estimation of the typical value of the order
parameter, $\varepsilon_{D}$ is the typical scale of energy dependence
of the matrix element $D$, and $E_{F}$ is the energy scale of the
disorder distribution $\nu\left(\xi\right)$. Finally, we define the
dimensionless Cooper attraction constant, which is assumed to be small:
\begin{equation}
\lambda:=2\nu_{0}ZD\left(0\right)\ll1.
\end{equation}
In what follows, we will also make use of the following parameter:
\begin{equation}
\kappa=\frac{\lambda}{Z_{\text{eff}}}\approx\frac{D\left(0\right)}{\Delta_{0}},\label{eqapp:kappa-def}
\end{equation}
which turns out be the only qualitatively important parameter of the
low-energy theory as long as all the assumption outlined above are
fulfilled. 

\subsection{Equation on the cumulant generating function\label{subsecapp:Eq-on-the-char-function-derivation}}

We start by performing the Fourier transform on the target distribution:
\begin{equation}
R\left(T|\xi,\Delta\right):=\intop_{\mathbb{R}}d\xi_{1}\intop_{0}^{\infty}d\Delta_{1}\cdot\mathcal{P}\left(\xi_{1},\Delta_{1}|\xi,\Delta\right)\cdot\exp\left\{ iTf\left(\xi_{1},\Delta_{1}\right)\right\} ,\label{eqapp:R-definition}
\end{equation}
which corresponds to calculating the moment generating function of
the quantity $f\left(\xi_{1},\Delta_{1}\right)$ over the distribution
$\mathcal{P}\left(\xi_{1},\Delta_{1}|\xi,\Delta\right)$. The normalization
of the $\mathcal{P}$ function translates to
\begin{equation}
R\left(0|\xi,\Delta\right)=1.\label{eqapp:R-normalization}
\end{equation}
There are several important properties of the $R$ function. First
of all, since the $f$ function assumes positive values, the $R$~function
is analytical in the upper half-plane of the complex variable~$T$.
Moreover, because the probability density function of the $f$~variable
does not diverge at~$f\rightarrow0$, the corresponding moment generating
function~$R$ decays at least as $1/T$ for $\text{Im}T\rightarrow\infty$.
This fact secures convergence of the Inverse Laplace Transform required
to restore the distribution function. For example, consider the following
integral
\[
I=\intop_{\mathbb{R}+i0}dT\cdot R\left(T|\xi,\Delta\right)\cdot\phi\left(T\right),
\]
where $\phi\left(T\right)$ is some analytical function in the upper
half-plane with a decaying behavior as~$\text{Im}T\rightarrow\infty$.
Due to the described analytical properties of~$R$, one can close
the integration contour in the upper half-plane and subsequently apply
the Cauchy theorem to obtain~$I=0$. Note, however, that the \emph{typical}
value of $f$ is given by~$f_{\text{typ}}\sim D\left(0\right)/Z$,
where $Z$ is a large quantity by assumption. Therefore, there exists
a large region $\text{Im}T\ll f_{\text{\text{typ}}}^{-1}$ where the
$R$ function might demonstrate nontrivial behavior. In particular,
we will observe that in this intermediate region the $R$ function
demonstrates growth of the form
\[
R\left(i\tau|\xi,\Delta\right)\sim\tau\ln\tau.
\]
In what follows, we will be primarily interested in the intermediate
region, while bearing in mind that at large scales the expected analytical
behavior is properly restored.

After applying the integral transformation of Eq.~\eqappref{R-definition}
to the right hand side of Eq.~\eqappref{equation-on-modified-distribution}
one arrives to

\begin{align}
R\left(T|\xi,\Delta\right) & =\intop_{\mathbb{R}}d\xi_{1}\intop_{0}^{\infty}d\Delta_{1}\cdot\exp\left\{ iTf\left(\xi_{1},\Delta_{1}\right)\right\} \cdot\nu\left(\xi_{1}\right)\cdot\intop_{\mathbb{R}-i0}\frac{dt}{2\pi}\cdot\exp\left\{ itf\left(\xi,\Delta\right)\right\} \nonumber \\
 & \times\frac{\partial}{\partial\Delta_{1}}\left\{ \left[\intop_{\infty}^{\Delta_{1}}d\Delta_{1}^{'}\exp\left\{ -it\Delta_{1}^{'}\right\} \right]\cdot\left[R\left(t|\xi_{1},\Delta_{1}\right)\right]^{Z-1}\right\} .\label{eqapp:equation-on-R-function}
\end{align}
To obtain this expression, we have shifted the integration contour
over $t$ to guarantee the convergence of the integral over $\Delta_{1}$
and made an explicit choice of the lower limit of integration over
$\Delta_{1}^{'}$, see explanation under \eqappref{expression-for-delta-function-of-equations-2}.
We now seek the solution in the following form:
\begin{equation}
R\left(T|\xi,\Delta\right)=\exp\left\{ 2\nu_{0}\Delta_{0}\cdot r\left(\Delta_{0}T\left|\frac{\xi}{\Delta_{0}},\frac{\Delta}{\Delta_{0}}\right.\right)\right\} ,\label{eqapp:substitution-for-R}
\end{equation}
where $r$ is the rescaled cumulant generating function of the order
parameter distribution, as will also be demonstrated later. This function
obeys the normalization condition
\begin{equation}
r\left(0|x,y\right)=0\label{eqapp:r-normalization}
\end{equation}
as a consequence of \ref{eqapp:R-normalization}. In this substitution,
we have also introduced proper dimensionless quantities for this problem:
\begin{equation}
S=\Delta_{0}T,\,\,\,\,\,x=\xi/\Delta_{0},\,\,\,\,\,y=\Delta/\Delta_{0}.\label{eqapp:dimensionless-parametrization}
\end{equation}

The next idea is to exploit the exponential smallness of the order
parameter by treating it as the only finite energy scale in the problem.
Formally, the value of $r$ turns out to be of order unity, which
allows us to use the relation~\eqappref{small-delta-assumption}
in the form $\nu_{0}\Delta_{0}\sim\Delta_{0}/E_{F}\ll1$ and expand
the expression~\eqappref{substitution-for-R} for $R$ as
\[
\exp\left\{ 2\nu_{0}\Delta_{0}\cdot r\left(S|x,y\right)\right\} =1+2\nu_{0}\Delta_{0}\cdot r\left(S|x,y\right)+O\left(\left[\nu_{0}\Delta_{0}\right]^{2}\right).
\]
The last term is exponentially small and can be safely neglected.
This allows one to express the new $r$ function as
\begin{equation}
r\left(S|x,y\right)=\frac{R\left(S\Delta_{0}^{-1}|x\Delta_{0},y\Delta_{0}\right)-R\left(0|x\Delta_{0},y\Delta_{0}\right)}{2\nu_{0}\Delta_{0}},\label{eqapp:expression-for-r-via-R}
\end{equation}
where we have also used the normalization relation \eqappref{R-normalization}.
In order to obtain the equation on the $r$ function, we note that
the initial equation~\eqappref{equation-on-R-function} respects
the normalization condition~\eqappref{R-normalization} as a special
case for $S=\Delta_{0}T=0$. That is why, one uses the right hand
side of~\eqappref{equation-on-R-function} to express both instances
of $R$ in~\eqappref{expression-for-r-via-R} and obtains the following
equation on the $r$ function:
\begin{align}
r\left(S|x,y\right) & =\intop_{\mathbb{R}}dx_{1}\intop_{0}^{\infty}dy_{1}\cdot\left[\exp\left\{ iS\frac{f\left(\Delta_{0}x_{1},\Delta_{0}y_{1}\right)}{\Delta_{0}}\right\} -1\right]\cdot\frac{\nu\left(\Delta_{0}x_{1}\right)}{2\nu_{0}}\nonumber \\
 & \times\intop_{\mathbb{R}-i0}\frac{ds}{2\pi}\cdot\exp\left\{ is\frac{f\left(\Delta_{0}x,\Delta_{0}y\right)}{\Delta_{0}}\right\} \frac{\partial}{\partial y_{1}}\left\{ \left[\intop_{\infty}^{y_{1}}dy_{1}^{'}\exp\left\{ -isy_{1}^{'}\right\} \right]\cdot\exp\left\{ Z_{\text{eff}}\cdot r\left(s|x_{1},y_{1}\right)\right\} \right\} .\label{eqapp:equation-on-r-function}
\end{align}
In this equation, we have used the dimensionless parametrization~\eqappref{dimensionless-parametrization}
for all dummy integration variables.

\subsubsection{Excluding high-energies from the problem}

The next step is to quantify the role of $\Delta_{0}$ as the only
relevant energy scale in the low-energy physics. Indeed, as long as
the hierarchy of energy scales~\eqappref{small-delta-assumption}
takes place, the sole role of higher energy scales of order $\varepsilon_{D}$
or $E_{F}$ is to define the value of the typical scale of the order
parameter $\Delta_{0}$ via the mean-field equation~\eqref{mean-field_order-parameter_eq}.
Therefore, only the behavior of all functions at scales $\xi\sim\Delta\sim\Delta_{0}$
and $T\sim\Delta_{0}^{-1}$ should be important. However, naively
letting $\nu\left(\xi\right)\approx\nu\left(0\right)=\nu_{0}$ in
Eq.~\eqappref{equation-on-r-function} would eventually result in
the following divergence in the integral over $x_{1}$ w.r.t the upper
limit:
\begin{equation}
\intop_{\#}^{\infty}dx_{1}\left[\frac{iSc_{1}}{x_{1}/y_{1}}+O\left(\frac{1}{x_{1}^{2}}\right)\right],
\end{equation}
where $c_{1}$ is some smooth function of $x$ and $y$, but not $y_{1}$,
$x_{1}$ or $S$, and $\#$ denotes some low-energy cutoff. The origin
of this logarithmic divergence lies solely in the $1/\xi$-behavior
of the $f$ function at intermediate scales $\Delta\ll\xi\ll\omega_{D}$.
In particular, this singularity is of the very same nature as that
in the standard BCS mean-field theory. To demonstrate this, let us
rewrite the mean-field equation~\eqref{mean-field_order-parameter_eq}
in terms of dimensionless variables:
\begin{equation}
1=Z_{\text{eff}}\cdot\intop_{\mathbb{R}}dx_{1}\cdot\frac{\nu\left(\Delta_{0}x_{1}\right)}{2\nu_{0}}\cdot\frac{f\left(\Delta_{0}x_{1},\Delta_{0}\right)}{\Delta_{0}},\label{eqapp:mean-field-equation-dimensionless}
\end{equation}
where we have neglected the difference between $Z$ and $Z-1$ in
the definition of $Z_{\text{eff}}$. One can observe that due to the
form of the $f$ function, neglecting high-energy dispersion of $\nu$
in this integral produces the same type of logarithmic divergence.
From this example, we also infer that the divergence is regularized
only at $x_{1}\sim E_{F}/\Delta_{0}\gg1$ due to the properties of
the $\nu$ function in the full problem. The first idea is thus to
extract all contributions in Eq.~\eqappref{equation-on-r-function}
that are linear in~$f\left(\Delta_{0}x_{1},\Delta_{0}y_{1}\right)$
and to compensate the associated $1/\xi$~divergence by adding and
subtracting a suitable modification of the mean-field equation~\eqappref{mean-field-equation-dimensionless}.
Because $f$ enters the expression only in combination with $S$,
the exact part of~\eqappref{equation-on-r-function} containing the
divergence is linear in $S$ and reads:
\begin{align}
\Delta r & =\intop_{\mathbb{R}}dx_{1}\intop_{0}^{\infty}dy_{1}\cdot iS\frac{f\left(\Delta_{0}x_{1},\Delta_{0}y_{1}\right)}{\Delta_{0}}\frac{\nu\left(\Delta_{0}x_{1}\right)}{2\nu_{0}}\nonumber \\
 & \times\intop_{\mathbb{R}-i0}\frac{ds}{2\pi}\cdot\exp\left\{ is\frac{f\left(\Delta_{0}x,\Delta_{0}y\right)}{\Delta_{0}}\right\} \frac{\partial}{\partial y_{1}}\left\{ \left[\intop_{\infty}^{y_{1}}dy_{1}^{'}\exp\left\{ -isy_{1}^{'}\right\} \right]\cdot\exp\left\{ Z_{\text{eff}}\cdot r\left(s|x_{1},y_{1}\right)\right\} \right\} .\label{eqapp:diverging-term-1}
\end{align}
One cannot, however, immediately use the mean-field equation \eqappref{mean-field-equation-dimensionless}
to calculate this expression because it contains some residual dependence
on $x_{1}$ in $r\left(s|x_{1},y_{1}\right)$. To move forward, we
exploit the fact that the $r$ function is expected to be a smooth
function of $x_{1}$. One expects that $r\left(s|x_{1},y_{1}\right)$
quickly approaches a constant finite value at $x_{1}\gg1$ regardless
of the value of $y_{1}$, as long as the latter is of order unity.
Such an expectation stems from the fact that the right hand side of
the target equation~\eqappref{equation-on-r-function} on the $r$
function does indeed indicate such a behavior for the region $1\ll x_{1}\ll E_{D}/\Delta_{0}$.
Let us denote the corresponding value as $r\left(s|\infty,0\right)$.
The second idea is then to exploit the following asymptotic formula:
\begin{equation}
\exp\left\{ Z_{\text{eff}}\cdot r\left(s|x_{1},y_{1}\right)\right\} -\exp\left\{ Z_{\text{eff}}\cdot r\left(s|0,\infty\right)\right\} \approx Z_{\text{eff}}\cdot\left[r\left(s|x_{1},y_{1}\right)-r\left(s|0,\infty\right)\right]\sim\frac{1}{x_{1}},\,\,\,\,\,1\ll x_{1}\ll\left(\nu_{0}\Delta_{0}\right)^{-1}.
\end{equation}
Because the difference of exponents produces an extra power of $1/x_{1}$,
we conclude that the difference between $\Delta r$ and a modified
version with the additional dependence on $x_{1}$ neglected is already
a quickly converging integral. In other words, it suffices to analyze
the following modification of $\Delta r$:
\begin{align}
\Delta r_{r} & =\intop_{\mathbb{R}}dx_{1}\intop_{0}^{\infty}dy_{1}\cdot iS\frac{f\left(\Delta_{0}x_{1},\Delta_{0}y_{1}\right)}{\Delta_{0}}\frac{\nu\left(\Delta_{0}x_{1}\right)}{2\nu_{0}}\nonumber \\
 & \times\intop_{\mathbb{R}-i0}\frac{ds}{2\pi}\cdot\exp\left\{ is\frac{f\left(\Delta_{0}x,\Delta_{0}y\right)}{\Delta_{0}}\right\} \frac{\partial}{\partial y_{1}}\left\{ \left[\intop_{\infty}^{y_{1}}dy_{1}^{'}\exp\left\{ -isy_{1}^{'}\right\} \right]\cdot\exp\left\{ Z_{\text{eff}}\cdot r\left(s|\infty,0\right)\right\} \right\} ,\label{eqapp:diverging-term-2}
\end{align}
where the additional regular dependence on $x_{1}$ is neglected.
The difference between this expression and the original diverging
term~\eqappref{diverging-term-1} already gains its value in the
region $x_{1}\apprle1$, while the error term from such an approximation
is of order $\Delta_{0}/E_{D}$ and can be neglected:
\begin{equation}
\Delta r-\Delta r_{r}=\text{low-energy part}+\cancelto{\Delta_{0}/\omega_{D}}{\intop_{\#}^{\infty}dx_{1}\left[\frac{iSc_{2}}{x_{1}^{2}}+O\left(\frac{1}{x_{1}^{3}}\right)\right].}
\end{equation}
Here the notation ``low-energy part'' corresponds to the part of
the expression resulting from ignoring the high-energy behavior of
$\nu$ function. As a result, the new high-energy term \eqappref{diverging-term-2}
now reads
\begin{align}
\Delta r_{r} & =\intop_{\mathbb{R}}dx_{1}\intop_{0}^{\infty}dy_{1}\cdot iS\cdot\frac{f\left(\Delta_{0}x_{1},\Delta_{0}y_{1}\right)}{\Delta_{0}}\cdot\frac{\nu\left(\Delta_{0}x_{1}\right)}{2\nu_{0}}\nonumber \\
 & \times\intop_{\mathbb{R}-i0}\frac{ds}{2\pi}\cdot\exp\left\{ is\frac{f\left(\Delta_{0}x,\Delta_{0}y\right)}{\Delta_{0}}\right\} \cdot\exp\left\{ -isy_{1}\right\} \cdot\exp\left\{ Z_{\text{eff}}\cdot r\left(s|\infty,0\right)\right\} .\label{eqapp:diverging-term-3}
\end{align}
In order to use the exact definition \eqappref{mean-field-equation-dimensionless}
to evaluate this integral, one has to modify the second argument of
$f$. However, naively putting $y_{1}=1$ in this expression will
modify the asymptotic behavior of $f$:
\begin{equation}
\frac{f\left(\Delta_{0}x_{1},\Delta_{0}y_{1}\right)}{\Delta_{0}}\approx\kappa\frac{y_{1}}{x_{1}},\,\,\,\,\,1\ll x_{1}\ll E_{F}/\Delta_{0},\,\,\,y_{1}\sim1,\label{eqapp:f-function_large-x1-asymptotic}
\end{equation}
and the term \eqappref{diverging-term-3} would no longer be able
to serve for a counter-term to the logarithmic divergence in the target
equation \eqappref{equation-on-r-function}. One thus has to manually
fix the emerging discrepancy in the leading coefficient by considering
a yet another modification in the high-energy term:
\begin{align}
\Delta r_{rf} & =\intop_{\mathbb{R}}dx_{1}\intop_{0}^{\infty}dy_{1}\cdot iS\cdot y_{1}\frac{f\left(\Delta_{0}x_{1},\Delta_{0}\right)}{\Delta_{0}}\cdot\frac{\nu\left(\Delta_{0}x_{1}\right)}{2\nu_{0}}\nonumber \\
 & \times\intop_{\mathbb{R}-i0}\frac{ds}{2\pi}\cdot\exp\left\{ is\frac{f\left(\Delta_{0}x,\Delta_{0}y\right)}{\Delta_{0}}\right\} \cdot\exp\left\{ -isy_{1}\right\} \cdot\exp\left\{ Z_{\text{eff}}\cdot r\left(s|\infty,0\right)\right\} ,
\end{align}
which results from $\Delta r_{r}$ after putting $y_{1}=1$ in the
second argument of $f$ and subsequent multiplication by $y_{1}$.
And again, the difference between $\Delta r_{r}$ and $\Delta r_{rf}$
is already controlled by low-energies by construction:
\begin{equation}
\Delta r_{r}-\Delta r_{rf}=\text{low energy part}+\cancel{\intop_{\#}^{\infty}dx_{1}\left[iSc_{3}\cdot\left[\frac{f\left(\Delta_{0}x_{1},\Delta_{0}y_{1}\right)}{\Delta_{0}}-y_{1}\frac{f\left(\Delta_{0}x_{1},\Delta_{0}\right)}{\Delta_{0}}\right]+O\left(\frac{1}{x_{1}^{3}}\right)\right]},
\end{equation}
where the last term vanishes due to the asymptotic form~\eqappref{f-function_large-x1-asymptotic}
of the $f$ function. The value of $\Delta r_{rf}$ can already be
calculated exactly by virtue of the mean-field equation~\eqappref{mean-field-equation-dimensionless}:
\begin{align}
\Delta r_{rf} & =\intop_{0}^{\infty}dy_{1}\cdot iS\cdot y_{1}\boxed{\intop_{\mathbb{R}}dx_{1}\frac{f\left(\Delta_{0}x_{1},\Delta_{0}\right)}{\Delta_{0}}\cdot\frac{\nu\left(\Delta_{0}x_{1}\right)}{2\nu_{0}}}\nonumber \\
 & \times\intop_{\mathbb{R}-i0}\frac{ds}{2\pi}\cdot\exp\left\{ is\frac{f\left(\Delta_{0}x,\Delta_{0}y\right)}{\Delta_{0}}\right\} \cdot\exp\left\{ -isy_{1}\right\} \cdot\exp\left\{ Z_{\text{eff}}\cdot r\left(s|\infty,0\right)\right\} \nonumber \\
 & =\frac{iS}{Z_{\text{eff}}}\frac{f\left(\Delta_{0}x,\Delta_{0}y\right)}{\Delta_{0}}+S\frac{\partial r}{\partial S}\left(0|\infty,0\right),
\end{align}
where the box highlights the part that is equal to $1/Z_{\text{eff}}$
due to \eqappref{mean-field-equation-dimensionless}. To simplify
the expression after using the mean-field equation, we have used the
infinitesimal imaginary part of $s$ to calculate the integral over
$y_{1}$. After that we have used the fact that $r$ decays exponentially
in the upper half-plane of $s$ variable to close the integration
contour and evaluate the remaining integral as a residue at its only
singularity at $s=0$.

\subsubsection{Extracting the low-energy behavior}

Let us now summarize the procedure described in the previous subsection.
We have first identified a logarithmically large contribution in Eq.~\eqappref{equation-on-r-function}
produced by the $1/\xi_{1}$-behavior of the BCS root in Eq.~\eqappref{saddle-point-equation_no-D-dependence}.
The exact value of this contribution is accumulated from all energy
scales up to $E_{F}$, which prevents us from analyzing low-energy
physics right away. The problem is solved as follows:
\begin{itemize}
\item first, one identifies the contribution that produces the divergence
if the high-energy regularization is ignored. The strategy is to come
up with a proper counter-term that can be calculated with the help
of some exact identities in the theory. Remarkably, the problematic
contribution turns out to be linear in $S$ variable which indicates
that that it is responsible for the exact value of the average order
parameter, while all other moments of the distribution are completely
determined by low-energy physics.
\item The contribution in question is divergent due to the presence of the
term proportional to the integral of the $f$ function over its first
argument. In the original theory, its finite value is delivered by
mean-field self-consistency equation~\eqappref{mean-field-equation-dimensionless},
so that the latter is a viable candidate to counter the discussed
divergence. However, the target contribution also contains some residual
dependence on its arguments which forbids direct usage of the mean-field
equation. 
\item One then has to strip off the residual regular dependence of the integrand
on $\xi_{1}$ and $\Delta_{1}$ and nonessential part of the dependence
on $\xi$, and the error term from this step is already controlled
by low energies. This is achieved by successive extraction of sub-leading
terms in formal $1/\xi_{1}$ expansion of the integrand.
\item After a chain of additions and subtractions, one is left with an expression
that can be computed exactly due to the mean-field equation~\eqappref{mean-field-equation-dimensionless}. 
\end{itemize}
As a result of this manipulations, the target expression \eqappref{equation-on-r-function}
can be rewritten as a sum of four terms:
\begin{equation}
r\left(S|x,y\right)=\left[\text{target equation}-\Delta r\right]+\left[\Delta r-\Delta r_{r}\right]+\left[\Delta r_{r}-\Delta r_{rf}\right]+\Delta r_{rf},\label{eqapp:r-function_expression-after-regularization-0}
\end{equation}
\begin{align}
\left[\text{target equation}-\Delta r\right] & =\intop_{\mathbb{R}}dx_{1}\intop_{0}^{\infty}dy_{1}\cdot\left[\exp\left\{ iS\frac{f\left(\Delta_{0}x_{1},\Delta_{0}y_{1}\right)}{\Delta_{0}}\right\} -1-iS\frac{f\left(\Delta_{0}x_{1},\Delta_{0}y_{1}\right)}{\Delta_{0}}\right]\cdot\frac{\nu\left(\Delta_{0}x_{1}\right)}{2\nu_{0}}\nonumber \\
 & \times\intop_{\mathbb{R}-i0}\frac{ds}{2\pi}\cdot\exp\left\{ is\frac{f\left(\Delta_{0}x,\Delta_{0}y\right)}{\Delta_{0}}\right\} \frac{\partial}{\partial y_{1}}\left\{ \left[\intop_{\infty}^{y_{1}}dy_{1}^{'}e^{-isy_{1}^{'}}\right]\cdot\exp\left\{ Z_{\text{eff}}\cdot r\left(s|x_{1},y_{1}\right)\right\} \right\} ,\label{eqapp:r-function_expression-after-regularization-1}
\end{align}
 
\begin{align}
\left[\Delta r-\Delta r_{r}\right] & =\intop_{\mathbb{R}}dx_{1}\intop_{0}^{\infty}dy_{1}\cdot iS\frac{f\left(\Delta_{0}x_{1},\Delta_{0}y_{1}\right)}{\Delta_{0}}\frac{\nu\left(\Delta_{0}x_{1}\right)}{2\nu_{0}}\cdot\intop_{\mathbb{R}-i0}\frac{ds}{2\pi}\cdot\exp\left\{ is\frac{f\left(\Delta_{0}x,\Delta_{0}y\right)}{\Delta_{0}}\right\} \nonumber \\
 & \times\frac{\partial}{\partial y_{1}}\left[\left\{ \left[\intop_{\infty}^{y_{1}}dy_{1}^{'}e^{-isy_{1}^{'}}\right]\cdot\exp\left\{ Z_{\text{eff}}\cdot r\left(s|x_{1},y_{1}\right)\right\} \right\} -\left\{ \left[\intop_{\infty}^{y_{1}}dy_{1}^{'}e^{-isy_{1}^{'}}\right]\cdot\exp\left\{ Z_{\text{eff}}\cdot r\left(s|\infty,0\right)\right\} \right\} \right],\label{eqapp:r-function_expression-after-regularization-2}
\end{align}
\begin{align}
\left[\Delta r_{r}-\Delta r_{rf}\right] & =\intop_{\mathbb{R}}dx_{1}\intop_{0}^{\infty}dy_{1}\cdot iS\cdot\left[\frac{f\left(\Delta_{0}x_{1},\Delta_{0}y_{1}\right)}{\Delta_{0}}-y_{1}\frac{f\left(\Delta_{0}x_{1},\Delta_{0}\right)}{\Delta_{0}}\right]\cdot\frac{\nu\left(\Delta_{0}x_{1}\right)}{2\nu_{0}}\nonumber \\
 & \times\intop_{\mathbb{R}-i0}\frac{ds}{2\pi}\cdot\exp\left\{ is\frac{f\left(\Delta_{0}x,\Delta_{0}y\right)}{\Delta_{0}}\right\} \cdot e^{-isy_{1}}\cdot\exp\left\{ Z_{\text{eff}}\cdot r\left(s|\infty,0\right)\right\} ,\label{eqapp:r-function_expression-after-regularization-3}
\end{align}
\begin{equation}
\Delta r_{rf}=\frac{iS}{Z_{\text{eff}}}\frac{f\left(\Delta_{0}x,\Delta_{0}y\right)}{\Delta_{0}}+S\frac{\partial r}{\partial S}\left(0|\infty,0\right).\label{eqapp:r-function_expression-after-regularization-4}
\end{equation}
In the resulting expression, all integrals over $x_{1}$ and $y_{1}$
in Eq.~\eqappref{equation-on-r-function} are now forced to gain
their value in the region $x_{1},y_{1}\sim1$. As a result, the typical
energy scales of $\nu\left(\xi\right)$ and $D\left(\xi\right)$ are
rendered irrelevant, so that one can safely replace
\begin{equation}
\intop_{\mathbb{R}}dx_{1}\cdot\frac{\nu\left(\Delta_{0}x_{1}\right)}{2\nu_{0}}\mapsto\intop_{0}^{\infty}dx_{1},\,\,\,\,\,\frac{f\left(\Delta_{0}x_{1},\Delta_{0}y_{1}\right)}{\Delta_{0}}\mapsto\frac{D\left(0\right)}{\Delta_{0}}\omega\left(\frac{x_{1}}{y_{1}}\right)=\kappa\cdot\omega\left(\frac{x_{1}}{y_{1}}\right),
\end{equation}
where $\omega$ is the BCS root given by Eq.~\eqappref{saddle-point-equation_no-D-dependence},
and $\kappa$ is the low energy control parameter given by Eq.~\eqappref{kappa-def}.
Applying these replacements to Eq.\nobreakdash-s~(\hphantom{}\ref{eqapp:r-function_expression-after-regularization-1}\nobreakdash-\ref{eqapp:r-function_expression-after-regularization-3}\hphantom{}),
one arrives to
\begin{align}
\left[\text{target equation}-\Delta r\right] & =\intop_{0}^{\infty}dx_{1}\intop_{0}^{\infty}dy_{1}\cdot\left[\exp\left\{ iS\kappa\cdot\omega\left(\frac{x_{1}}{y_{1}}\right)\right\} -1-iS\kappa\cdot\omega\left(\frac{x_{1}}{y_{1}}\right)\right]\nonumber \\
 & \times\intop_{\mathbb{R}-i0}\frac{ds}{2\pi}\cdot\exp\left\{ is\kappa\cdot\omega\left(\frac{x}{y}\right)\right\} \frac{\partial}{\partial y_{1}}\left\{ \left[\intop_{\infty}^{y_{1}}dy_{1}^{'}e^{-isy_{1}^{'}}\right]\cdot\exp\left\{ Z_{\text{eff}}\cdot r\left(s|x_{1},y_{1}\right)\right\} \right\} ,\label{eqapp:r-function-terms_low-energy-expr-1}
\end{align}
\begin{align}
\left[\Delta r-\Delta r_{r}\right] & =\intop_{0}^{\infty}dx_{1}\intop_{0}^{\infty}dy_{1}\cdot iS\kappa\cdot\omega\left(\frac{x_{1}}{y_{1}}\right)\cdot\intop_{\mathbb{R}-i0}\frac{ds}{2\pi}\cdot\exp\left\{ is\kappa\cdot\omega\left(\frac{x}{y}\right)\right\} \nonumber \\
 & \times\frac{\partial}{\partial y_{1}}\left[\left\{ \left[\intop_{\infty}^{y_{1}}dy_{1}^{'}e^{-isy_{1}^{'}}\right]\cdot\exp\left\{ Z_{\text{eff}}\cdot r\left(s|x_{1},y_{1}\right)\right\} \right\} -\left\{ \left[\intop_{\infty}^{y_{1}}dy_{1}^{'}e^{-isy_{1}^{'}}\right]\cdot\exp\left\{ Z_{\text{eff}}\cdot r\left(s|\infty,0\right)\right\} \right\} \right],\label{eqapp:r-function-terms_low-energy-expr-2}
\end{align}
\begin{align}
\left[\Delta r_{r}-\Delta r_{rf}\right] & =\intop_{0}^{\infty}dx_{1}\intop_{0}^{\infty}dy_{1}\cdot iS\cdot\left[\omega\left(\frac{x_{1}}{y_{1}}\right)-y_{1}\omega\left(x_{1}\right)\right]\nonumber \\
 & \times\intop_{\mathbb{R}-i0}\frac{ds}{2\pi}\cdot\exp\left\{ is\kappa\cdot\omega\left(\frac{x}{y}\right)\right\} \cdot e^{-isy_{1}}\cdot\exp\left\{ Z_{\text{eff}}\cdot r\left(s|\infty,0\right)\right\} .\label{eqapp:r-function-terms_low-energy-expr-3}
\end{align}
These equations can further be simplified by noting that the dependence
$r$ on $x$ and $y$ in the right hand side of all equations is now
expressed via a single variable $w=\omega\left(x/y\right)$. This
allows one to carry out all integrals over $y_{1}$ explicitly by
making a change of variables 
\[
\left(x_{1},y_{1}\right)\mapsto\left(y_{1},z_{1}=x_{1}/y_{1}\right)\Rightarrow\frac{\partial}{\partial y_{1}}\mapsto\frac{\partial}{\partial y_{1}}-\frac{1}{y_{1}}\cdot z_{1}\frac{\partial}{\partial z_{1}},\,\,\,\,\,\intop_{0}^{\infty}dx_{1}\intop_{0}^{\infty}dy_{1}\mapsto\intop_{0}^{\infty}dz_{1}\intop_{0}^{\infty}dy_{1}\cdot y_{1}.
\]
The result reads:
\begin{align}
\left[\text{target equation}-\Delta r\right] & =\intop_{0}^{\infty}dz_{1}\cdot\left[\exp\left\{ iS\kappa\cdot\omega\left(z_{1}\right)\right\} -1-iS\kappa\cdot\omega\left(z_{1}\right)\right]\cdot\nonumber \\
 & \times\intop_{\mathbb{R}-i0}\frac{ds}{2\pi}\cdot\exp\left\{ is\kappa w\right\} \frac{1}{\left(is\right)^{2}}\left[1+z_{1}\frac{\partial}{\partial z_{1}}\right]\exp\left\{ Z_{\text{eff}}\cdot r\left(s|\omega\left(z_{1}\right)\right)\right\} ,\label{eqapp:r-function-terms_holomorphic-variable-1}
\end{align}
\begin{align}
\left[\Delta r-\Delta r_{r}\right] & =iS\kappa\cdot\intop_{0}^{\infty}dz_{1}\omega\left(z_{1}\right)\cdot\intop_{\mathbb{R}-i0}\frac{ds}{2\pi}\cdot\exp\left\{ is\kappa w\right\} \nonumber \\
 & \times\frac{1}{\left(is\right)^{2}}\left[1+z_{1}\frac{\partial}{\partial z_{1}}\right]\left[\exp\left\{ Z_{\text{eff}}\cdot r\left(s|\omega\left(z_{1}\right)\right)\right\} -\exp\left\{ Z_{\text{eff}}\cdot r\left(s|0\right)\right\} \right].\label{eqapp:r-function-terms_holomorphic-variable-2}
\end{align}
To simplify the third term~\eqappref{r-function-terms_low-energy-expr-3},
we choose to evaluate the integral over $x_{1}$ instead:
\begin{equation}
\left[\Delta r_{r}-\Delta r_{rf}\right]=iS\cdot\intop_{0}^{\infty}dy_{1}\cdot y_{1}\ln\frac{1}{y_{1}}\cdot\intop_{\mathbb{R}-i0}\frac{ds}{2\pi}\cdot\exp\left\{ is\kappa\cdot w\right\} \cdot\exp\left\{ -isy_{1}\right\} \cdot\exp\left\{ Z_{\text{eff}}\cdot r\left(s|0\right)\right\} .\label{eqapp:r-function-terms_holomorphic-variable-3}
\end{equation}
The remaining integrals over $s$ in Eq.-s \eqappref{r-function-terms_holomorphic-variable-1}
and \eqappref{r-function-terms_holomorphic-variable-2} are expressed
via a single residue at $s=0$ by means of Cauchy theorem, rendering:
\begin{equation}
\left[\text{target equation}-\Delta r\right]=\intop_{0}^{\infty}dz_{1}\cdot\left[\exp\left\{ iS\kappa\cdot\omega\left(z_{1}\right)\right\} -1-iS\kappa\cdot\omega\left(z_{1}\right)\right]\cdot\left[1+z_{1}\frac{\partial}{\partial z_{1}}\right]\left(\kappa w-iZ_{\text{eff}}\frac{\partial r}{\partial S}\left(0|\omega\left(z_{1}\right)\right)\right),\label{eqapp:r-function-terms_integrated-s-1}
\end{equation}
\begin{equation}
\left[\Delta r-\Delta r_{r}\right]=iS\kappa\cdot\intop_{0}^{\infty}dz_{1}\omega\left(z_{1}\right)\cdot\left[1+z_{1}\frac{\partial}{\partial z_{1}}\right]\left(-iZ_{\text{eff}}\frac{\partial r}{\partial S}\left(0|\omega\left(z_{1}\right)\right)+iZ_{\text{eff}}\frac{\partial r}{\partial S}\left(0|0\right)\right).\label{eqapp:r-function-terms_integrated-s-2}
\end{equation}
The integral over $s$ in Eq.~\eqappref{r-function-terms_holomorphic-variable-3}
can also be evaluated explicitly, although for now we opt to keep
it in the unevaluated form. One then performs the following simplifying
substitutions
\begin{equation}
w=\omega\left(z\right),\,\,\,\,\,Z_{\text{eff}}r\left(S|w\right)=:m\left(S|w\right).
\end{equation}
In these terms, the transformed system (\hphantom{}\ref{eqapp:r-function_expression-after-regularization-0}\nobreakdash-\ref{eqapp:r-function_expression-after-regularization-4}\hphantom{})
reads:
\begin{align}
m\left(S|w\right) & =\lambda\intop_{0}^{1}\frac{dw_{1}}{w_{1}^{2}\sqrt{1-w_{1}^{2}}}\cdot\left[\exp\left\{ iS\kappa w\right\} -1-iS\kappa w_{1}\right]\cdot\left[1-w_{1}\left(1-w_{1}^{2}\right)\frac{\partial}{\partial w_{1}}\right]\left(w-i\frac{1}{\kappa}\frac{\partial m}{\partial S}\left(0|w_{1}\right)\right)\nonumber \\
 & +iS\lambda\cdot\intop_{0}^{1}\frac{dw_{1}}{w_{1}^{2}\sqrt{1-w_{1}^{2}}}\cdot w_{1}\cdot\left[1-w_{1}\left(1-w_{1}^{2}\right)\frac{\partial}{\partial w_{1}}\right]\left(-i\frac{\partial m}{\partial S}\left(0|w_{1}\right)+i\frac{\partial m}{\partial S}\left(0|0\right)\right)\nonumber \\
 & +iS\lambda\intop_{\mathbb{R}-i0}\frac{ds}{2\pi}\cdot\exp\left\{ is\cdot\kappa w\right\} \cdot\intop_{0}^{\infty}dy_{1}\cdot y_{1}\ln\frac{1}{y_{1}}\cdot\exp\left\{ -isy_{1}\right\} \cdot\exp\left\{ m\left(s|0\right)\right\} \nonumber \\
 & +iS\kappa w+S\frac{\partial m}{\partial S}\left(0|0\right).\label{eqapp:m-function_full-expression}
\end{align}
For further analysis, it is convenient to extract the linear in $S$
term in $m$:
\begin{equation}
m\left(S|w\right):=iSm_{1}\left(w\right)+m_{2}\left(S|w\right),\label{eqapp:m-split-into-m1-and-m2}
\end{equation}
where 
\begin{equation}
m_{1}\left(w\right):=\frac{\partial m}{i\partial S}\left(0|w\right),
\end{equation}
so that $m_{2}$ has vanishing first derivative at $S=0$. The equation~\eqappref{m-function_full-expression}
then splits into two:
\begin{align}
m_{1}\left(w\right) & =m_{1}\left(0\right)+\kappa w+\lambda\cdot\intop_{0}^{1}dw_{1}\cdot\sqrt{1-w_{1}^{2}}\cdot\frac{m_{1}\left(w_{1}\right)-m_{1}\left(0\right)}{w_{1}}\nonumber \\
 & +\lambda\intop_{\mathbb{R}-i0}\frac{ds}{2\pi}\cdot\exp\left\{ is\kappa w\right\} \cdot\intop_{0}^{\infty}dy_{1}\cdot y_{1}\ln\frac{1}{y_{1}}\cdot\exp\left\{ m\left(s|0\right)-isy_{1}\right\} ,\label{eqapp:equation-on-m1}
\end{align}
\begin{equation}
m_{2}\left(S|w\right)=\lambda\cdot\intop_{0}^{1}\frac{dw_{1}}{w_{1}^{2}\sqrt{1-w_{1}^{2}}}\cdot\left[\exp\left\{ iS\kappa w_{1}\right\} -1-iS\kappa w_{1}\right]\cdot\left[1-w_{1}\left(1-w_{1}^{2}\right)\frac{\partial}{\partial w_{1}}\right]\frac{\kappa w+m_{1}\left(w_{1}\right)}{\kappa}.\label{eqapp:equation-on-m2}
\end{equation}
To simplify equation \eqappref{equation-on-m1}, the second term in
Eq.~\eqappref{m-function_full-expression} was additionally integrated
by parts with respect to $w_{1}$. This is the final stage of the
transformation. As intended, it contains only dimensionless variables
and thus describes the low-energy physics. Higher energy scales enter
the problem only via the values of the control parameters $\kappa$~and~$\lambda$.
One now has to solve the resulting pair of integral equations with
respect to $m_{1}$, after which the value of $m_{2}$ is restored
via the integral representation~\eqappref{equation-on-m2}.

As it follows from the derivation, the expressions are valid for $x,y<\varepsilon_{\max}/\Delta_{0}$,
where $\varepsilon_{\max}$ is the high-energy cut-off of the mean
field equation. In the simplistic model considered above with no energy
dependence of the interaction matrix element $D$, one has $\varepsilon_{\max}=E_{F}$
as governed by the single-particle density of states $\nu$. As discussed
in \appref{Effect-of-fluctuating-coupling}, the actual value of $\varepsilon_{\max}$
is given by the characteristic scale $\varepsilon_{D}$ of the $D$~function.
We also emphasize that it is the value of $m_{1}$ that collects all
information about the high-energy physics. Indeed, all high-energy
terms and their regulators eventually found their way only into the
expression for~$m_{1}$, while the remaining part of the $m$~function
is restored from the form of~$m_{1}$.

\subsection{Expressions for distribution functions\label{subsecapp:Expressions-for-distributions}}

One can restore the probability distributions of interest by using
the original set of integral equations~(\hphantom{}\ref{eqapp:equation-on-onsite-distribution}\nobreakdash-\ref{eqapp:equation-on-modified-distribution}\hphantom{}).
After some algebra, the results read:

\begin{align}
\mathcal{P}\left(x_{1},y_{1}|w_{0}\right) & =P\left(x_{1}\right)\cdot\intop_{\mathbb{R}-i0}\frac{ds}{2\pi}\cdot\exp\left\{ is\kappa w_{0}\right\} \cdot\frac{\partial}{\partial y_{1}}\left\{ \left[\intop_{\infty}^{y_{1}}dy_{1}^{'}\exp\left\{ -isy_{1}^{'}\right\} \right]\cdot\exp\left\{ m\left(s|\omega\left(\frac{x_{1}}{y_{1}}\right)\right)\right\} \right\} ,\label{eqapp:expression-for-modified-distribution_small-delta}
\end{align}

\begin{equation}
P\left(x,y\right)=P\left(x\right)\cdot\intop_{\mathbb{R}-i0}\frac{ds}{2\pi}\cdot\frac{\partial}{\partial y}\left\{ \left[\intop_{\infty}^{y}dy^{'}\exp\left\{ -isy^{'}\right\} \right]\cdot\exp\left\{ m\left(s|\omega\left(\frac{x}{y}\right)\right)\right\} \right\} ,\label{eqapp:expression-for-onsite-probability_small-delta}
\end{equation}
where $P\left(x\right)=\Delta_{0}\cdot\nu\left(\xi=\Delta_{0}x\right)$.
In this approximation, we are using $x$ and $y$ as the arguments
of the distribution functions in the sense that the corresponding
probability measure is given by $P\left(x,y\right)dxdy$. According
to the original equation~\eqappref{equation-on-modified-distribution},
the $\mathcal{P}$ function contains dependence on both $x_{0}$ and
$y_{0}$, but for $x_{0},y_{0}\sim1$ it can be collected into a single
variable $w_{0}=\omega\left(x_{0}/y_{0}\right)$, in the same fashion
as it takes place for functions $r_{1}$ and $r_{2}$.

One can further notice that the difference of $P\left(x,y\right)$
with $P\left(\infty,y\right)$ is only present in a small vicinity
of $x\sim1$ and decays quickly, so that the marginal distribution
of the order parameter alone formally coincide with the $x=\infty$
limit of the conditional distribution:

\begin{equation}
P\left(y\right):=\intop_{\mathbb{R}}P\left(\xi,\Delta_{0}y\right)d\xi=\intop_{\mathbb{R}-i0}\frac{dS}{2\pi}\cdot\exp\left\{ m\left(S|0\right)-iSy\right\} .\label{eqapp:marginal-probability-distribution_small-dela}
\end{equation}
Alternatively, the same result can be demonstrated by a direct calculation.
It is still implied that the probability measure to be used in any
sort of averaging is $P\left(y\right)dy$, so that the probability
density function of the dimension-full order parameter is restored
as
\begin{equation}
P\left(\Delta\right)=\frac{1}{\Delta_{0}}P\left(y=\frac{\Delta}{\Delta_{0}}\right).
\end{equation}
Note that in general it is not correct to let $x=\infty$ directly
in the expression~\eqappref{expression-for-onsite-probability_small-delta}
for the joint probability distribution. Despite the fact that the
onsite correlation between $\xi$ and $\Delta$ is only visible in
a small region $\xi\sim\Delta$, this is the defining region for all
quantities with a typical energy scale of the order of $\Delta$.

With expression \eqappref{marginal-probability-distribution_small-dela}
at hand, one can now observe that the value of $m_{1}\left(0\right)$
is directly connected to the mean value of the order parameter:
\begin{equation}
\left\langle y\right\rangle =m_{1}\left(0\right),\label{eqapp:m10-expression_average-form}
\end{equation}
where $\left\langle \bullet\right\rangle $ now stands for the average
over the distribution of the order parameter as given by \eqappref{marginal-probability-distribution_small-dela}.
More generally, the function $m\left(S|0\right)$ is the cumulant
generating function of the dimensionless order parameter:
\begin{equation}
m\left(S|0\right)=\ln\left\langle e^{iSy}\right\rangle .
\end{equation}
Going further, the equation \eqappref{equation-on-m1} on $m_{1}\left(w\right)$
can also be rewritten as
\begin{equation}
m_{1}\left(w\right)=\left\langle y\right\rangle +\kappa w+\lambda\cdot\left\langle \left(y+\kappa w\right)\ln\frac{1}{y+\kappa w}-y\ln\frac{1}{y}\right\rangle .\label{eqapp:m1-expression_average-form}
\end{equation}
Letting $w=0$ provides the following self-consistency equation for
the value of $\left\langle y\right\rangle $:
\begin{equation}
0=\lambda\cdot\intop_{0}^{1}dw_{1}\cdot\sqrt{1-w_{1}^{2}}\cdot\frac{m_{1}\left(w_{1}\right)-m_{1}\left(0\right)}{w_{1}}+\lambda\cdot\left\langle y\ln\frac{1}{y}\right\rangle ,\label{eqapp:equation-on-the-mean-value-1}
\end{equation}
or, after substituting the functional form of $m_{1}\left(w\right)$:
\begin{equation}
0=\frac{\pi}{4}\left(1-\lambda\ln\kappa\right)+\left\langle \frac{y}{\kappa}\right\rangle \ln\frac{1}{\kappa}+\left\langle \frac{y}{\kappa}\ln\frac{\kappa}{y}+\lambda\cdot G\left(y/\kappa\right)\right\rangle ,\label{eqapp:self-consistency-equation}
\end{equation}
where $G\left(u=y/\kappa\right)$ is the following special function:
\begin{align}
G\left(u\right) & =\intop_{0}^{1}dw\cdot\sqrt{1-w^{2}}\cdot\frac{\left(u+w\right)\ln\frac{1}{u+w}-u\ln\frac{1}{u}}{w}\nonumber \\
 & =\frac{\pi}{4}\left(\ln2+\frac{1}{2}\right)+\frac{1}{8}\left(2\pi u^{2}-\left(4+\pi^{2}\right)u+4u\left[\text{arccos}\frac{1}{u}\right]^{2}-\left[u\sqrt{u^{2}-1}-\ln\left(u-\sqrt{u^{2}-1}\right)\right]\text{arccos}\frac{1}{u}\right)\nonumber \\
 & +\text{Im}\left\{ \text{Li}_{2}\left(\frac{-i+\sqrt{u^{2}-1}}{u}\right)-\text{Li}_{2}\left(\frac{i-\sqrt{u^{2}-1}}{u}\right)-\frac{i\pi}{4}\ln\left(\frac{-u^{2}+2\sqrt{1-u^{2}}+2}{u^{2}}\right)\right\} ,\label{eqapp:G-function-definition}
\end{align}
with $\text{Li}_{a}\left(z\right)$ being the generalized polylogarithm
function defined as
\begin{equation}
\text{Li}_{a}\left(z\right)=\sum_{k=1}^{\infty}\frac{z^{k}}{k^{a}}.\label{eqapp:polylog-function_definition}
\end{equation}
The presented expression for $G$ is valid for $u>0,$ while the integral
representation for $G$ is a holomorphic function of $u\in\mathbb{C}$
with a branch cut along $\left[-\infty,0\right]$.

We conclude this subsection by noting that the self-consistency equation~\eqappref{self-consistency-equation}
is the only trace of high energy physics. Indeed, the only role of
this equation is to define the exact value of $m_{1}\left(0\right)\equiv\left\langle y\right\rangle $.
Our derivation indicates that the high-energy physics takes essential
part in the formation of this mean value, so that equation~\eqappref{self-consistency-equation}
is a counterpart of the mean field self-consistency equation~\eqappref{mean-field_order-parameter-value}
in the conventional BCS theory. However, once the exact value of $\left\langle y\right\rangle $
is specified by whatever mechanism, equations \eqappref{m1-expression_average-form}~and~\eqappref{equation-on-m2}
define the entire cumulant generating function without any influence
of large energy scales. In other words, all statistical properties
of the order parameters are entirely defined by the parameters $\lambda$,
$\kappa$ and the value of $\left\langle y\right\rangle $. 

\subsection{Gaussian limit $Z_{\text{eff}}\gg1$\label{subsecapp:Gaussian-limit}}

Within our model, the limit of $Z_{\text{eff}}=2\nu_{0}\Delta_{0}\cdot\left(Z-1\right)\gg1$
corresponds to conventional BCS-like theory with relatively weak disorder
and nearly homogeneous order parameter. In this limit, the distribution
of the order parameter is nearly Gaussian with mean value close to
the mean-field order parameter defined by Eq.~\eqref{mean-field_order-parameter_eq},
and the fluctuations are suppressed as $Z_{\text{eff}}^{-1}$. Below
we demonstrate how our results reduce to a simple Gaussian distribution
for the case of large $Z_{\text{eff}}$ as a manifestation of the
central limit theorem applied to the original self-consistency equation~\eqappref{saddle-point-equation_no-D-dependence}.

At its heart, the analysis of this case amounts to applying the saddle-point
approximation to all integrals over $s$. By doing so, we essentially
replace all distributions of the order parameter with some version
of a Gaussian distribution, which is entirely consistent with the
central limit theorem applied to the original set of equations~\eqref{saddle-point_order-parameter}.
In the limit of large $Z_{\text{eff}}$, the position of the saddle
point is in some sense close to $S=0$ and is thus governed by the
behavior of the leading terms of $m$ in its expansion in powers of
$S$. The whole theory thus reduces to a set of algebraic equations
on the values of leading moments of the distribution, which are precisely
the leading Taylor coefficients of $m\left(S|0\right)$. 

From the formal point of view, we can seek the solution in the following
form:
\begin{equation}
m_{2}\left(S|w\right)=\frac{1}{2}\lambda\kappa\cdot\mu_{2}\left(w\right)\cdot\left(iS\right)^{2}+O\left(\left(iS\right)^{3}\right),\label{eqapp:m2-small-kappa_expansion}
\end{equation}
where $\mu_{2}$ is some function of order unity, which is to be verified
later. We now substitute this ansatz in the equation~\eqappref{equation-on-m1}
on $m_{1}$ and evaluate the integral over $s$ in the third term:
\begin{align}
m_{1}\left(w\right) & =m_{1}\left(0\right)+\kappa w+\lambda\cdot\intop_{0}^{1}dw_{1}\cdot\sqrt{1-w_{1}^{2}}\cdot\frac{m_{1}\left(w_{1}\right)-m_{1}\left(0\right)}{w_{1}}\nonumber \\
 & +\lambda\intop_{0}^{\infty}dy_{1}\cdot y_{1}\ln\frac{1}{y_{1}}\cdot\frac{1}{\sqrt{2\pi\lambda\kappa\mu_{2}\left(0\right)}}\exp\left\{ -\frac{\left[y_{1}-\left(m_{1}\left(0\right)+\kappa w\right)\right]^{2}}{2\lambda\kappa\mu_{2}\left(0\right)}\right\} .s\label{eqapp:equation-on-m1_Gaussian-limit}
\end{align}
Because $\mu_{1}$ is of order unity and $\kappa=\lambda/Z_{\text{eff}}\ll1$,
the remaining integral over $y_{1}$ is governed by a small region
$\left|y-m_{1}\left(0\right)-\kappa w\right|\sim\sqrt{\lambda\kappa\mu_{2}}\ll1$,
so that one can expand the logarithmic part of the integrand around
the center of this region. Additionally, it becomes obvious that the
function $m_{1}\left(w\right)$ has a typical scale $w\sim\kappa^{-1}\gg1$.
However, the value of $w$ itself only assumes values in the interval
$\left[0,1\right]$. That is why, one can only retain the leading
powers of $\kappa w\ll1$ by replacing $m_{1}\left(w\right)$ with
a linear function:
\begin{equation}
m_{1}\left(w\right)=m_{10}+m_{11}\cdot\kappa w,
\end{equation}
where the coefficients $m_{10},m_{11}\sim1$ are determined directly
from Eq.~\eqappref{equation-on-m1_Gaussian-limit}:
\begin{align}
m_{11} & =1-\lambda,\,\,\,\,\,m_{10}=1+\frac{\pi}{4}\kappa m_{11}.\label{eqapp:m1_large-Keff-limit}
\end{align}
The value of $\mu_{2}\left(w\right)$ is now deduced from the direct
expansion of the equation~\eqappref{equation-on-m2} on $m_{2}\left(S|w\right)$
with the use of approximate expression for $m_{1}$:
\begin{align}
\mu_{2}\left(w\right) & \equiv-\frac{1}{\lambda\kappa}\frac{\partial^{2}m_{2}}{\partial S^{2}}\left(0|w\right)=-\kappa\cdot\intop_{0}^{1}\frac{dw_{1}}{w_{1}^{2}\sqrt{1-w_{1}^{2}}}\cdot\left(iw_{1}\right)^{2}\cdot\left[1-w_{1}\left(1-w_{1}^{2}\right)\frac{\partial}{\partial w_{1}}\right]\frac{\kappa w+m_{1}\left(w_{1}\right)}{\kappa}\nonumber \\
 & =\left(m_{10}+\kappa w\right)\cdot\intop_{0}^{1}\frac{dw_{1}}{\sqrt{1-w_{1}^{2}}}+\kappa m_{11}\cdot\intop_{0}^{1}\frac{w_{1}^{3}dw_{1}}{\sqrt{1-w_{1}^{2}}}=\left(m_{10}+\kappa w\right)\cdot\frac{\pi}{2}+\kappa m_{11}\cdot\frac{2}{3}.\label{eqapp:m2_large-Keff-limit}
\end{align}
Evidently, the assumptions about the values of $\mu_{2}$ and $m_{1}$
turned out to be correct.

The region of applicability of this result is governed by the behavior
of the corresponding integral~\eqappref{marginal-probability-distribution_small-dela}
over $S$. The latter gains its value in the region described by the
condition that the $m$ functions reaches the value of order unity,
viz:
\begin{equation}
\left|\lambda\kappa\cdot\mu_{2}\left(w\right)\cdot\left(iS\right)^{2}\right|\sim1\Leftrightarrow S\sim\frac{1}{\sqrt{\lambda\kappa}},
\end{equation}
where we have taken into account that $\mu_{2}\sim1$. The proposed
expansion~\eqappref{m2-small-kappa_expansion} is applicable whenever
higher order corrections to it are small in the relevant region. The
latter can be estimated in a way similar to that for $\mu_{2}$:
\begin{equation}
m_{2}\left(S|w\right)=\frac{1}{2}\frac{\lambda}{\kappa}\cdot\mu_{2}\left(w\right)\cdot\left(i\kappa S\right)^{2}+\frac{1}{6}\frac{\lambda}{\kappa}\cdot\left(i\kappa S\right)^{3}\cdot\mu_{3}\left(w\right)+O\left(S^{4}\right),
\end{equation}
\begin{align}
\mu_{3}\left(w\right) & \equiv\frac{1}{i^{3}\cdot\lambda\kappa^{2}}\cdot\frac{\partial^{3}m_{2}}{\partial S^{3}}\left(0|w\right)=\kappa\cdot\intop_{0}^{1}\frac{dw_{1}}{w_{1}^{2}\sqrt{1-w_{1}^{2}}}\cdot w_{1}^{3}\cdot\left[1-w_{1}\left(1-w_{1}^{2}\right)\frac{\partial}{\partial w_{1}}\right]\frac{\kappa w+m_{1}\left(w_{1}\right)}{\kappa}\nonumber \\
 & =\left(m_{10}+\kappa w\right)\cdot\intop_{0}^{1}\frac{w_{1}dw_{1}}{\sqrt{1-w_{1}^{2}}}+\kappa m_{11}\cdot\intop_{0}^{1}\frac{w_{1}^{4}dw_{1}}{\sqrt{1-w_{1}^{2}}}=m_{10}+\frac{3\pi}{16}\cdot\kappa m_{11}+\kappa w.
\end{align}
One observes that $\mu_{3}\sim1$, and we conclude that the criteria
of applicability reads:
\begin{equation}
\left|\frac{\lambda}{\kappa}\cdot\left(i\kappa S\right)^{3}\right|_{S\sim\frac{1}{\sqrt{\lambda\kappa}}}\ll1\Leftrightarrow\kappa\ll\lambda,
\end{equation}
which is consistent with the purely physical argument based on the
effective number of neighbors:
\begin{equation}
\kappa\ll\lambda\Leftrightarrow Z_{\text{eff}}=Z\cdot2\nu_{0}\Delta_{0}\gg1.
\end{equation}

We conclude this subsection by noting that the values of $m_{1}\left(0\right)$
and $\mu_{2}\left(0\right)$ are consistent with a direct perturbative
expansion of the initial self-consistency equation~\eqappref{saddle-point-equation_no-D-dependence}
around the mean value defined by Eq.~\eqappref{mean-field-equation-dimensionless}.
For instance, one can calculate the dispersion in the leading order
of small parameter $\kappa$ as
\begin{equation}
\frac{\left\langle \left\langle \Delta^{2}\right\rangle \right\rangle }{\Delta_{0}^{2}}=\frac{\left\langle \Delta^{2}\right\rangle -\left\langle \Delta\right\rangle ^{2}}{\Delta_{0}^{2}}=\frac{1}{\Delta_{0}^{2}}\sum_{k=1}^{Z}\left\langle \left\langle f^{2}\left(\xi,\Delta_{0}|\xi_{0}\right)\right\rangle \right\rangle _{\xi}\approx\frac{\pi}{2}\frac{\lambda^{2}}{Z\nu_{0}\Delta_{0}}=\frac{\pi}{2}\kappa\lambda,
\end{equation}
where $\left\langle \left\langle \bullet\right\rangle \right\rangle $
denotes the corresponding cumulant. Within our approach, one can trivially
verify the same value of the standard deviation:
\begin{equation}
\frac{\left\langle \left\langle \Delta^{2}\right\rangle \right\rangle }{\Delta_{0}^{2}}=\left\langle \left\langle y^{2}\right\rangle \right\rangle \equiv-\frac{\partial^{2}m_{2}}{\partial S^{2}}\left(0|w\right)=\lambda\kappa\mu_{2}\left(0\right)=\lambda\kappa\frac{\pi}{2}\left(1+O\left(\kappa\right)\right).\label{eqapp:Gaussian-limit-dispersion}
\end{equation}
In order to extract sub-leading in $\kappa$ effects directly from
the self-consistency equation, additional technical effort is required.
On the other hand, our approach provides a straightforward procedure
in the form of direct expansion of equations~(\hphantom{}\ref{eqapp:equation-on-m1}\nobreakdash-\ref{eqapp:equation-on-m2}\hphantom{})
in powers of $\kappa\ll1$ up to the required order.

\section{Extreme value statistics\label{app:Extreme-value-statistics}}

In this Appendix, we analyze the asymptotic behavior of the probability
density function in order to obtain extreme value statistics. This
is done by means of saddle-point analysis of the integral in Eq.~\eqappref{marginal-probability-distribution_small-dela}.
The corresponding estimation is composed as
\begin{equation}
P\left(y\right)\approx\sum_{n}\frac{1}{\sqrt{2\pi\cdot\left[-\partial^{2}m_{2}/\partial^{2}s\left(s_{n}|0\right)\right]}}\cdot\exp\left\{ m_{2}\left(s_{n}|0\right)-is_{n}\cdot\left(y-\left\langle y\right\rangle \right)\right\} ,\label{eqapp:P-integral_saddle-point-estimation}
\end{equation}
where we have used that $\left\langle y\right\rangle =m_{1}\left(0\right)$
and $\partial^{2}m/\partial s^{2}\equiv\partial^{2}m_{2}/\partial s^{2}$,
and each saddle point $s_{n}$ is a solution of the following equation:
\begin{equation}
\frac{\partial m_{2}}{\partial s}\left(s_{n}|0\right)=i\left(y-\left\langle y\right\rangle \right).\label{eqapp:P-integral_saddle-point-equation}
\end{equation}
The applicability of such an approximation is controlled by the subleading
terms in expansion of $m_{2}$ around $s_{0}$. The corresponding
criteria reads
\begin{equation}
\left|\frac{\partial^{3}m_{2}}{\partial s^{3}}\cdot\left(\frac{\partial^{2}m_{2}}{\partial s^{2}}\right)^{-3/2}\right|_{s=s_{n}}\ll1.\label{eqapp:P-integral_saddle-point-criteria}
\end{equation}
In practice, it always corresponds to some requirement on the value
of $\left|y-\left\langle y\right\rangle \right|$.

\begin{figure}
\begin{centering}
\includegraphics[scale=0.4]{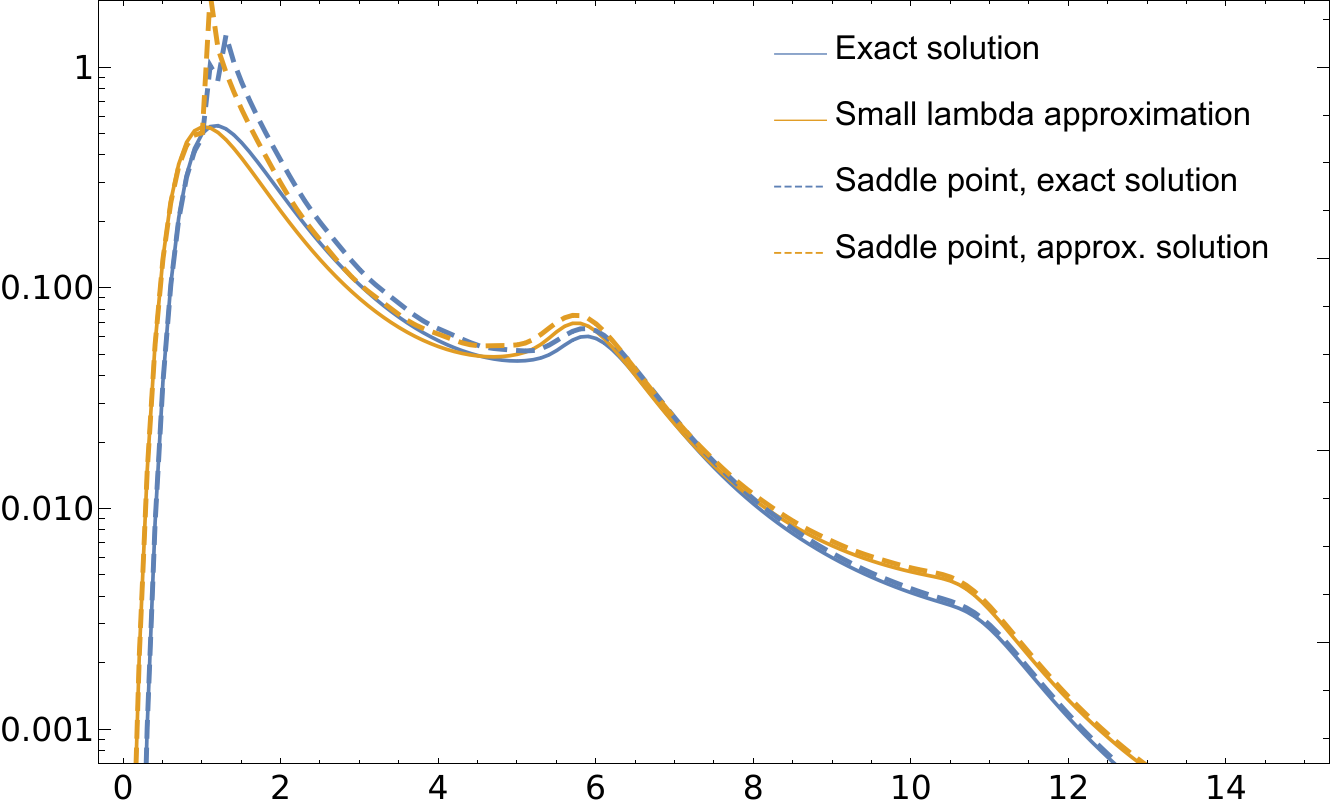}
\par\end{centering}
\caption{A log-scaled plot of the probability density function $P\left(y\right)$
calculated by four different methods. The solid lines represent direct
numerical integration of Eq.~\eqappref{marginal-probability-distribution_small-dela}
with the use of the exact (blue) and approximate (orange) solution
to the integral equations~(\protect\hphantom{}\ref{eqapp:equation-on-m1}\protect\nobreakdash-\ref{eqapp:equation-on-m2}\protect\hphantom{}).
The approximate solution is described in detail in \appref{Solution_small-lambda}.
The dashed lines correspond to evaluation of integral~\eqappref{marginal-probability-distribution_small-dela}
within the saddle-point approximation, with the exact numerical solution
for the $m$~function used to determine the position of all the saddle
points and the associated contribution to the integral. The peak at
$y\sim1$ results from vanishing of the second derivative of the integrand
at the saddle points for $y=\left\langle y\right\rangle $, where
the saddle-point approximation ceases to be applicable. For $y\apprle\left\langle y\right\rangle $
the exact and approximate curves coincide, thus demonstrating a perfect
agreement. For $y\apprge\left\langle y\right\rangle $, the saddle-point
approximation comprises multiple saddle points, as described in \subsecappref{Probability-large-y-asymptotic}.
The presented curves correspond to taking into account $n=7$ leading
contributions. Using a smaller number produces oscillations in the
region $2\apprle y\apprle4$, while further increase of $n$ does
not lead to any noticeable change in the form of the curve. For $y\apprge6$,
the the approximate result is in good agreement with the exact value.
The parameters of the model are $\lambda=0.12$ and $Z=51$, rendering
$\kappa\approx5.0$. \label{figapp:saddle-point-approx}}
\end{figure}

In both cases of large and small values of $y$, the leading contribution
corresponds to a saddle point $s_{0}$ that turns out to be purely
imaginary and large in absolute value. This fact allows one extract
the corresponding asymptotic behavior from the integral equations~(\hphantom{}\ref{eqapp:equation-on-m1}\nobreakdash-\ref{eqapp:equation-on-m2}\hphantom{})
on $m_{1}$,~$m_{2}$. A representative result of such approach is
presented on \figappref{saddle-point-approx}, and below we extract
analytical behavior of the large- and low-$y$ tails of the distribution
and overview the key qualitative features of the result.

\subsection{Probability function for $y\apprle\left\langle y\right\rangle $\label{subsecapp:Probability-small-y-asymptotic}}

For large arguments $y\gg1$, the only relevant saddle point is in
the upper half-plane. The asymptotic behavior of the $m$~function
for arguments with large positive imaginary part can be calculated
directly from Eq.~\eqappref{equation-on-m2}. The corresponding integral
over $w_{1}$ gains its value in the region $w_{1}\sim1/\left|i\kappa S\right|$
and reads:
\begin{align}
\frac{m_{2}\left(S|w\right)}{\lambda} & =\frac{\kappa w+m_{1}\left(0\right)}{\kappa}\cdot\intop_{0}^{1}dw_{1}\cdot\left[\frac{\exp\left\{ iS\kappa w_{1}\right\} -1-iS\kappa w_{1}}{w_{1}^{2}\sqrt{1-w_{1}^{2}}}\right]+O\left(\frac{1}{iS\kappa}\right)\nonumber \\
 & +\intop_{0}^{1}\frac{dw_{1}}{w_{1}^{2}\sqrt{1-w_{1}^{2}}}\left[\left(-iS\kappa w_{1}\right)-1\right]\left[1-w_{1}\left(1-w_{1}^{2}\right)\frac{\partial}{\partial w_{1}}\right]\left(\frac{m_{1}\left(w_{1}\right)-m_{1}\left(0\right)}{\kappa}\right).
\end{align}
The second term can be further transformed via integration by parts
to render
\begin{equation}
m_{2}\left(S|w\right)=Z_{\text{eff}}\left(\kappa w+\left\langle y\right\rangle \right)\cdot\left(-iS\kappa\right)\left(\ln\left(-2iS\kappa\right)+\gamma-1\right)-\left(-iS\kappa\right)\cdot Z_{\text{eff}}\left\langle y\ln\frac{1}{y}\right\rangle +O\left(\frac{1}{\kappa S}\right),\,\,\,\,\text{\text{Im}}\left\{ \kappa S\right\} \gg1,\label{eqapp:m2-asymptotic_upper-half-plane}
\end{equation}
where $\gamma\approx0.577...$ is the Euler--Mascheroni constant,
the value of $m_{1}\left(0\right)$ was replaced with $\left\langle y\right\rangle $,
and the second term $\propto\left\langle y\ln1/y\right\rangle $ was
obtained after using Eq.~\eqappref{equation-on-the-mean-value-1}.
The saddle-point equation~\eqappref{P-integral_saddle-point-equation}
possesses a single purely imaginary solution:
\begin{equation}
s_{0}=\frac{i}{2\kappa}\exp\left\{ \frac{1}{\lambda}\left(1-\frac{y}{\left\langle y\right\rangle }\right)-\frac{\left\langle y\ln y\right\rangle }{\left\langle y\right\rangle }-\gamma\right\} .
\end{equation}
The corresponding estimation~\eqappref{P-integral_saddle-point-estimation}
for the probability reads:
\begin{equation}
P\left(y\right)=\sqrt{\frac{\zeta\left(y\right)}{2\pi\cdot\left[\lambda\left\langle y\right\rangle \right]^{2}}}\cdot\exp\left\{ -\zeta\left(y\right)\right\} ,\label{eqapp:marginal-probability_small-value-asymptotic}
\end{equation}
\begin{equation}
\zeta\left(y\right)=-i\lambda s_{0}\left\langle y\right\rangle =\frac{Z_{\text{eff}}\left\langle y\right\rangle }{2}\exp\left\{ \frac{1}{\lambda}\left(1-\frac{y}{\left\langle y\right\rangle }\right)-\frac{\left\langle y\ln y\right\rangle }{\left\langle y\right\rangle }-\gamma\right\} .\label{eqapp:marginal-probability_small-value-asympt_zeta-def}
\end{equation}
The expression turns out to be an excellent approximation for the
true value of the probability, as seen e.g. on \figref{P0-log-plot_with-asymptotics}.

We also note that the double-exponential behavior is to be expected
as the observed profile of the distribution is delivered by a certain
kind of disorder configurations. Indeed, as it can be seen from the
original self-consistency equation~\eqref{saddle-point_order-parameter},
the only way to produce a low value of the order parameter in a given
site is to have the values of $\xi$ on \emph{all neighboring sites}
much larger than the typical order parameter $\Delta_{0}$. In this
case, one can linearize the self-consistency equation~\eqref{saddle-point_order-parameter}
by neglecting the order parameter in the denominator of the right
hand side. For the purpose of estimations, we will also estimate the
order parameter $\Delta_{j}$ in the numerator as $\left\langle \Delta\right\rangle $,
as the latter has a narrow distribution compared to that of $1/\left|\xi\right|$
and can thus be approximated by its mean value. As a result, the order
parameter is roughly given by the following expression:
\begin{equation}
\Delta_{i}\sim D\left(0\right)\sum_{j=1}^{Z}\frac{\left\langle \Delta\right\rangle }{\left|\xi_{j}\right|}.\label{eqapp:extreme-value-statistics_small-Delta-approx-expr}
\end{equation}
Because of the large number of terms in this sum, it can be estimated
by replacing the sum with the average over the distribution of $\xi$:
\begin{equation}
\Delta\sim D\left(0\right)\sum_{j=1}^{Z}\frac{\left\langle \Delta\right\rangle }{\left|\xi_{j}\right|}\sim2Z\left\langle \Delta\right\rangle D\left(0\right)\intop_{\left|\xi\right|_{\min}}^{\infty}\frac{\nu\left(\xi\right)d\xi}{\xi}=\left\langle \Delta\right\rangle \cdot2\nu_{0}ZD\left(0\right)\cdot\ln\frac{E_{1}}{\left|\xi\right|_{\min}},
\end{equation}
where $\left|\xi\right|_{\min}=\min\left|\xi_{j}\right|$ is the minimum
absolute value of the onsite energy among the neighboring sites, and
$E_{1}$ is some high energy cutoff. It can be estimated by plugging
the average order parameter in both the left hand side and instead
of $\left|\xi\right|_{\min}$ in the right hand side:
\[
\left\langle \Delta\right\rangle =\left\langle \Delta\right\rangle \cdot2\nu_{0}ZD\left(0\right)\cdot\ln\frac{E_{1}}{\left\langle \Delta\right\rangle }.
\]
The resulting estimation for the order parameter then reads:
\begin{equation}
\Delta\sim\left\langle \Delta\right\rangle \lambda\cdot\left[\ln\frac{\left\langle \Delta\right\rangle }{\left|\xi\right|_{\min}}+\frac{1}{\lambda}\right],
\end{equation}
where we have also taken into account that $\lambda=2\nu_{0}ZD\left(0\right)$.
In order for the resulting value of order parameter to be smaller
than $\Delta$, one thus needs
\begin{equation}
\left|\xi\right|_{\min}\apprge\left\langle \Delta\right\rangle \exp\left\{ \frac{1}{\lambda}\left(1-\frac{\Delta}{\left\langle \Delta\right\rangle }\right)\right\} .
\end{equation}
The distribution of the quantity $\left|\xi\right|_{\min}$ can approximately
be described by a Poisson distribution for sufficiently small values
of $\xi$:
\begin{equation}
P\left(\min\left|\xi_{j}\right|>E\right)=\left[1-F\left(E\right)\right]^{Z}\approx\exp\left\{ -ZF\left(E\right)\right\} ,
\end{equation}
where $F\left(E\right)=\text{Prob}\left(\left|\xi\right|<E\right)$
is the distribution function of the disorder field $\xi$ that can
be approximated as
\begin{equation}
F\left(E\right)=2\intop_{0}^{E}\nu\left(\xi\right)d\xi\approx2\nu_{0}E,\,\,\,\,\,E\ll E_{F}.
\end{equation}
Here we have taken into account that $\left|\xi\right|_{\min}$ is
still much smaller than $E_{F}$, so that the density of states $\nu\left(\xi\right)$
was replaced with a constant value. As a result, one obtains the estimation
for the probability:
\begin{equation}
\text{Prob}\left(\Delta<\overline{\Delta}\right)\sim\text{Prob}\left(\min\xi>\Delta_{0}\exp\left\{ \frac{1}{\lambda}\left(1-\frac{\overline{\Delta}}{\left\langle \Delta\right\rangle }\right)\right\} \right)\approx\exp\left[-Z_{\text{eff}}\exp\left\{ \frac{1}{\lambda}\left(1-\frac{\overline{\Delta}}{\left\langle \Delta\right\rangle }\right)\right\} \right],
\end{equation}
which thus reproduces the asymptotic result~(\hphantom{}\ref{eqapp:marginal-probability_small-value-asymptotic}\nobreakdash-\ref{eqapp:joint-probability_small-value-asymptotic}\hphantom{})
up to prefactor $1/2$ in the exponent and other sub-exponential prefactors
arising due to a crude estimation of the sum in Eq.~\eqappref{extreme-value-statistics_small-Delta-approx-expr}.

In a similar vein, one can use Eq.~\eqappref{expression-for-onsite-probability_small-delta}
to obtain the asymptotic form of the joint probability distribution:
\begin{equation}
P\left(x,y\right)=P\left(x\right)\cdot\frac{\partial}{\partial y}\left\{ \intop_{\infty}^{y}dy_{1}^{'}\cdot\sqrt{\frac{\zeta\left(y|\omega\left(x/y\right)\right)}{2\pi\cdot\left[\lambda m_{1}\left(w\right)\right]^{2}}}\cdot\exp\left\{ -\zeta\left(y|\omega\left(x/y\right)\right)\right\} \right\} ,\label{eqapp:joint-probability_small-value-asymptotic}
\end{equation}
\begin{equation}
\zeta\left(y|w\right)=\frac{Z_{\text{eff}}m_{1}\left(w\right)}{2}\exp\left\{ \frac{1}{\lambda}\left(1-\frac{y}{m_{1}\left(w\right)}\right)-\frac{\left\langle y\ln y\right\rangle }{m_{1}\left(w\right)}-\gamma\right\} .
\end{equation}

Note that because $\lambda\ll1$ and $y<\left\langle y\right\rangle $,
one has $x_{0}\gg1$, and the increase of the probability function
with $y$ is very steep. As a result, the onsite joint probability
distribution $P\left(x,y\right)$ and other joint distributions of
various quantities on neighboring sites all feature strong correlations
between $\xi$ and $\Delta$ for small values of $\xi$. Indeed, the
relation between the value of $\xi$ and $\Delta$ determines the
value of the $w$ argument in the expression for $\zeta$, thus determining
the exact position of the onset of the exponential tail. 

The region of applicability of approximate expressions~(\hphantom{}\ref{eqapp:marginal-probability_small-value-asymptotic}\nobreakdash-\ref{eqapp:joint-probability_small-value-asymptotic}\hphantom{})
is controlled by two factors. First of all, the width of the relevant
region around the saddle point of the integral over $s$ has to be
small, i. e. the general condition~\eqappref{P-integral_saddle-point-criteria}
must be fulfilled:
\begin{equation}
\left|\frac{d^{3}m_{2}}{dS^{3}}\cdot\left(\frac{d^{2}m_{2}}{dS^{2}}\right)^{-3/2}\right|\ll1\Leftrightarrow\sqrt{\zeta\left(y|w\right)}\gg1.
\end{equation}
Secondly, the value of the saddle point $s_{0}$ has to be within
the region of applicability of the asymptotic expansion \eqappref{m2-asymptotic_upper-half-plane},
i. e.
\begin{equation}
-i\kappa s_{0}\gg1\Leftrightarrow\zeta\left(y|w\right)\gg Z_{\text{eff}}m_{1}\left(w\right).
\end{equation}
The two requirements above can thus be summarized in the following
criteria of applicability:
\begin{equation}
\zeta\left(y|w\right)\gg\max\left\{ Z_{\text{eff}}m_{1}\left(w\right),1\right\} \Leftrightarrow\frac{1}{\lambda}\left(1-\frac{y}{m_{1}\left(w\right)}\right)\apprge-\frac{\left\langle y\ln y\right\rangle }{m_{1}\left(w\right)}+\gamma+\ln2+\max\left\{ 0,\ln\frac{1}{Z_{\text{eff}}m_{1}\left(w\right)}\right\} \label{eqapp:small-value-asymptotic_criteria}
\end{equation}

At this point, we also note that this result is consistent with the
Gaussian limit described previously. Indeed, for the case of large
$Z_{\text{eff}}$, the criteria of applicability~\eqappref{small-value-asymptotic_criteria}
is only satisfied for sufficiently large deviations of $y$ from the
mean value $\left\langle y\right\rangle $. As a result, the corresponding
asymptotic behavior is rendered effectively unobservable due to small
absolute value of the probability.

\subsection{Probability function for $y>\left\langle y\right\rangle $: leading
dependence\label{subsecapp:Probability-large-y-asymptotic}}

\begin{figure}[h]
\begin{centering}
\includegraphics[scale=0.4]{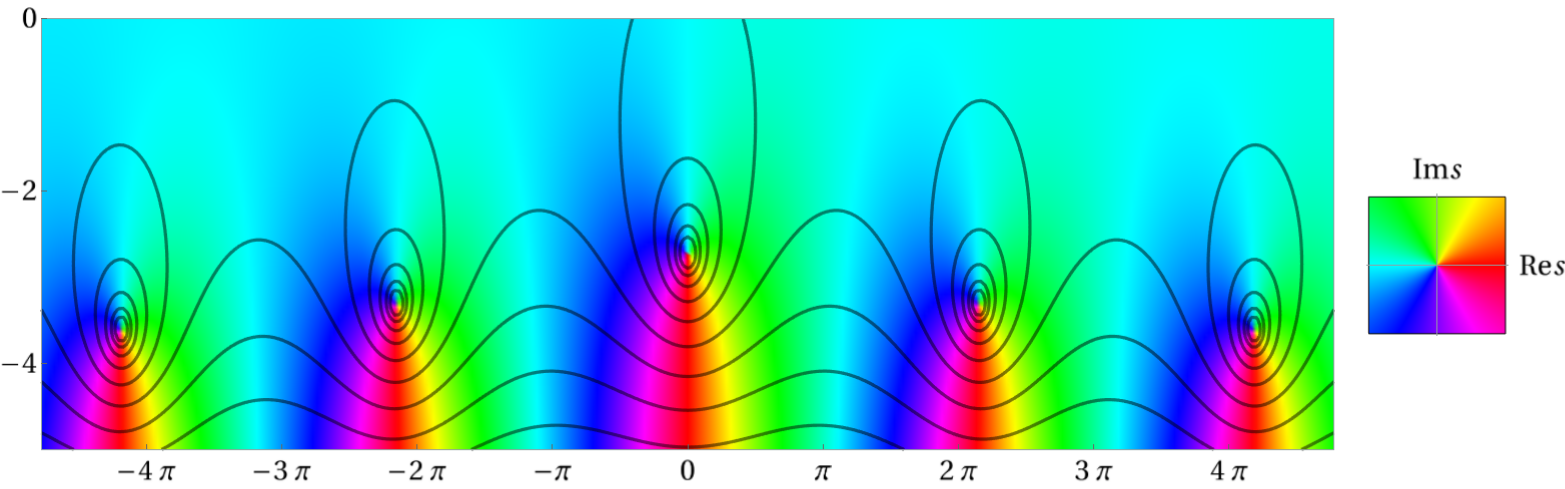}
\par\end{centering}
\raggedright{}\caption{The complex plot of the derivative $f\left(s\right)=-i\partial m/\partial s-y$
of the integrand of Eq.~\eqappref{marginal-probability-distribution_small-dela}
for $y=10$ computed for the model with $\lambda=0.12$ and $Z=51$.
The color encodes the value of $\arg f$ according to the legend to
the right, which stands for the complex plot of the function $f\left(s\right)=s$.
One can observe that $f$ has 5 zeroes in the presented plot region.
The imaginary parts of zeros are close to a corresponding multiple
of $2\pi$. The black solid lines represent the contours of constant
value of $\log_{10}\left|f\right|$ with the step of $0.2$, with
the smallest value corresponding to $0.1$. The function grows exponentially
in the lower half-plane, as evident from the fact that for $\text{Im}s\apprle3.5$
the contours are approximately equidistant. \label{figapp:probability-integrand-derivative_complex-plot}}
\end{figure}

For large values of $y$, the integral~\eqappref{marginal-probability-distribution_small-dela}
possess a whole series of saddle points in the lower half-plane of
$s$ variable, see \figappref{probability-integrand-derivative_complex-plot}
for an illustration. In the relevant region, the asymptotic expression
for the $m_{2}$ function reads:
\begin{equation}
m_{2}\left(S|0\right)=Z_{\text{eff}}\cdot m_{1}\left(1\right)\sqrt{\frac{\pi}{2i\kappa S}}e^{i\kappa S}\left[1+O\left(\frac{1}{i\kappa S}\right)\right]+O\left(i\kappa S\ln i\kappa S\right),\,\,\,\,\left(-\text{Im}\kappa S\right)\gg1.\label{eqapp:m2-asymptotic_lower-half-plane}
\end{equation}
The last error term reflects the presence of one more asymptotic series
of the same type as the one for $\text{Im}S>0$. The leading term
is exponentially large, and we can neglect this sub-leading series,
although it is valid to retain it for any $S$ with nonzero real part.

The integral over $s$ thus has a series of saddle points that can
be described by the following equation:
\begin{equation}
S_{n}:=\frac{z_{n}}{i\kappa},\,\,\,\,\psi:=\frac{y-\left\langle y\right\rangle }{\lambda m_{1}\left(1\right)\sqrt{\frac{\pi}{2}}}\Rightarrow\frac{e^{z}}{\sqrt{z}}\left[1+O\left(\frac{1}{z}\right)\right]+O\left(\ln z\right)=\psi.\label{eqapp:secondary-saddle-points_expression}
\end{equation}
This equation possesses an infinite set of solutions $\left\{ z_{n},n\in\mathbb{Z}\right\} $
that come in conjugate pairs, i. e. $z_{-n}=\overline{z_{n}}$, with
each pair having an imaginary part of order $2\pi in$, as can be
seen e.g. on \figappref{probability-integrand-derivative_complex-plot}.
The asymptotic behavior of $z_{n}$ is captured by the following expression:
\begin{equation}
z_{n}\approx\begin{cases}
-\frac{1}{2}W_{-1}\left(-\frac{2}{\psi^{2}}\right), & n=0,\\
-\frac{1}{2}W_{-\left(2n+1\right)}\left(-\frac{2}{\psi^{2}}\right), & n\in\mathbb{N},\\
-\frac{1}{2}W_{-2n}\left(-\frac{2}{\psi^{2}}\right), & -n\in\mathbb{N},
\end{cases}\label{eqapp:secondary-saddle-points_estimation}
\end{equation}
where $W_{k}\left(u\right)$ is the $k$-th branch of Lambert\textquoteright s
$W$-function. To obtain this approximation, we have neglected the
error terms in Eq.~\eqappref{secondary-saddle-points_expression}.
Each of the resulting saddle points produces a contribution to the
value of the probability according to Eq.~\eqappref{P-integral_saddle-point-estimation}:
\begin{equation}
P^{\left(n\right)}\left(y\right)\sim\frac{1}{\sqrt{2\pi\kappa\left(y-\left\langle y\right\rangle \right)}}\cdot\exp\left\{ -\frac{y-\left\langle y\right\rangle }{\kappa}\left[z_{n}-1+O\left(\frac{1}{z_{n}}\right)\right]+O\left(\ln z_{n}\right)\right\} ,\label{eqapp:marginal-probability_single-saddle-contribution}
\end{equation}
rendering the following result for the total probability:
\begin{align}
P\left(y\right) & \sim\frac{\exp\left\{ -\frac{y-\left\langle y\right\rangle }{\kappa}\left[z_{0}-1+O\left(\frac{1}{z_{0}}\right)\right]+O\left(\ln z_{0}\right)\right\} }{\sqrt{2\pi\kappa\left(y-\left\langle y\right\rangle \right)}}\nonumber \\
 & \times\left[1+\sum_{n\neq0}\exp\left\{ -\frac{y-\left\langle y\right\rangle }{\kappa}\left[z_{n}-z_{0}+O\left(\frac{1}{z_{n}}-\frac{1}{z_{0}}\right)\right]+O\left(\ln\frac{z_{n}}{z_{0}}\right)\right\} \right].\label{eqapp:marginal-probability_large-value-asymptotic}
\end{align}
Because $z_{n}$ for $n\neq0$ come in conjugate pairs, the whole
expression is real as it should be. The range of applicability of
this result is again controlled by the region of convergence of the
Gaussian integral, i. e.
\begin{equation}
\left|\frac{d^{3}m_{2}}{dS^{3}}\cdot\left(\frac{d^{2}m_{2}}{dS^{2}}\right)^{-3/2}\right|\ll1\Leftrightarrow\sqrt{\frac{\kappa}{y-\left\langle y\right\rangle }}\ll1.\label{eqapp:large-value-asymptotic_criteria}
\end{equation}

Due to the exponential behavior~\eqappref{m2-asymptotic_lower-half-plane}
of the $m_{2}$ function, each term of the resulting series is only
available up to logarithmic accuracy and we cannot provide an estimation
for the sum in Eq.~\eqappref{marginal-probability_large-value-asymptotic}.
In addition, this analysis is unable to provide the exact value of
$m\left(1\right)$, which is controlled by the form of the distribution
$P\left(y\right)$ near its maximum. In order to complete both of
these tasks, one has to use the exact expressions for $m_{2}$ (e.
g. those available in the limit of weak coupling, see \appref{Solution_small-lambda}).
Nevertheless, this analysis provides us with insights into the asymptotic
behavior of the probability $P\left(y\right)$ for large values of
the argument.

According to Eq.~\eqappref{secondary-saddle-points_expression},
the real part of $z_{n}$ grows with $n$. While the exact rate of
decay depends on the particular form of the errors terms that we have
neglected, one expects that for sufficiently large value of $y$ only
several first terms contribute to the sum in Eq.~\eqappref{marginal-probability_large-value-asymptotic}.
In particular, one can establish the general form of the asymptotic
tail of the distribution by using only the main saddle point:
\begin{equation}
\ln P\left(y\right)\sim\ln P^{\left(0\right)}\left(y\right)\sim-\frac{y-\left\langle y\right\rangle }{\kappa}\left[z_{0}-1+O\left(\frac{1}{z_{0}}\right)\right]+\ln\left[\frac{1}{\sqrt{2\pi\kappa\left(y-\left\langle y\right\rangle \right)}}\right]+O\left(\ln z_{0}\right),\label{eqapp:marginal-probability_large-value-leading-term}
\end{equation}
where $z_{0}$ can be estimated according to Eq.~\eqappref{secondary-saddle-points_expression}
as
\begin{equation}
z_{0}\approx\ln\psi\cdot\left(1+O\left(\frac{1}{\ln\psi}\right)\right)+\frac{1}{2}\ln\ln\psi\cdot\left(1+O\left(\frac{1}{\ln\psi}\right)\right)+O\left(\frac{1}{\ln\psi}\right).
\end{equation}
The precision of our calculations allows us to provide the final result
in the following form:
\begin{equation}
\ln P\left(y\right)\sim-\frac{y-\left\langle y\right\rangle }{\kappa}\left[\ln\psi+\frac{1}{2}\ln\ln\psi-1\right]+\ln\frac{1}{\sqrt{2\pi\kappa\left(y-\left\langle y\right\rangle \right)}},\,\,\,\,\,\psi=\frac{y-\left\langle y\right\rangle }{\lambda m_{1}\left(1\right)\sqrt{\frac{\pi}{2}}}.\label{eqapp:extreme-value-statistics_large-y-leading-approx}
\end{equation}
A practically important observation is that in a broad region of parameters
this distribution is numerically close to a ``squashed exponent''
distribution of the form $\exp\left\{ -Ay^{1+\beta}+\text{const}\right\} $
for some small parameter $\beta$, which is consistent with some experimental
observations.

This result is also consistent with the Gaussian limit discussed in
\subsecappref{Gaussian-limit}, as the latter corresponds to $\kappa\ll1$.
Thus, according to the criteria~\eqappref{large-value-asymptotic_criteria},
the asymptotic behavior~\eqappref{marginal-probability_large-value-leading-term}
starts at 
\begin{equation}
y-\left\langle y\right\rangle \gg\kappa=\sqrt{\left\langle \left\langle y^{2}\right\rangle \right\rangle /\lambda},
\end{equation}
where the last equation is due to Eq.~\eqappref{Gaussian-limit-dispersion}.
Consequently, the probability at these values is already exponentially
small, rendering the corresponding regime unobservable.

\subsection{Probability function for $y>\left\langle y\right\rangle $: sub-leading
corrections and secondary maxima\label{subsecapp:Probability_large-y-asymptotic_secondary-peaks}}

For moderately large values of $y$ and $\kappa$, the secondary saddle
points bring in additional oscillatory behavior as seen on \figappref{saddle-point-approx}.
The qualitative origin of these oscillations lies in the fact that
each secondary saddle point has imaginary part of the order of $2\pi n,\,\,n\in\mathbb{Z}$.
As a result of the latter, the contribution~\eqappref{marginal-probability_single-saddle-contribution}
of each saddle point exhibits damping oscillations w.r.t $y$ with
period close to $\kappa/n$. One thus expects that all secondary contributions
in the total sum~\eqappref{marginal-probability_large-value-asymptotic}
will demonstrate constructive interference as $y$ approaches 
\begin{equation}
y_{n}=\left\langle y\right\rangle +\kappa n,\,\,\,\,\,n\in\mathbb{N}.\label{eqapp:position-of-secondary-peaks}
\end{equation}

\begin{figure}[h]
\begin{centering}
\includegraphics[scale=0.21]{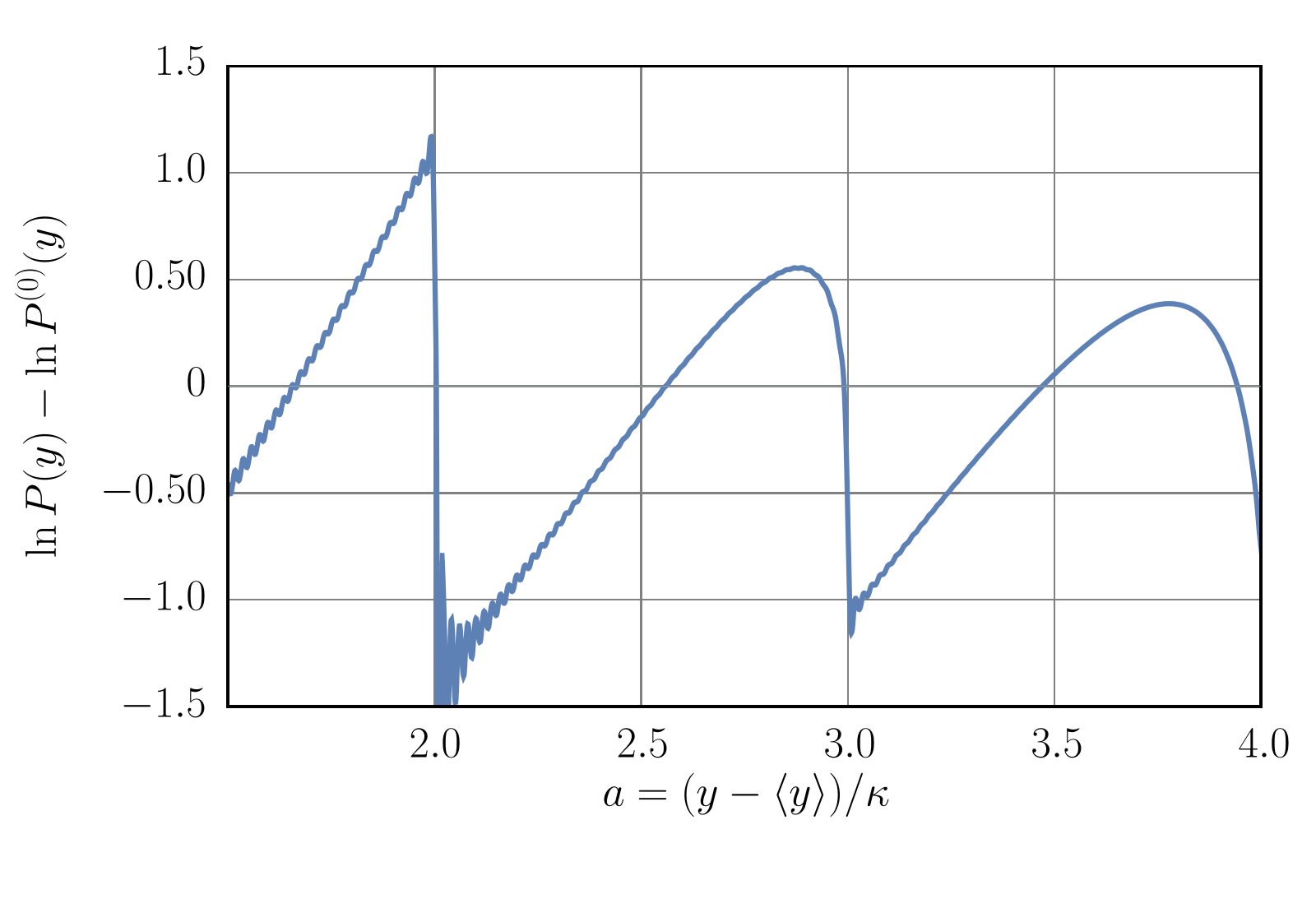}\hspace*{\fill}\includegraphics[viewport=0bp 15.00457bp 1200.75bp 797.493bp,clip,scale=0.21]{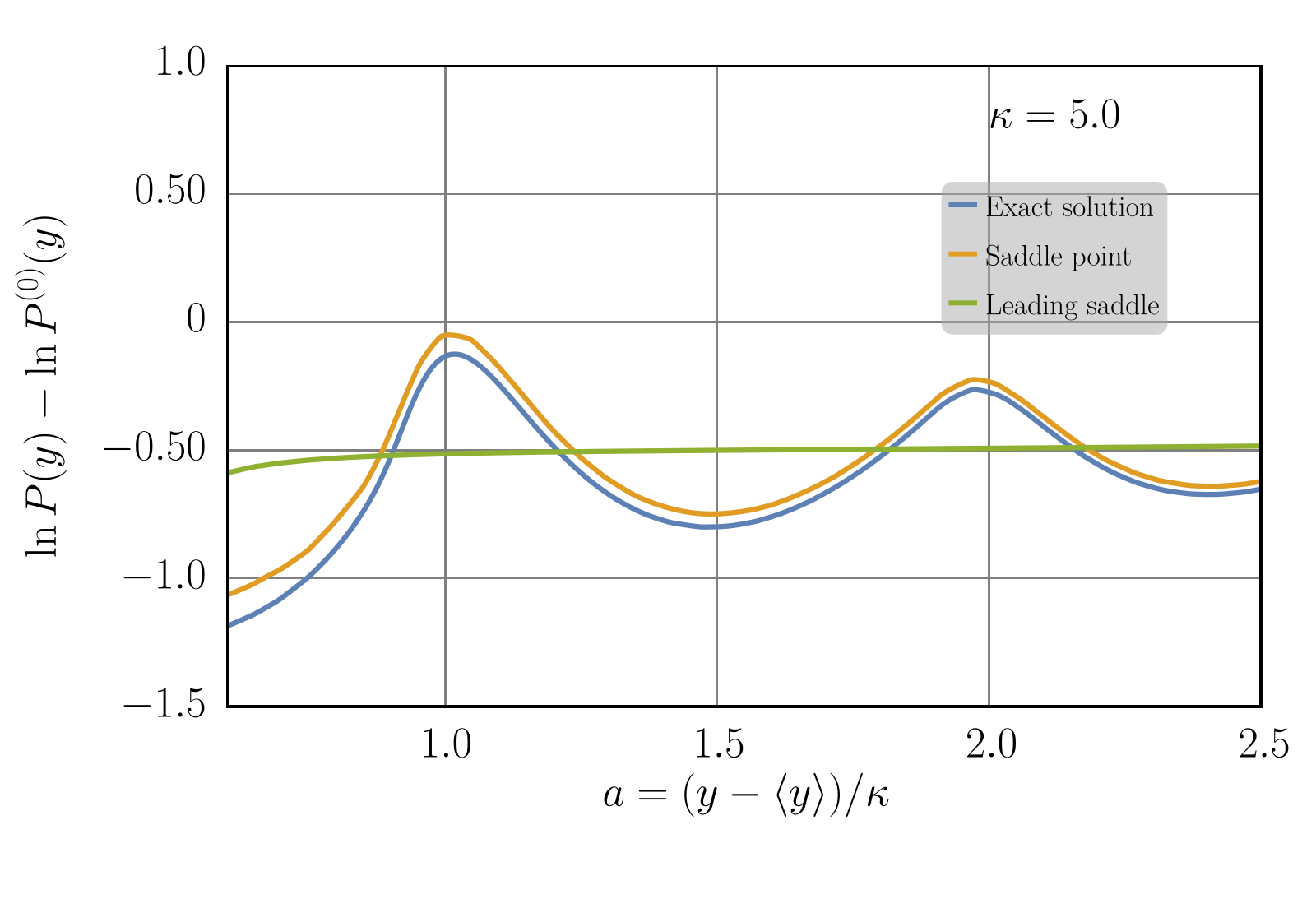}
\par\end{centering}
\caption{Plots of the logarithm of the multiplicative factor $f\left(y\right)$
distinguishing the leading asymptotic $P_{0}^{\left(0\right)}\left(y\right)$
behavior at large $y$ given by Eq.~\eqappref{extreme-value-statistics_large-y-leading-approx}
and the true distribution of the dimensionless order parameter $P\left(y\right)=\exp\left\{ f\left(y\right)\right\} \cdot P_{0}^{\left(0\right)}\left(y\right)$.
The argument is given by $a=\left(y-\left\langle y\right\rangle \right)/\kappa$.
The microscopical parameters of the model are $\lambda\approx0.12$,
$Z=51$ and $\kappa\approx5.0$. \emph{Left}. The multiplicative correction
according to the qualitative estimation~\eqappref{marginal-probability_subleading_estimation}.
The noise at moderate values of $a$ results from cutting the sum
in Eq.~\eqappref{marginal-probability_subleading_estimation} at
a finite number of terms for the purposes of numerical evaluation.
\emph{Right}. The multiplicative correction according to various theoretical
approximations for the true distribution. The blue curve represents
the value of the integral~\eqappref{marginal-probability-distribution_small-dela}
obtained by direct numerical integration. The orange line corresponds
to saddle-point approximation of the integral~\ref{eqapp:marginal-probability-distribution_small-dela}
with all saddle point taken into account for $y>\left\langle y\right\rangle $.
The green line reflects contribution of the leading purely imaginary
saddle point only. When required, the exact $m$~function is used.
The quantitative difference between the two plots is explained by
the subleading corrections to the exponent of each term in Eq.~\eqappref{marginal-probability_subleading_estimation}
that are beyond the accuracy of the used expansions. \label{figapp:large-y_multiplicative-correction-plots}}
\end{figure}

While the precision of our calculations does not allow us to demonstrate
these oscillations explicitly, we can still use the leading asymptotic
form to construct a meaningful model that illustrates such a behavior.
By using the leading approximation~\eqappref{secondary-saddle-points_estimation}
for $z_{n}$ and discarding the corrections despite their growing
nature we can estimate the multiplicative correction to the probability
as:
\begin{equation}
\ln\frac{P\left(y\right)}{P^{\left(0\right)}\left(y\right)}\sim\ln\left[1+2\text{Re}\sum_{n=1}^{\infty}\exp\left\{ -\frac{y-\left\langle y\right\rangle }{\kappa}\left[-\frac{1}{2}W_{-\left(2n+1\right)}\left(-\frac{2}{\psi^{2}}\right)+\frac{1}{2}W_{-1}\left(-\frac{2}{\psi^{2}}\right)\right]\right\} \right],\,\,\,\,\,\psi=\frac{y-\left\langle y\right\rangle }{\lambda m_{1}\left(1\right)\sqrt{\frac{\pi}{2}}},\label{eqapp:marginal-probability_subleading_estimation}
\end{equation}
where we have taken into account that $z_{-n}=\overline{z}_{n}$ and
thus the whole sum is real. It is worth specifying the asymptotic
expression of the exponent at large $y$:
\begin{equation}
-\frac{1}{2}W_{-\left(2n+1\right)}\left(-\frac{2}{\psi^{2}}\right)+\frac{1}{2}W_{-1}\left(-\frac{2}{\psi^{2}}\right)=2\pi in\cdot\left[1+\frac{1}{2\ln\psi}+O\left(\frac{\ln\ln\psi}{\ln^{2}\psi}\right)\right]+\left[\left(\frac{\pi n}{\ln\psi}\right)^{2}+O\left(\frac{1}{\ln^{3}\psi}\right)\right].
\end{equation}
We once again underscore that such an estimation cannot guarantee
any sort of qualitative convergence and thus serves only illustrative
purposes (at least because the correct estimation~\eqappref{marginal-probability_single-saddle-contribution}
contains the error term of the order $O\left(1/z_{n}\right)$ which
is disregarded in the expression above). The plot of the resulting
function for some values of the parameters is shown on the left plot
of \figappref{large-y_multiplicative-correction-plots}. Crucially,
one can observe that the resulting expression exhibits oscillations
with a slowly drifting period close to $\Delta y=\kappa$. The exact
answer for the distribution function confirms this qualitative result,
as evident from the neighboring right plot of \figappref{large-y_multiplicative-correction-plots}
as well as from the data presented in the Main Text.

From the physics point of view, the secondary peaks are delivered
by a certain spatial configurations of the disorder. Namely, the $n$\nobreakdash-th
secondary maximum of the distribution corresponds to the sites with
exactly $n$ neighbors with small value of onsite disorder $\left|\xi_{k}\right|\apprle\Delta_{0}$.
Indeed, suppose a given site $i$ has $n\sim1$ such neighbors. Then
the self-consistency equation~\eqappref{saddle-point-equation_no-D-dependence}
for this site $i$ reads:
\[
\Delta_{i}=\sum_{j\in\partial i}f\left(\xi_{j},\Delta_{j}|\xi_{i}\right)\sim\sum_{j=1}^{n}\frac{\lambda}{\nu_{0}Z}+\sum_{j=n+1}^{Z}f\left(\xi_{j},\Delta_{j}|\xi_{i}\right)\sim\frac{n\lambda}{\nu_{0}Z}+\frac{n+1}{Z}\cdot\Delta_{0}
\]
where we have used $\lambda/\nu_{0}Z$ as an approximation for the
values of the right hand side for the chosen sites with small values
of $\xi$, while the remaining sum was estimated by its mean field
value. Rescaling this estimation to the units of $\Delta_{0}$ immediately
leads us to Eq.~\eqappref{position-of-secondary-peaks}.

The apparent sharpness of the peaks can be perceived as a consequence
of Van Hove singularity in the distribution of the right hand side
of the self-consistency equation. Indeed, at small values of $\xi$,
the BSC root in Eq.~\eqappref{saddle-point-equation_no-D-dependence}
features a maximum, leading to a square root singularity in its distribution.
The latter is subsequently contracted with the distribution of the
order parameter itself, thus producing a shifted replica the main
maximum of the distribution.

\begin{figure}[h]
\begin{centering}
\includegraphics[scale=0.23]{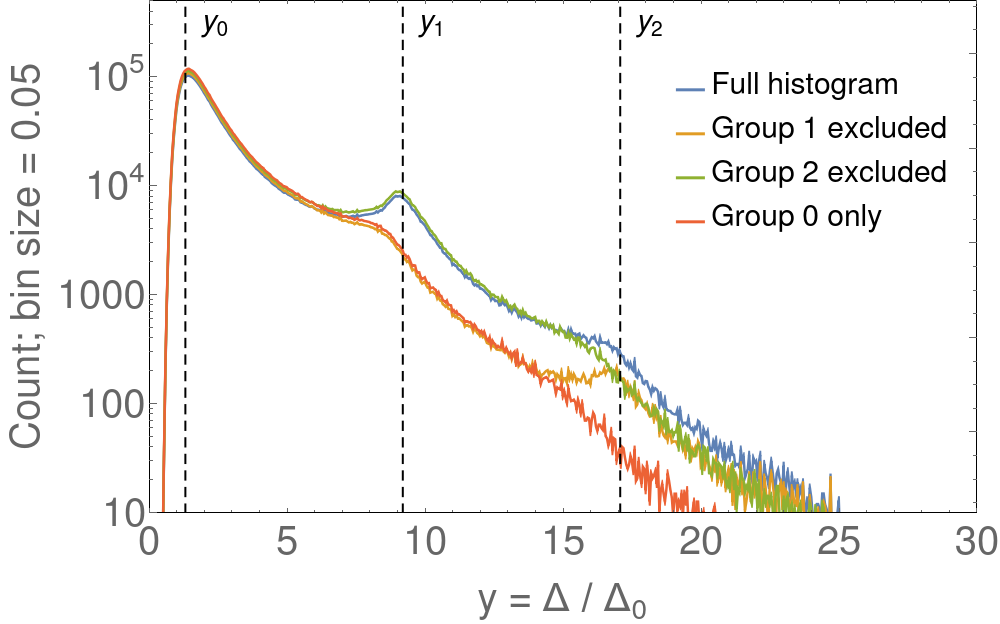}\hspace*{\fill}\includegraphics[scale=0.23]{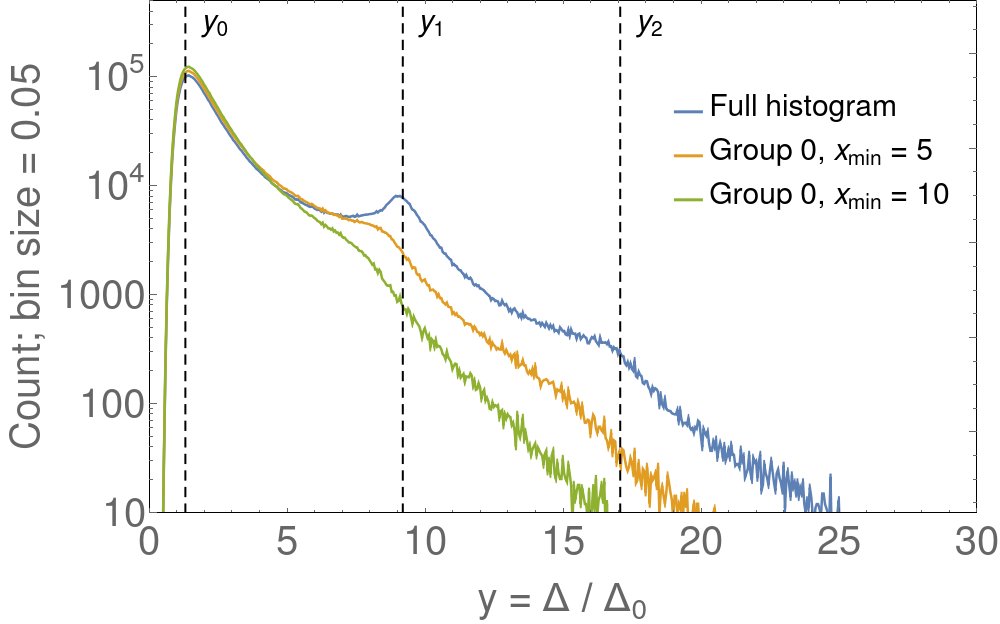}
\par\end{centering}
\raggedright{}\caption{The plots of \textquotedblleft filtered\textquotedblright{} histograms
of the order parameter in a particular disorder realization with $N=2^{22}\approx4.2\cdot10^{6}$,
$Z=51$, $\lambda=0.113$, corresponding to $Z_{\text{eff}}=1.43\cdot10^{-2}$,
$\kappa=7.88$. The curves are offset by a constant multiplier close
to unity so that all curves are visible (otherwise they coincide with
high precision). The vertical dashed lines denote the expected position
of the maximum of the distribution according to Eq.~\eqappref{position-of-secondary-peaks},
but with the value of $\left\langle y\right\rangle $ replaced with
the actual position of main maximum $y_{\text{max}}\approx1.3$. The
noise present on both plots at large values of $y$ is due to statistical
uncertainty, as can be deduced from the \textquotedblleft count\textquotedblright{}
value for these values of $y$. \emph{Left.} Histograms obtained after
excluding sets of sites with the corresponding number of neighbors
satisfying $\left|\xi\right|<\xi_{\min}=5\Delta_{0}$. For instance,
\textquotedblleft Group 1\textquotedblright{} refers to the histogram
obtained after excluding the sites with exactly one nearest neighbor
that has $\left|\xi\right|<5\Delta_{0}$. One can see how the corresponding
secondary peaks are suppressed in accordance to the fact that Group
$n$ contributes to formation of the $n$-th peak. \emph{Right}. Demonstration
of the effect of various threshold values $\xi_{\min}=x_{\min}\cdot\Delta_{0}$
on the behavior of the histogram. One can observe how increasing the
threshold suppresses all peaks except the main one. \label{figapp:secondary-peak-filtration}}
\end{figure}

The presented explanation for the secondary peaks also admits a straightforward
verification for each particular realization of the disorder. Given
a solution to the self-consistency equation~\eqappref{saddle-point-equation_no-D-dependence}
in a particular realization of a random graph and disorder fields,
one classifies all sites according to the exact number $k$ of neighbors
with $\left|\xi\right|<\xi_{\min}$ where $\xi_{\min}$ is some threshold
of order of several $\Delta_{0}$. By removing the $k$\nobreakdash-th
group from the complete set of $\Delta$ values, one expects to flatten
out the corresponding secondary peak, while leaving all other peaks
intact. The results of this procedure are presented on \figappref{secondary-peak-filtration}
and are rather confirmatory. The secondary peaks are not eliminated
entirely by the described numerical classification because the distribution
of the right hand side of the self-consistency equation exhibits a
broad power-law tail away from the Van Hove singularity. As a result,
there is no exact scale for the threshold parameter $\xi_{\min}$.
This fact is also demonstrated on \figappref{secondary-peak-filtration}:
depending on the exact value of the threshold, one can observe different
degrees of deterioration of the secondary peaks.

The proposed explanation is also apparent from the theoretical analysis
presented thus far. Indeed, the exponential behavior of the $m_{2}$
function originates from the vicinity of $w_{1}=1$ point in the integral~\eqappref{equation-on-m2}.
This region, in turn, corresponds to the values of the $f$ function
achieved in the limit $\xi\ll\Delta$. As a result, each secondary
peak effectively represents configurations with $n$ neighbors with
small values of $\xi$, while the remaining neighbors form a background
value of $\left\langle y\right\rangle $ in a mean field fashion.

\begin{figure}[h]
\begin{centering}
\includegraphics[scale=0.3]{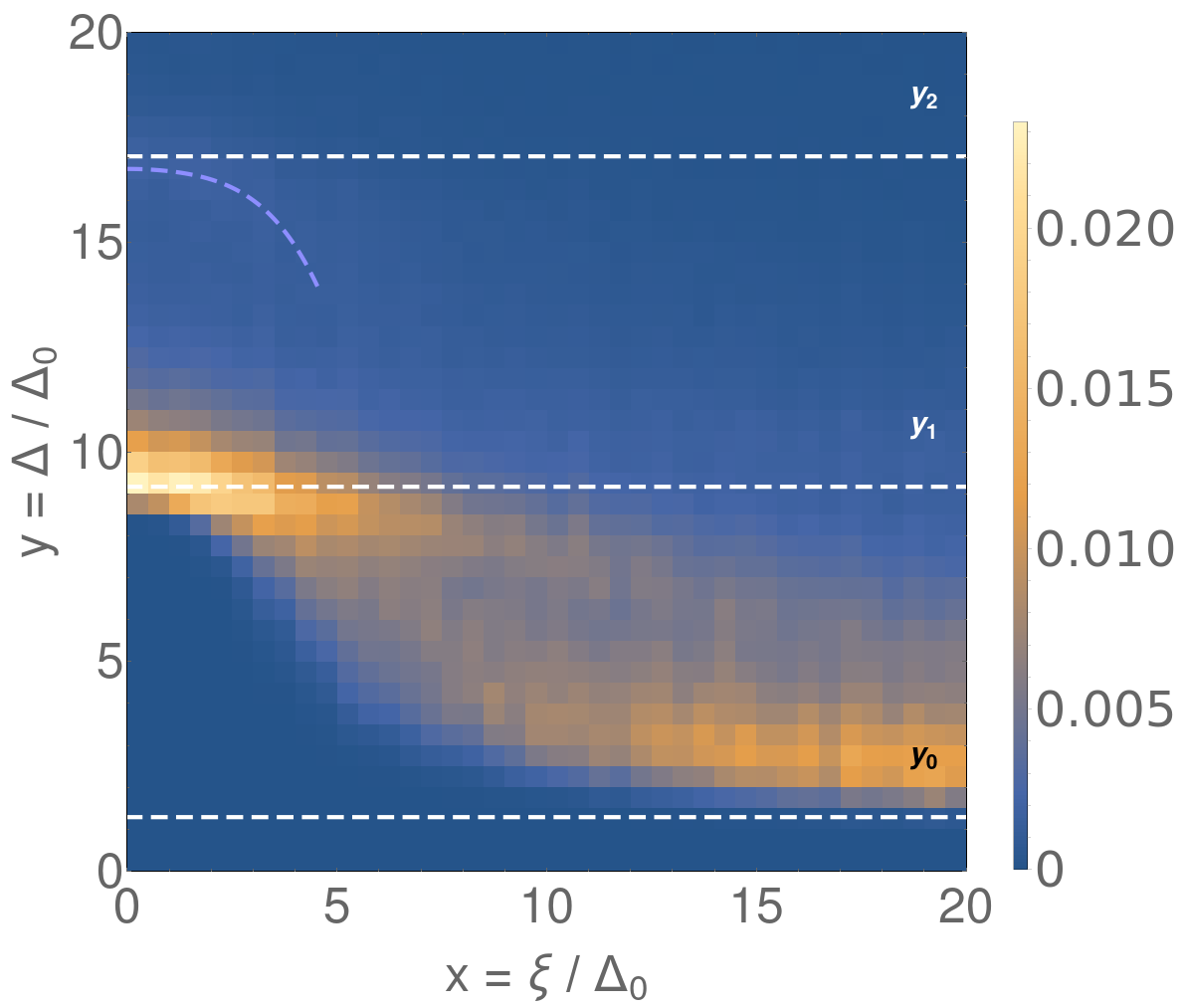}
\par\end{centering}
\raggedright{}\caption{Density histogram of the joint probability distribution $P\left(\xi_{j},\Delta_{i}\right)$
for all pairs $\left(i,j\right)$ of neighboring sites in a particular
disorder realization with $N=2^{22}\approx4.2\cdot10^{6}$, $Z=51$,
$\lambda=0.113$, corresponding to $Z_{\text{eff}}=1.43\cdot10^{-2}$,
$\kappa=7.88$. The color represents the value of the probability
density function according to the legend on the right. Horizontal
dashed lines denote the expected position of the maximum of the marginal
distribution $P\left(y\right)$ according to Eq.~\eqappref{position-of-secondary-peaks},
but with the value of $\left\langle y\right\rangle $ replaced with
the actual position of main maximum $y_{\text{max}}=1.3$. Note that
the main maximum is situated at $y_{1}$ for small $x$ and approaches
$y_{0}$ as $x$ grows. The light blue dashed line at the top left
corner serves as a guide for the position of the secondary maximum
wheres the latter can be resolved, although barely visible on the
plot itself. The secondary maximum at small values of $x$ is thus
situated close to $y_{2}$, in full agreement with the proposed interpretation.
\label{figapp:Delta-neighboring-xi_joint-distribution}}
\end{figure}

A yet another consistency test is to examine the joint probability
distribution $P\left(\xi_{j},\Delta_{i}\right)$ for some pair of
neighboring sites $\left(i,j\right)$. Both numerical and analytical
means then clear indicate that for small values of $\xi_{j}$ the
conditional distribution experiences a well pronounced shift by the
value of $f\left(\Delta,\xi\right)$, in full agreement with the proposed
explanation. A representative example of numerical data is shown on
\figappref{Delta-neighboring-xi_joint-distribution}.

We conclude this appendix by noting that the observed behavior of
both the joint distribution $P\left(\xi_{i},\Delta_{j}\right)$ and
the distribution of the order parameter $P\left(y\right)$ is expected
to persist in the presence of the Onsager reaction term. As our analysis
suggests, the positions of the maxima in both these distributions
are determined solely by the direct influence of neighboring sites.
On the other hand, the Onsager reaction term eliminates the ``self-action''
of the order parameter, i. e. its response to its own value via value
of the order parameter on the neighboring sites. Therefore, such a
term is incapable of altering the positions and strengths of the observed
maxima of the distribution behavior qualitatively. For instance, one
expects such an effect to be present if the self-consistency equations
are solved on an indefinite directed Caley tree instead of a finite
random regular graph, as visible e. g. on Figs~2~and~3 of Ref.~\citep{Lemarie2013}.

\section{Solution for the cumulant generating function $m$ in the limit of
small $\lambda$\label{app:Solution_small-lambda}}

In this Appendix, we derive the analytical solution for the integral
equations~(\hphantom{}\ref{eqapp:equation-on-m1}\nobreakdash-\ref{eqapp:equation-on-m2}\hphantom{})
in the limit of small BCS coupling constant $\lambda$. The resulting
solution appears to be valid in the entire region of applicability
of the self-consistency equation~\eqappref{zero-temperature-saddle-point-equation}.

We seek the solution for both $m_{1}$ and $m_{2}$ in the form of
a formal expansion in powers of small $\lambda$ up to the leading
order. We start by discussing the magnitude of the functions in question.
Let us consider the equation on $m_{1}$ first. As we will see later,
the first two terms in Eq.~\eqappref{equation-on-m1} are both of
order $O\left(\lambda^{0}\right)$. Assuming those to be the leading
terms, one can see that the remaining contributions are all of order
$O\left(\lambda^{1}\right)$ or smaller. Having these estimates hold,
one can infer from Eq.~\eqappref{equation-on-m2} that the whole
function $m_{2}$ is of order $O\left(\lambda^{1}\right)$ or smaller.
As a result, we assert the following estimations for the functions
in question:
\begin{equation}
m_{1}\left(w\right)=\kappa\cdot\left(w_{0}+w\right)+O\left(\lambda\right),\,\,\,\,\,w_{0}=O\left(\lambda^{0}\right),\,\,\,\,\,m_{2}\left(S|w\right)=O\left(\lambda^{1}\right)
\end{equation}
where we have denoted 
\begin{equation}
w_{0}:=m_{1}\left(0\right)/\kappa,
\end{equation}
which is a suitable variable for further calculations.

This expansion can now be substituted in \eqappref{equation-on-m2}
to obtain $m_{2}$:
\begin{equation}
m_{2}\left(S|w\right)=\lambda\cdot\left[\left(w+w_{0}\right)\Phi_{0}\left(\kappa S\right)+\Phi_{1}\left(\kappa S\right)\right]+O\left(\lambda^{2}\right),\label{eqapp:m2-expression_small-lambda-limit}
\end{equation}
\begin{align}
\Phi_{0}\left(\sigma\right) & =\intop_{0}^{1}\frac{dw_{1}}{w_{1}^{2}\sqrt{1-w_{1}^{2}}}\left\{ \exp\left\{ i\sigma w_{1}\right\} -1-i\sigma w_{1}\right\} \nonumber \\
 & =-\frac{1}{12}\sigma^{2}\left[3\pi\cdot\,_{1}F_{2}\left(\frac{1}{2};\frac{3}{2},2;-\frac{\sigma^{2}}{4}\right)+2i\sigma\cdot\,_{2}F_{3}\left(1,1;\frac{3}{2},2,\frac{5}{2};-\frac{\sigma^{2}}{4}\right)\right],\label{eqapp:Phi-0-definition}
\end{align}
\begin{equation}
\Phi_{1}\left(\sigma\right)=\intop_{0}^{1}\frac{w_{1}dw_{1}}{\sqrt{1-w_{1}^{2}}}\left\{ \exp\left\{ i\sigma w_{1}\right\} -1-i\sigma w_{1}\right\} =i\frac{d^{3}\Phi_{0}\left(\sigma\right)}{d\sigma^{3}}-\frac{\pi}{4}i\sigma-1.\label{eqapp:Phi-1-definition}
\end{equation}
where $\,_{p}F_{q}$ is the generalized hypergeometric series:
\begin{equation}
\,_{p}F_{q}\left(a_{1},..a_{p};b_{1}...b_{q};z\right)=\sum_{k=0}^{\infty}\frac{\left(a_{1}\right)_{k}...\left(a_{p}\right)_{k}}{\left(b_{1}\right)_{k}...\left(b_{q}\right)_{k}}\frac{z^{k}}{k!},
\end{equation}
with $\left(x\right)_{k}=\Gamma\left(x+k\right)/\Gamma\left(x\right)$
being the Pochhammer symbol.

With this expression at hand, one can turn back to the equation~\eqappref{equation-on-m1}
for the $m_{1}$~function. It is convenient to use it in the form
of expressions \eqappref{m1-expression_average-form}~and~\eqappref{self-consistency-equation},
rendering:
\begin{equation}
m_{1}\left(w\right)=\kappa\left(w+w_{0}\right)+\lambda\cdot\left\langle \left(y+\kappa w\right)\ln\frac{1}{y+\kappa w}-y\ln\frac{1}{y}\right\rangle +O\left(\lambda^{2}\right),
\end{equation}
\begin{equation}
0=\frac{\pi}{4}\left(1+\lambda\ln\frac{1}{\kappa}\right)+\left\langle \frac{y}{\kappa}\right\rangle \ln\frac{1}{\kappa}+\left\langle \frac{y}{\kappa}\ln\frac{\kappa}{y}\right\rangle +\lambda\cdot\left\langle G\left(y/\kappa\right)\right\rangle .
\end{equation}
where we have also retained $O\left(\lambda\right)$ contributions
in the first equation in order to match with the precision of Eq.~\eqappref{m2-expression_small-lambda-limit}.
The average values $\left\langle \bullet\right\rangle $ in this expressions
are calculated as:
\[
\left\langle \phi\left(y\right)\right\rangle =\intop_{0}^{\infty}dy\cdot\phi\left(y\right)P\left(y\right)\approx\intop_{\mathbb{R}-i0}\frac{d\sigma}{2\pi}\cdot\intop_{0}^{\infty}da\cdot\phi\left(\kappa a\right)\cdot\exp\left\{ i\sigma w_{0}-i\sigma a+\lambda\cdot\left[w_{0}\Phi_{0}\left(\sigma\right)+\Phi_{1}\left(\sigma\right)\right]\right\} .
\]
To obtain the last expression, we used the result~\eqappref{m2-expression_small-lambda-limit}
for the $m_{2}$~function and carried out substitutions $\sigma=\kappa s$,
$a=y/\kappa$. It is thus convenient to rewrite the result as
\begin{equation}
m_{1}\left(w\right)=\kappa\left(w+w_{0}+\lambda\left[g_{1}\left(w_{0},w;\lambda\right)-g_{1}\left(w_{0},0;\lambda\right)\right]\right),
\end{equation}
\begin{equation}
0=\frac{\pi}{4}\left(1+\lambda\ln\frac{1}{\kappa}\right)+w_{0}\ln\frac{1}{\kappa}+g_{1}\left(w_{0},0;\lambda\right)+\lambda\cdot g_{2}\left(w_{0};\lambda\right)+O\left(\lambda^{2}\right),
\end{equation}
where we have used the exact identity $\left\langle y\right\rangle =m_{1}\left(0\right)=\kappa w_{0}$,
and the $g$ functions are defined as
\begin{equation}
g_{1}\left(w_{0},w;\lambda\right)=\intop_{\mathbb{R}-i0}\frac{d\sigma}{2\pi}\cdot\intop_{0}^{\infty}da\cdot\left(a+w\right)\ln\frac{1}{a+w}\cdot\exp\left\{ i\sigma w_{0}-i\sigma a+\lambda\cdot\left[w_{0}\Phi_{0}\left(\sigma\right)+\Phi_{1}\left(\sigma\right)\right]\right\} ,\label{eqapp:g1-function-def}
\end{equation}
\begin{equation}
g_{2}\left(w_{0};\lambda\right)=\intop_{0}^{1}dw\cdot\sqrt{1-w_{1}^{2}}\cdot\frac{g_{1}\left(w_{0},w_{1};\lambda\right)-g_{1}\left(w_{0},0;\lambda\right)}{w_{1}}.
\end{equation}
To obtain the expression for $g_{2}$, we have used the integral representation~\eqappref{G-function-definition}
for the $G$ function. We now need to develop a proper approximation
for $g_{1}$ function. Because we expect $w_{0}$ to be large, as
it follows from the previous subsection, it is valid to seek a formal
expansion of $g_{1}$ in powers of $1/w_{0}$. It is convenient to
use the following integral representation for the $a$-dependence
of the integrand in Eq.~\eqappref{g1-function-def}:
\begin{equation}
x\ln\frac{1}{x}=-x-\intop_{0}^{\infty}dt\cdot\frac{e^{-xt}+e^{-t}xt-1}{t^{2}}.
\end{equation}
With the use of this representation, we can rewrite the expression
$g_{1}$ as:
\begin{equation}
g_{1}\left(w_{0},w;\lambda\right)=-\left(w_{0}+w\right)-\intop_{0}^{\infty}dt\left[\frac{\exp\left\{ -t\left(w_{0}+w\right)+\lambda\cdot\left[w_{0}\Phi_{0}\left(it\right)+\Phi_{1}\left(it\right)\right]\right\} +e^{-t}\left(w_{0}+w\right)t-1}{t^{2}}\right].\label{eqapp:g1-expression_integral-representation}
\end{equation}
To obtain this result, the integral over $a$ was taken exactly, and
the subsequent integration over $\sigma$ was carried out by means
of Cauchy theorem. Before we move on, it should be noted, that the
integral over $t$ should be formally cut at $t_{0}\sim\exp\left\{ 1/\lambda\right\} $.
Indeed, according to asymptotic expression~\eqappref{m2-asymptotic_upper-half-plane}
derived earlier, the $\Phi_{0}$ term brings in a contribution whose
real part grows as $+\lambda w_{0}\cdot t\ln t$ at large $t$, so
that it eventually dominates the linear decay provided by the first
term in the exponent. However, this happens only at $t\sim t_{0}$,
where the expression for $m_{2}$ itself ceases to work, as discussed
in \appref{equation-on-distribution_small-Delta}. The integral itself
gains its value at $t\sim\left(w_{0}+w\right)^{-1}$ due to the first
term, thus allowing one to treat the expression as convergent. 

The expression~\eqappref{g1-expression_integral-representation}
is thus suitable for expanding in powers of $\lambda$, rendering:
\begin{align}
g_{1}\left(w_{0},w;\lambda\right) & =-\left(w_{0}+w\right)-\intop_{0}^{\infty}dt\left[\frac{e^{-t\left(w_{0}+w\right)}+e^{-t}\left(w_{0}+w\right)t-1}{t^{2}}\right]\nonumber \\
 & -\lambda\intop_{0}^{\infty}dt\cdot e^{-t\left(w_{0}+w\right)}\cdot\frac{w_{0}\Phi_{0}\left(it\right)+\Phi_{1}\left(it\right)}{t^{2}}+O\left(\lambda^{2}\right).
\end{align}
Finally, integrals over $t$ can be evaluated, resulting in:
\begin{equation}
g_{1}\left(w_{0},w;\lambda\right)=\left(w_{0}+w\right)\ln\frac{1}{w_{0}+w}-\lambda\cdot H\left(w_{0},w\right)+O\left(\lambda^{2}\right),\label{eqapp:g1-expression_first-order-in-lambda}
\end{equation}
\begin{equation}
H\left(w_{0},w\right)=\intop_{0}^{1}\frac{dw_{1}}{\sqrt{1-w_{1}^{2}}}\cdot\left(w_{0}+w_{1}^{3}\right)\cdot\frac{\left(w+w_{0}+w_{1}\right)\ln\left(\frac{w_{1}+w+w_{0}}{w+w_{0}}\right)-w_{1}}{w_{1}^{2}}.
\end{equation}
Now, the result can be used to calculate the $g_{2}$ term. Within
the required precision, we only need to use the leading term in Eq.~\eqappref{g1-expression_first-order-in-lambda},
rendering:
\begin{equation}
g_{2}\left(w_{0};\lambda\right)=\intop_{0}^{1}dw\cdot\sqrt{1-w_{1}^{2}}\cdot\frac{\left(w_{0}+w_{1}\right)\ln\frac{1}{w_{0}+w_{1}}-w_{0}\ln\frac{1}{w_{0}}}{w_{1}}\equiv G\left(w_{0}\right)+O\left(\lambda\right),
\end{equation}
where $G$ is the function defined in Eq.~\eqappref{G-function-definition}.

At this point, it is worth noting that the derived expression~\eqappref{g1-expression_first-order-in-lambda}
for $g_{1}$ possesses a transparent qualitatively interpretation.
According to~\eqappref{g1-function-def}, the $g_{1}$~function
represents the mean value of the form
\begin{equation}
g_{1}\left(w_{0},w;\lambda\right)=\left\langle \frac{y+\kappa w}{\kappa}\ln\frac{\kappa}{y+\kappa w}\right\rangle ,
\end{equation}
where the average is taken over the exact distribution $P\left(y\right)$
function of the order parameter. On the other hand the value of $w_{0}$
also corresponds to the average value of the form $\left\langle y/\kappa\right\rangle $.
One can thus see that the leading term in the approximation~\eqappref{g1-expression_first-order-in-lambda}
for the $g_{1}$ function corresponds to replacing the full distribution
$P\left(y\right)$ with a $\delta$-function centered at the mean
value, i. e. $\delta\left(y-\kappa w_{0}\right)$. The sub-leading
corrections to $g_{1}$ are then obtained by treating the actual form
of the distribution as a perturbation on top of the trial distribution.
This appears to be sufficient to determine the value of $w_{0}$ with
the required accuracy solely because the averaged functions $y$ and
$y\ln1/y$ differ by a slow function that does not contribute substantially
to the result unless the underlying distribution possesses fat tails.
That is why, it would be qualitatively wrong to perform such an approximation
for other averages. Consider, for instance, low order cumulants: they
vanish identically for a $\delta$-like distribution, but from the
form of the cumulant generating function given by $m$ itself one
can tell that in our problem they are all of the same order.

Collecting everything together, one obtains the following set of equations
for the $m_{1}$ function:
\begin{equation}
m_{1}\left(w\right)=\kappa\left(w+w_{0}\right)+\lambda\cdot\left[\left(w_{0}+w\right)\ln\frac{1}{w_{0}+w}-w_{0}\ln\frac{1}{w_{0}}\right]+O\left(\lambda^{2}\right),\label{eqapp:m1-expression_first-order-in-lambda}
\end{equation}
\begin{equation}
0=\frac{\pi}{4}\left(1+\lambda\ln\frac{1}{\kappa}\right)+w_{0}\ln\frac{1}{\kappa}+w_{0}\ln\frac{1}{w_{0}}+\lambda\cdot F\left(w_{0}\right)+O\left(\lambda^{2}\right),\label{eqapp:w0-equation_first-order-in-lambda}
\end{equation}
\begin{align}
F\left(w_{0}\right) & =G\left(w_{0}\right)-H\left(w_{0},0\right)=\intop_{0}^{1}dw_{1}\cdot\left[\frac{w_{1}^{2}+\left(1-w_{1}^{2}\right)\ln\frac{1}{w_{0}}}{\sqrt{1-w_{1}^{2}}}+\frac{\left(w_{0}+w_{1}\right)^{2}\ln\frac{w_{0}}{w_{0}+w_{1}}+w_{0}w_{1}}{w_{1}^{2}\sqrt{1-w_{1}^{2}}}\right]\nonumber \\
 & =\frac{\pi}{4}\left(1+\ln\frac{1}{w_{0}}\right)+\frac{\pi}{4}\left(2w_{0}^{2}+3+2\ln2-\left(\frac{4}{\pi}+\pi\right)w_{0}-2\ln\frac{1}{w_{0}}\right)\nonumber \\
 & +w_{0}\left[\text{arccos}\frac{1}{w_{0}}\right]^{2}-w_{0}\left[w_{0}\sqrt{w_{0}^{2}-1}-\ln\left(w_{0}-\sqrt{w_{0}^{2}-1}\right)\right]\text{arccos}\frac{1}{w_{0}}\nonumber \\
 & +\left[i\cdot\text{Li}_{2}\left(\frac{i-\sqrt{w_{0}^{2}-1}}{w_{0}}\right)-i\cdot\text{Li}_{2}\left(\frac{-i+\sqrt{w_{0}^{2}-1}}{w_{0}}\right)-\frac{i\pi}{4}\ln\left(\frac{-w_{0}^{2}+2\sqrt{1-w_{0}^{2}}+2}{w_{0}^{2}}\right)\right],\label{eqapp:F-function-def}
\end{align}
where $\text{Li}_{n}\left(z\right)$ is the polylogarithm function~\eqappref{polylog-function_definition}.
Similarly to expression~\eqappref{G-function-definition} for the
$G$ function, the $F$ function is purely real for all $w_{0}>0$
and represents a holomorphic function of $w_{0}\in\mathbb{C}$ with
a branch cut along $\left[-\infty,w_{0}\right]$. 

From this result, we can also infer the actual limit of applicability
of the proposed procedure. Indeed, at large $w_{0}$ the sub-leading
term behaves as
\begin{equation}
F\left(w_{0}\gg1\right)=\frac{\pi}{4}\left(1+\ln\frac{1}{w_{0}}\right)-\frac{1}{3w_{0}}+O\left(\frac{1}{w_{0}^{2}}\right),
\end{equation}
so that the obtained series for $g_{1}+\lambda g_{2}$ is governed
by the parameter $\lambda/w_{0}\ll1$. Quite conveniently, this also
happens to coincide with the criteria of applicability of the expression~\eqappref{m2-expression_small-lambda-limit}
for the $m_{2}$ function.

As a by-product of the presented calculation, one can infer the explicit
result of the typical value of the distribution. The latter is defined
as
\begin{equation}
y_{\text{typ}}:=\exp\left\{ \left\langle \ln y\right\rangle \right\} .
\end{equation}
One can observe that it is connected to the derivative of $g_{1}$
function as:
\begin{equation}
\frac{\partial g_{1}}{\partial w}\left(w_{0},0;\lambda\right)\equiv\left[\frac{\partial}{\partial w}\left\langle \frac{y+\kappa w}{\kappa}\ln\frac{\kappa}{y+\kappa w}\right\rangle \right]_{w=0}=\left\langle \ln\frac{\kappa}{y}\right\rangle -1\equiv\ln\kappa-\ln y_{\text{typ}}-1.
\end{equation}
Consequently, the typical value reads:
\begin{equation}
y_{\text{typ}}=\kappa\cdot\exp\left\{ -\left[\frac{\partial g_{1}}{\partial w}\left(w_{0},0;\lambda\right)+1\right]\right\} =\kappa w_{0}\cdot\exp\left\{ -\lambda\cdot\frac{\partial H}{\partial w}\left(w_{0},0\right)+O\left(\lambda^{2}\right)\right\} =\kappa w_{0}\cdot\exp\left\{ -\frac{\pi\lambda}{4w_{0}}+O\left(\lambda^{2}\right)\right\} .
\end{equation}
To obtain the last expression, we have evaluated the required expression
for the $H$ function.

Solving the resulting equation~\eqappref{w0-equation_first-order-in-lambda}
for $w_{0}$ up to the available precision renders
\begin{equation}
w_{0}=w_{0}^{\left(0\right)}+\lambda\cdot w_{0}^{\left(1\right)}+O\left(\lambda^{2}\right),\,\,\,\,\,w_{0}^{\left(0\right)}=\frac{\pi/4}{W\left(\pi\kappa/4\right)},\,\,\,\,\,w_{0}^{\left(1\right)}=\frac{\frac{\pi}{4}\ln\frac{1}{\kappa}+F\left(w_{0}^{\left(0\right)}\right)}{\ln\kappa w_{0}^{\left(0\right)}+1}.\label{eqapp:solution-to-w0-aka-mean-value}
\end{equation}
The result suggests that the presented derivation is applicable when
\begin{equation}
w_{0}^{\left(0\right)}\apprge\lambda\Leftrightarrow\kappa\apprle\frac{4}{\pi}\exp\left\{ \frac{\pi}{4}\frac{1}{\lambda}\right\} .\label{eqapp:small-lamda-solution_criteria-of-appilicability}
\end{equation}
This corresponds to the following limitation on the value of $Z$:
\begin{equation}
Z\apprge Z^{*}=\frac{\lambda}{2\nu_{0}\Delta_{0}\cdot\frac{4}{\pi}\exp\left\{ \frac{\pi}{4}\frac{1}{\lambda}\right\} }\sim\frac{\pi}{4}\cdot\frac{\lambda}{4\nu_{0}\varepsilon_{D}}\cdot\exp\left\{ \frac{1}{\lambda}\left(1-\frac{\pi}{4}\right)\right\} .
\end{equation}
Remarkably, this scale is exponentially smaller than $Z_{1}=\lambda\cdot\exp\left\{ 1/2\lambda\right\} $,
with the latter being the scale suggested by \citep{Feigelman_SIT_2010}
as the lower limit below which the original self-consistency equation
is rendered inapplicable, see the discussion after Eq.~\eqref{saddle-point_order-parameter}
for details. Consequently, the developed approximation covers the
entire region of applicability of the proposed model for sufficiently
small $\lambda$.

Let us now summarize the obtained results: we have derived the following
analytical solution to the integral equations~(\hphantom{}\ref{eqapp:equation-on-m1}\nobreakdash-\ref{eqapp:equation-on-m2}\hphantom{}):
\begin{equation}
m_{1}\left(w\right)=\kappa\left(w+w_{0}\right)+\lambda\cdot\left[\left(w_{0}+w\right)\ln\frac{1}{w_{0}+w}-w_{0}\ln\frac{1}{w_{0}}\right],\label{eqapp:m1-expression_first-order-in-lambda-1}
\end{equation}
\begin{equation}
m_{2}\left(S|w\right)=\lambda\cdot\left[\left(w+w_{0}\right)\Phi_{0}\left(\kappa S\right)+\Phi_{1}\left(\kappa S\right)\right],\label{eqapp:m2-expression_small-lambda-limit-1}
\end{equation}
\begin{equation}
w_{0}=w_{0}^{\left(0\right)}+\lambda\cdot w_{0}^{\left(1\right)},\,\,\,\,\,w_{0}^{\left(0\right)}=\frac{\pi/4}{W\left(\pi\kappa/4\right)},\,\,\,\,w_{0}^{\left(1\right)}=\frac{\frac{\pi}{4}\ln\frac{1}{\kappa}+F\left(w_{0}^{\left(0\right)}\right)}{\ln\kappa w_{0}^{\left(0\right)}+1},\label{eqapp:w0-solution-small-lambda-limit}
\end{equation}
where $\Phi_{i}$ are special functions given by Eq.\nobreakdash-s~(\hphantom{}\ref{eqapp:Phi-0-definition}\nobreakdash-\ref{eqapp:Phi-1-definition}\hphantom{}),
and $F$ is given by Eq.~\eqappref{F-function-def}. The solution
is controlled by the parameter $\lambda/w_{0}\ll1$, which turns out
to be small in the entire limit of applicability of the proposed model.
The presented solution comprises all previously discussed special
cases. For instance, one can formally expand these equations in the
limit $\kappa\ll1$, corresponding to the Gaussian regime. The result
then reduces back to expressions~(\hphantom{}\ref{eqapp:m1_large-Keff-limit}\nobreakdash-\ref{eqapp:m2_large-Keff-limit}\hphantom{}),
thus successfully reproducing the large $Z_{\text{eff}}$ limit. In
the opposite limit of $\kappa\apprge1$, the expressions~(\hphantom{}\ref{eqapp:m1-expression_first-order-in-lambda-1}\nobreakdash-\ref{eqapp:w0-solution-small-lambda-limit}\hphantom{})
provide a quantitative demonstration for the asymptotic behavior described
in \appref{Extreme-value-statistics}.

\section{The effect of smooth energy dependence of the matrix element\label{app:Effect-of-energy-dependence-of-matrix-element}}

In the simplified model discussed in \appref{equation-on-distribution_small-Delta},
we have ignored the dependence of the matrix element $D\left(\xi\right)$
on the energy difference, replacing it with a constant value given
by $D\left(0\right)$. In this section, we restore this dependence
and discuss its effect on our results.

\subsection{Qualitative discussion\label{subsecapp:qualitative-discussion-of-xi-dependence}}

Let us first briefly discuss the role of the $D$ function within
the mean-field approach described in \appref{Saddle-point-equation_derivation}.
In the simplified version of the theory we have been discussing thus
far, the logarithmically divergent integral in Eq.~\eqappref{saddle-point_mean-field}
was cut off at energies of order $E_{F}$ originating purely from
the behavior of the density of states $\nu\left(\xi\right)$. The
main role of the energy dependence $D\left(\xi\right)$ is to provide
a more physical upper limit cut-off of the order of $\varepsilon_{D}$,
as it was shown in \appref{Saddle-point-equation_derivation}. However,
the difference between the effect of $\nu\left(\xi\right)$ and $D\left(\xi\right)$
is that the latter endows the order parameter with and additional
dependence on the onsite value of the disorder field $\xi_{0}$. In
other words, one now has to describe not just the value of $\Delta_{0}$,
but the whole function $\Delta\left(\xi_{0}\right)$. As it is demonstrated
in \appref{Saddle-point-equation_derivation}, the profile of $\Delta\left(\xi_{0}\right)$
resembles that of $D\left(\xi_{0}\right)$ itself, and thus the emergent
typical scale of $\xi_{0}$-dependence in also given by $\xi_{0}\sim\varepsilon_{D}$.

Connected to the mean-field equation is the question of the actual
number of sites participating in the superconducting order. Indeed,
in the current version of the model, each site develops its own value
of the order parameter of order $\Delta_{0}$, regardless of the onsite
value of $\xi$. On the other hand, already at the level of the mean-field
equation it is clear that only sites with energies $\left|\xi\right|\apprle\varepsilon_{D}$
can participate in the formation of superconducting state due to the
limitation on maximum energy transfer. In other words, we expect the
actual joint probability distribution of $\xi$ and $\Delta$ on a
given site to behave as
\begin{equation}
P\left(\xi,\Delta\right)\sim\begin{cases}
\frac{\varepsilon_{D}}{E_{F}}\cdot P_{0}\left(\xi,\Delta\right), & \left|\xi\right|\apprle\varepsilon_{D},\\
\left(1-\frac{\varepsilon_{D}}{E_{F}}\right)\cdot\nu\left(\xi\right)\cdot\delta\left(\Delta\right), & \left|\xi\right|\apprge\varepsilon_{D},
\end{cases}
\end{equation}
where $P_{0}$ is the low-energy joint distribution~\eqappref{marginal-probability-distribution_small-dela}
found in the previous section. The leading prefactors in both expressions
ensure normalization and reflect the fact that only $\sim N\cdot\varepsilon_{D}/E_{F}$
of all $N$ sites in the system actually develop superconducting ordering.
Therefore, the role of the $\eta$ function in a more accurate version
of the theory is to exclude the sites deep within the Fermi sea from
superconducting correlations.

\subsection{Equations for the $m$ function}

Let us now turn to quantitative description of the outlined differences.
From the considerations above one concludes that the $m$~function
and all associated objects should also exhibit a slow dependence on
the onsite value of $x=\xi/\Delta_{0}$. The corresponding counterpart
of the equation~\eqappref{equation-on-r-function} for the $r$ function
now reads:
\begin{align}
r\left(S|x,y\right) & =\intop_{\mathbb{R}}dx_{1}\intop_{0}^{\infty}dy_{1}\cdot\left[\exp\left\{ iS\frac{f\left(\Delta_{0}x_{1},\Delta_{0}y_{1}|\Delta_{0}x\right)}{\Delta_{0}}\right\} -1\right]\cdot\frac{\nu\left(\Delta_{0}x_{1}\right)}{2\nu_{0}}\nonumber \\
 & \times\intop_{\mathbb{R}-i0}\frac{ds}{2\pi}\cdot\exp\left\{ is\frac{f\left(\Delta_{0}x,\Delta_{0}y|\Delta_{0}x_{1}\right)}{\Delta_{0}}\right\} \frac{\partial}{\partial y_{1}}\left\{ \left[\intop_{\infty}^{y_{1}}dy_{1}^{'}\exp\left\{ -isy_{1}^{'}\right\} \right]\cdot\exp\left\{ Z_{\text{eff}}\cdot r\left(s|x_{1},y_{1}\right)\right\} \right\} .
\end{align}
It is convenient to parametrize the $f$ function as
\begin{equation}
f\left(\xi,\Delta|\xi_{0}\right)=\Delta_{0}\cdot\kappa\cdot\omega\left(\xi/\Delta\right)\cdot\eta\left(\frac{\xi-\xi_{0}}{\Delta_{0}}\right),
\end{equation}
where $\eta$ is the function determining the energy dependence of
the matrix element (c.f. with \eqappref{saddle-point-equation-derivation_envelope-functions-defs}):
\begin{equation}
\eta\left(x\right)\equiv u\left(\Delta_{0}x\right)=D\left(\Delta_{0}x\right)/D\left(0\right).\label{eqapp:smooth-energy-dependece_eta-definition}
\end{equation}
One can also introduce the following parametrization for the $r$
function:
\begin{equation}
r\left(S|x,y\right):=r\left(S|w=\omega\left(x/y\right),x\right),
\end{equation}
so that the $x$-dependence is explicitly factorized into the low-energy
part corresponding to $w$ argument and high-energy dependence originating
from the presence of the $\eta$ function. Extracting the main logarithmic
divergence according to the procedure outlined in \subsecappref{Expressions-for-distributions}
produces the following set of equations:

\begin{equation}
r\left(S|w,\underline{x}\right)=\left[\text{target eq.}-\Delta r\right]+\left[\Delta r-\Delta r_{f}\right]+\Delta r_{f},
\end{equation}
\begin{align}
\left[\text{target eq.}-\Delta r\right] & =\intop_{\mathbb{R}}dx_{1}\intop_{0}^{\infty}dy_{1}\cdot\left[\exp\left\{ iS\kappa\cdot\omega\left(x_{1}/y_{1}\right)\cdot\underline{\eta\left(x-x_{1}\right)}\right\} -1-iS\kappa\cdot\omega\left(x_{1}/y_{1}\right)\cdot\underline{\eta\left(x-x_{1}\right)}\right]\cdot\frac{\nu\left(\Delta_{0}x_{1}\right)}{2\nu_{0}}\nonumber \\
 & \times\intop_{\mathbb{R}-i0}\frac{ds}{2\pi}\cdot\exp\left\{ is\kappa\cdot w\cdot\underline{\eta\left(x_{1}-x\right)}\right\} \frac{\partial}{\partial y_{1}}\left\{ \left[\intop_{\infty}^{y_{1}}dy_{1}^{'}e^{-isy_{1}^{'}}\right]\cdot\exp\left\{ Z_{\text{eff}}\cdot r\left(s|\omega\left(x_{1}/y_{1}\right),\underline{x_{1}}\right)\right\} \right\} ,
\end{align}
 
\begin{align}
\left[\Delta r-\Delta r_{r}\right] & =\intop_{\mathbb{R}}dx_{1}\intop_{0}^{\infty}dy_{1}\cdot iS\kappa\cdot\omega\left(x_{1}/y_{1}\right)\cdot\underline{\eta\left(x-x_{1}\right)}\cdot\frac{\nu\left(\Delta_{0}x_{1}\right)}{2\nu_{0}}\cdot\intop_{\mathbb{R}-i0}\frac{ds}{2\pi}\cdot\exp\left\{ is\kappa\cdot w\cdot\underline{\eta\left(x_{1}-x\right)}\right\} \nonumber \\
 & \times\frac{\partial}{\partial y_{1}}\left\{ \left[\intop_{\infty}^{y_{1}}dy_{1}^{'}e^{-isy_{1}^{'}}\right]\left(\exp\left\{ Z_{\text{eff}}\cdot r\left(s|\omega\left(x_{1}/y_{1}\right),\underline{x_{1}}\right)\right\} -\exp\left\{ Z_{\text{eff}}\cdot r\left(s|0,\underline{x_{1}}\right)\right\} \right)\right\} ,
\end{align}
 
\begin{align}
\left[\Delta r_{r}-\Delta r_{rf}\right] & =\intop_{\mathbb{R}}dx_{1}\intop_{0}^{\infty}dy_{1}\cdot iS\kappa\cdot\left[\omega\left(x_{1}/y_{1}\right)-y_{1}\omega\left(x_{1}\right)\right]\cdot\underline{\eta\left(x-x_{1}\right)}\cdot\frac{\nu\left(\Delta_{0}x_{1}\right)}{2\nu_{0}}\nonumber \\
 & \times\intop_{\mathbb{R}-i0}\frac{ds}{2\pi}\cdot\exp\left\{ is\kappa\cdot w\cdot\underline{\eta\left(x_{1}-x\right)}\right\} \cdot\frac{\partial}{\partial y_{1}}\left\{ \left[\intop_{\infty}^{y_{1}}dy_{1}^{'}e^{-isy_{1}^{'}}\right]\cdot\exp\left\{ Z_{\text{eff}}\cdot r\left(s|0,\underline{x_{1}}\right)\right\} \right\} ,
\end{align}
 
\begin{align}
\Delta r_{rf} & =\intop_{\mathbb{R}}dx_{1}\intop_{0}^{\infty}dy_{1}\cdot iS\kappa\cdot y_{1}\omega\left(x_{1}\right)\cdot\underline{\eta\left(x-x_{1}\right)}\cdot\frac{\nu\left(\Delta_{0}x_{1}\right)}{2\nu_{0}}\nonumber \\
 & \times\intop_{\mathbb{R}-i0}\frac{ds}{2\pi}\cdot\exp\left\{ is\kappa\cdot w\cdot\underline{\eta\left(x_{1}-x\right)}\right\} \cdot\frac{\partial}{\partial y_{1}}\left\{ \left[\intop_{\infty}^{y_{1}}dy_{1}^{'}e^{-isy_{1}^{'}}\right]\cdot\exp\left\{ Z_{\text{eff}}\cdot r\left(s|0,\underline{x_{1}}\right)\right\} \right\} ,
\end{align}
where underlined are the differences of these expressions to their
counterparts in \subsecappref{Eq-on-the-char-function-derivation}.
Similarly to the calculations of \subsecappref{Eq-on-the-char-function-derivation},
the next step is to treat the functions $\eta\left(x\right)$, $\nu\left(x\right)$
and $r\left(...,x\right)$ as constant in all expressions where the
corresponding integral is convergent at the scale $x\sim1$. This
allows one to rewrite the equations above as
\begin{align}
\left[\text{target eq.}-\Delta r\right] & =\frac{1}{2}\intop_{\mathbb{R}}dx_{1}\intop_{0}^{\infty}dy_{1}\cdot\left[\exp\left\{ iS\kappa\cdot\omega\left(x_{1}/y_{1}\right)\cdot\underline{\eta\left(x\right)}\right\} -1-iS\kappa\cdot\omega\left(x_{1}/y_{1}\right)\cdot\underline{\eta\left(x\right)}\right]\nonumber \\
 & \times\intop_{\mathbb{R}-i0}\frac{ds}{2\pi}\cdot\exp\left\{ is\kappa\cdot w\cdot\underline{\eta\left(x\right)}\right\} \frac{\partial}{\partial y_{1}}\left\{ \left[\intop_{\infty}^{y_{1}}dy_{1}^{'}e^{-isy_{1}^{'}}\right]\cdot\exp\left\{ Z_{\text{eff}}\cdot r\left(s|\omega\left(x_{1}/y_{1}\right),\underline{0}\right)\right\} \right\} ,
\end{align}
 
\begin{align}
\left[\Delta r-\Delta r_{r}\right] & =iS\kappa\underline{\eta\left(x\right)}\cdot\frac{1}{2}\intop_{\mathbb{R}}dx_{1}\intop_{0}^{\infty}dy_{1}\cdot\omega\left(x_{1}/y_{1}\right)\cdot\intop_{\mathbb{R}-i0}\frac{ds}{2\pi}\cdot\exp\left\{ is\kappa\cdot w\cdot\underline{\eta\left(x\right)}\right\} \nonumber \\
 & \times\frac{\partial}{\partial y_{1}}\left\{ \left[\intop_{\infty}^{y_{1}}dy_{1}^{'}e^{-isy_{1}^{'}}\right]\left(\exp\left\{ Z_{\text{eff}}\cdot r\left(s|\omega\left(x_{1}/y_{1}\right),\underline{0}\right)\right\} -\exp\left\{ Z_{\text{eff}}\cdot r\left(s|0,\underline{0}\right)\right\} \right)\right\} ,
\end{align}
 
\begin{align}
\left[\Delta r_{r}-\Delta r_{rf}\right] & =iS\kappa\underline{\eta\left(x\right)}\cdot\frac{1}{2}\intop_{\mathbb{R}}dx_{1}\intop_{0}^{\infty}dy_{1}\cdot\left[\omega\left(x_{1}/y_{1}\right)-y_{1}\omega\left(x_{1}\right)\right]\nonumber \\
 & \times\intop_{\mathbb{R}-i0}\frac{ds}{2\pi}\cdot\exp\left\{ is\kappa\cdot w\cdot\underline{\eta\left(x\right)}\right\} \cdot\frac{\partial}{\partial y_{1}}\left\{ \left[\intop_{\infty}^{y_{1}}dy_{1}^{'}e^{-isy_{1}^{'}}\right]\cdot\exp\left\{ Z_{\text{eff}}\cdot r\left(s|0,\underline{0}\right)\right\} \right\} ,
\end{align}
 
\begin{align}
\Delta r_{rf} & =\intop_{\mathbb{R}}dx_{1}\intop_{0}^{\infty}dy_{1}\cdot iS\kappa\cdot y_{1}\omega\left(x_{1}\right)\cdot\underline{\eta\left(x-x_{1}\right)}\cdot\frac{\nu\left(\Delta_{0}x_{1}\right)}{2\nu_{0}}\nonumber \\
 & \times\intop_{\mathbb{R}-i0}\frac{ds}{2\pi}\cdot\exp\left\{ is\kappa\cdot w\cdot\underline{\eta\left(x_{1}-x\right)}\right\} \cdot\frac{\partial}{\partial y_{1}}\left\{ \left[\intop_{\infty}^{y_{1}}dy_{1}^{'}e^{-isy_{1}^{'}}\right]\cdot\exp\left\{ Z_{\text{eff}}\cdot r\left(s|0,\underline{x_{1}}\right)\right\} \right\} .
\end{align}
Note that the last expression remained intact, as it still contains
a logarithmic integral over $x_{1}$. One then proceeds to simplifying
these expressions in a way similar to that presented in \subsecappref{Eq-on-the-char-function-derivation}.
The result then reads:
\begin{equation}
m\left(S|w,\underline{x}\right):=Z_{\text{eff}}\cdot r\left(S|w,\underline{x}\right)=iSm_{1}\left(w,\underline{x}\right)+m_{1}\left(S|w,\underline{x}\right),
\end{equation}
\begin{equation}
m_{2}\left(S|w,x\right)=\lambda\intop_{0}^{1}dw_{1}\cdot\frac{\exp\left\{ iS\kappa\underline{\eta\left(x\right)}\cdot w_{1}\right\} -1-iS\kappa\underline{\eta\left(x\right)}\cdot w_{1}}{w_{1}^{2}\sqrt{1-w_{1}^{2}}}\cdot\left[1-w_{1}\left(1-w_{1}^{2}\right)\frac{\partial}{\partial w_{1}}\right]\cdot\left[\frac{\kappa w\underline{\eta\left(x\right)}+m_{1}\left(w_{1},\underline{0}\right)}{\kappa}\right],
\end{equation}
\begin{align}
m_{1}\left(w,\underline{x}\right) & =\underline{\eta\left(x\right)}\cdot\lambda\intop_{0}^{1}dw_{1}\cdot\sqrt{1-w_{1}^{2}}\cdot\frac{m_{1}\left(w_{1},\underline{0}\right)-m_{1}\left(0,\underline{0}\right)}{w_{1}}\nonumber \\
 & +\underline{\eta\left(x\right)}\cdot\lambda\intop_{0}^{\infty}dy_{1}\cdot y_{1}\ln\frac{1}{y_{1}}\cdot\intop_{\mathbb{R}-i0}\frac{ds}{2\pi}\cdot\exp\left\{ is\kappa\underline{\eta\left(x\right)}\cdot w\right\} \cdot\exp\left\{ m\left(s|0,\underline{0}\right)-isy_{1}\right\} \nonumber \\
 & +\lambda\intop_{\mathbb{R}}dx_{1}\cdot\omega\left(x_{1}\right)\cdot\underline{\eta\left(x-x_{1}\right)}\cdot\frac{\nu\left(\Delta_{0}x_{1}\right)}{2\nu_{0}}\cdot\left\{ \kappa w\cdot\underline{\eta\left(x_{1}-x\right)}+m_{1}\left(0,\underline{x_{1}}\right)\right\} .\label{eqapp:m1-equation-1}
\end{align}

In order to further simply the last term in the equation on $m_{1}$,
one considers the following identity:
\begin{equation}
m_{1}\left(0,x\right)-\eta\left(x\right)m_{1}\left(0,0\right)=\lambda\intop_{\mathbb{R}}dx_{1}\cdot\omega\left(x_{1}\right)\cdot\frac{\nu\left(\Delta_{0}x_{1}\right)}{2\nu_{0}}\cdot m_{1}\left(0,x_{1}\right)\cdot\left(\eta\left(x-x_{1}\right)-\eta\left(x\right)\eta\left(x_{1}\right)\right),
\end{equation}
which is obtained by using Eq.~\eqappref{m1-equation-1} for both
instance of $m$ in the right hand side. As it follows from the qualitative
considerations in the beginning of this Appendix, we expect $m_{1}\left(0,x_{1}\right)$
to depend on $x_{1}$ only at the scale $x_{1}\sim\varepsilon_{D}/\Delta_{0}$.
The resulting integral over $x_{1}$ is thus governed by $x_{1}\sim\varepsilon_{D}/\Delta_{0}$,
so that one can replace $\omega\left(x_{1}\right)$ with $1/\left|x_{1}\right|$,
similarly to the derivation of Eq.~\eqappref{m1-equation-1}. We
then observe that the value of $m_{1}\left(0,x\right)$ is given by
the mean-field answer corresponding to Eq.~\eqappref{mean-field-equation_dimensionless-form}:
\begin{equation}
m_{1}\left(0,x\right)=m_{1}\left(0,0\right)\cdot d\left(\Delta_{0}\cdot x\right),
\end{equation}
where $d\left(\xi\right)$ is the function obeying Eq.~\eqappref{mean-field_d-function-equation}.
Because the~$d$~function also happens to describe the exact energy
dependence of the mean-field order parameter $\Delta_{0}\left(\xi\right)$,
one can evaluated explicitly the value of the integral containing
$m_{1}$ in Eq.~\eqappref{m1-equation-1} due to the self-consistency
equation~\eqappref{mean-field_order-parameter-value}:
\begin{equation}
\lambda\intop_{\mathbb{R}}dx_{1}\cdot\omega\left(x_{1}\right)\cdot\eta\left(x-x_{1}\right)\cdot\frac{\nu\left(\Delta_{0}x_{1}\right)}{2\nu_{0}}m_{1}\left(0,x_{1}\right)=m_{1}\left(0,0\right)\cdot d\left(\Delta_{0}x\right)
\end{equation}
The equation on $m_{1}$ can then be rewritten as
\begin{align}
m_{1}\left(w,\underline{x}\right) & =\underline{\eta\left(x\right)}\cdot\lambda\intop_{0}^{1}dw_{1}\cdot\sqrt{1-w_{1}^{2}}\cdot\frac{m_{1}\left(w_{1},\underline{0}\right)-m_{1}\left(0,\underline{0}\right)}{w_{1}}\nonumber \\
 & +\underline{\eta\left(x\right)}\cdot\lambda\intop_{0}^{\infty}dy_{1}\cdot y_{1}\ln\frac{1}{y_{1}}\cdot\intop_{\mathbb{R}-i0}\frac{ds}{2\pi}\cdot\exp\left\{ is\kappa\underline{\eta\left(x\right)}\cdot w\right\} \cdot\exp\left\{ m\left(s|0,\underline{0}\right)-isy_{1}\right\} \nonumber \\
 & +m_{1}\left(0,\underline{0}\right)\cdot\underline{d\left(\Delta_{0}x\right)}+\kappa w\cdot\lambda\intop_{\mathbb{R}}dx_{1}\cdot\frac{\nu\left(\Delta_{0}x_{1}\right)}{2\nu_{0}}\cdot\omega\left(x_{1}\right)\cdot\underline{\eta^{2}\left(x-x_{1}\right)}.\label{eqapp:m1-equation-2}
\end{align}
It only remains to simplify the last term. In a simpler model with
no $\xi$ dependence of the matrix element, this integral evaluated
to unity due to the self-consistency mean-field equation. Guided by
the analysis in \appref{Saddle-point-equation_derivation}, we rewrite
it as
\begin{equation}
\lambda\intop_{\mathbb{R}}dx_{1}\cdot\omega\left(x_{1}\right)\cdot\frac{\nu\left(\Delta_{0}x_{1}\right)}{2\nu_{0}}\cdot\eta^{2}\left(x_{1}-x\right)=\alpha\cdot\eta^{2}\left(x\right)+\lambda\cdot\psi\left(\Delta_{0}x\right),
\end{equation}
where we have denoted
\begin{equation}
\alpha=\lambda\intop_{\mathbb{R}}dx_{1}\cdot\omega\left(x_{1}\right)\cdot\frac{\nu\left(\Delta_{0}x_{1}\right)}{2\nu_{0}}\cdot\eta^{2}\left(x_{1}\right),\label{eqapp:alpha-definition}
\end{equation}
\begin{equation}
\psi\left(\xi\right)=\intop_{\mathbb{R}}d\xi_{1}\cdot\frac{\nu\left(\xi\right)}{2\nu_{0}}\cdot\frac{u^{2}\left(\xi_{1}-\xi\right)-u^{2}\left(\xi_{1}\right)u^{2}\left(\xi\right)}{\left|\xi_{1}\right|}.\label{eqapp:smooth-energy-dependence_psi-def}
\end{equation}
The equation~\eqappref{alpha-definition} for $\alpha$ can be further
simplified to exclude low-energy scales:
\begin{equation}
\alpha=1+\lambda\intop_{\mathbb{R}}d\xi_{1}\cdot\frac{\nu\left(\xi_{1}\right)}{2\nu_{0}}\cdot u\left(\xi_{1}\right)\cdot\frac{u\left(\xi_{1}\right)-d\left(\xi_{1}\right)}{\left|\xi_{1}\right|},
\end{equation}
where we have again made use of the mean-field equation~\eqappref{mean-field_order-parameter-value}
in its dimensionless form as well as the dimensionfull counterpart
$u\left(\xi\right)$ of the $\eta$ function. For the low-energy physics
it is important that $\alpha$ is close to unity. Indeed, as it is
shown in \appref{Saddle-point-equation_derivation}, the difference
between $g$ and $\eta$ is of order $\lambda$, so that $\alpha$
differs from unity by a quantity the of order $O\left(\lambda^{2}\right)$.

One can thus write down the equations for the $m$ function in their
final form:
\begin{equation}
m\left(S|w,\underline{x}\right)=iSm_{1}\left(w,\underline{x}\right)+m_{1}\left(S|w,\underline{x}\right),
\end{equation}
\begin{equation}
m_{2}\left(S|w,\underline{x}\right)=\lambda\intop_{0}^{1}dw_{1}\cdot\frac{\exp\left\{ iS\kappa\underline{\eta\left(x\right)}\cdot w_{1}\right\} -1-iS\kappa\underline{\eta\left(x\right)}\cdot w_{1}}{w_{1}^{2}\sqrt{1-w_{1}^{2}}}\cdot\left[1-w_{1}\left(1-w_{1}^{2}\right)\frac{\partial}{\partial w_{1}}\right]\cdot\left[\frac{\kappa w\underline{\eta\left(x\right)}+m_{1}\left(w_{1},\underline{0}\right)}{\kappa}\right],\label{eqapp:smooth-energy-dependence_m2-equation}
\end{equation}
\begin{align}
m_{1}\left(w,\underline{x}\right) & =\underline{\eta\left(x\right)}\cdot\lambda\intop_{0}^{1}dw_{1}\cdot\sqrt{1-w_{1}^{2}}\cdot\frac{m_{1}\left(w_{1},\underline{0}\right)-m_{1}\left(0,\underline{0}\right)}{w_{1}}\nonumber \\
 & +\underline{\eta\left(x\right)}\cdot\lambda\intop_{0}^{\infty}dy_{1}\cdot y_{1}\ln\frac{1}{y_{1}}\cdot\intop_{\mathbb{R}-i0}\frac{ds}{2\pi}\cdot\exp\left\{ is\kappa\underline{\eta\left(x\right)}\cdot w\right\} \cdot\exp\left\{ m\left(s|0,\underline{0}\right)-isy_{1}\right\} \nonumber \\
 & +m_{1}\left(0,\underline{0}\right)\cdot\underline{d\left(\Delta_{0}x\right)}+\kappa w\cdot\left[\underline{\eta^{2}\left(x\right)\alpha+\lambda\cdot\psi\left(\Delta_{0}x\right)}\right].\label{eqapp:smooth-energy-dependence_m1-equation}
\end{align}
Here, the $d\left(\xi=\Delta_{0}x\right)$ function is given by the
solution to Eq.~\eqappref{mean-field_d-function-equation}, the function
$\psi\left(\xi\right)$ is defined as
\begin{equation}
\psi\left(\xi\right)=\intop_{\mathbb{R}}d\xi_{1}\cdot\frac{\nu\left(\xi_{1}\right)}{2\nu_{0}}\cdot\frac{u^{2}\left(\xi_{1}-\xi\right)-u^{2}\left(\xi_{1}-\xi\right)u^{2}\left(\xi_{1}-\xi\right)}{\left|\xi_{1}\right|},\label{eqapp:smooth-energy-dependence_psi-def-2}
\end{equation}
and the value of the coefficient $\alpha$ is given by
\begin{equation}
\alpha=1+\lambda\intop_{\mathbb{R}}d\xi_{1}\cdot\frac{\nu\left(\xi_{1}\right)}{2\nu_{0}}\cdot u\left(\xi_{1}\right)\cdot\frac{\eta\left(\xi_{1}\right)-d\left(\xi_{1}\right)}{\left|\xi_{1}\right|}.\label{eqapp:smooth-energy-dependence_alpha-def}
\end{equation}
Note that $\psi\left(\xi\right)$ obeys the condition $\psi\left(0\right)=0$
and depends on $\xi$ at the scale of the Debye energy $\left|\xi\right|\sim\varepsilon_{D}$
as it follows from the definition of the $\eta$ function. Similarly
to equations~(\hphantom{}\ref{eqapp:equation-on-m1}\nobreakdash-\ref{eqapp:equation-on-m2}\hphantom{})
of a simpler model, one has to determine the form of the $m_{1}$~function
by solving the system of coupled integro-differential equations. 

Let us consider this system of equations at $x=0$:
\begin{align}
m_{2}\left(S|w,0\right) & =\lambda\cdot\intop_{0}^{1}dw_{1}\cdot\frac{\exp\left\{ iS\kappa\cdot w_{1}\right\} -1-iS\kappa\cdot w_{1}}{w_{1}^{2}\sqrt{1-w_{1}^{2}}}\cdot\left[1-w_{1}\left(1-w_{1}^{2}\right)\frac{\partial}{\partial w_{1}}\right]\cdot\left[\frac{\kappa w+m_{1}\left(w_{1},0\right)}{\kappa}\right],
\end{align}
\begin{align}
m_{1}\left(w,0\right) & =\lambda\intop_{0}^{1}dw_{1}\cdot\sqrt{1-w_{1}^{2}}\cdot\frac{m_{1}\left(w_{1},0\right)-m_{1}\left(0,0\right)}{w_{1}}\nonumber \\
 & +\lambda\intop_{0}^{\infty}dy_{1}\cdot y_{1}\ln\frac{1}{y_{1}}\cdot\intop_{\mathbb{R}-i0}\frac{ds}{2\pi}\cdot\exp\left\{ is\kappa\cdot w\right\} \cdot\exp\left\{ m\left(s|0,0\right)-isy_{1}\right\} \nonumber \\
 & +m_{1}\left(0,0\right)+\kappa w\cdot\alpha,
\end{align}
where we have used the normalization $\eta\left(0\right)=d\left(0\right)=1$.
The only difference between these equations and the ones obtained
for the simple model with no $D\left(\xi\right)$ dependence is that
$\alpha$ now differs from unity. Remarkably, however, the difference
is small as $\alpha-1=O\left(\lambda^{2}\right)$. In particular,
the approximate solution developed in \appref{Solution_small-lambda}
applies to these equations unchanged. The value of $m$ function at
$x=0$ remains quantitatively correct up until $\left|x\right|\sim\varepsilon_{D}/\Delta_{0}$,
where it quickly decays to zero.

\subsection{Distribution functions}

As discussed in \subsecappref{qualitative-discussion-of-xi-dependence},
the energy dependence of the matrix element of the Cooper attraction
leads to a modification of the expressions for the distribution functions.
Similarly to \subsecappref{Expressions-for-distributions}, one can
use the exact equations~(\hphantom{}\ref{eqapp:equation-on-onsite-distribution}\nobreakdash-\ref{eqapp:equation-on-modified-distribution}\hphantom{})
to obtain the following expressions for the distribution functions
of interest:
\begin{align}
\mathcal{P}\left(x_{1},y_{1}|x_{0},y_{0}\right) & =P\left(x_{1}\right)\cdot\intop_{\mathbb{R}}\frac{ds}{2\pi}\cdot\exp\left\{ is\kappa\cdot\omega\left(x_{0}/y_{0}\right)\cdot\underline{\eta\left(x_{1}-x_{0}\right)}\right\} \nonumber \\
 & \times\frac{\partial}{\partial y_{0}}\left\{ \left[\intop^{y_{0}}dy_{1}^{'}\exp\left\{ -isy_{1}\right\} \right]\cdot\exp\left\{ m\left(s|\omega\left(x_{1}/y_{1}\right),\underline{x_{1}}\right)\right\} \right\} ,
\end{align}
\begin{equation}
P\left(x,y\right)=P\left(x\right)\cdot\intop_{\mathbb{R}}\frac{ds}{2\pi}\cdot\frac{\partial}{\partial y}\left\{ \left[\intop^{y}dy^{'}\exp\left\{ -isy^{'}\right\} \right]\cdot\exp\left\{ m\left(s|\omega\left(x/y\right),\underline{x}\right)\right\} \right\} .
\end{equation}
For $x_{i}\sim1$ one can neglect the slow explicit dependence of
$m$ and $\eta$ on $x$ and simplify the expressions above to
\begin{align}
\mathcal{P}\left(x_{1},y_{1}|x_{0},y_{0}\right) & =P\left(x_{1}\right)\cdot\intop_{\mathbb{R}}\frac{ds}{2\pi}\cdot\exp\left\{ is\kappa\cdot\omega\left(x_{0}/y_{0}\right)\right\} \nonumber \\
 & \times\frac{\partial}{\partial y_{0}}\left\{ \left[\intop^{y_{0}}dy_{1}^{'}\exp\left\{ -isy_{1}\right\} \right]\cdot\exp\left\{ m\left(s|\omega\left(x_{1}/y_{1}\right),0\right)\right\} \right\} .
\end{align}
\begin{equation}
P\left(x,y\right)=P\left(x\right)\cdot\intop_{\mathbb{R}}\frac{ds}{2\pi}\cdot\frac{\partial}{\partial y}\left\{ \left[\intop^{y}dy^{'}\exp\left\{ -isy^{'}\right\} \right]\cdot\exp\left\{ m\left(s|\omega\left(x/y\right),0\right)\right\} \right\} .
\end{equation}
Finally, one can neglect the difference between $\alpha$ and unity
e.g. when using the approximate solution from \appref{Solution_small-lambda},
in which case these expressions are identical to the results~(\hphantom{}\ref{eqapp:expression-for-modified-distribution_small-delta}\nobreakdash-\ref{eqapp:expression-for-onsite-probability_small-delta}\hphantom{})
in the previous section. 

We note that for a trivial choice $\eta\equiv1$, the new results
properly reduce back to Eq.~\appref{equation-on-distribution_small-Delta}.
One can also consider a simplistic model function of the form
\begin{equation}
\eta\left(x\right)=\begin{cases}
1, & \left|x\right|<\varepsilon_{D}/\Delta_{0},\\
0, & \left|x\right|>\varepsilon_{D}/\Delta_{0},
\end{cases}
\end{equation}
which corresponds to a hard cut-off of the $D$ function at the Debye
energy $\varepsilon_{D}$. It is not exactly physical, but it helps
to illustrate the result of our calculations. In this case, $\alpha=1$
exactly and the solution for $m$ is given by
\begin{equation}
m\left(S|w,x\right)=\begin{cases}
m_{0}\left(S|w\right), & \left|x\right|\le\varepsilon_{D}/\Delta_{0},\\
0, & \left|x\right|>\varepsilon_{D}/\Delta_{0},
\end{cases}
\end{equation}
where $m_{0}$ is the solution for the case of no $D\left(\xi\right)$
dependence. The values for the probabilities then read:
\begin{equation}
P\left(x,y\right)=\begin{cases}
P_{0}\left(x,y\right), & \left|x\right|<\varepsilon_{D}/\Delta_{0},\\
\nu\left(x\right)\cdot\delta\left(y\right), & \left|x\right|>\varepsilon_{D}/\Delta_{0},
\end{cases}
\end{equation}
\begin{equation}
P\left(y\right)=P_{0}\left(y\right)\cdot\intop_{0}^{\varepsilon_{D}/\Delta_{0}}dx\cdot P\left(x\right)+\delta\left(y\right)\cdot\intop_{\varepsilon_{D}/\Delta_{0}}^{\infty}dx\cdot P\left(x\right)\approx P_{0}\left(y\right)\cdot\frac{\varepsilon_{D}}{\Delta_{0}}+\delta\left(y\right)\cdot\left(1-\frac{\varepsilon_{D}}{\Delta_{0}}\right),
\end{equation}
where $P_{0}\left(x,y\right)$ and $P_{0}\left(y\right)$ are the
values for the case of no $\xi$ dependence of the matrix element.
The results are thus consistent with the expectations outlined in
the beginning of this section.

We conclude this section by noting that the exact marginal probability
distribution $P\left(y\right)$ ceases to be physically important
for the case of nontrivial $\eta\left(x\right)$ dependence. Indeed,
as it is apparent from the discussion above, the value of $P\left(y\right)$
does not discriminate between physically important sites close to
the Fermi level and those deep within the Fermi sea. It is physically
more sensible to consider the conditional probability function $P\left(y|x\right)=P\left(x,y\right)/P\left(x\right)$
for $x\sim1$, which contains the actual behavior of the order parameter.
That is why, it is valid to claim that the distribution $P_{0}\left(y\right)$
in the naive model without the energy dependence is a proper quantity
describing the statistics the order parameter. Consequently, the $m$
function still characterizes the cumulants of this distribution. Finally,
the average value denoted by $\left\langle \bullet\right\rangle $
in \subsecappref{Expressions-for-distributions} should be interpreted
as those over $P_{0}\left(y\right)$ rather than the full distribution
$P\left(y\right)$.

\section{The model with fluctuations of the matrix element of the Cooper attraction\label{app:Effect-of-fluctuating-coupling}}

One of the most drastic simplifications of the model thus far is our
complete disregard to the  fluctuations of the matrix element of Cooper
attractions between the localized single-particle states. Not only
we have neglected the fluctuations of the sheer number $Z$ of effectively
interacting neighbors, but we have also treated the value of this
matrix element between each pair of interacting states as constant.
As discussed previously in~\subsecappref{Probability_large-y-asymptotic_secondary-peaks},
this results in physically improbable secondary maxima in the distribution
of the order parameter. In this Appendix, we present a more realistic
model that takes into account the described fluctuations and eventually
provides a more complete picture for the distribution of the order
parameter.

The model can be summarized by representing the value $D_{ij}$ of
the matrix element between the two single-particle states by the following
combination:
\begin{equation}
D_{ij}=c_{ij}\cdot D\left(\xi_{i}-\xi_{j}\right),
\end{equation}
where $c_{ij}$ are independent random variables distributed according
to some distribution $P\left(c\right)$, and $D\left(\xi\right)$
is the energy dependence of the interaction discussed in~\appref{Effect-of-energy-dependence-of-matrix-element}.
In this way, the matrix element now contains two types of fluctuations:
explicit  fluctuations due to $c$ and implicit ones due to the $\xi$-dependence.
The the model analyzed previously is reproduced by letting $P\left(c\right)=\delta\left(c-1\right)$,
i. e. setting all couplings $c_{ij}$ equal to unity. The corresponding
self-consistency equation for the value of the order parameter then
reads
\begin{equation}
\Delta_{i}=\sum_{j}c_{ij}\cdot D\left(\xi_{i}-\xi_{j}\right)\cdot\frac{\Delta_{j}}{\sqrt{\Delta_{j}^{2}+\xi_{j}^{2}}}.\label{eqapp:saddle-point-equation_fluctuating-coupling}
\end{equation}
One has to solve this system of equations for each realization of
the disorder field $\xi$ \emph{and }random couplings $c_{ij}$. 

Below we present both numerical and analytical study of this extended
model. Sections \ref{subsecapp:fluctuating-coupling_mean-field}~through~\ref{subsecapp:fluctuating-coupling_weak-coupling-approximation}
provide a concise derivation of the generalized theory, which includes
the mean-field approximation, the equations on the modified distribution
function, the $m$~function and the solution for the $m$~function
in the limit of weak coupling. In \subsecappref{fluctuating-coupling_extreme-value-statistics}
we then present detailed results for two specific choices of the coupling
distribution $P\left(c\right)$. We first analyze the effect of weakly
fluctuating $c$ by choosing $P\left(c\right)$ to be a narrow Gaussian-like
distribution of mean value $1$ and standard deviation $\delta\ll1$.
We then touch on the effect of the fluctuating \emph{number of neighbors
$Z$ }by exploring the model with $P\left(c\right)=p\delta\left(1-c\right)+\left(1-p\right)\delta\left(c\right)$.
The outcomes of our analysis substantiate the qualitative claims made
in \subsecref{Extreme-value-statistics_fluctuating-coupling}.

\subsection{The mean-field approximation\label{subsecapp:fluctuating-coupling_mean-field}}

Within a simple mean-field approximation, the self-consistency equation~\eqappref{saddle-point-equation_fluctuating-coupling}
reduces to the following:
\begin{equation}
\Delta\left(\xi_{0}\right)=Z\left\langle c\right\rangle \cdot\left\langle D\left(\xi_{0}-\xi\right)\cdot\frac{\Delta\left(\xi\right)}{\sqrt{\Delta\left(\xi\right)^{2}+\xi^{2}}}\right\rangle .
\end{equation}
Performing the same type analysis as the one presented in~\appref{Saddle-point-equation_derivation}
results in the following answers for the mean-field order parameter:
\begin{equation}
\Delta\left(\xi\right)=2E_{D}\exp\left\{ -\frac{1}{\lambda}\right\} \cdot d\left(\xi\right),
\end{equation}
where $E_{D}$ and $d\left(\xi\right)$ are still defined by equations~\eqappref{E0-definition}~and~\eqappref{mean-field_d-function-equation},
respectively, but the dimensionless coupling constant $\lambda$ in
all expressions is now defined as
\begin{equation}
\lambda=2\nu_{0}\cdot D\left(0\right)Z\left\langle c\right\rangle .
\end{equation}
Here, $\left\langle c\right\rangle $ is the mean coupling constant.
In this way, the fluctuations do not affect the mean-field behavior,
as the extra multiplier $\left\langle c\right\rangle $ can absorbed
into the $D$ function.

\subsection{Equations on the modified distribution function $P\left(\xi,\Delta|\xi_{0},\Delta_{0}\right)$\label{subsecapp:fluctuating-coupling_eq-on-the-distribution}}

The next step is to derive the generalization of the equation on the
distribution according to the program described in \appref{equation-on-distribution}.
Similarly to previous cases, we introduce the following shorthand
notation for the right hand side of the new self-consistency equation~\eqappref{saddle-point-equation_fluctuating-coupling}:
\begin{equation}
f\left(\xi_{j},\Delta_{j}|\xi_{i},c_{ij}\right)=c_{ij}\cdot D\left(\xi_{i}-\xi_{j}\right)\cdot\frac{\Delta_{j}}{\sqrt{\Delta_{j}^{2}+\xi_{j}^{2}}}.
\end{equation}
While the disorder is not restricted to a single site anymore, it
is still local in a sense that the configuration on a given site $i$
is completely determined by quantities $\xi_{i},\xi_{j},\Delta_{j},c_{ij}$
in the nearest neighborhood only. One then considers the joint probability
distribution of the nearest neighborhood of the chosen site $i$.
The latter now also describes the couplings $c_{ij}$ and is explicitly
defined as:
\begin{equation}
P_{i}\left(\left\{ \overline{\xi}_{k},\Delta_{k}\right\} _{k\in N\left(i\right)},\left\{ \overline{c}_{ij}\right\} _{j\in\partial i}\right):=\left\langle \prod_{k\in N\left(i\right)}\delta\left(\overline{\xi}_{k}-\xi_{k}\right)\delta\left(\Delta_{k}-S_{k}\right)\cdot\prod_{j\in\partial i}\delta\left(\overline{c}_{ij}-c_{ij}\right)\right\rangle _{\xi,c}
\end{equation}
where $k$ runs through $N\left(i\right)=\partial i\cup\left\{ i\right\} $,
$S_{k}$ is the solution to the self-consistency equations~\eqappref{saddle-point-equation_fluctuating-coupling}
on site $k$ for a given disorder realization (thus depending on the
values of $\xi_{j}$ and $c_{ij}$ in the whole system), and the average
$\left\langle \bullet\right\rangle $ is performed over all values
of $\xi_{j},c_{ij}$ in the whole system. Similarly to the simpler
case of \appref{equation-on-distribution}, we introduce the modified
problem, where a chosen site $i$ has $\Delta_{i},\xi_{i}$ and $c_{ij}$
for all $j\in\partial i$ specified externally. We denote the solution
to this modified problem as $S_{j}^{i}\left(\left\{ \xi_{k}\right\} |\xi_{i},\Delta_{i},\left\{ c_{ij}\right\} _{j\in\partial i}\right)$.
The corresponding modified probability distribution for a site $j$
neighboring with $i$ is defined as
\begin{equation}
P_{j}^{i}\left(\overline{\xi}_{j},\Delta_{j}|\xi_{i},\Delta_{i},c_{ij}\right):=\left\langle \delta\left(\overline{\xi}_{j}-\xi_{j}\right)\delta\left(\Delta_{j}-S_{j}^{i}\right)\right\rangle _{\xi,c},
\end{equation}
where the average is now performed over the values of $\xi_{j}$ on
all sites except $i$ and over the values of $c_{ij}$ on all edges
except those incident with $i$. By following the derivation identical
to that of \appref{equation-on-distribution}, one arrives to the
following relation between the introduced joint probability distribution
$P_{i}$ and the distribution $P_{j}^{i}$ in the modified problem:

\begin{align}
P_{i}\left(\left\{ \xi_{k},\Delta_{k}\right\} ,\left\{ c_{ij}\right\} \right) & =\prod_{k\in N\left(i\right)}P_{\xi}\left(\xi_{k}\right)\cdot\prod_{j\in\partial i}P_{c}\left(c_{ij}\right)\cdot\prod_{j\in\partial i}P_{j}^{i}\left(\xi_{j},\Delta_{j}|\xi_{i},\Delta_{i},c_{ij}\right)\cdot\delta\left(\Delta_{0}-s\right)\nonumber \\
 & -\prod_{k\in N\left(i\right)}P_{\xi}\left(\xi_{k}\right)\cdot\prod_{j\in\partial i}P_{c}\left(c_{ij}\right)\cdot\sum_{j}\prod_{k\in\partial i,k\neq j}P_{k}^{i}\left(\xi_{k},\Delta_{k}|\xi_{i},\Delta_{i},c_{ik}\right)\nonumber \\
 & \times\left[\frac{\partial}{\partial\Delta_{0}}\intop^{\Delta_{j}}d\Delta_{j}^{'}\cdot P_{j}^{i_{0}}\left(\xi_{j},\Delta_{j}^{'}|\xi_{i},\Delta_{i},c_{ij}\right)\right]\cdot\left[\frac{\partial}{\partial\Delta_{j}}\intop^{\Delta_{0}}d\Delta_{0}^{'}\cdot\delta\left(\Delta_{0}^{'}-s\right)\right],
\end{align}
where 
\begin{equation}
s=\sum_{j\in\partial i}f\left(\xi_{j},\Delta_{j}|\xi_{i},c_{ij}\right),
\end{equation}
and $P_{\xi}\left(\xi\right),\,P_{c}\left(c\right)$ are the distributions
of local disorder fields $\xi$ and $c$, respectively. For simplicity
we have assumed $c_{ij}$ to be uncorrelated and independent on $\xi$.
Integrating out fields $\Delta_{j},\xi_{j}$ on neighboring sites
$j\in\partial i$ and the corresponding couplings $c_{ij}$ then renders
the following equation for the onsite probability distribution:

\begin{align}
P_{i}\left(\xi_{i},\Delta_{i}\right) & =P_{\xi}\left(\xi_{i}\right)\cdot\prod_{j\in\partial i}\intop d\xi_{j}d\Delta_{j}dc_{ij}\cdot P_{\xi}\left(\xi_{j}\right)P_{c}\left(c_{ij}\right)\cdot P_{j}^{i}\left(\xi_{j},\Delta_{j}|\xi_{i},\Delta_{i},c_{ij}\right)\cdot\delta\left(\Delta_{0}-s\right)\nonumber \\
 & -P_{\xi}\left(\xi_{i}\right)\cdot\prod_{j\in\partial i}\intop d\xi_{j}d\Delta_{j}dc_{ij}\cdot P_{\xi}\left(\xi_{j}\right)P_{c}\left(c_{ij}\right)\cdot\sum_{j}\prod_{k\in\partial i,k\neq j}P_{k}^{i}\left(\xi_{k},\Delta_{k}|\xi_{i},\Delta_{i},c_{ik}\right)\nonumber \\
 & \times\left[\frac{\partial}{\partial\Delta_{0}}\intop^{\Delta_{j}}d\Delta_{j}^{'}\cdot P_{j}^{i_{0}}\left(\xi_{j},\Delta_{j}^{'}|\xi_{i},\Delta_{i},c_{ij}\right)\right]\cdot\left[\frac{\partial}{\partial\Delta_{j}}\intop^{\Delta_{0}}d\Delta_{0}^{'}\cdot\delta\left(\Delta_{0}^{'}-s\right)\right].
\end{align}
In a similar vein one then derives the recursive equation for $P_{j}^{i}$:

\begin{align}
P_{j}^{i}\left(\xi_{1},\Delta_{1}|\xi_{0},\Delta_{0},c_{0}\right) & =\prod_{k\in\partial j,k\neq i}\intop d\xi_{k}d\Delta_{k}dc_{jk}\cdot P_{k}^{j}\left(\xi_{k},\Delta_{k}|\xi_{1},\Delta_{1},c_{jk}\right)\cdot P_{c}\left(c_{0}\right)P_{\xi}\left(\xi_{1}\right)\cdot\delta\left(\Delta_{1}-s^{i}\right)\nonumber \\
 & -\prod_{k\in\partial j,k\neq i}\intop d\xi_{k}d\Delta_{k}dc_{jk}\cdot\sum_{k}\prod_{r\in\partial j,r\neq i,k}P_{r}^{j}\left(\xi_{r},\Delta_{r}|\xi_{1},\Delta_{1},c_{jr}\right)\nonumber \\
 & \times\left[\frac{\partial}{\partial\Delta_{1}}\intop^{\Delta_{k}}d\Delta_{k}^{'}\cdot P_{k}^{j}\left(\xi_{k},\Delta_{k}^{'}|\xi_{1},\Delta_{1},c_{jk}\right)\right]\cdot\left[P_{c}\left(c_{0}\right)P_{\xi}\left(\xi_{1}\right)\cdot\frac{\partial}{\partial\Delta_{k}}\intop^{\Delta_{1}}d\Delta_{1}^{'}\cdot\delta\left(\Delta_{1}^{'}-s^{i}\right)\right],
\end{align}
where
\begin{equation}
s^{i}=\sum_{k\in\partial j,k\neq i}f\left(\xi_{k},\Delta_{k}|\xi_{1},c_{jk}\right)+f\left(\xi_{0},\Delta_{0}|\xi_{1},c_{0}\right).
\end{equation}
Finally, the arguments of translational and rotational invariance
on the graph allows one expect identical distributions on all sites,
so that one obtains the following equations after a proper Fourier
transform:

\begin{equation}
P\left(\xi_{0},\Delta_{0}\right)=P_{\xi}\left(\xi_{0}\right)\cdot\intop_{\mathbb{R}}\frac{dt}{2\pi}\cdot\frac{\partial}{\partial\Delta_{0}}\left\{ \left[\intop^{\Delta_{0}}d\Delta^{'}\exp\left\{ -it\Delta^{'}\right\} \right]\cdot\left[\intop d\xi d\Delta dc\cdot\mathcal{P}\left(\xi,\Delta|\xi_{0},\Delta_{0},c\right)\cdot\exp\left\{ itf\left(\xi,\Delta|\xi_{0},c\right)\right\} \right]^{Z}\right\} ,
\end{equation}
\begin{align}
\mathcal{P}\left(\xi_{1},\Delta_{1}|\xi_{0},\Delta_{0},c_{0}\right) & =P_{\xi}\left(\xi_{1}\right)P_{c}\left(c_{1}\right)\cdot\intop_{\mathbb{R}}\frac{dt}{2\pi}\cdot\exp\left\{ itf\left(\xi_{0},\Delta_{0}|\xi_{1},c_{1}\right)\right\} \nonumber \\
 & \times\frac{\partial}{\partial\Delta_{1}}\left\{ \left[\intop^{\Delta_{1}}d\Delta_{1}^{'}\exp\left\{ -it\Delta_{1}^{'}\right\} \right]\cdot\left[\intop d\xi d\Delta dc\cdot\mathcal{P}\left(\xi,\Delta|\xi_{1},\Delta_{1},c\right)\cdot\exp\left\{ itf\left(\xi,\Delta|\xi_{1},c\right)\right\} \right]^{Z-1}\right\} ,\label{eqapp:fluctuating-coupling_P1-equation}
\end{align}
which are direct generalizations of Eq.\nobreakdash-s~(\hphantom{}\ref{eqapp:equation-on-onsite-distribution}\nobreakdash-\ref{eqapp:equation-on-modified-distribution}\hphantom{}).
By employing a procedure similar to that described in \appref{equation-on-distribution},
one can express all other local joint distributions in terms of $\mathcal{P}$. 

\subsection{Equations for the $m$ function}

Upon deriving a closed set of equations on the joint distribution
functions, we proceed to simplifying them in the limit $\nu_{0}\Delta_{0}\ll1$,
$Z\gg1$. One starts with the following definition of the $m$~function:
\[
m\left(S|x,y\right):=\ln\left\{ \left[\intop d\xi_{1}d\Delta_{1}dc_{1}\cdot P_{1}\left(\xi_{1},\Delta_{1}|\xi,\Delta,c_{1}\right)\cdot\exp\left\{ iS\cdot f\left(\xi_{1},\Delta_{1}|\xi,c_{1}\right)/\Delta_{0}\right\} \right]^{Z-1}\right\} ,\,\,\,\,\,\xi=\Delta_{0}x,\,\,\,\,\Delta=\Delta_{0}y,
\]
so that it satisfies the integral equation obtained from~\eqappref{fluctuating-coupling_P1-equation}:
\begin{align}
\frac{1}{Z_{\text{eff}}}m\left(S|x,y\right) & =\intop_{\mathbb{R}}\frac{dx_{1}\nu\left(\Delta_{0}x_{1}\right)}{2\nu_{0}}\cdot\boxed{\intop P\left(c\right)dc}\cdot\intop_{0}^{\infty}dy_{1}\cdot\left[\exp\left\{ iS\frac{f\left(\Delta_{0}x_{1},\Delta_{0}y_{1}|\Delta_{0}x,\boxed{c}\right)}{\Delta_{0}}\right\} -1\right]\nonumber \\
 & \times\intop_{\mathbb{R}-i0}\frac{ds}{2\pi}\cdot\exp\left\{ is\frac{f\left(\Delta_{0}x,\Delta_{0}y|\Delta_{0}x_{1},\boxed{c}\right)}{\Delta_{0}}\right\} \frac{\partial}{\partial y_{1}}\left\{ \left[\intop_{\infty}^{y_{1}}dy_{1}^{'}\exp\left\{ -isy_{1}^{'}\right\} \right]\cdot\exp\left\{ m\left(s|x_{1},y_{1}\right)\right\} \right\} .
\end{align}
For the purpose of visualization, we have highlighted the modifications
due to the presence of fluctuating coupling by $c$ by a box.

The next step is to exclude high-energy scales while carefully treating
the emerging logarithmic divergencies. Note that in our model the
$f$~function contains $c$ in a simple multiplicative form, i. e.
$f\left(\xi_{1},\Delta_{1},c_{1}|\xi\right)=c_{1}\cdot f\left(\xi_{1},\Delta_{1}|\xi\right)$,
with the latter term multiplier being of the same form as the one
used in the previous \appref{Effect-of-energy-dependence-of-matrix-element}.
The solution can be seen as a straightforward modification of the
derivation presented earlier in \appref{Effect-of-energy-dependence-of-matrix-element}.
It is still convenient to represent the $m$~function as a sum of
two terms:
\begin{equation}
m\left(S|x,y\right)=iSm_{1}\left(x,y\right)+m_{2}\left(S|x,y\right).
\end{equation}
The equation for $m_{2}$ then readily reads
\begin{align}
m_{2}\left(S|w,x\right) & =\lambda\cdot\boxed{\intop dcP\left(c\right)}\cdot\intop_{0}^{1}dw_{1}\cdot\frac{\exp\left\{ iS\kappa\eta\left(x\right)\cdot\boxed{c}w_{1}\right\} -1-iS\kappa\eta\left(x\right)\cdot\boxed{c}w_{1}}{w_{1}^{2}\sqrt{1-w_{1}^{2}}}\nonumber \\
 & \times\left[1-w_{1}\left(1-w_{1}^{2}\right)\frac{\partial}{\partial w_{1}}\right]\cdot\left[\frac{\boxed{c}\kappa w\eta\left(x\right)+m_{1}\left(w_{1},0\right)}{\kappa}\right],\label{eqapp:fluctuating-coupling_m2-equation}
\end{align}
while the equation for $m_{1}$ is obtained after the procedure identical
to that of \appref{Effect-of-energy-dependence-of-matrix-element}
and reads:
\begin{align}
m_{1}\left(w,x\right) & =\boxed{\left\langle c\right\rangle }\cdot\eta\left(x\right)\cdot\lambda\intop_{0}^{1}dw_{1}\cdot\sqrt{1-w_{1}^{2}}\cdot\frac{m_{1}\left(w_{1},0\right)-m_{1}\left(0,0\right)}{w_{1}}\nonumber \\
 & +\eta\left(x\right)\cdot\lambda\intop_{0}^{\infty}dy_{1}\cdot y_{1}\ln\frac{1}{y_{1}}\cdot\intop_{\mathbb{R}-i0}\frac{ds}{2\pi}\cdot\boxed{\intop dcP\left(c\right)}\cdot\boxed{c}\exp\left\{ i\boxed{c}s\kappa\eta\left(x\right)w\right\} \cdot\exp\left\{ m\left(s|0,0\right)-isy_{1}\right\} \nonumber \\
 & +m_{1}\left(0,0\right)\cdot d\left(\Delta_{0}x\right)+\boxed{\left\langle c^{2}\right\rangle }\cdot\kappa w\left[\alpha\cdot\eta^{2}\left(x\right)+\lambda\cdot\psi\left(\Delta_{0}x\right)\right],\label{eqapp:fluctuating-coupling_m1-equation}
\end{align}
where the constant $\alpha$ and the functions $d\left(\xi\right)$,
$\psi\left(\xi\right)$ are defined in Eq.\nobreakdash-s~\eqappref{smooth-energy-dependence_alpha-def},
\eqappref{mean-field_d-function-equation}~and~\eqappref{smooth-energy-dependence_psi-def-2},
respectively. The equations~(\hphantom{}\ref{eqapp:fluctuating-coupling_m1-equation}\nobreakdash-\ref{eqapp:fluctuating-coupling_m2-equation}\hphantom{})
are the direct counterparts of Eq.\nobreakdash-s~(\hphantom{}\ref{eqapp:smooth-energy-dependence_m1-equation}\nobreakdash-\ref{eqapp:smooth-energy-dependence_m2-equation}\hphantom{})
discussed previously in \appref{Effect-of-energy-dependence-of-matrix-element}.

Similarly to the case of \appref{Effect-of-energy-dependence-of-matrix-element},
the distribution of the order parameter for the states participating
in the superconducting order (i. e. within the energy strip of width
$\sim2\varepsilon_{D}$ around the Fermi surface) is retains its original
expression:
\begin{equation}
P_{0}\left(y\right)=\intop_{\mathbb{R}-i0}\frac{ds}{2\pi}\exp\left\{ m\left(s|0,0\right)-isy\right\} .\label{eqapp:fluctuating-coupling_P0-equation}
\end{equation}

\subsection{Weak coupling approximation\label{subsecapp:fluctuating-coupling_weak-coupling-approximation}}

The obtained equations~(\hphantom{}\ref{eqapp:fluctuating-coupling_m1-equation}\nobreakdash-\ref{eqapp:fluctuating-coupling_m2-equation}\hphantom{})
admit a solution in terms of expansion in powers of $\lambda\ll1$.
The procedure is completely analogous to that of \appref{Solution_small-lambda}.
One starts with calculating $m_{2}$ function by approximating the
value of $m_{1}$ with the leading $O\left(\lambda^{0}\right)$ term
and immediately finds:
\begin{equation}
m_{2}\left(S|w,x\right)=\lambda\cdot\boxed{\intop dcP\left(c\right)}\left[\left(\boxed{c}w\eta\left(x\right)+\frac{m_{1}\left(0,0\right)}{\kappa}\right)\Phi_{0}\left(\boxed{c}\kappa S\eta\left(x\right)\right)+\alpha\boxed{\left\langle c^{2}\right\rangle }\cdot\Phi_{1}\left(\boxed{c}\kappa S\eta\left(x\right)\right)\right]+O\left(\lambda^{2}\right),
\end{equation}
where $\Phi_{0}\left(s\right),\,\Phi_{1}\left(s\right)$ are defined
in equations~\eqappref{Phi-0-definition}~and~\eqappref{Phi-1-definition}.
The equation on the $m_{1}$~function can be rewritten as
\begin{equation}
m_{1}\left(0,x\right)=m_{1}\left(0,0\right)\cdot d\left(\Delta_{0}x\right)+\lambda\boxed{\left\langle c\right\rangle }\cdot\eta\left(x\right)\cdot\boxed{\left\langle c^{2}\right\rangle }\alpha\cdot\frac{\pi}{4}\kappa+\eta\left(x\right)\cdot\lambda\boxed{\left\langle c\right\rangle }\cdot\left\langle y\ln\frac{1}{y}\right\rangle +O\left(\lambda^{2}\right),\label{eqapp:fluctuating-coupling_m1-small-lambda-1}
\end{equation}
\begin{align}
m_{1}\left(w,x\right)-m_{1}\left(0,x\right) & =\boxed{\left\langle c^{2}\right\rangle }\kappa w\left[\alpha\eta^{2}\left(x\right)+\lambda\cdot\psi\left(\Delta_{0}x\right)\right]\nonumber \\
 & +\lambda\cdot\eta\left(x\right)\cdot\left\langle \boxed{c}\left(y+\boxed{c}\kappa\eta\left(x\right)w\right)\ln\frac{1}{y+\boxed{c}\kappa\eta\left(x\right)w}-\boxed{\left\langle c\right\rangle }y\ln\frac{1}{y}\right\rangle ,\label{eqapp:fluctuating-coupling_m1-small-lambda-2}
\end{align}
where we have denoted
\begin{equation}
\left\langle f\left(y,c\right)\right\rangle =\intop dcP\left(c\right)\cdot\intop dyP_{0}\left(y\right)\cdot f\left(y,c\right),
\end{equation}
with $P\left(c\right)$ being the distribution of coupling, and $P_{0}\left(y\right)$
being the distribution of the order parameter near the Fermi surface
given by the standard expression~\eqappref{fluctuating-coupling_P0-equation}.
The value of $m_{1}\left(0,0\right)$ is found self-consistently from
the following equation:
\begin{equation}
0=\frac{\pi}{4}\alpha\kappa\boxed{\left\langle c^{2}\right\rangle }+\lambda\cdot\frac{1}{\left\langle c\right\rangle }\intop_{0}^{1}dw_{1}\cdot\sqrt{1-w_{1}^{2}}\cdot\left\langle c\frac{\left(y+\boxed{c}\kappa w_{1}\right)\ln\frac{1}{y+\boxed{c}\kappa w_{1}}-y\ln\frac{1}{y}}{w_{1}}\right\rangle +\boxed{\left\langle y\ln\frac{1}{y}\right\rangle }.\label{eqapp:fluctuating-coupling_equation-on-average}
\end{equation}

Guided by the calculation of \appref{Solution_small-lambda}, we introduce
\begin{equation}
g_{1}\left(\mu,w;\lambda\right)=\intop dcP\left(c\right)\cdot\intop_{\mathbb{R}-i0}\frac{ds}{2\pi}\cdot\intop_{0}^{\infty}dy\cdot c\left(y+\kappa wc\right)\ln\frac{1}{y+\kappa wc}\cdot\exp\left\{ is\mu-isy+m_{2}\left(s|0,0\right)\right\} .
\end{equation}
The system of equations \eqappref{fluctuating-coupling_m1-small-lambda-1},~\eqappref{fluctuating-coupling_m1-small-lambda-2}~and~\eqappref{fluctuating-coupling_equation-on-average}
can be rewritten as
\begin{equation}
m_{1}\left(0,x\right)=m_{1}\left(0,0\right)\cdot d\left(\Delta_{0}x\right)+\lambda\left\langle c\right\rangle \cdot\eta\left(x\right)\cdot\left\langle c^{2}\right\rangle \alpha\cdot\frac{\pi}{4}\kappa+\eta\left(x\right)\cdot\lambda\cdot g_{1}\left(\mu,0;\lambda\right)+O\left(\lambda^{2}\right),
\end{equation}
\begin{equation}
m_{1}\left(w,x\right)-m_{1}\left(0,x\right)=\left\langle c^{2}\right\rangle \kappa w\left[\alpha\eta^{2}\left(x\right)+\lambda\cdot\psi\left(\Delta_{0}x\right)\right]+\lambda\cdot\eta\left(x\right)\cdot\left[g_{1}\left(\mu,\eta\left(x\right)w;\lambda\right)-g_{1}\left(\mu,0;\lambda\right)\right],
\end{equation}
\begin{equation}
0=\frac{\pi}{4}\kappa\alpha\left\langle c^{2}\right\rangle +\lambda\cdot\intop_{0}^{1}dw_{1}\cdot\sqrt{1-w_{1}^{2}}\cdot\frac{g_{1}\left(\mu,w;\lambda\right)-g_{1}\left(\mu,0;\lambda\right)}{\left\langle c\right\rangle w_{1}}+\frac{1}{\left\langle c\right\rangle }g_{1}\left(\mu,0;\lambda\right)+O\left(\lambda^{2}\right).
\end{equation}
Upon substituting the explicit form of $m_{2}$ and expanding in powers
of small $\lambda$ one obtains
\[
g_{1}\left(\mu,w;\lambda\right)=\left\langle c\cdot\mu_{w}\ln\frac{1}{\mu_{w}}\right\rangle -\lambda\cdot\intop_{0}^{1}\frac{dw_{1}}{\sqrt{1-w_{1}^{2}}}\cdot\left\langle c\left[\frac{\mu_{w}}{\kappa}+\alpha\left\langle c^{2}\right\rangle \cdot w_{1}^{3}\right]\frac{\left(\mu_{w}+c'\kappa w_{1}\right)\ln\left(\frac{\mu_{w}+c'\kappa w_{1}}{\mu_{w}}\right)-c'\kappa w_{1}}{w_{1}^{2}}\right\rangle +O\left(\lambda^{2}\right),
\]
where we have denoted $\mu_{w}=\mu+\kappa wc$ for brevity, and the
average in now performed over both $c$ and $c'$ independently (another
instance $c'$ emerges after substituting the expression for $m_{2}$
that contains its own, independent integration over $c$). In particular,
one observes that
\begin{equation}
g_{1}\left(\mu,0;\lambda\right)=\left\langle c\right\rangle \mu\ln\frac{1}{\mu}-\lambda\cdot\intop_{0}^{1}\frac{dw_{1}}{\sqrt{1-w_{1}^{2}}}\cdot\left\langle c\left[\frac{\mu}{\kappa}+\alpha\left\langle c^{2}\right\rangle \cdot w_{1}^{3}\right]\frac{\mu_{w_{1}}\ln\frac{\mu_{w_{1}}}{\mu}-c\kappa w_{1}}{w_{1}^{2}}\right\rangle +O\left(\lambda^{2}\right).
\end{equation}
Similarly to \appref{Solution_small-lambda}, the integrals over $w_{1}$
in $g_{1}$ and $\mu_{1}$ can be evaluated in terms of special functions,
but we choose to leave it in an unevaluated form as the subsequent
average over the distribution of $c$ cannot be performed for arbitrary
$P\left(c\right)$ anyway. 

The self-consistency equation~\eqappref{fluctuating-coupling_equation-on-average}
for $\mu=m_{1}\left(0,0\right)$ can still be solved within the perturbation
theory in powers of $\lambda$. For brevity, here we will present
only the leading order:
\begin{equation}
m_{1}\left(0,0\right)=\frac{\pi\kappa/4\cdot\alpha\left\langle c^{2}\right\rangle }{W\left(\pi\kappa/4\cdot\alpha\left\langle c^{2}\right\rangle \right)}+O\left(\lambda\right),
\end{equation}
where $W$ is the Lambert's $W$-function. Higher orders are expressed
in terms of $g_{1}$ function, similarly to \appref{Solution_small-lambda}.

\subsection{Extreme value statistics\label{subsecapp:fluctuating-coupling_extreme-value-statistics}}

\subsubsection{Fluctuating $D$ model}

We start with the simplest model describing small fluctuations in
the value of the matrix element around its mean value. The corresponding
distribution can be chosen in a form of a uniform distribution around
$c=1$ with a small width $\text{\ensuremath{\delta}}\ll1$:
\begin{equation}
P\left(c\right)=\begin{cases}
N\cdot\frac{1}{\sqrt{2\pi\delta^{2}}}\exp\left\{ -\frac{\left(c-1\right)^{2}}{2\delta^{2}}\right\} , & c>0,\\
0, & c<0,
\end{cases}
\end{equation}
where $N$ is the normalization constant close to $1$. The distribution
is characterized by unit expectation $\left\langle c\right\rangle =1$
and small standard deviation $\left\langle c^{2}-1\right\rangle =\delta^{2}\ll1$.
Due to truncation of negative values, corrections of order $\exp\left\{ -\frac{1}{2\delta^{2}}\right\} \ll1$
exists to $N$, $\left\langle c\right\rangle $ and $\left\langle c^{2}\right\rangle $
but we are going to discard them in what follows. Within such a model,
it is possible to analyze the asymptotic behavior qualitatively in
the same spirit as done in \appref{Extreme-value-statistics}. 

Let us start with the region $y\apprle\left\langle y\right\rangle $
first. The only saddle point contributing to the integral~\eqappref{fluctuating-coupling_P0-equation}
for the probability density still lies on the imaginary axis in the
upper half-plane. There exists a large region $\kappa\left|S\right|\ll1/\delta^{2}$
where the modified asymptotic expression for $m_{2}$ can be obtained
by direct perturbation theory, i. e. by formally treating deviation
of $c$ from one as a small correction. By repeating the calculation
of~\subsecappref{Probability-small-y-asymptotic} one the obtains
the following asymptotic expression:
\begin{equation}
m_{2}\left(S|w\right)=Z_{\text{eff}}\frac{\left\langle y\right\rangle }{\kappa}\cdot a\left(\ln2a+\gamma-1\right)-a\cdot Z_{\text{eff}}\left\langle y\ln\frac{1}{y}\right\rangle -a\cdot\frac{Z_{\text{eff}}\left\langle y\right\rangle \delta^{2}}{2}+O\left(\frac{1}{a}\right),\,\,\,\,a=-i\kappa S\gg1.\label{eqapp:fluctuating-coupling_m2-upper-half-plane}
\end{equation}
The saddle-point estimation of the $P_{0}$ then reads:
\begin{equation}
P_{0}\left(y\right)\approx\sqrt{\frac{\zeta\left(y\right)}{2\pi\cdot\left[\lambda\left\langle y\right\rangle \right]^{2}}}\exp\left\{ -\zeta\left(y\right)\right\} ,\label{eqapp:fluctuating-coupling_P0-small-y}
\end{equation}
\begin{equation}
\zeta\left(y\right)=\frac{\lambda\left\langle y\right\rangle }{2\kappa}\exp\left\{ \frac{1}{\lambda}\left(1-\frac{y}{\left\langle y\right\rangle }\right)-\frac{\left\langle y\ln y\right\rangle }{\left\langle y\right\rangle }-\gamma\boxed{+\frac{\delta^{2}}{2}}\right\} .\label{eqapp:fluctuating-coupling_zeta-expr}
\end{equation}
This result is valid while
\begin{equation}
\kappa\left|S\right|=\frac{\kappa\zeta\left(y\right)}{\lambda\left\langle y\right\rangle }\ll\frac{1}{\delta^{2}},
\end{equation}
which imposes a lower bound on the available values of $y$:
\begin{equation}
1-\lambda\left(\ln\frac{2}{\delta^{2}}+\gamma-\frac{\left\langle y\ln y\right\rangle }{\left\langle y\right\rangle }\right)\le\frac{y}{\left\langle y\right\rangle }.\label{eqapp:fluctuating-coupling_small-y-lower-bound}
\end{equation}
Because of the smallness of $\lambda$ this region might turn out
to be narrow. This does not imply, however, that the corresponding
asymptotic behavior is unobservable. What matters is the change in
the value of the probability density. The lowest value of the probability
attained with this asymptotic regime can be estimated as:
\begin{equation}
P\left(y\right)\apprge\sqrt{\frac{1}{2\pi\cdot\left[\lambda\left\langle y\right\rangle \kappa\delta^{2}\right]}}\exp\left\{ -\frac{\lambda\left\langle y\right\rangle }{\kappa\delta^{2}}\right\} .\label{eqapp:fluctuationg-coupling_double-exp-criteria}
\end{equation}
If this value is small enough compared to unity (the value of $P\left(y\right)$
for $y\sim\left\langle y\right\rangle $), the corresponding sharp
profile will be well observed. Moreover, the profile will not differ
from the one with no fluctuations of the coupling constant as the
expression~\eqappref{fluctuating-coupling_zeta-expr} suggests. Demanding
the lower value of the probability to be much smaller than unity results
in the following criteria for the value of $\delta$:
\begin{equation}
\delta\apprle\sqrt{\frac{\lambda\left\langle y\right\rangle }{\kappa}}\sim\sqrt{\frac{\lambda}{\kappa}}=\sqrt{Z_{\text{eff}}}.
\end{equation}
Note that the latter quantity is small everywhere in the non-Gaussian
region of interest $\kappa\apprge\lambda$. 

When $\delta$ is not small enough to satisfy this criteria, the logarithmic
asymptotic expression~\eqappref{fluctuating-coupling_m2-upper-half-plane}
for the $m_{2}$~function ceases to be applicable. Indeed, such a
behavior originates from the region near $w=0$ of the integral~\eqappref{fluctuating-coupling_m2-equation},
but for $\kappa\left|S\right|\delta^{2}>1$ this contribution is clearly
superseded by an exponential one originating from the region $w_{1}\sim1$.
The resulting saddle-point estimation of the integral~\eqappref{fluctuating-coupling_P0-equation}
for the probability density is a topic for a separate study. For our
model it can be shown that beyond the limit of applicability~\eqappref{fluctuationg-coupling_double-exp-criteria}
of the double-exponential asymptotic behavior given by Eq.\nobreakdash-s~(\hphantom{}\ref{eqapp:fluctuating-coupling_P0-small-y}\nobreakdash-\ref{eqapp:fluctuating-coupling_zeta-expr}\hphantom{})
the probability distribution is described by a much slower dependence
of the form
\begin{equation}
\ln P\left(y\right)\sim-\frac{\left\langle y\right\rangle -y}{\kappa}\left[\ln\left[\frac{\left\langle y\right\rangle -y}{\lambda\sqrt{\pi/2}\cdot m_{1}\left(1,0\right)}\right]+\ln\frac{1}{\delta}\right],
\end{equation}
which can be obtained by a technique similar to the one used in \subsecappref{Probability-large-y-asymptotic}
of \appref{Extreme-value-statistics} for \emph{large }values of $y$.
From the physics point of view, it corresponds to the fact that the
distribution now rests on a different type of optimal fluctuation
in real space. As explained e.g. in \subsecappref{Probability-small-y-asymptotic},
the observed double-exponential profile corresponds to sites with
all neighbors exhibiting large value of the disorder field $\xi\gg\Delta_{0}$.
For the case with constant matrix element of the interaction this
fluctuation is the only way to deliver a small value of the order
parameter. On the other hand, with fluctuating coupling constant one
can suppress the order parameter on a given site by picking diminished
values of the coupling matrix elements on sufficiently large fraction
of incident edges. These two mechanisms compete with each other, providing
a transition to different type of the asymptotic behavior of the probability
as $y$ approaches the value defined by Eq.~\eqappref{fluctuating-coupling_small-y-lower-bound}.
This also implies that the low-$y$ behavior of the distribution with
sufficiently small fluctuations of the coupling constant will be sensitive
to fine qualitative details of the distribution of the coupling constants
$c_{ij}$, such as the exact form of the distribution presence of
local correlations.

In the opposite limit $y\ge\left\langle y\right\rangle $ one has
to analyze multiple saddle points. Within the region $kS\ll1/\delta$,
one can again treat the correction arising from $\delta$ perturbatively.
This can be done by using the following operator representation:
\begin{equation}
m_{2}\left(S|w,x\right)=\boxed{\left\langle \exp\left\{ \frac{c-\left\langle c\right\rangle }{\left\langle c\right\rangle }\cdot S\frac{\partial}{\partial S}\right\} \left[1+\frac{c-\left\langle c\right\rangle }{\left\langle c\right\rangle }\cdot w\frac{\partial}{\partial w}\right]\right\rangle }\cdot m_{2}^{\text{clean}}\left(\left\langle c\right\rangle S|w,x\right),
\end{equation}
where the boxed operator is understood as its formal power series,
with each term being averaged over distribution of the coupling constant
$c$, and $m_{2}^{\text{clean}}\left(S|w,x\right)$ is the value of
the $m_{2}$~function obtained for the case with no fluctuations
of the coupling constant. By formally expanding this expression up
to leading powers of the coupling fluctuation $c-\left\langle c\right\rangle $
one arrives at
\begin{equation}
m_{2}\left(S|w,x\right)\approx\left(1+\boxed{\frac{\left\langle \left(c-\left\langle c\right\rangle \right)^{2}\right\rangle }{\left\langle c\right\rangle ^{2}}\left\{ S\frac{\partial}{\partial S}w\frac{\partial}{\partial w}+\frac{1}{2}\left(S\frac{\partial}{\partial S}\right)^{2}\right\} }\right)\cdot m_{2}^{\text{clean}}\left(\left\langle c\right\rangle S|w,x\right),
\end{equation}
thus obtaining the exact equation for the leading perturbative correction
to the $m_{2}$~function.

For the sake of brevity, let us now analyze the case $w=0$, $x=0$
sufficient to determine the value of the probability density of the
order parameter. Upon using the available asymptotic expression for
$m_{2}^{\text{clean}}$, the expression for the $m_{2}$~function
then evaluates to
\begin{equation}
m_{2}\left(S|0,0\right)\approx\left[1+\frac{\left(i\kappa S\delta\right)^{2}}{2}Z_{\text{eff}}\right]\cdot m_{1}\left(1,0\right)\sqrt{\frac{\pi}{2i\kappa S}}e^{i\kappa S}\left(1+O\left(\frac{1}{i\kappa S}\right)\right),
\end{equation}
where we have also used that $\left\langle c\right\rangle =1$ and
$\left(c-\left\langle c\right\rangle \right)^{2}=\delta^{2}$ without
loss of generality. The region of applicability of such an approximation
is defined by the converge radius of the used expansion:
\begin{equation}
\kappa\left|S\right|\delta\ll1.
\end{equation}
For the relevant values of $S$, the criteria evaluates to
\begin{equation}
\kappa\left|S\right|\sim\ln\frac{y-\left\langle y\right\rangle }{\lambda m_{1}\left(1,0\right)\sqrt{\pi/2}}\Leftrightarrow y-\left\langle y\right\rangle \ll\lambda m_{1}\left(1,0\right)\sqrt{\frac{\pi}{2}}\cdot\exp\left\{ \frac{1}{\delta}\right\} 
\end{equation}
and appears to specify an exponentially large region. Because the
saddle-point analysis essentially requires performing the Legendre
transform on the $m_{2}$ function, the leading effect of the perturbation
is delivered solely by the change of the $m_{2}$~function itself.
One can thus approximate the contribution of each saddle point as
\begin{equation}
P^{\left(n\right)}\left(y\right)\sim P^{\left(n\right)}\left(y,\delta=0\right)\cdot\exp\left\{ +\frac{\left(z_{n}\delta\right)^{2}}{2}\frac{y-\left\langle y\right\rangle }{\kappa}\right\} ,\label{eqapp:fluctuating-coupling_large-y-saddle-point-contribution}
\end{equation}
where $P_{n}\left(y,\delta=0\right)$ stands for the magnitude of
the contribution without fluctuations of the matrix element, and the
value of the exponential part in $m_{2}$ was approximated with a
proper linear function of $y$ according to the unperturbed saddle-point
equation~\eqappref{secondary-saddle-points_expression} of \appref{Extreme-value-statistics}.
One immediately observes that the main asymptotic behavior of the
probability density given by Eq.~\eqappref{extreme-value-statistics_large-y-leading-approx}
remains intact up to $\delta\sim1$, since only at this point do the
correction to the contribution of main saddle point become significant.
Another particular consequence of this result is that the contribution
$P_{n}$ of the $n$-th secondary saddle point acquires additional
multiplier the form $\exp\left\{ -\frac{\left(2\pi n\delta\right)^{2}}{2}\frac{y-\left\langle y\right\rangle }{\kappa}\right\} $
due to the imaginary part of $z_{n}$ that can be estimated as $\text{Im}z_{n}\sim2\pi n$.
This has a certain influence on the secondary maxima of the probability
density observed in the case with no fluctuations (see e.g. \figappref{large-y_multiplicative-correction-plots}).
The $m$-th secondary maximum located close to $y_{m}=\left\langle y\right\rangle +\kappa m$
will thus be smeared for $\left(2\pi\delta\right)^{2}m\sim1$. In
particular, for $\delta\sim1/2\pi$ all of the secondary maxima will
disappear.

\begin{figure}[h]
\begin{centering}
\includegraphics[scale=0.2]{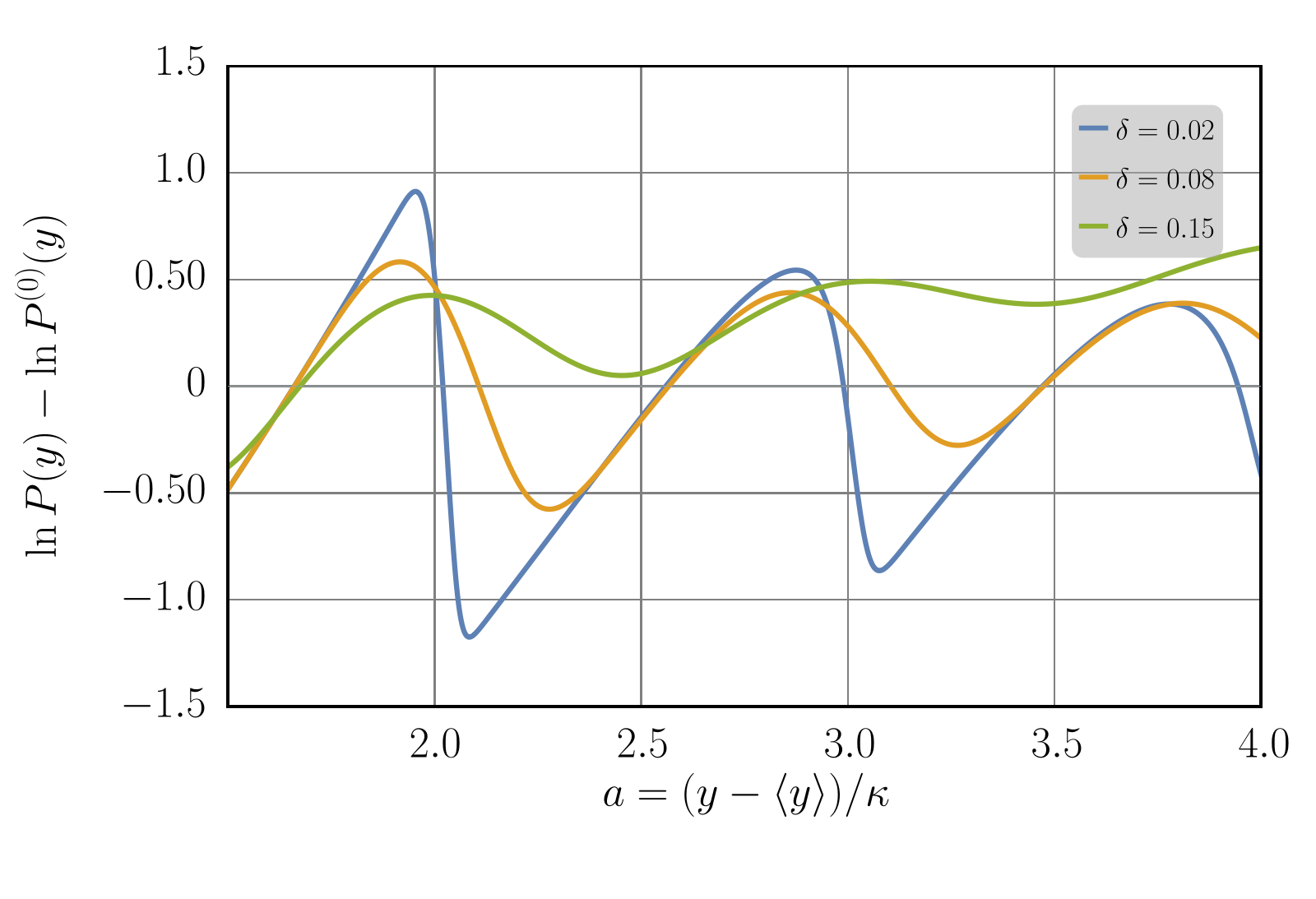}\hspace*{\fill}\includegraphics[viewport=0bp 0bp 1204.5bp 820bp,clip,scale=0.2]{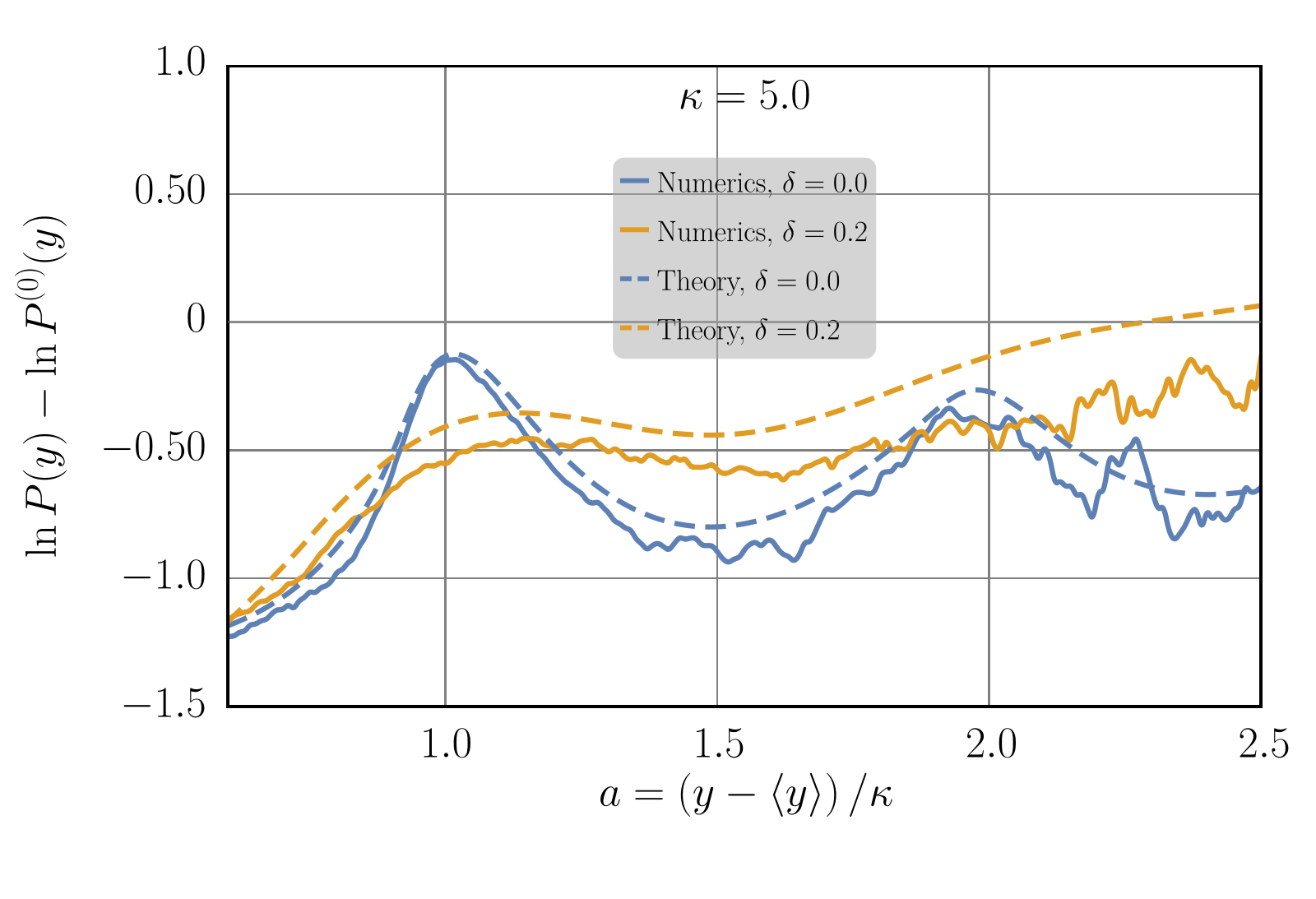}
\par\end{centering}
\raggedright{}\caption{Plots of the logarithm of the multiplicative factor $f\left(y\right)$
distinguishing the leading asymptotic $P_{0}^{\left(0\right)}\left(y\right)$
behavior at large $y$ given by Eq.~\eqappref{extreme-value-statistics_large-y-leading-approx}
(i. e. with $\delta=0$) and the true distribution of the dimensionless
order parameter $P\left(y\right)=\exp\left\{ f\left(y\right)\right\} \cdot P_{0}^{\left(0\right)}\left(y\right)$.
On both plots, various curves correspond to various values of the
standard deviation $\delta$ of the coupling matrix element. The argument
is given by $a=\left(y-\left\langle y\right\rangle \right)/\kappa$.
The microscopical parameters of the model are $\lambda\approx0.12$,
$Z=51$ and $\kappa\approx5.0$. \emph{Left}. The multiplicative correction
estimated by the \textquotedblleft model sum\textquotedblright ~
\eqappref{marginal-probability_subleading_estimation}, but with each
term adjusted according to Eq.~\eqappref{fluctuating-coupling_large-y-saddle-point-contribution}\emph{.
Right}. The plots for the multiplicative correction according to the
direct numerical solution of the self-consistency equation~\eqappref{saddle-point-equation_fluctuating-coupling}
(solid lines) and the theoretical value for the PDF (dashed lines).
The data from \figref{P0-log-plot_with-weak-coupling-fluctuations}
was used, and discrepancies between the numerical and theoretical
plots are also addressed under \figref{P0-log-plot_with-weak-coupling-fluctuations}.
The quantitative difference between the two plots is explained by
the subleading corrections to the exponent of each term in Eq.~\eqappref{marginal-probability_subleading_estimation}
that are beyond the accuracy of the used expansions. \label{figapp:multiplcative-correction-estimation-plot_various-delta}}
\end{figure}

For the purposes of qualitative demonstration, the left plot on \figappref{multiplcative-correction-estimation-plot_various-delta}
shows a set of plots resulting from using the properly modified ``model''
sum~\eqappref{marginal-probability_subleading_estimation}. The latter
is composed of the leading asymptotic estimations~(\hphantom{}\ref{eqapp:secondary-saddle-points_estimation}\nobreakdash-\ref{eqapp:marginal-probability_single-saddle-contribution}\hphantom{})
for the contributions of each secondary saddle point with the correction~\eqappref{fluctuating-coupling_large-y-saddle-point-contribution}
taken into account. The right plot of \figappref{multiplcative-correction-estimation-plot_various-delta}
demonstrates this behavior in the true distribution of the order parameter
found both theoretically and by direct numerical solution of Eq.~\eqappref{saddle-point-equation_fluctuating-coupling}
in a number of disorder realizations. One can indeed note that two
major effects are induced by a finite value of $\delta$. Firstly,
one observes smearing of the secondary maxima as $\delta$ increases
in accordance with the described mechanism. Secondly, the expression~\eqappref{fluctuating-coupling_large-y-saddle-point-contribution}
for $n=0$ suggests that $\delta$~introduces an additional nearly
linear growth of the exponent of the actual leading contribution.
This growth is then observed as an upward tendency on both plots.

\subsubsection{Fluctuating $Z$ model}

Another simple yet informative model is the one that reproduces fluctuations
of number of neighbors $Z$. Within this model, one chooses
\begin{equation}
P\left(c\right)=p\cdot\delta\left(1-c\right)+\left(1-p\right)\cdot\delta\left(c\right),\,\,\,\,0<p<1.
\end{equation}
Each neighbor then has a fluctuating number of neighbors because each
edge is either turned on with probability $p$, or turned off with
probability $1-p$. The first moments of the actual number of neighbors
are given by
\begin{equation}
\left\langle Z\right\rangle =pZ,\,\,\,\,\left\langle Z^{2}\right\rangle -\left\langle Z\right\rangle ^{2}=Zp\left(1-p\right).
\end{equation}

This simple model turns out to be very similar to the original model
without the coupling disorder. Let us introduce the following renormalized
values of the microscopical quantities:
\begin{equation}
\lambda_{R}=p\lambda,\,\,\,\,\,Z_{R}=pZ,\,\,\,\,\,\alpha_{R}=p\alpha,\,\,\,\,\,\kappa_{R}=\frac{\lambda_{R}}{\Delta_{0}Z_{R}},
\end{equation}
where $\Delta_{0}$ is evaluated with the renormalized dimensionless
Cooper constant $\lambda_{R}$, as described in \subsecappref{fluctuating-coupling_mean-field}.
Upon such renormalization, the equations on both $m_{1}$ and $m_{2}$
are \emph{exactly }mapped on those for constant $Z$ presented in
\appref{Effect-of-energy-dependence-of-matrix-element}:
\begin{align}
m_{2}\left(S|w,x\right) & =\lambda_{R}\cdot\intop_{0}^{1}dw_{1}\cdot\frac{\exp\left\{ iS\kappa_{R}\eta\left(x\right)w_{1}\right\} -1-iS\kappa_{R}\eta\left(x\right)\cdot w_{1}}{w_{1}^{2}\sqrt{1-w_{1}^{2}}}\nonumber \\
 & \times\left[1-w_{1}\left(1-w_{1}^{2}\right)\frac{\partial}{\partial w_{1}}\right]\cdot\left[\frac{\kappa_{R}w\eta\left(x\right)+m_{1}\left(w_{1},0\right)}{\kappa_{R}}\right],
\end{align}
\begin{align}
m_{1}\left(w,x\right) & =\eta\left(x\right)\cdot\lambda_{R}\intop_{0}^{1}dw_{1}\cdot\sqrt{1-w_{1}^{2}}\cdot\frac{m_{1}\left(w_{1},0\right)-m_{1}\left(0,0\right)}{w_{1}}\nonumber \\
 & +\eta\left(x\right)\cdot\lambda_{R}\intop_{0}^{\infty}dy_{1}\cdot y_{1}\ln\frac{1}{y_{1}}\cdot\intop_{\mathbb{R}-i0}\frac{ds}{2\pi}\cdot\exp\left\{ is\kappa_{R}\eta\left(x\right)w\right\} \cdot\exp\left\{ m\left(s|0,0\right)-isy_{1}\right\} \nonumber \\
 & +m_{1}\left(0,0\right)\cdot d\left(\Delta_{0}x\right)+\kappa_{R}w\left[\alpha_{R}\cdot\eta^{2}\left(x\right)+\lambda_{R}\cdot\psi\left(\Delta_{0}x\right)\right].
\end{align}
As a result, the sole effect of the fluctuation of the number of neighbors
within such a model is pure renormalization of the microscopic parameters
$\Delta_{0},\kappa,\lambda,\alpha$.

\end{document}